\documentclass{emulateapj}
\usepackage{lscape}
\usepackage{color}

\shorttitle{Testing the Unification Model in the Infrared}
\shortauthors{Ramos Almeida et al.}

\begin{document}

\title{Testing the Unification Model for AGN in the Infrared: are the obscuring tori of Type 1 and 2 Seyferts different?}

\author{C. Ramos Almeida\altaffilmark{1}, N. A. Levenson\altaffilmark{2}, 
A. Alonso-Herrero\altaffilmark{3}, A. Asensio Ramos\altaffilmark{4,5}, J. M. Rodr\'\i guez Espinosa\altaffilmark{4,5}, 
A. M. P\'erez Garc\' ia\altaffilmark{4,5}, C. Packham\altaffilmark{6}, R. Mason\altaffilmark{7}, J. T. Radomski\altaffilmark{2}, 
and T. D\' iaz-Santos\altaffilmark{8}}

\altaffiltext{1}{University of Sheffield, Department of Physics \& Astronomy, S3 7RH, UK; C.Ramos@sheffield.ac.es}
\altaffiltext{2}{Gemini Observatory, Casilla 603, La Serena, Chile.}
\altaffiltext{3}{Centro de Astrobiolog\' ia, INTA-CSIC, E-28850, Madrid, Spain.}
\altaffiltext{4}{Instituto de Astrof\'\i sica de Canarias (IAC), Tenerife, Spain}
\altaffiltext{5}{Departamento de Astrof\' isica, Universidad de La Laguna, Tenerife, Spain}
\altaffiltext{6}{Astronomy Department, University of Florida, Gainesville, FL 32611-2055.}
\altaffiltext{7}{Gemini Observatory, Northern Operations Center, 670 North A\`{o}hoku Place, Hilo, HI 96720.}
\altaffiltext{8}{Department of Physics, University of Crete, GR-71003, Heraklion, Greece.}

\begin{abstract}
We present new mid-infrared (MIR) imaging data for three Type-1 Seyfert galaxies
obtained with T-ReCS on the Gemini-South Telescope at subarcsecond resolution. Our aim is to enlarge the sample studied in a 
previous work to compare the properties of Type-1 and Type-2 Seyfert tori
using clumpy torus models and a Bayesian approach to fit the infrared nuclear spectral energy distributions (SEDs). 
Thus, the sample considered here comprises 7 Type-1, 11 Type-2, and 3 intermediate-type Seyferts. 
The unresolved IR emission of the Seyfert 1 galaxies can be reproduced by a combination of dust heated by the central engine and 
direct AGN emission, while for the Seyfert 2 nuclei only dust emission is considered. 
These dusty tori have physical sizes smaller than 6 pc radius, as derived from our fits. 
Unification schemes of AGN account for a variety of observational differences in terms of viewing geometry. However, we find
evidence that strong unification may not hold,
and that the immediate dusty surroundings of Type-1 and Type-2 Seyfert nuclei are intrinsically different. 
The Type-2 tori studied here are broader, have more clumps, and these clumps have lower optical depths than those of Type-1 tori.  
The larger the covering factor of the torus, the smaller the probability of having direct view of the AGN, and vice-versa. In our sample, Seyfert 2 tori have 
larger covering factors (C$_T$=0.95$\pm$0.02) and smaller escape probabilities (P$_{esc}$=0.05$\pm^{0.08}_{0.03}$ \%) than those of Seyfert 1 
(C$_T$=0.5$\pm$0.1; P$_{esc}$=18$\pm$3 \%). All the previous differences are significant according to
the Kullback-Leibler divergence. 
Thus, on the basis of the results presented here, the classification of a Seyfert galaxy as a Type-1 or Type-2 depends more 
on the intrinsic properties of the torus rather than on its mere inclination towards us, in contradiction with the simplest unification model.
\end{abstract}


\section{Introduction}
\label{intro}
 
Observational evidence in the X-rays and the MIR indicates that the strong AGN continuum source must be absorbed by 
obscuring material over a wide solid angle (see e.g., \citealt{Antonucci85,Maiolino98,Risaliti02}).
According to observed spectra of different AGN types, the obscuring structure has to block the emission 
of the subparsec-scale Broad-Line Region (BLR) where the broad lines are produced,
but not that of the kiloparsec-scale Narrow-Line Region (NLR). 
 
The unified model for active galaxies \citep{Antonucci93,Urry95} is based on the existence of a dusty toroidal structure 
surrounding the central region of AGN. This toroidal geometry 
explains the biconical shapes observed in Hubble Space Telescope (HST) imaging of several AGNs \citep{Tadhunter89,Malkan98,Tadhunter99} and 
also the polarimetric observations \citep{Antonucci85,Packham97}. 
Thus, considering this geometry of the obscuring material, the central engines of Type-1 AGN can be seen directly, 
resulting in typical spectra with both narrow and broad emission lines, whereas in Type-2 AGN the BLR is obscured.

Pioneering work in modelling the dusty torus \citep{Pier92,Pier93,Granato94,Efstathiou95,Granato97,Siebenmorgen04} 
assumed a uniform dust density distribution to simplify the modelling, although from the start, \citet{Krolik88} realized
that smooth dust distributions cannot survive within the AGN vicinity. They proposed instead that the material in the torus must be 
distributed in a clumpy structure, in order to prevent the dust grains from being destroyed by the hot surrounding gas. 

The IR range (and particularly the MIR) is key to set constraints on the torus models, 
since the reprocessed radiation from the dust in the torus is re-emitted in this range.
However, in comparing the predictions of any
torus model with observations, its small-scale emission must be isolated. 
High angular resolution is then essential to separate torus emission  
from stellar emission and star-heated dust in the near-IR (NIR) and MIR, respectively. 
Indeed, starlight dominates the nuclear NIR emission of Seyfert 2 galaxies when using large aperture 
data (see e.g., \citealt{Alonso96}) and still has a significant contribution for Seyfert 1 galaxies
\citep{Kotilainen92}. Similar contamination problems can be present in the MIR with the star-heated dust and 
dust in the ionization cones \citep{Alonso06,Mason06}.

Another controversial issue about the torus structure is its typical dimensions. \citet{Pier93} and \citet{Granato94} reproduced
the infrared observations of nearby Seyfert galaxies with $\sim$100 pc scale tori. However, hard X-ray observations showed that about
half of nearby Type-2 Seyferts are Compton-thick (i.e., they are obscured by a column density higher than 10$^{24}~cm^{-2}$; 
\citealt{Risaliti99}). For these highly obscured sources the torus dimensions are expected to be of a few parsecs, because otherwise 
the dynamical mass of the obscuring material would be too large to be realistic \citep{Risaliti99}.
In addition, recent ground-based MIR observations of nearby Seyferts reveal that the torus size is likely restricted to a few parsecs.
\citet{Packham05} and \citet{Radomski08} established upper limits of 2 and 1.6 pc for the outer radii of the Circinus galaxy and 
Centaurus A tori, respectively. Besides, interferometric observations obtained with the MIR Interferometric Instrument (MIDI) 
at the Very Large Telescope Interferometer (VLTI) of Circinus, NGC 1068, and Centaurus A suggest a 
scenario where the torus emission would only extend out to R = 1 pc \citep{Tristram07}, 
R = 1.7 - 2 pc \citep{Jaffe04,Raban09}, and R = 0.3 pc \citep{Meisenheimer07}, respectively.

In order to solve the discrepancies between observations and previous models, 
an intensive search for an alternative torus geometry has been carried out in the last decade. 
The first results of radiative transfer calculations of a clumpy rather than a smooth medium were reported 
by \citet{Nenkova02} and \citet{Elitzur06}, and further work was done by \citet{Dullemond05}.
The clumpy dusty torus models \citep{Nenkova02,Nenkova08a,Nenkova08b,Honig06,Schartmann08}
propose that the dust is distributed in clumps, instead of homogeneously filling the torus volume. These models are making
significant progress in accounting for the MIR emission of AGNs (\citealt{Mason06,Mason09,Mor09,Horst08,Horst09,Nikutta09}; Ramos Almeida et al.~2009a;
\citealt{Honig10}).

In our previous work (Ramos Almeida et al.~2009a; hereafter \citealt{Ramos09a}), we constructed subarcsecond resolution 
IR SEDs for eighteen Seyfert galaxies, mostly Type-2 Seyferts. From the comparison between 
our high angular resolution MIR fluxes and large aperture data, such as those from ISO, IRAS, or Spitzer, we confirmed
that the former provide a spectral shape that is substantially different from that of the large aperture data \citep{Rodriguez97}. 
Since our nuclear measurements allowed us to better characterize the torus emission, we modelled our SEDs with clumpy
torus models. In general, we found that Type-2 views are more inclined than those of Type-1s, and more importantly, 
we derive larger covering factors for the Type-2 tori (i.e., more clumps and wider torus angular distributions). This would imply that
the observed differences between Type-1 and Type-2 AGN would not be due to orientation effects only, but to intrinsic 
differences in their tori. However, due to the limited size of the sample analyzed by \citealt{Ramos09a}, and 
in particular of Type-1 Seyferts compared with Type-2s, our aim is to enlarge the sample studied in the
previous work with new Seyfert 1 infrared data to compare the properties of Type-1 and Type-2 Seyfert tori.



In this work, we report new subarcsecond MIR imaging data for the 3 nearby Type-1 Seyfert galaxies NGC 7469, NGC 6221, 
and NGC 6814, for which we estimate unresolved nuclear MIR fluxes.
We enlarge the sample by including the galaxies NGC 1097, NGC 1566, NGC 3227, and NGC 4151, which have similar MIR data, and 
we compile NIR nuclear fluxes from the literature of similar resolution to construct nuclear SEDs for all the galaxies. 
We fit these SEDs with clumpy torus models which we interpolate from the \citet{Nenkova08a,Nenkova08b} database, 
and compare them with the larger sample studied by \citealt{Ramos09a}.
Table \ref{sources} summarizes key observational properties of the sources in the sample. 
Section \ref{observations} describes the observations, data reduction, and compilation
of NIR and MIR fluxes. Sections \ref{extended} and \ref{sed} present the main observational results, 
and in \S \ref{modelling} we report the modelling results.
We discuss differences between Type-1 and Type-2 Seyferts and draw conclusions about the clumpy torus models and AGN obscuration
in general  in \S \ref{discussion}.  Finally, Section \ref{final} summarizes the main conclusions of this work.

\begin{deluxetable*}{lcccccc}
\tablewidth{0pt}
\tablecaption{Basic Galaxy Data}
\tablehead{
\colhead{Galaxy} & \colhead{Seyfert Type} & Ref. & \colhead{$z$} & \colhead{Distance} & \colhead{Scale} & Ref.  \\
 & & & & \colhead{(Mpc)} & \colhead{(pc~arcsec$^{-1}$)} & } 
\startdata
NGC 1097    &  Sy1\tablenotemark{\dag}& A1&  0.0042 &   19   & 92    & B1   \\  
NGC 1566    &  Sy1		     & A2 &  0.0050 &   20   & 97    & B2   \\  
NGC 6221    &  Sy1		     & A3 &  0.0050 &   18   & 87    & B3   \\ 
NGC 6814    &  HII/Sy1.5	     & A4 &  0.0052 &   21   & 102   & B4   \\ 
NGC 7469    &  Sy1   		     & A5 &  0.0163 &   65   & 315   & B5   \\
\hline
NGC 3227    &  Sy1.5                 & A6 & 0.0039  &   17   & 82    & B6   \\
NGC 4151    &  Sy1.5                 & A7 & 0.0033  &   13   & 64    & B7   \\
\enddata
\tablecomments{\footnotesize{Classification and distance are taken from the literature (references below) 
and spectroscopic redshift from the NASA/IPAC Extragalactic Database (NED).}}
\tablenotetext{\dag}{\footnotesize{Originally classified as a LINER by \citet{Keel83} and \citet{Phillips84}.}}
\tablerefs{\footnotesize{(A1) \citet{Storchi97}; (A2) \citet{Kriss90}; (A3) \citet{Levenson01}; (A4) \citet{Veron06}; (A5) \citet{Osterbrock93}; 
(A6) \citet{Rubin68}; (A7) \citet{Ayani91};
(B1) \citet{Willick97}; (B2) \citet{Sandage94}; (B3) \citet{Koribalski04}; (B4) \citet{Liszt95}; (B5) \citet{Heckman86};
(B6) \citet{Garcia93}; (B7) \citet{Radomski03}.}}
\label{sources}
\end{deluxetable*}

\section{Observations and Data Reduction}
\label{observations}

\subsection{MIR Imaging Observations}
\label{MIR}

In order to enlarge the sample of 18 Seyfert galaxies presented in \citealt{Ramos09a}, which comprises 
12 Seyfert 2 (Sy2), two Seyfert 1.9 (Sy1.9), one Seyfert 1.8 (Sy1.8), two Seyfert 1.5 (Sy1.5), and one Seyfert 1 galaxy (Sy1), 
we obtained new subarcsecond MIR observations of the Type-1 Seyferts NGC 6221, NGC 6814, and NGC 7469 (see Table \ref{sources}).

The observations were performed with the MIR camera/spectrograph T-ReCS (Thermal-Region Camera Spectrograph;
\citealt{Telesco98}) on the Gemini-South telescope during the Summer of 2009. T-ReCS uses a Raytheon 320x240 pixel Si:As IBC array, 
providing a plate scale of 0.089\arcsec~pixel$^{-1}$, corresponding to a FOV of 28.5\arcsec x21.4\arcsec. 
The filters employed for the observations were the narrow Si-2 filter ($\lambda_c$=8.74 \micron, $\Delta\lambda$=0.78 \micron, 
50\% cut-on/off) and the broad Qa filter ($\lambda_c$=18.3 \micron, $\Delta\lambda$=1.5 \micron, 50\% cut-on/off).
The resolutions obtained were 0.3\arcsec~at 8.7 \micron~and 0.5\arcsec~at 18.3 \micron, as measured from the observed Point Spread
Function (PSF) stars. A summary of the observations is reported in Table \ref{log}.

\begin{deluxetable*}{llccccc}
\tablewidth{0pt}
\tablecaption{Summary of MIR Observations}
\tablehead{
\colhead{Galaxy} & \colhead{Filters} & \colhead{Observation} & \multicolumn{2}{c}{On-Source Time (s)}  & \multicolumn{2}{c}{PSF FWHM}   \\
& & \colhead{epoch} & \colhead{N band} & \colhead{Q band} & \colhead{N band} & \colhead{Q band}}
\startdata
NGC 1097  & Si-5, Qa   & Sep 2005   &  456  & 912   & 0.41\arcsec & 0.52\arcsec      \\ 
NGC 1566  & Si-2, Qa   & Sep 2005   &  152  & 304   & 0.30\arcsec & 0.53\arcsec      \\  
NGC 6221  & Si-2, Qa   & Aug 2009   &  145  & 203   & 0.32\arcsec & 0.55\arcsec      \\        
NGC 6814  & Si-2, Qa   & Aug 2009   &  145  & 203   & 0.28\arcsec & 0.53\arcsec      \\      
NGC 7469  & Si-2, Qa   & Sep 2009   &  145  & 203   & 0.31\arcsec & 0.55\arcsec      \\    
\hline
NGC 3227  & N'         & Apr 2006   &  300  & \dots & 0.39\arcsec & \dots            \\
NGC 4151  & N, IHW18   & May 2001   &  360  & 480   & 0.53\arcsec & 0.58\arcsec      \\
\enddata       
\tablecomments{\footnotesize{Images were obtained in the 8.74 \micron~(Si-2, $\Delta\lambda$ = 0.78 \micron~at 50\% cut-on/off), 
11.66 \micron~(Si-5, $\Delta\lambda$ = 1.13 \micron), and 18.30 \micron~(Qa, $\Delta\lambda$ = 1.5 \micron) T-ReCS filters.
NGC 3227 was observed in the 11.29 \micron~(N', $\Delta\lambda$ = 2.4 \micron) Michelle/Gemini-North filter, and 
NGC 4151 in the 10.75 \micron~(N, $\Delta\lambda$ =5.2 \micron) and 18.17 \micron~(IHW18, $\Delta\lambda$ =1.7 \micron) OSCIR/Gemini-North filters.}}
\label{log}
\end{deluxetable*}

The standard chopping-nodding technique was used to remove the time-variable sky background,
the telescope thermal emission, and the so-called 1/f detector noise.
The chopping throw was 15\arcsec, and the telescope was nodded 15\arcsec~in the direction of the chop
every 45 s. 
The difference for each chopped pair for each given nodding set was calculated,
and the nod sets were then differenced and combined until a single image was created. 
The data were reduced using in-house-developed IDL routines.

Observations of flux standard stars were made for the flux calibration of each galaxy through the 
same filters. The uncertainties in the flux calibration are typically $\sim$5-10\% at 8.7 \micron~ 
and $\sim$15-20\% at 18.3 \micron. PSF star observations were also made immediately prior to or
after each galaxy observation to accurately sample the image quality. 
These images were employed to determine the unresolved (i.e., nuclear) component of each galaxy.
The PSF star, scaled to the peak of the galaxy emission, represents the maximum
contribution of the unresolved source (100\%), where we integrate flux within an aperture of 2\arcsec. 
The residual of the total emission minus the scaled PSF represents the host galaxy contribution, 
which is analyzed in Section \ref{extended}.
We require a flat profile in the residual for a realistic galaxy profile over the central pixels. 
Therefore, we reduce the scale of the PSF from matching the peak of the galaxy emission (100\%), when necessary, to 
obtain the unresolved fluxes reported in Table \ref{psf}. They include corresponding aperture corrections 
to take into account possible flux losses when integrating the scaled PSF flux in the 2\arcsec~aperture.

\begin{figure*}[!ht]
\centering
\includegraphics[width=15cm]{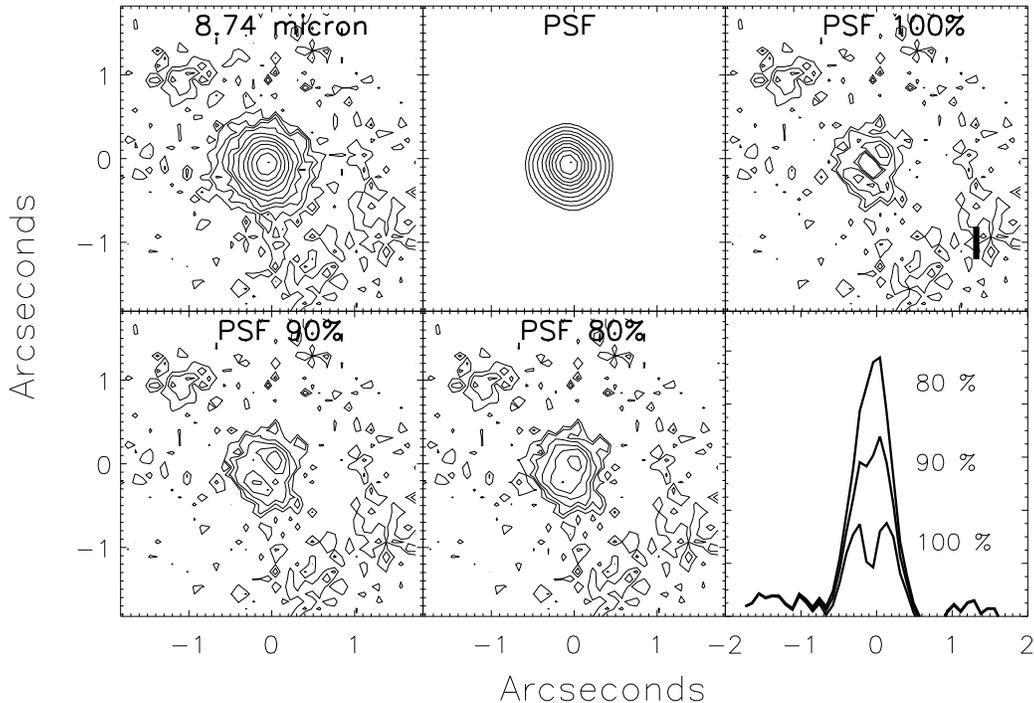}
\caption{\footnotesize{8.74 \micron~contour plots of NGC 7469 at the 3$\sigma$ level, the PSF star, and 
scaled subtraction of the PSF for this galaxy at various levels (100\%, 90\%, and 80\%).
The residuals of the subtraction in the lower right panel show the host galaxy profile.}
\label{psf1}}
\end{figure*}

Figure \ref{psf1} shows an example of PSF subtraction at various levels
(in contours) for NGC 7469 in the Si-2 T-ReCS filter, following 
\citet{Radomski02,Radomski03}. 
The residual profiles from the different scales demonstrate the best-fitting result.
The uncertainty in the unresolved fluxes determination from PSF subtraction is $\sim$10-15\%. Thus, 
we estimated the errors in the flux densities reported in Table \ref{psf} by adding quadratically the 
flux calibration and PSF subtraction uncertainties, resulting in $\sim$15\% at 8.7 \micron~and 
$\sim$25\% at 18.3 \micron. 

\begin{deluxetable}{lcccc}
\tablewidth{0pt}
\tablecaption{Unresolved MIR Fluxes}
\tablehead{
\colhead{Galaxy} & \multicolumn{2}{c}{Level of PSF subtraction} & \multicolumn{2}{c}{Flux Density (mJy)}    \\
& \colhead{N band} & \colhead{Qa band} & \colhead{N band} & \colhead{Qa band}}
\startdata
NGC 1097                    	&   80\% &   20\%	&   23    &  55   	\\
NGC 1566\tablenotemark{a}       &  100\% &  100\%	&   29    &  117    	\\  
NGC 6221	            	&   80\% &   30\%	&   48    &  314   	\\  
NGC 6814                       	&  100\% &   40\%	&   53    &  159   	\\
NGC 7469	    		&   85\% &  100\%	&   174   &  1354       \\
\hline
NGC 3227	    		&  100\% &  \nodata	&   401   & \nodata 	\\
NGC 4151	    		&   90\% &  100\%	&   1320  & 3200  	\\
\enddata        
\tablecomments{The percentages of PSF subtraction level are reported 
in the employed filters (listed in Table \ref{log}). Errors in flux densities  are dominated in 
general by uncertainties in the flux calibration and PSF subtraction ($\sim$15\% at N band and $\sim$25\% at Qa band).
Fluxes include aperture corrections.}
\tablenotetext{a}{MIR fluxes for NGC 1566 are from aperture photometry using 0.45\arcsec~aperture radii (\citealt{Ramos09a}).}
\label{psf}
\end{deluxetable}

\subsection{Compilation of NIR and MIR High Spatial Resolution Data}

In addition to the new T-ReCS observations described in Section \ref{MIR}, we include here two more Sy1 galaxies with 
already published T-ReCS data: NGC 1097 and NGC 1566\footnote{NGC 1097 and NGC 1566 were not included in the analysis of the SEDs 
presented in \citealt{Ramos09a} because of the lack of high angular resolution NIR fluxes for them at the time of publication.}. 
We reduced the images presented in \citet{Mason07} for NGC 1097 and obtained unresolved MIR fluxes by performing the same technique explained
in Section \ref{MIR} (see Table \ref{psf}). For NGC 1566 we compiled the MIR nuclear fluxes reported 
\citealt{Ramos09a}.
Both galaxies were observed in September 2005 with T-ReCS: NGC 1566 in the Si-2 and Qa filters, and NGC 1097 in the Si-5 
($\lambda_c$=11.66 \micron, $\Delta\lambda$=1.13 \micron, 50\% cut-on/off) and Qa filters (see Table \ref{log}). 
The resolutions of the images are 0.3\arcsec~at 8.7 \micron, $\sim$0.4\arcsec~at 11.66 \micron, 
and 0.5\arcsec~at 18.3 \micron.

NIR subarcsecond resolution nuclear fluxes compiled from the literature are reported in Table \ref{literature}.
For NGC 1097, NGC 1566, and NGC 7469, \citet{Prieto10} reported diffraction-limited and near-diffraction-limited adaptive 
optics NACO/VLT fluxes. The three galaxies are unresolved in the NIR down to the highest resolution achieved 
(FWHM$\sim$0.15\arcsec~for NGC 1097 in the L'-band, $\sim$0.12\arcsec~for NGC 1566 in the K-band, and $\sim$0.08\arcsec~for NGC 7469 in the H-band). 

For NGC 7469, \citet{Prieto10} reported NACO J-, H-, and K-band nuclear fluxes obtained from observations in Nov 2002, as well as in the 
L' and NB-4.05 \micron~filters, observed in this case in Dec 2005. First, we discarded the narrow-band filter, since it is designed
to collect the Br$\alpha$ emission, which in the case of this galaxy is important (as inferred from the Br$\gamma$ line detected 
with NIR spectroscopy; \citealt{Genzel95,Hicks08}). The L'-band filter also contains Br$\alpha$, which is very likely 
compromising the flux measurement, since the L' and NB-4.05 fluxes do not 
match the SED shape of the remaining J, H, K, Si-2 and Qa data. Indeed, \citet{Prieto10} also reported a NICMOS/HST flux measurement 
in the filter F187N, obtained in 2007. This data point lies in between the NACO H and K measurements in wavelength and flux, which
gives us extra-confidence in the reliability of the NACO J, H, and K measurements. Considering all the previous, we finally decided 
to consider the NACO L'-band flux as an upper limit\footnote{The nucleus of NGC 7469 had undergone different periods of activity, 
including a maximum in the optical happening in the period 1996 to 2000, followed by a relaxation epoch in the following years
\citep{Prieto10}. For this reason, here we only consider data obtained from 2002.}. 


For NGC 6221 and NGC 6814, which do not have any published subarcsecond resolution NIR data, we 
retrieved broad-band images from the Hubble Legacy Archive\footnote{http://archive.stsci.edu/} obtained 
with NICMOS on the HST. 
The two galaxies were observed in the F160W filter, using the NIC2 camera, as part of the program 7330 (PI: Mulchaey, J.). 
The typical FWHM for an unresolved PSF is $\sim$0.13\arcsec~using the F160W filter with NIC2.
Details of the observations can be found in \citet{scoville00} and \citet{regan99}.
For the analysis, we first separated the nuclear emission from the underlying host galaxy emission. We then
applied the two-dimensional image decomposition GALFIT program \citep{peng02} to fit
and subtract the unresolved component (PSF). PSF models were created using the TinyTim
software package, which includes the optics of HST plus the specifics of the camera and filter
system \citep{krist93}. We checked that no prominent emission lines are included in the wavelength range covered 
by the filter F160W. 
The resulting unresolved NIR fluxes for NGC 6221 and NGC 6814 are reported in Table \ref{literature}.

We finally include the Sy1.5 galaxies NGC 3227 and NGC 4151 in this study, which were also part of the 
\citealt{Ramos09a} sample. The MIR nuclear fluxes are the same reported in the latter work (see description 
of the observations in Section 2.1 in \citealt{Ramos09a}), whereas the NIR fluxes have been updated (see Table 
\ref{literature}).

Using the NIR nuclear fluxes reported in Table \ref{literature} in combination with our 
MIR unresolved measurements (Table \ref{psf}) we construct AGN-dominated SEDs for the five Sy1 galaxies. 

\begin{deluxetable*}{lcccccc}
\tabletypesize{\footnotesize}
\tablewidth{0pt}
\tablecaption{High Spatial Resolution NIR Fluxes}
\tablehead{
\colhead{Galaxy} & \multicolumn{4}{c}{Flux Density (mJy)} & \colhead{Filters} & \colhead{Reference(s)}  \\
& \colhead{J band} & \colhead{H band} & \colhead{K band} & \colhead{L band} &   &}
\startdata
NGC 1097  & 1.1$\pm$0.1   & 2.7$\pm$0.1  & 3.9$\pm$0.1  & 11$\pm$1                      & NACO J,H,K,L' & a  \\
NGC 1566  & 1.1$\pm$0.1   & \nodata	 & 2.1$\pm$0.1  & 7.8$\pm$0.1                   & NACO J,K,L'   & a  \\ 
NGC 6221  & \nodata       & 2.1$\pm$0.2  & \nodata	& \nodata                       & F160W	        & b  \\	 
NGC 6814  & \nodata       & 5.2$\pm$0.5  & \nodata	& \nodata                       & F160W	        & b  \\ 
NGC 7469\tablenotemark{a} & 8.0$\pm$0.1  & 15$\pm$1	& 20$\pm$1     & 84$\pm$1       & NACO J,H,K,L' & a  \\   
\hline
NGC 3227\tablenotemark{b} & \nodata       & 7.8$\pm$0.8 & 16.4$\pm$1.7 & 46.7$\pm$9.3   & F160W,F222M, NSFCam L & c,d \\
NGC 4151\tablenotemark{b} & 60$\pm$6      & 100$\pm$10  & 197$\pm$20   & 325$\pm$65     & F110W,F160W,F222M, NSFCam L & c,e \\
\enddata
\tablecomments{\footnotesize{Ground-based instruments and telescopes are NACO on the 8 m VLT and NSFCam on the 3 m NASA IRTF. 
Measurements in the F110W, F160W, and F222M filters are from NICMOS on HST.}}
\tablenotetext{a}{\footnotesize{\citet{Prieto10} also reported a flux measurement of 19$\pm$1 mJy obtained from 
a NICMOS/HST observation in the filter F187N.}}
\tablenotetext{a}{\footnotesize{\citet{Ward87} also reported M-band flux measurements of 72$\pm$27 mJy for NGC 3227 and 
449$\pm$34 mJy for NGC 4151 obtained with IRCAM3 on the 3.8 m UKIRT.}}
\tablerefs{\footnotesize{(a) \citet{Prieto10}; (b) This work; (c) \citet{Kishimoto07}; (d) \citet{Alonso03}; (e) \citet{Ward87}}}
\label{literature}
\end{deluxetable*}

\section{The MIR Extended Emission of NGC 7469 and NGC 6221}
\label{extended}

Our new MIR imaging data reveal complex extended emission for NGC 7469 and NGC 6221, which
is known to be intense and associated with emitting-dust heated by star formation. On the other hand, 
NGC 6814 lacks of any extended emission. 
In this section we present the MIR images of NGC 7469 and NGC 6221, and compare them 
with published data in different wavelength ranges.

\subsection{NGC 7469}
\label{ngc7469extended}

The most spectacular morphological feature of this Sy1 galaxy  
is a circumnuclear ring of powerful starbursts of $\sim$1.6 kpc diameter, which is deeply embedded in a large cloud of molecular gas
and dust. This ring contains several super star clusters and regions of star formation and it accounts for 
two-thirds of the galaxy bolometric luminosity \citep{Genzel95}. 
The star-forming ring has been the subject of study of several works in different wavelengths (see D\' iaz-Santos et al.~2007, 
hereafter \citealt{Diaz07}, and references therein), including the MIR \citep{Miles94,Soifer03,Horst09}.
Based on VISIR/VLT MIR observations at 12.3 \micron, \citet{Horst09} report the detection of distinct knots of star formation 
around the nucleus located at a distance of $\sim$1.3\arcsec~($\sim$400 pc), although with low signal-to-noise. 
Using NIR HST data, \citealt{Diaz07} identified 30 clusters of star formation 
in the ring at 1.1 \micron. By fitting the individual ultraviolet-to-NIR SEDs of the clusters, 
the authors reported the presence of a dominant intermediate age population (8-20 Myr) and a younger and more extinguished 
one ($\sim$1-3 Myr). The latter does not coincide with the optical/NIR 
continuum-emitting regions, but seems to be traced by the MIR/radio emission, less affected by extinction than the optical/NIR. 

\begin{figure*}[!ht]
\centering
\par{
\includegraphics[width=8cm]{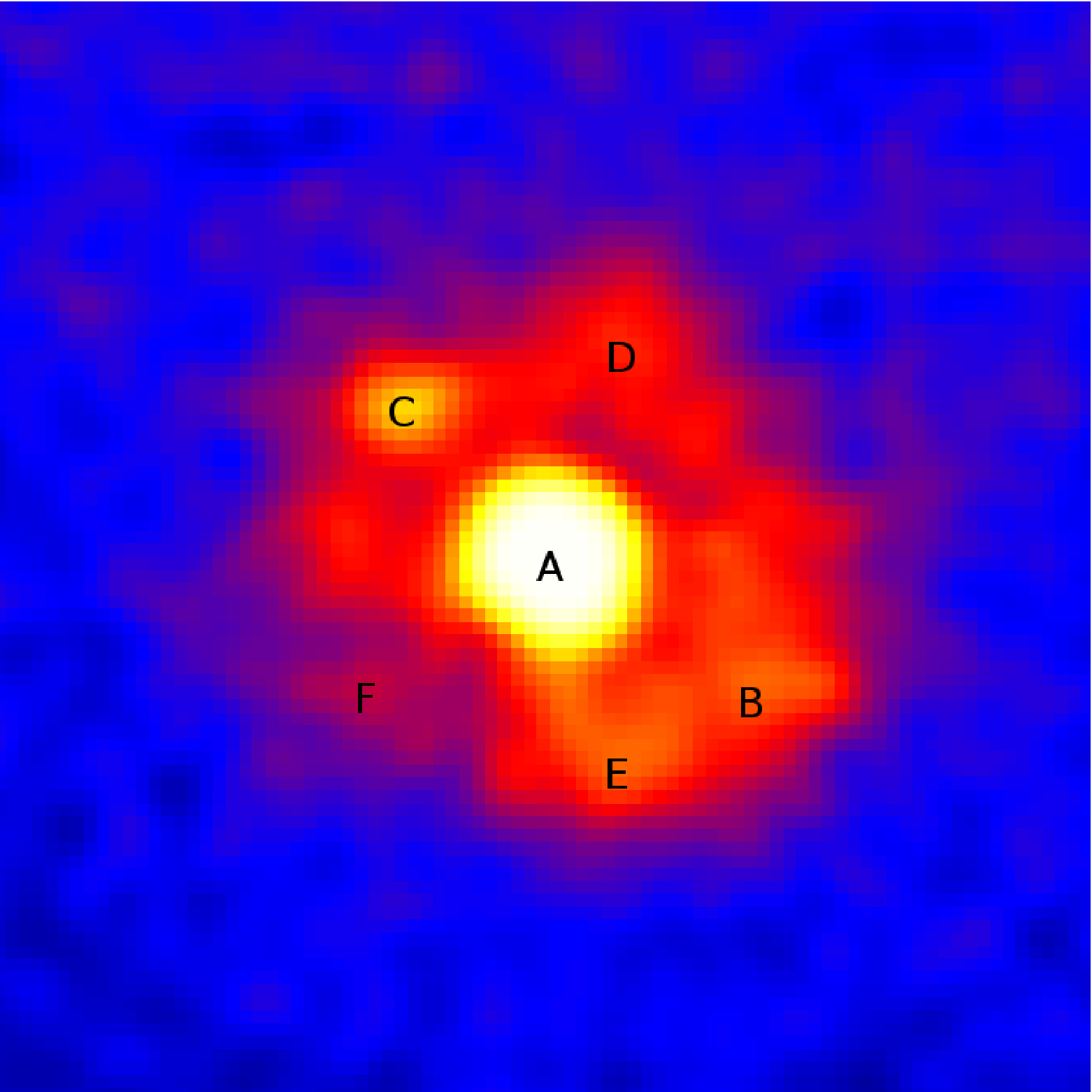}
\includegraphics[width=8cm]{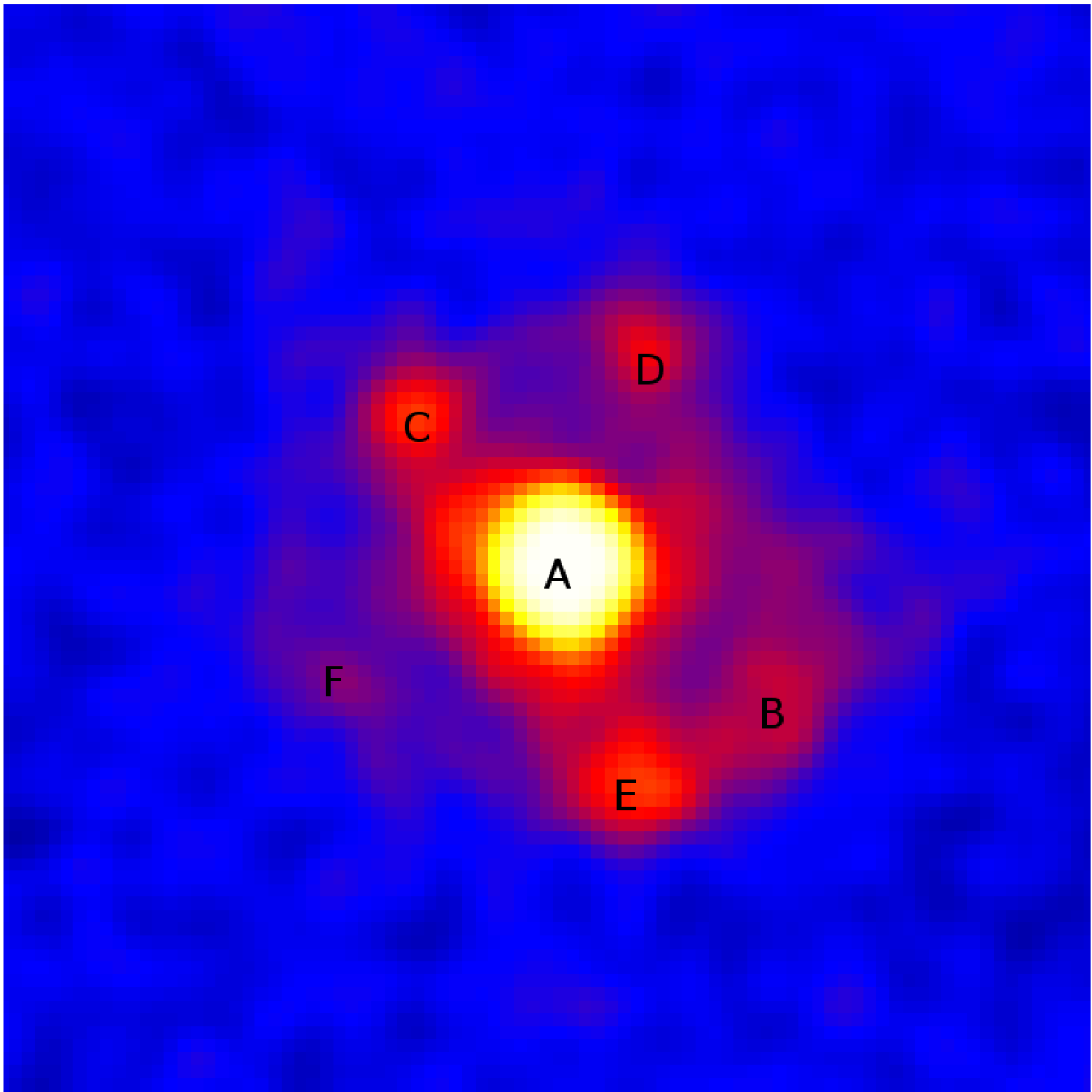}\par}
\caption{\footnotesize{8.74 \micron~(left) and 18.3 \micron~(right) T-ReCS images of NGC 7469 smoothed by using a 
moving box of 3 pixel size. The images size is 7.8 arcsec side. North is up, and East to the left. 
Emitting regions identified in both the Si-2 and Qa images, and also coincident
with the radio emission, are labelled from A to F, where A corresponds to the AGN.}
\label{ngc7469mid}}
\end{figure*}

Figure \ref{ngc7469mid} shows the high resolution 8.74 and 18.3 \micron~T-ReCS images of NGC 7469.    
At both wavelengths we detect extended emission with high signal-to-noise coincident with the star-forming ring. 
Indeed, our flux maps appears very similar to the high resolution 11.7 \micron~contours 
presented by \citet{Miles94} and to the high resolution (0.2\arcsec) VLA 8.4 GHz radio map presented in \citet{Colina01}. 
We identified the brightest knots in our MIR images in Figure \ref{ngc7469mid} using the same notation as in \citet{Miles94},
where the AGN is labelled A. The B and C knots in our images correspond to the brightest regions in radio, 
according to the 5 GHz \citep{Wilson91} and 8.4 GHz radio maps \citep{Colina01}.
We also identified the knots D, E, and F. In Table \ref{ring} we report the star clusters identified 
by \citealt{Diaz07} which better match the positions of the A to F MIR compact regions. The distances between these
knots and the AGN range from 1.4\arcsec~to 1.8\arcsec~(median distance of $\sim$480 pc).
All the knots appear more compact in the 18.3 \micron~image than in the 
8.74 \micron, where the ring emission is more extended. This is expected since the 8.74 \micron~filter contains the 
8.6 \micron~Polycyclic Aromatic Hydrocarbon (PAH) feature, whereas the 18.3 \micron~filter is mostly probing hot dust emission.
As found by Spitzer, the $\sim$8 \micron~PAH emission of nearby galaxies appears to be more extended than at $\sim$24 \micron~(e.g., 
\citealt{Helou04,Calzetti05}).

In Table \ref{ring} we report aperture fluxes for the six identified knots in both the 8.74 and 18.3 
\micron~images calculated using IRAF\footnote{IRAF is distributed by 
the National Optical Astronomy Observatory, which
is operated by the Association of Universities for the Research in Astronomy,
Inc., under cooperative agreement with the National Science Foundation
(http://iraf.noao.edu/).}.
The aperture radius was defined on the basis of the resolution element in the Qa band (0.55\arcsec), 
and corresponding aperture corrections were applied, since the knots are only barely resolved in both bands. 
Positions relative to the nucleus (A) in arcseconds are also given.

\begin{deluxetable*}{lccccccc}
\tablewidth{0pt}
\tablecaption{Positions and flux densities of NGC 7469 MIR knots}
\tablehead{
\colhead{Knot} & \multicolumn{1}{c}{ID} & \multicolumn{3}{c}{8.74 \micron} & \multicolumn{3}{c}{18.3 \micron}  \\
& DS07 & \colhead{(X)} & \colhead{(Y)} & \colhead{Flux (mJy)} & \colhead{(X)} & \colhead{(Y)} & \colhead{Flux (mJy)}}
\startdata
A  & \dots &  0.00         &  0.00         & 229     &  0.00         &  0.00         &  1320   \\
B  & C20   & -1.58\arcsec  & -0.89\arcsec  & 11.9    & -1.54\arcsec  & -0.93\arcsec  &  50     \\ 
C  & C6    &  1.09\arcsec  &  1.07\arcsec  & 13.4    &  0.95\arcsec  &  1.06\arcsec  &  79     \\	  
D  & C10   & -0.49\arcsec  &  1.31\arcsec  & 11.5    & -0.58\arcsec  &  1.38\arcsec  &  107    \\ 
E  & C7,C15& -0.40\arcsec  & -1.36\arcsec  & 15.8    & -0.52\arcsec  & -1.49\arcsec  &  104    \\  
F  & C12   &  1.25\arcsec  & -0.89\arcsec  & 3.7     &  1.47\arcsec  & -0.83\arcsec  &  14.8   \\  
\enddata
\tablecomments{\footnotesize{Columns (1) and (2) indicate the label asigned to each knot in Figure \ref{ngc7469mid} and in \citealt{Diaz07}. 
B, C, and D are also coincident with the R3, R2, and R1 regions in the VLA 8.4 GHz radio
maps in \citet{Colina01}. Columns (3,4) and (6,7) indicate
the position of each knot relative to A at 8.7 and 18.3 \micron, respectively. 
Columns (5) and (8) list the flux densities of the knots in a 0.55\arcsec~aperture radius.  Fluxes include aperture corrections 
of 19\% at 8.7 \micron~and 37\% at 18.3 \micron. Errors in flux densities are dominated 
by uncertainties in the flux calibration ($\sim$5-10\% at 8.7 \micron~and $\sim$15-20\% at 18.3 \micron).}}
\label{ring}
\end{deluxetable*}

\subsection{NGC 6221}
\label{ngc6221extended}

The galaxy NGC 6221 is a prototypical example of the so-called ``X-ray--loud composite galaxies'' 
(see e.g., \citealt{Moran96}). This classification comes from the comparison
between its optical spectrum, which is starburst-like, and its X-ray data, where the AGN is revealed \citep{Levenson01}.
According to the X-ray data, and in particular
the width of the Fe K$\alpha$ line, the orientation of the Seyfert nucleus is Type-1 \citep{Levenson01}. 
However, as mentioned above, the optical spectrum of the galaxy resembles more
that of a starburst galaxy, implying that there must be a big amount of dust (likely associated with the starburst) 
along the line of sight (LOS) hiding the BLR. A nuclear optical extinction of 
A$_V$=3.0 mag was measured from the optical spectrum by \citet{Levenson01}. 
They also presented HST images of the central region of NGC 6221 in the optical 
(F606W/WFPC2) and in the NIR (F160W/NICMOS) and identified the bright central NIR source as the AGN. 
At optical wavelengths, the nucleus is diffuse and weaker than other bright knots, identified as 
star clusters. 

\begin{figure*}[!ht]
\centering
\par{
\includegraphics[width=8cm]{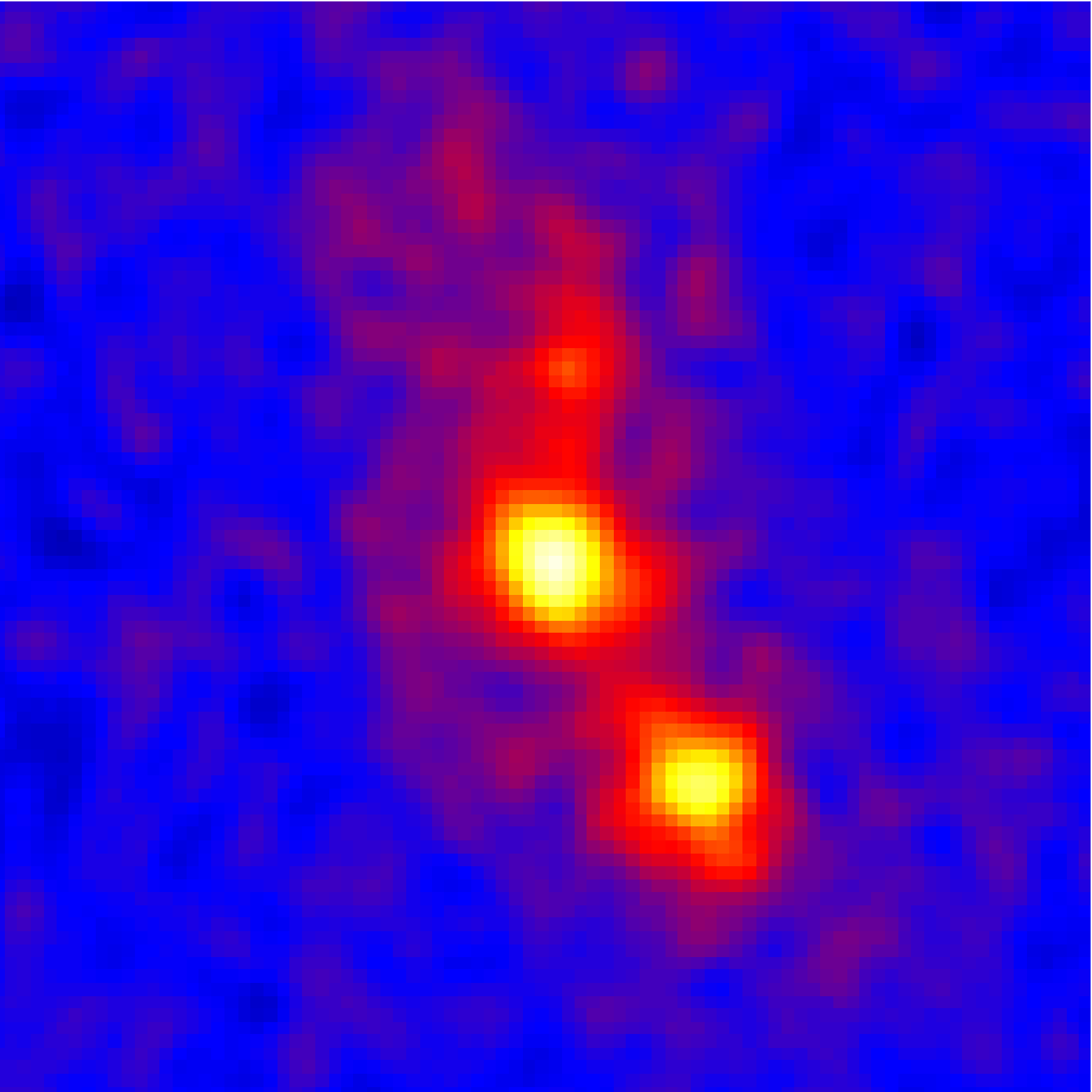}
\includegraphics[width=8cm]{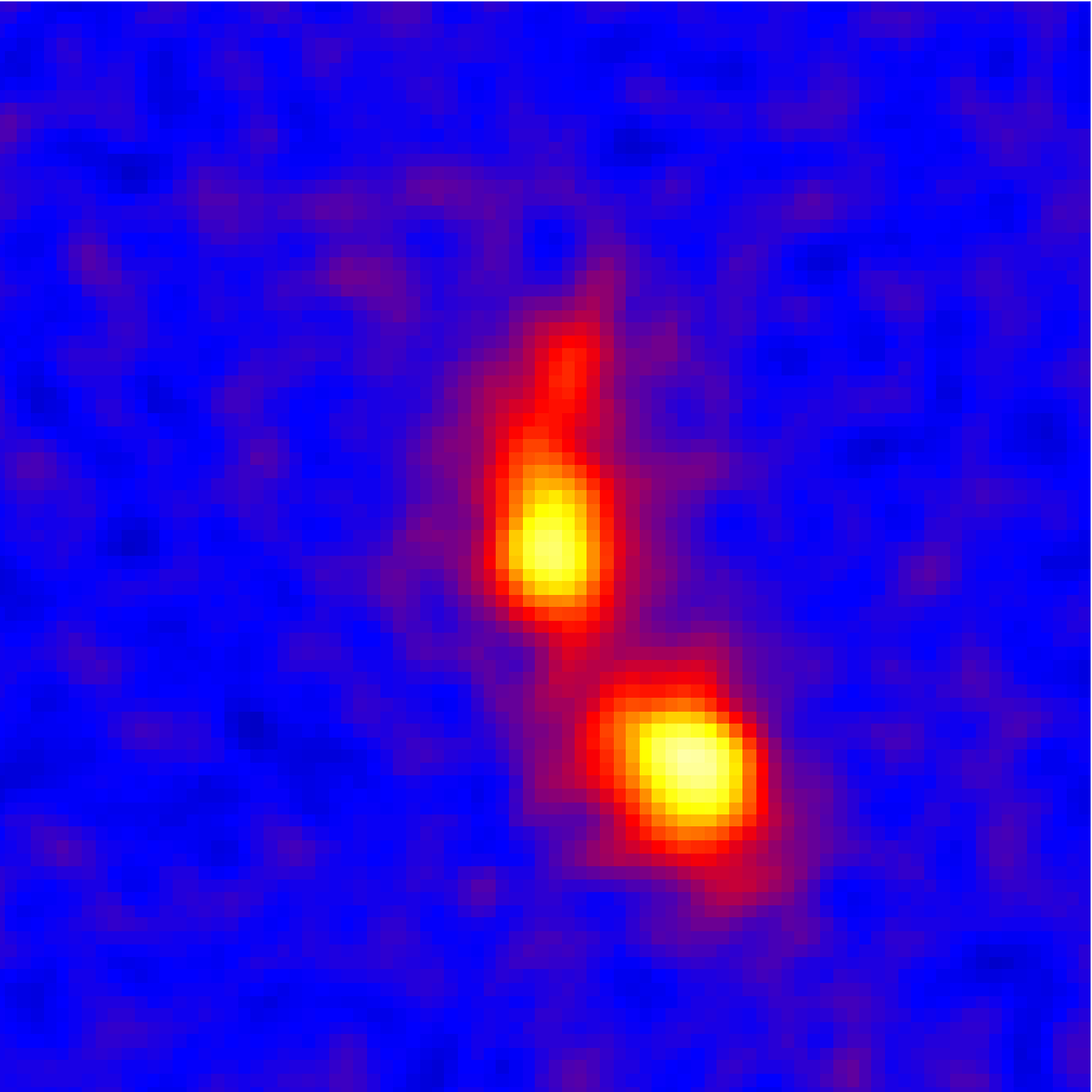}\par}
\caption{\footnotesize{8.74 \micron~(left) and 18.3 \micron~(right) T-ReCS images of NGC 6221 smoothed by 
using a moving box of 3 pixel size. The images size is 7.8 arcsec side. North is up, and East to the left.
The AGN is centered in both images.}
\label{ngc6221mid}}
\end{figure*}

Our 8.7 and 18.3 \micron~images of NGC 6221 are shown in Figure \ref{ngc6221mid}. They reveal spectacular extended emission 
with two bright knots. The AGN is centered in both images. The other bright knot, at $\sim$1.9\arcsec~SW
from the nucleus, which is roughly coincident with 
the SW starburst region shown in the F606W optical image in Figure 3 of \citet{Levenson01}, 
reaches practically the same intensity as the AGN at 8.7 \micron~and appears brighter at 18.3 \micron. 
We measured aperture fluxes in a 0.7\arcsec~radius for the SW knot, selected to collect the bulk 
of its MIR emission, and obtain fluxes of 42 mJy at 8.7 \micron~and 422 mJy at 18.3 \micron. 
Surrounding the AGN there is more diffuse emission extending towards the North, which in this case is more
intense in the 8.7 \micron~image than in the 18.3 \micron~one. As already mentioned in Section \ref{ngc7469extended}, the 
$\sim$8 \micron~PAH emission of nearby galaxies appears more extended than the $\sim$24 \micron~emission (e.g., 
\citealt{Helou04,Calzetti05}).

\section{SED Observational Properties}
\label{sed}

\subsection{Average Seyfert 1 Spectral Energy Distribution}
\label{averageSy1}

\begin{figure*}[!ht]
\centering
\includegraphics[width=15cm]{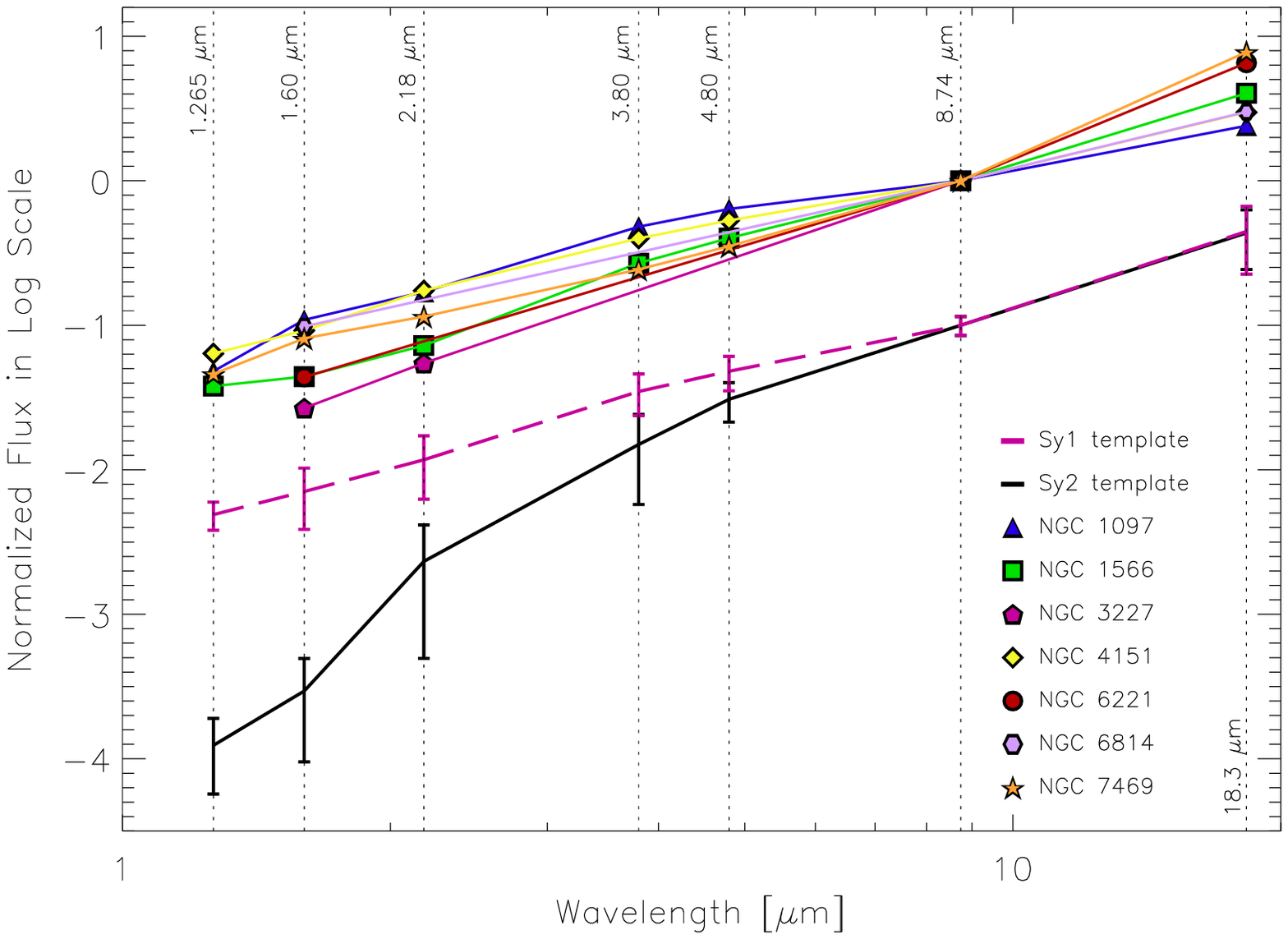}
\caption{\footnotesize{Observed IR SEDs for the seven Type-1 Seyfert galaxies (in color and with different 
symbols) used for the construction of the Sy1 average template (dashed pink). The Sy2 template from \citealt{Ramos09a} (solid black)
is also plotted for comparison. 
The SEDs have been normalized at 8.74 \micron, and the average Sy1 and Sy2 SEDs have been shifted in the Y-axis for clarity.}
\label{template}}
\end{figure*}

Using the MIR and NIR data reported in Tables \ref{psf} and \ref{literature} we construct 
subarcsecond resolution nuclear SEDs in the wavelength range from $\sim$1 to 18 \micron~for the five Type-1 Seyferts 
analyzed here. Figure \ref{template} shows a comparison between their spectral shapes 
and the average Sy2 SED from \citealt{Ramos09a}. This mean template was constructed using individual Sy2 data of 
the same angular resolution as that achieved in this work ($\lesssim$0.55\arcsec).  
In the same way, we have constructed an average Type-1 Seyfert template using the IR nuclear SEDs of the seven 
galaxies studied here\footnote{We have not considered
the NIR ground-based data reported in Table 4 of \citealt{Ramos09a} for NGC 3227 and NGC 4151 in the construction of the mean template
to be consistent with the angular resolutions of the other SEDs.}. 
The spectral shape of Sy1 and Sy1.5 galaxies is practically identical (as can be seen from Figure \ref{template}).
Based on the previous, and on the fact that both types of nuclei present broad lines in their optical spectra, 
in this work we consider them as Type-1 Seyferts.

The Sy2 template defines the wavelength grid, and we performed a quadratic interpolation of nearby measurements of the individual Sy1 
galaxies onto its scale (1.265, 1.60, 2.18, 3.80, 4.80, 8.74, and 18.3 \micron).
We did not interpolate the sparse observations of NGC 6221 and NGC 6814, which only have NICMOS 1.6 \micron~data in 
addition to the MIR measurements.
The interpolated fluxes were used solely for the purpose of deriving the average Sy1 template.
The error bars correspond to the standard deviation 
of each averaged point, except for the 8.74 \micron~point (the wavelength chosen for the normalization). In this case, 
we assigned a 15\% error, which is the nominal percentage considered for the N-band flux measurements (see Section \ref{observations}).

We measured the 1.265--18.3 \micron~IR slope
(f$_{\nu}~\alpha~\nu^{-\alpha_{IR}}$) of the Sy1 template, which is representative of 
the individual Sy1 SEDs: $\alpha_{IR} = 1.7\pm0.3$. We also calculated the NIR 
($\alpha_{NIR} = 1.6\pm0.2$, from 1.265 to 8.74 \micron)
and MIR spectral indexes ($\alpha_{MIR} = 2.0\pm0.2$, using the 8.7 and 18.3 \micron~points). A flat NIR slope
indicates an important contribution of hot dust emission (up to $\sim$1000-1200 K;
\citealt{Rieke81,Barvainis87}) that comes from the immediate vicinity of the AGN. 
$\alpha_{IR}$, $\alpha_{NIR}$, and $\alpha_{MIR}$ values for the individual Type-1 Seyfert galaxies and for the mean Sy1 and Sy2 SEDs 
are shown in Table \ref{slopes}.
The shape of the Sy2 mean SED is very steep ($\alpha_{IR} = 3.1\pm0.9$, $\alpha_{NIR}=3.6\pm0.8$, and $\alpha_{MIR}=2.0\pm0.2$) 
compared with those of the Type-1 Seyferts. In general, Sy2 have steeper 1--10 \micron~SEDs than Sy1 
\citep{Rieke78,Edelson87,Ward87,Fadda98,Alonso01,Alonso03}.
On the contrary, the MIR slope results to be the same ($\alpha_{MIR}=2.0\pm0.2$) for both 
the Sy1 and Sy2 templates.

\citet{Alonso03} also reported IR spectral indices measured from 1 to 16 \micron~for Sy1 
and Sy1.5 galaxies ($\alpha_{IR}^{Alonso}$=1.5-1.6).
On the other hand, the NIR slopes of the Sy1.8 and Sy1.9 in \citealt{Ramos09a} have intermediate values between 
those of Sy2 and Sy1 (mean slope $\alpha_{IR} = 2.0\pm0.4$), also in agreement with the IR slopes reported in 
\citet{Alonso03} for Sy1.8 and Sy1.9 ($\alpha_{IR}^{Alonso}$=1.8-2.6).
In summary, the slope of the IR nuclear SED is generally correlated with Seyfert type: Type-2 nuclei show steeper SEDs, 
whereas Type-1 and intermediate Seyferts are flatter. The NIR excess responsible for flattening the SED of the Type-1 nuclei 
would come from the contribution of hot dust in the directly-illuminated faces of the clouds in the torus, as well as 
from the direct AGN emission (i.e., the tail of the optical power-law continuum).

\begin{deluxetable}{lccccc}
\tablewidth{0pt}
\tablecaption{Spectral Shape Information}
\tablehead{
\colhead{Galaxy} & \colhead{$\alpha_{IR}$} & \colhead{$\alpha_{NIR}$} & \colhead{$\alpha_{MIR}$} & \colhead{H/N} & \colhead{N/Q}}
\startdata
Average Sy2   & 3.1$\pm$0.9  & 3.6$\pm$0.8 & 2.0$\pm$0.2 &  0.003		      &  0.23	  \\  
Average Sy1   & 1.7$\pm$0.3  & 1.6$\pm$0.2 & 2.0$\pm$0.2 &  0.071		      &  0.22	  \\
NGC 1097      & 1.4$\pm$0.2  & 1.6$\pm$0.2 & 1.2$\pm$0.1 &  0.117		      &  0.42	  \\
NGC 1566      & 1.8$\pm$0.3  & 1.8$\pm$0.3 & 1.9$\pm$0.2 &  0.040\tablenotemark{a}    &  0.25	  \\
NGC 3227      & 2.0$\pm$0.4  & 2.0$\pm$0.4 & \dots	 &  0.019		      & \dots	  \\
NGC 4151      & 1.4$\pm$0.2  & 1.4$\pm$0.1 & 1.5$\pm$0.1 &  0.076		      &  0.41	  \\
NGC 6221      & 2.0$\pm$0.5  & 1.8$\pm$0.4 & 2.5$\pm$0.3 &  0.044		      &  0.15	  \\
NGC 6814      & 1.4$\pm$0.2  & 1.4$\pm$0.2 & 1.5$\pm$0.1 &  0.098		      &  0.33	  \\
NGC 7469      & 1.8$\pm$0.3  & 1.5$\pm$0.2 & 2.8$\pm$0.4 &  0.086		      &  0.13	  \\
\enddata        
\tablecomments{\footnotesize{Columns 2, 3, and 4 show the fitted spectral indexes (f$_{\nu}~\alpha~\nu^{-\alpha}$) measured from the average Sy2 and Sy1 templates and
for the individual Sy1 SEDs in the whole range 
($\alpha_{IR}$, from $\sim$1 to 18 \micron), in the NIR ($\alpha_{NIR}$, from $\sim$1 to $\sim$9 \micron), and in the MIR ($\alpha_{MIR}$, 
using the N and Q band data points) respectively. Columns 5 and 6 give the values of the H/N and N/Q band ratios.}}
\tablenotetext{a}{\footnotesize{Due to the lack of H-band data for NGC 1566, we used the interpolated value at 1.6 $\micron$ for calculating the
H/N.}}
\label{slopes}
\end{deluxetable}

A similar comparison between Type-1 and Type-2 spectral shapes can be done by using the H/N and N/Q flux ratios.  
H/N is larger for Type-1 Seyferts (0.07$\pm$0.03) than for Sy2 (0.003$\pm$0.002), as measured from the 
individual values of the Sy1 considered here and the Sy2 in \citealt{Ramos09a}. 
The difference in H/N between Sy1 and Sy2 galaxies is significantly different at the 100\% confidence level, 
according to the Kolmogorov-Smirnov (K-S) test. This ratio depends on both the torus inclination and covering factor, 
as we will discuss in Section \ref{nirflux}.
On the other hand, N/Q is very similar for Type-1 and Type-2 Seyferts (mean values of 0.27$\pm$0.11 and 
0.23$\pm$0.14 respectively). Values of H/N and N/Q for the individual Sy1 galaxies and the average templates are reported in Table \ref{slopes}.

\section{SED Modelling}
\label{modelling}
\subsection{Clumpy Dusty Torus Models and Bayesian approach}

\begin{figure}[!ht]
\centering
\includegraphics[width=10cm]{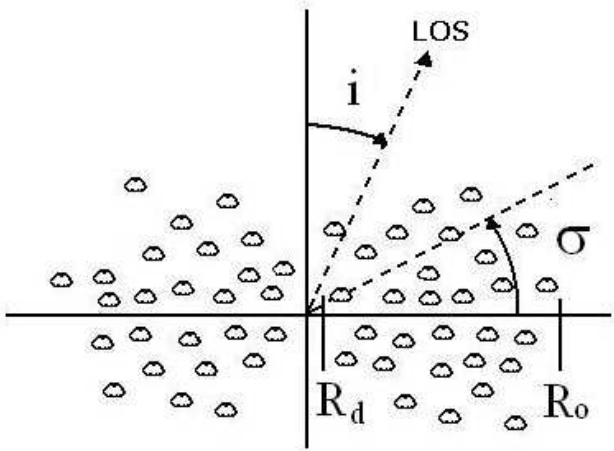}
\caption{\footnotesize{Scheme of the clumpy torus described in \citet{Nenkova08a,Nenkova08b}. The radial extent of the torus is defined by the outer radius
($R_o$) and the dust sublimation radius ($R_d$). All the clouds are supposed to have the same $\tau_{V}$, and $\sigma$ 
characterizes the width of the angular distribution. The number of cloud encounters is function of the viewing angle, $i$.}
\label{clumpy_scheme}}
\end{figure}

The clumpy dusty torus models of \citet{Nenkova02} hold that the dust
surrounding the central engine of an AGN is distributed in clumps, instead of homogeneously filling the torus volume.  
These clumps are distributed with a radial extent $Y = R_{o}/R_{d}$, where 
$R_{o}$ and $R_{d}$ are the outer and inner radius of the toroidal distribution, respectively (see Figure \ref{clumpy_scheme}). 
The inner radius is defined by the dust sublimation temperature ($T_{sub} \approx 1500$ K),
with $R_{d} = 0.4~(1500~K~T_{sub}^{-1})^{2.6}(L / 10^{45}\,\mathrm{erg ~s^{-1}})^{0.5}$ pc.  
Within this geometry, each clump has the same optical depth ($\tau_{V}$, defined at the $V$-band).
The average number of clouds along a radial equatorial ray is $N_0$. The radial density profile is a
power-law ($\propto r^{-q}$). A width parameter, $\sigma$, characterizes the angular distribution of the clouds, which has
a smooth edge.  The number of clouds along the LOS 
at an inclination angle $i$ is $N_{LOS}(i) = N_0~e^{(-(i-90)^2/\sigma^2)}$. 
Finally, the optical extinction produced by the torus along the LOS is computed as 
$A_{V}^{LOS} = 1.086~N_0~\tau_{V}~e^{(-(i-90)^{2}/\sigma^{2})}$ mag.
For a detailed description of the clumpy models see \citet{Nenkova02,Nenkova08a,Nenkova08b}.

The clumpy database now contains 1.2$\times$10$^6$ models, calculated for a fine grid of model
parameters. The inherent degeneracy between these parameters has to be taken into account
when fitting the observables. To this end, we recently developed a Bayesian inference tool (BayesClumpy),
that extracts as much information as possible from the observations. 
Details on the interpolation methods and algorithms employed can be found in \citet{Asensio09}. 
Thus, for the following analysis of the Seyfert SEDs, we are not using the original set of models 
described in \citet{Nenkova08a,Nenkova08b},
but an interpolated version of them (See Figures 3 and 4 in \citealt{Asensio09} 
for a comparison between the original and interpolated models). In this work 
we use the most up-to-date version of the models that are corrected for a
mistake in the torus emission calculations, which in principle only affects 
the AGN scaling factor (see erratum by \citealt{Nenkova10}).
The present version of BayesClumpy has also been
updated to use the Multinest algorithm of \citet{Feroz09} for sampling.
This algorithm is very robust and efficient when sampling from complex posterior distributions.

The prior distributions for the model parameters are assumed to be truncated uniform
distributions in the intervals reported in Table \ref{parametros}. Therefore, we give the same weight
to all the values in each interval. Apart from
the six parameters that characterize the models, there is an additional parameter
that accounts for the vertical displacement required to match the fluxes of a chosen model to an observed 
SED, which we allow to vary freely. This vertical shift scales with
the AGN bolometric luminosity (see Section \ref{discussion}). In order to compare with the observations,
BayesClumpy simulates the effect of the employed filters on the simulated SED by integrating
the product of the synthetic SED and the filter transmission curve. Observational errors
are assumed to be Gaussian or upper/lower limit detections. A detailed description of the
Bayesian inference applied to the clumpy models can
be found in \citet{Asensio09}. Additionally, to see an example of
the use of clumpy model fitting to IR SEDs using BayesClumpy, see \citealt{Ramos09a}.


\begin{deluxetable*}{lcl}
\tabletypesize{\footnotesize}
\tablewidth{0pt}
\tablecaption{Clumpy Model Parameters and Considered Intervals}
\tablehead{
\colhead{Parameter} & \colhead{Abbreviation} & \colhead{Interval}}
\startdata
Width of the angular distribution of clouds            & $\sigma$        & [15\degr, 70\degr]  \\
Radial extent of the torus                             & $Y$             & [5, 30]        \\
Number of clouds along the radial equatorial direction & $N_0$           & [1, 15] \\
Power-law index of the radial density profile          & $q$             & [0, 3]    \\
Inclination angle of the torus                         & $i$             & [0\degr, 90\degr]   \\
Optical depth per single cloud                         & $\tau_{V}$      & [5, 150] \\
\enddata         
\label{parametros}
\end{deluxetable*}

\subsection{Model Results}

\subsubsection{Seyfert 1 Individual Fits}
\label{individual}

The results of the fitting process of the IR SEDs with the interpolated version of the
clumpy models of \citet{Nenkova08a,Nenkova08b} are the posterior distributions for the six free 
parameters that describe the models and the vertical shift. These are indeed the probability
distributions of each parameter, represented as histograms. 
When the observed data introduce sufficient information into the fit, 
the resulting posteriors will clearly differ from the input uniform priors, 
either showing trends or being centered at certain values within the intervals considered.
For all the Sy1 fits, we considered uniform priors in the intervals shown in Table \ref{parametros}.
The only exception is NGC 7469, for which we use a gaussian prior of 85\degr$\pm$2\degr
for the inclination angle of the torus, based on the value of the accretion disk viewing angle 
deduced from X-ray observations \citep{Nandra07}, assuming that the disk and the torus are coplanar.

We fit the individual Sy1 SEDs with BayesClumpy,
modelling the torus emission and the direct AGN contribution (the latter as a broken power law).
We also consider the IR extinction curve of \citet{Chiar06} to take into account any possible foreground extinction 
from the host galaxy.
The AGN scales self-consistently with the torus flux, and the foreground extinction
(separate from the clumpy torus) is another free parameter, which we set as a uniform prior ranging from A$_V$=0 to 10 
mag\footnote{Note that A$_V$ is the foreground extinction from the host galaxy, which is different from 
the A$_V^{LOS}$ value reported in Table \ref{clumpy_parameters}, corresponding to the extinction produced by the torus.
$A_V^{LOS} = 1.086~N_0~\tau_{V}~e^{(-(i-90)^{2}/\sigma^{2})}$ mag.}.

In addition to the Gemini MIR unresolved fluxes reported in Table \ref{psf}  
we consider MIR nuclear fluxes from VISIR compiled from the literature when available (see Table \ref{literature1} in 
Appendix \ref{appendixA}) We find good agreement between the VISIR and T-ReCS unresolved fluxes.  
We did not include measurements in the VISIR PAH filters (8.59, 11.25 and 11.88 \micron) for the galaxies NGC 7469 and NGC 1097
because of their intense star formation and their already well-sampled SED.



\begin{figure*}[!ht]
\centering
{\par
\includegraphics[width=8cm]{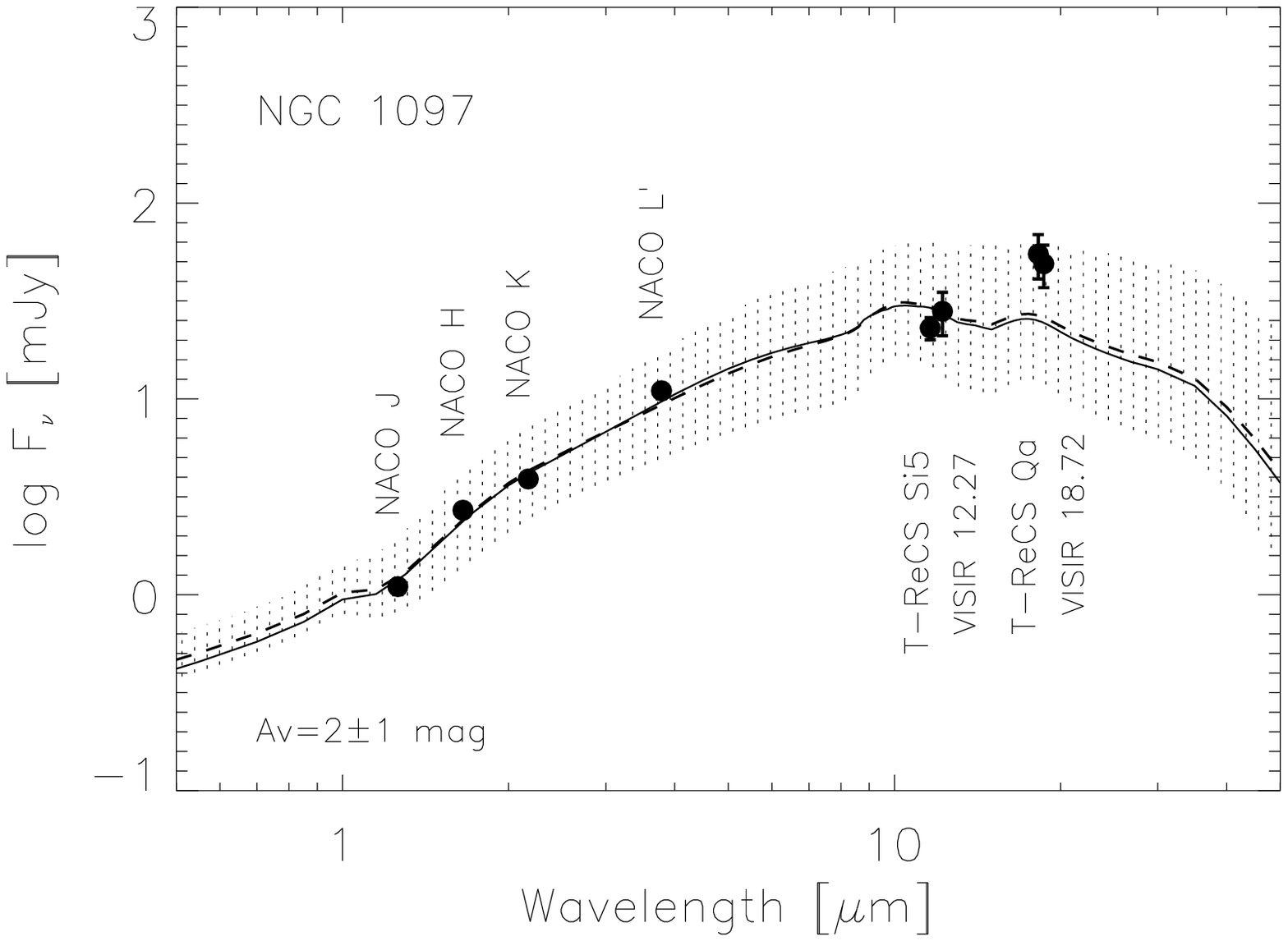}
\includegraphics[width=8cm]{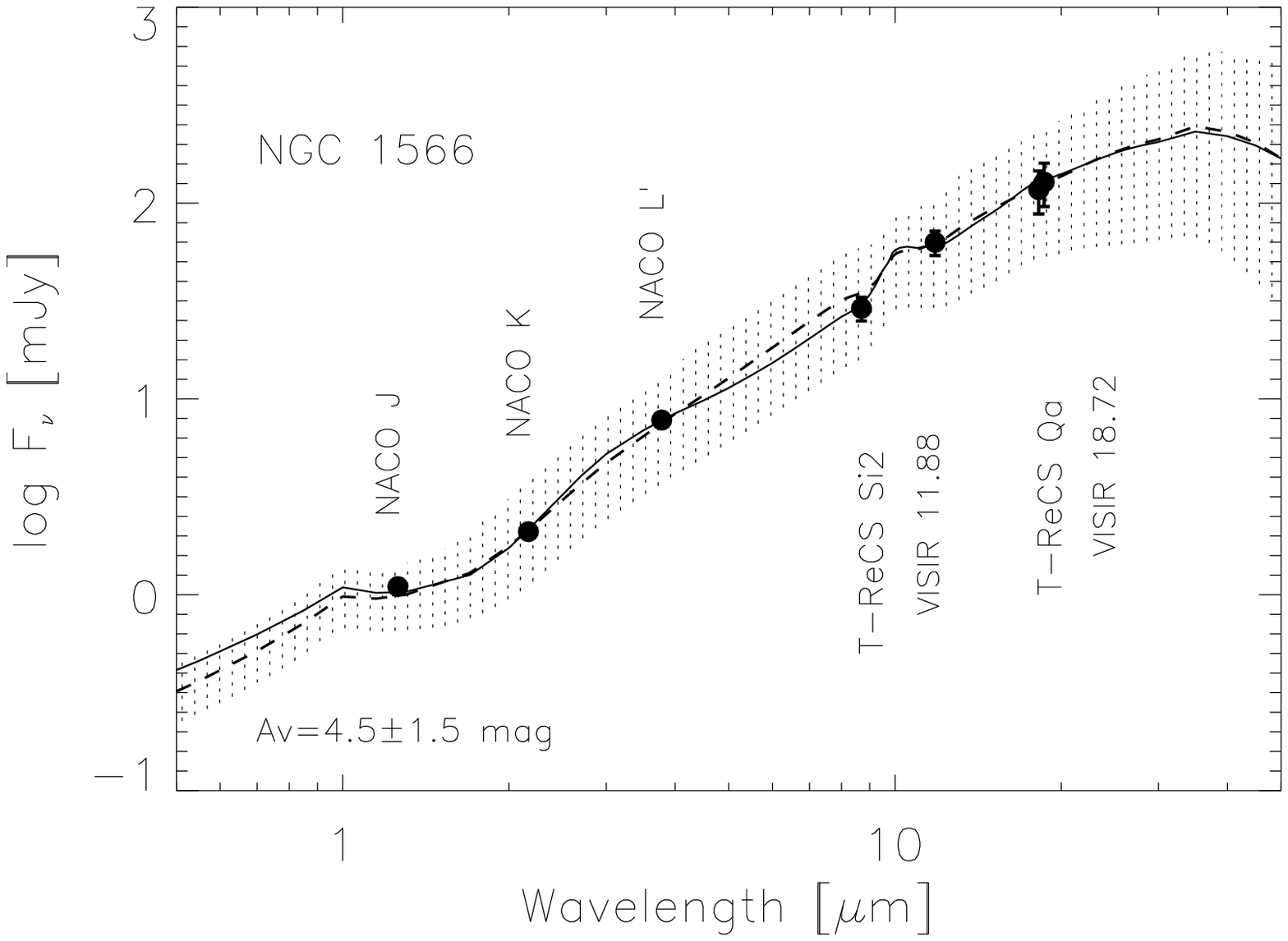}
\includegraphics[width=8cm]{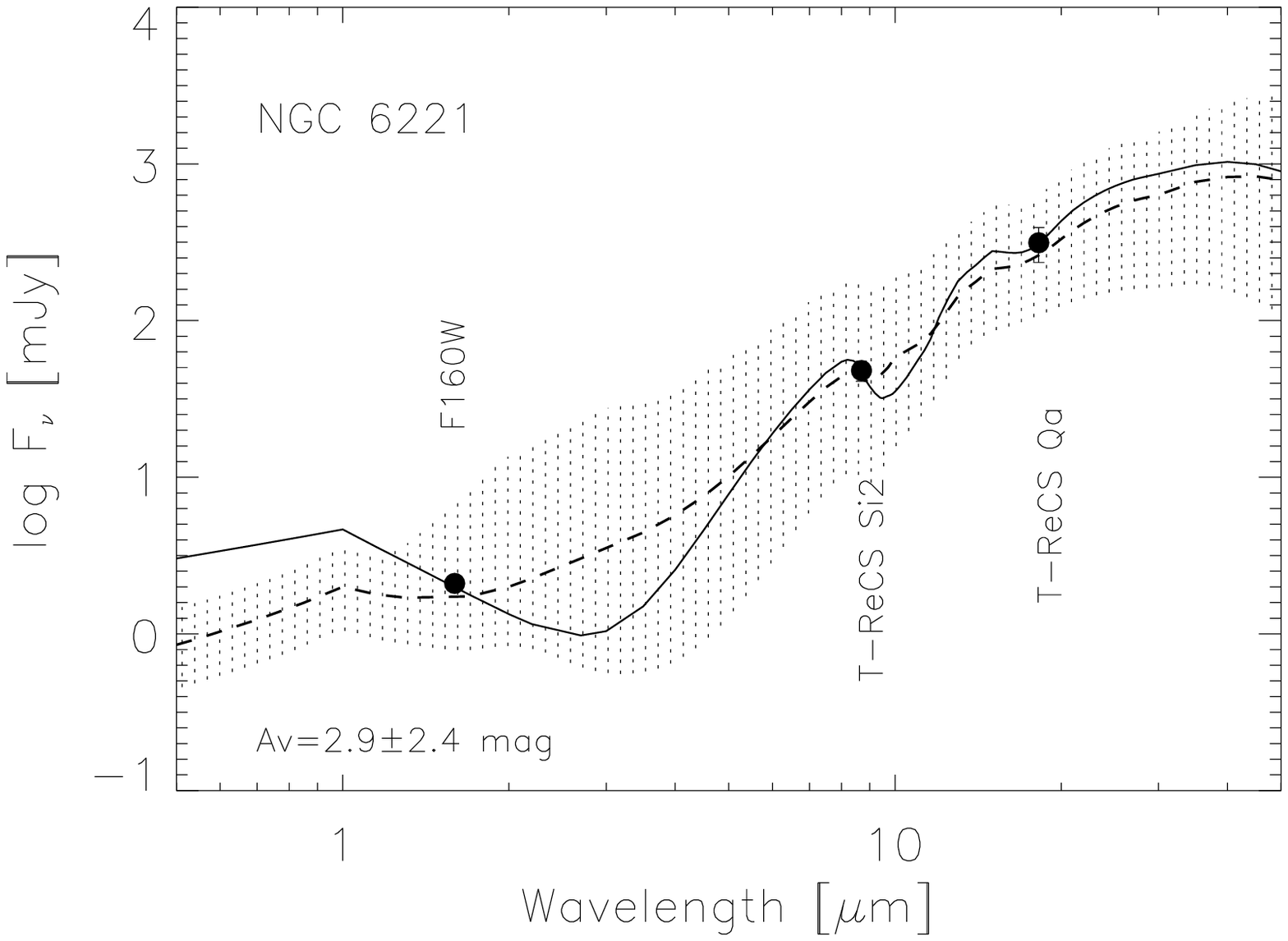}
\includegraphics[width=8cm]{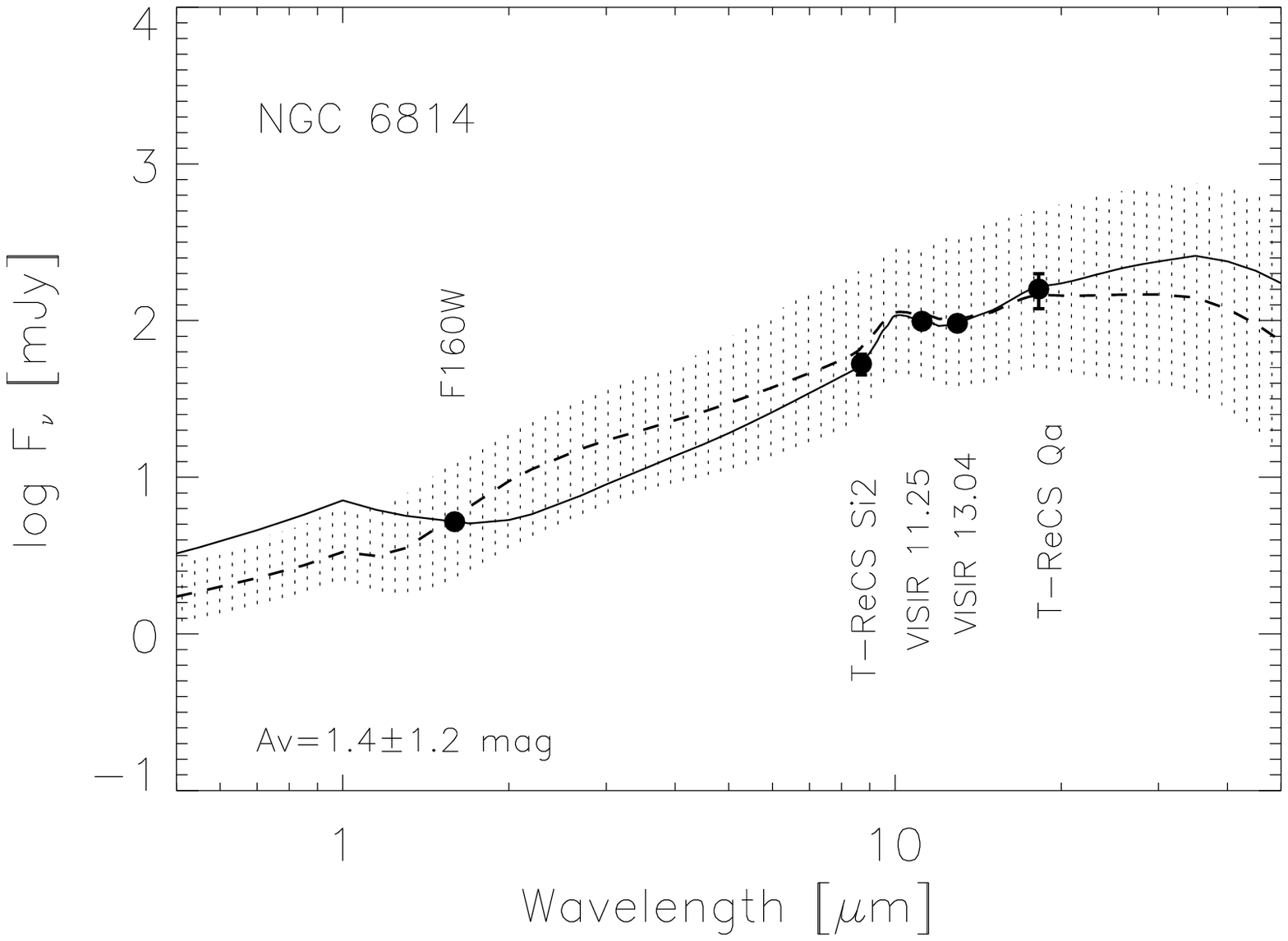}\par}
\caption{\footnotesize{High spatial resolution IR SEDs of the Sy1 galaxies NGC 1097, NGC 1566, NGC 6221, and
NGC 6814. Solid and dashed lines correspond 
to the MAP and median models, respectively. Shaded regions indicate the range of models compatible with the 
68\% confidence interval for each parameter around the median.}
\label{sy1_fits_a}}
\end{figure*}

\begin{figure*}[!ht]
\centering
{\par
\includegraphics[width=8cm]{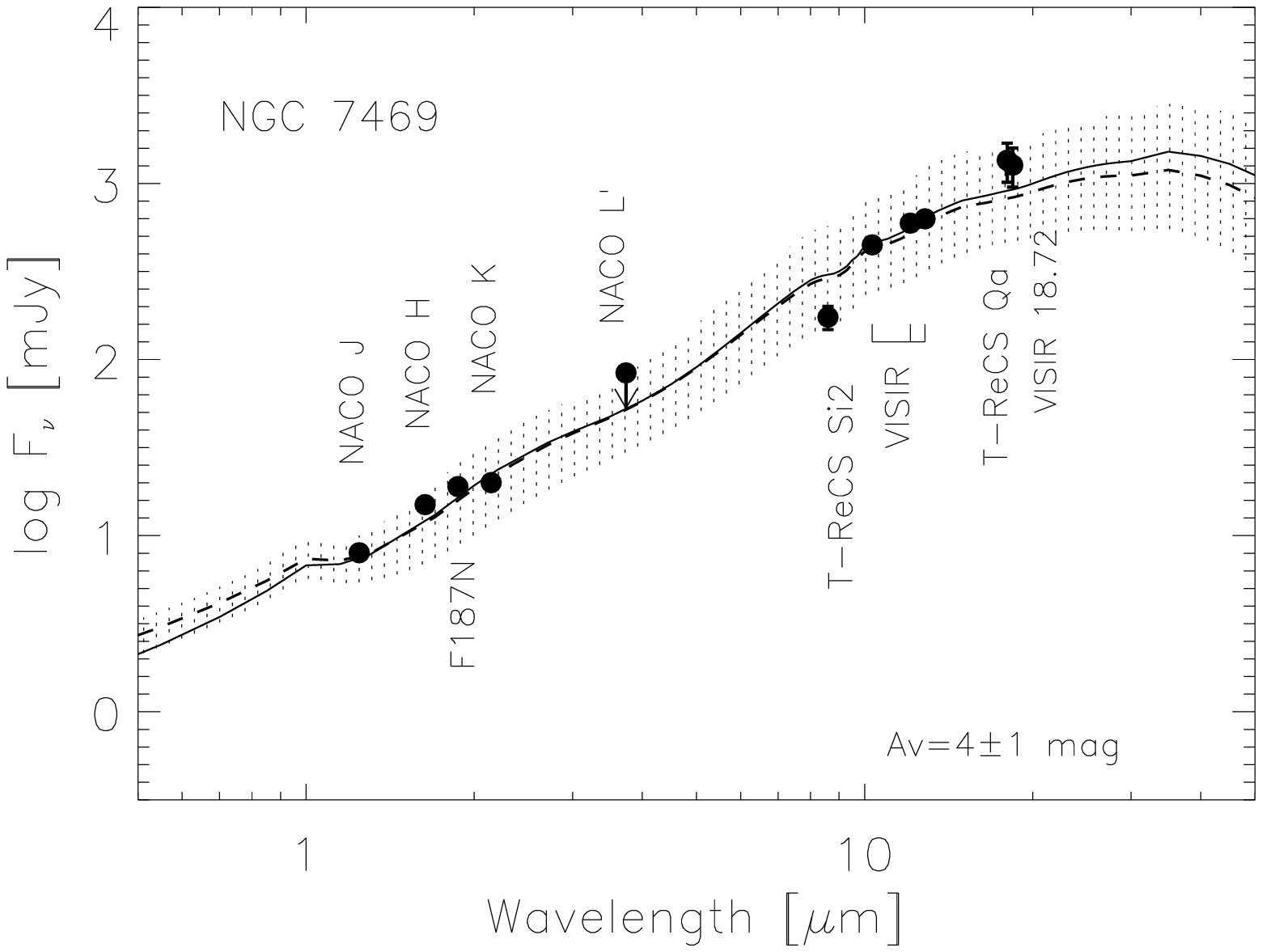}
\includegraphics[width=8cm]{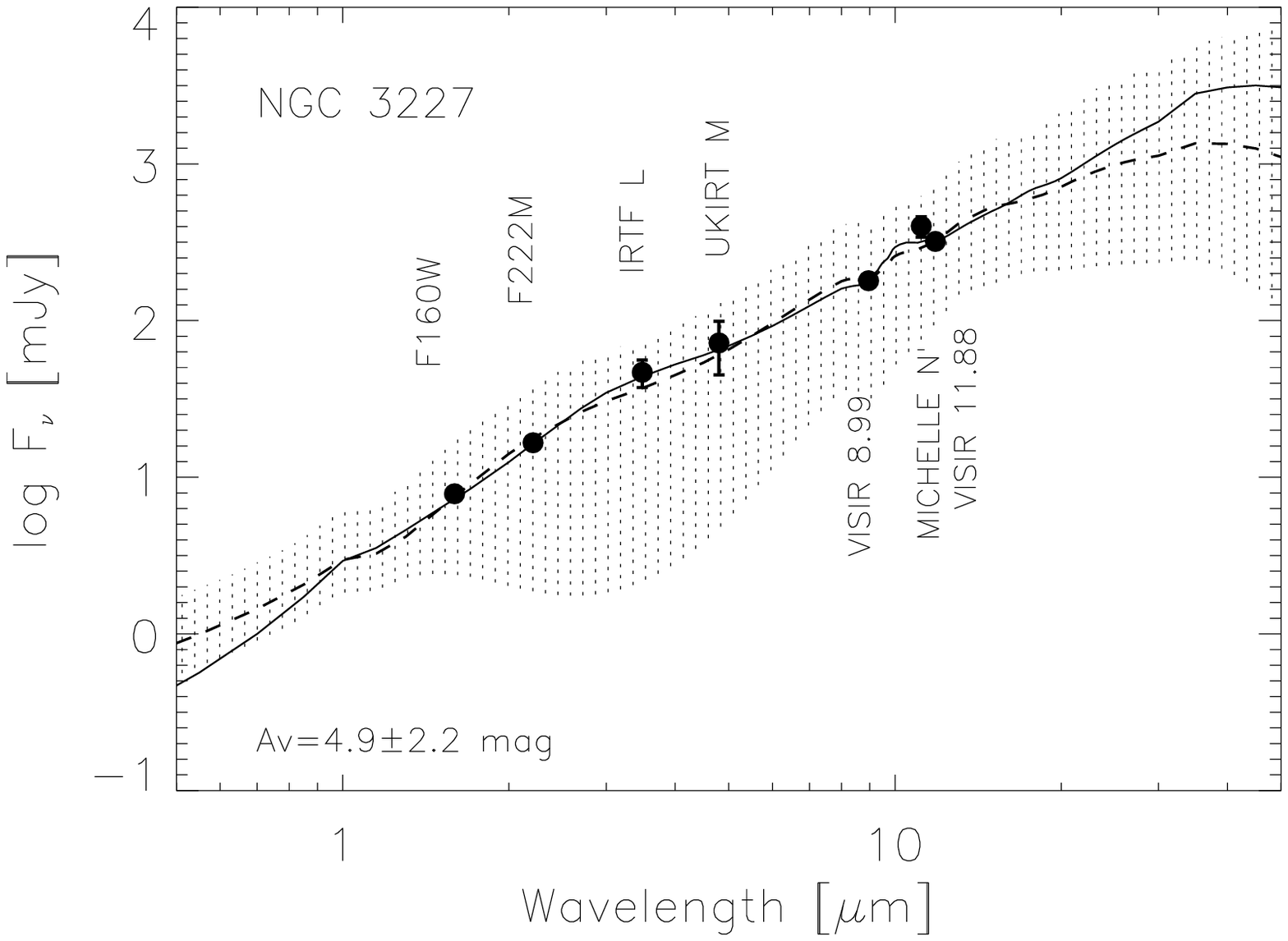}
\includegraphics[width=8cm]{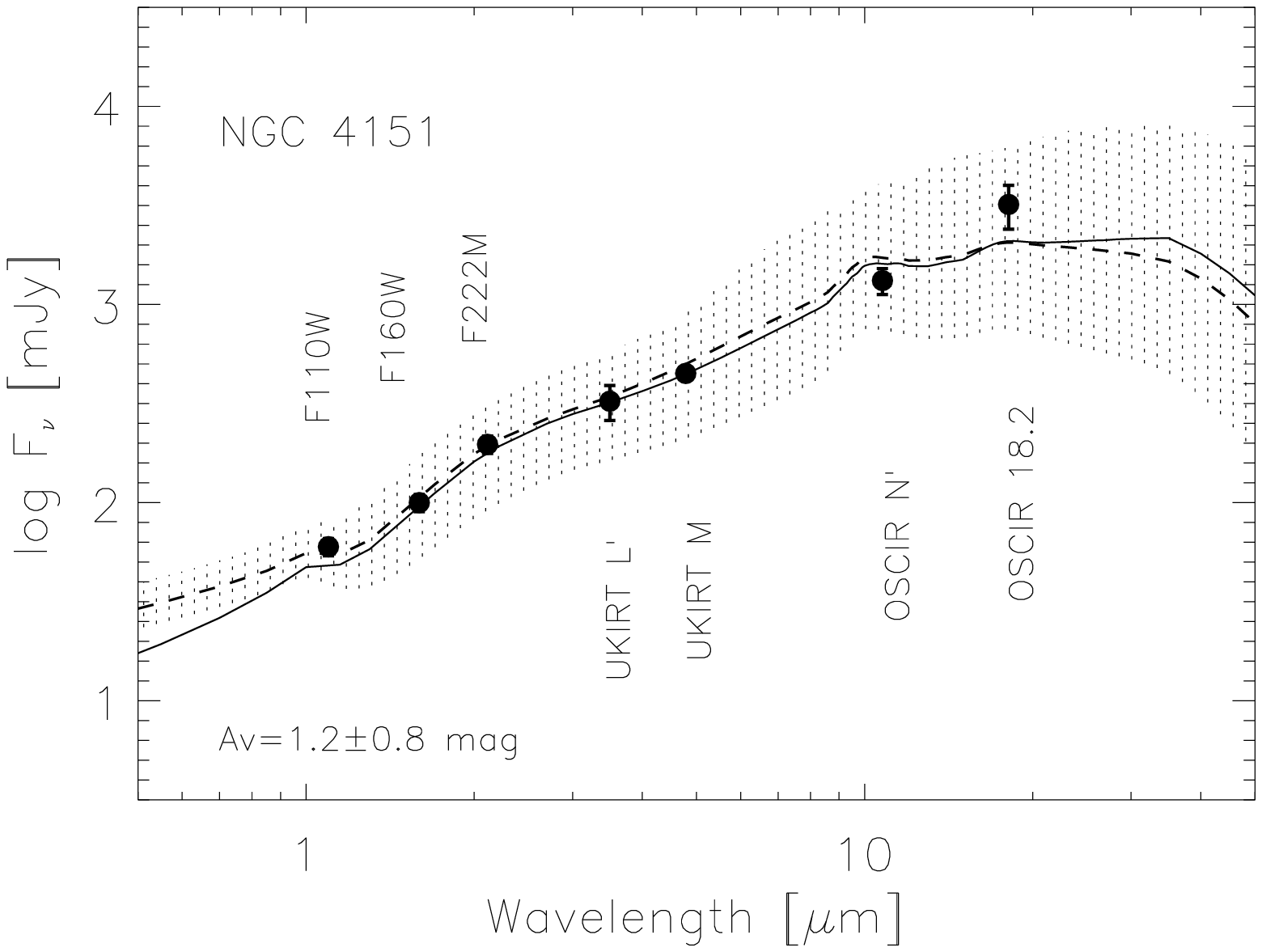}\par}
\caption{\footnotesize{Same as in Figure \ref{sy1_fits_a}, but for NGC 7469, NGC 3227, and NGC 4151.}
\label{sy1_fits_b}}
\end{figure*}

\begin{figure*}[!ht]
\centering
{\par
\includegraphics[width=5.3cm]{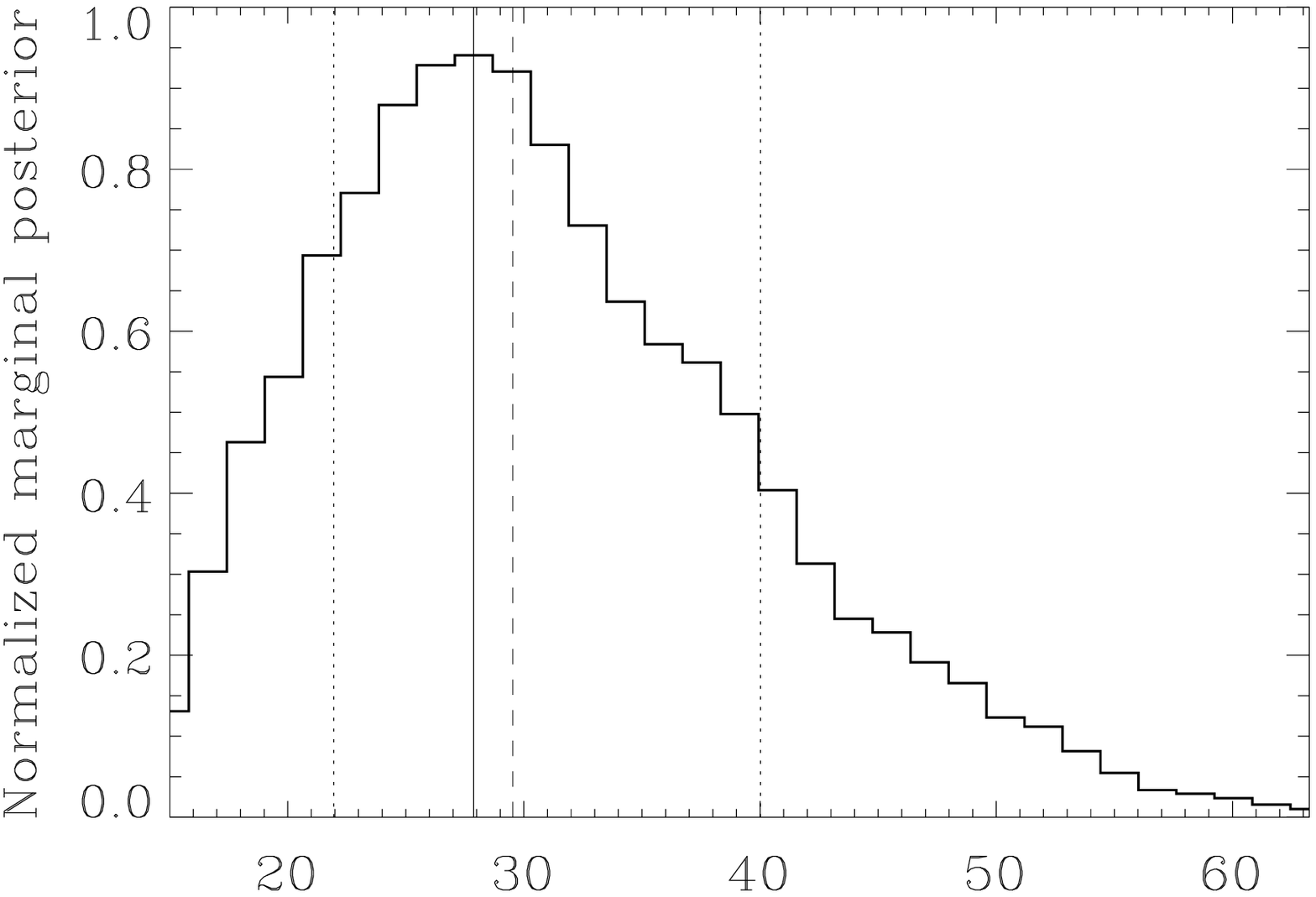}
\includegraphics[width=5.3cm]{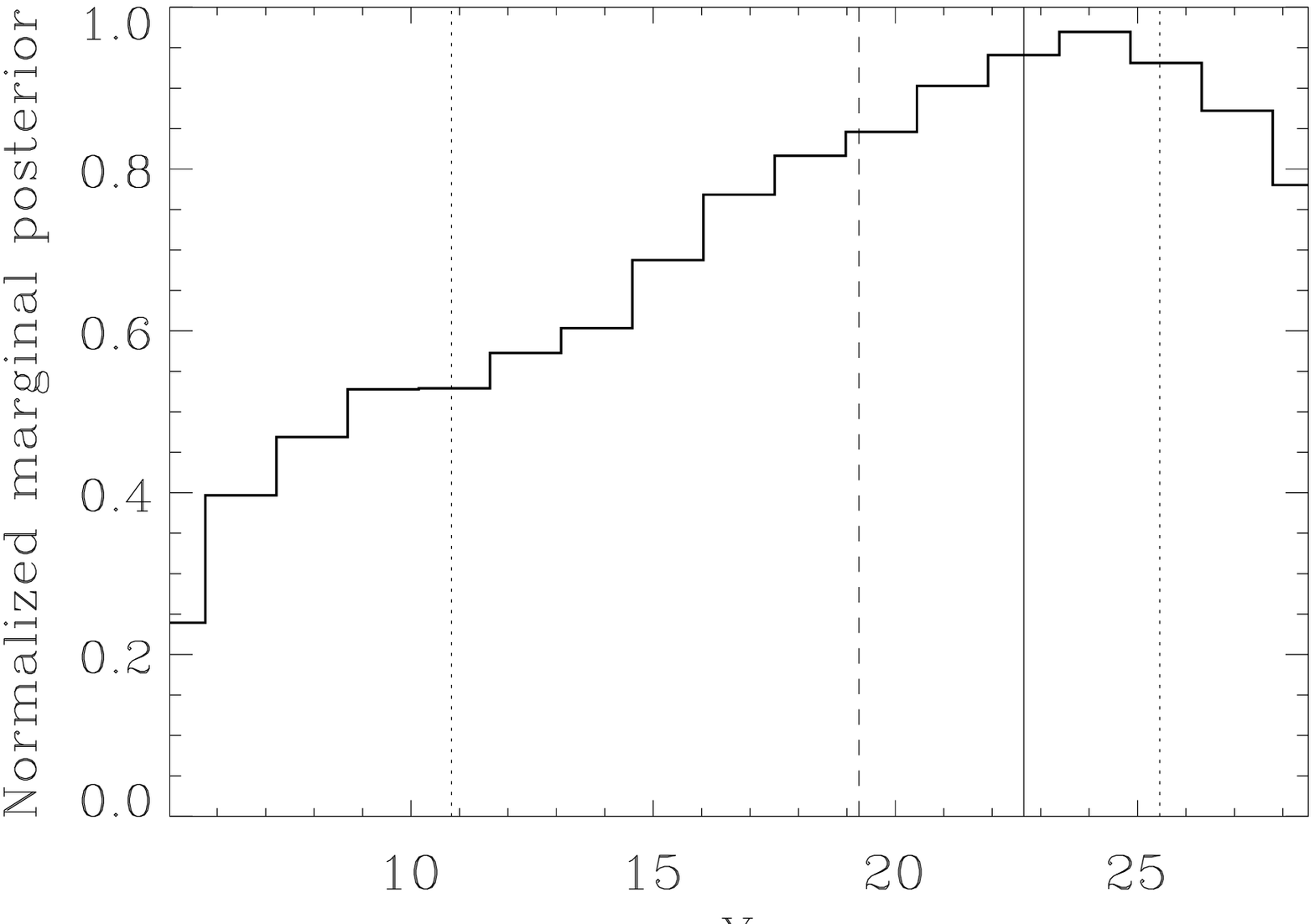}
\includegraphics[width=5.3cm]{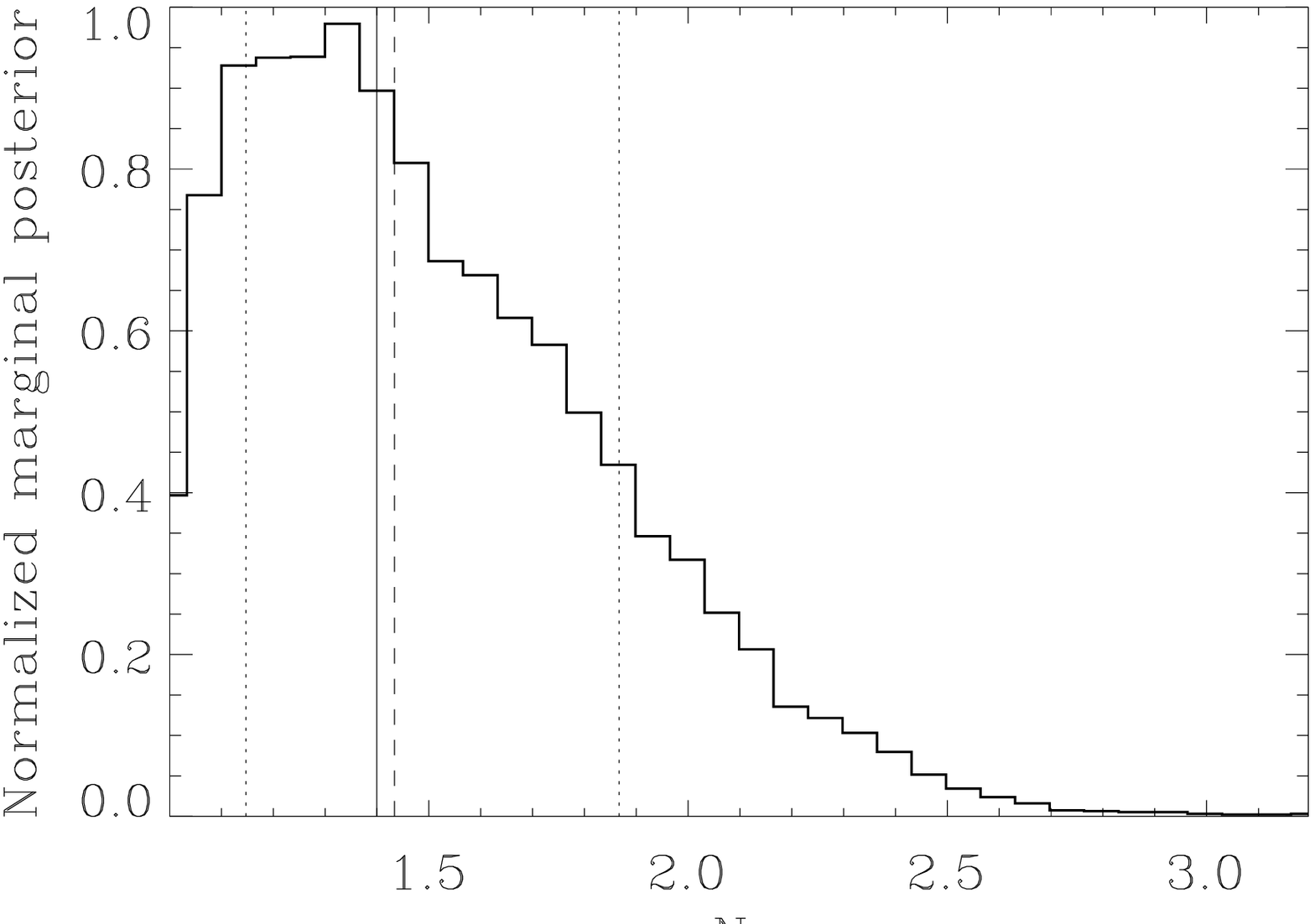}
\includegraphics[width=5.3cm]{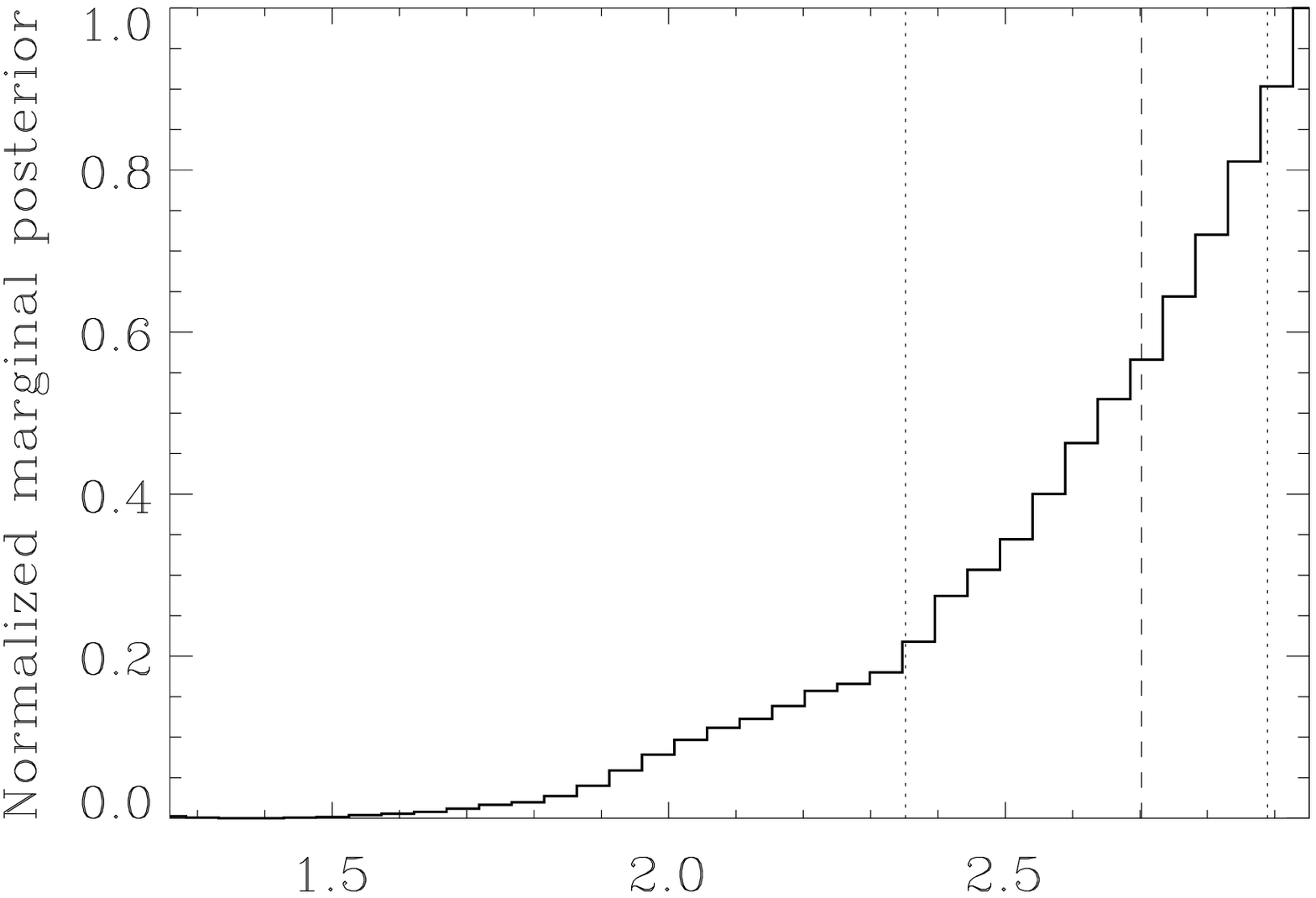}
\includegraphics[width=5.3cm]{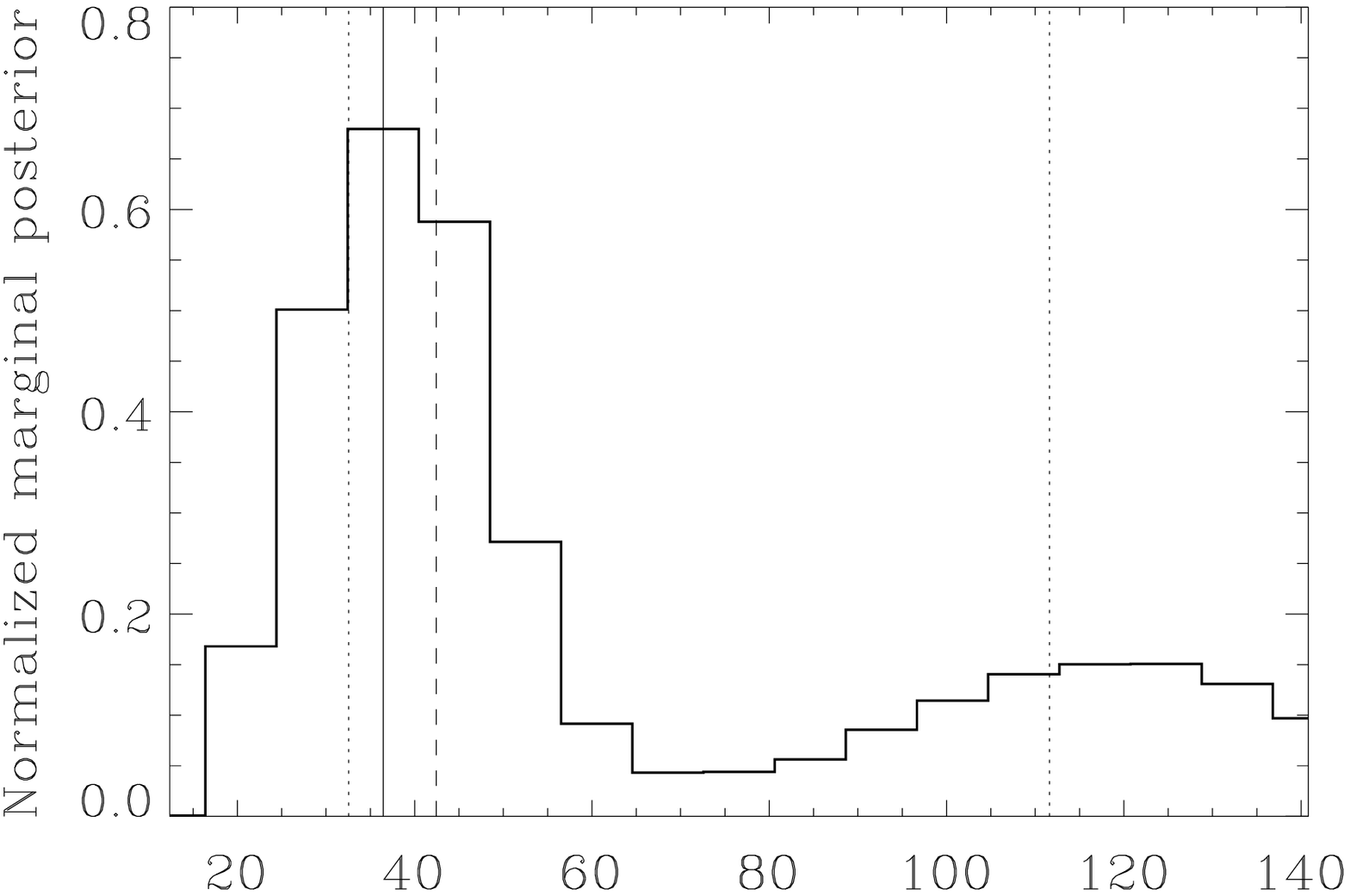}
\includegraphics[width=5.3cm]{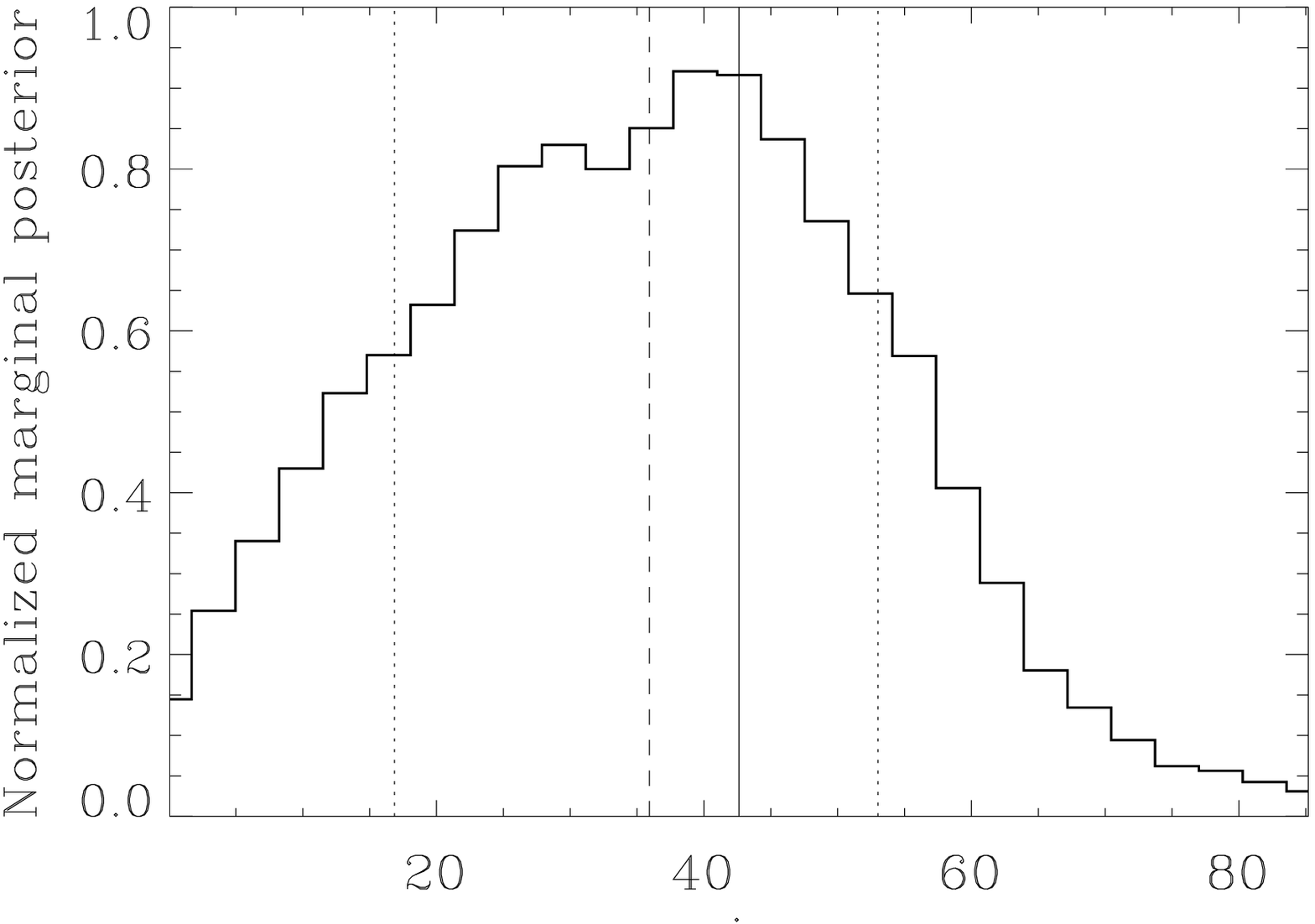}\par}
\caption{\footnotesize{Probability distributions resulting from the fit of NGC 1097.
Solid and dashed lines represent the mode and the median of each distribution and dotted lines indicate 
the 68\% confidence level around the median. The histograms have been smoothed for presentation 
purposes, occasionally leading to small offsets between the solid line corresponding to the mode and the position of the maximum of the histograms.}
\label{ngc1097}}
\end{figure*}

Although the solutions to the Bayesian inference problem are the
probability distributions of each parameter, we can translate the results
into corresponding spectra (Figures \ref{sy1_fits_a} and \ref{sy1_fits_b}).  
The solid lines correspond to the model described by 
the combination of parameters that maximizes their probability distributions (maximum-a-posteriori; MAP). 
Dashed lines represent the model computed with the median value of the probability distribution of 
each parameter. Shaded regions indicate the range of models compatible with the 
68\% confidence interval for each parameter around the median.
In Figure \ref{ngc1097} we show the posteriors of the six torus parameters
(the vertical shift and the foreground extinction have been marginalized) for the 
galaxy NGC 1097. Those for the rest of the Sy1 galaxies are presented in Appendix \ref{appendixA}
(Figures \ref{ngc1566} to \ref{ngc4151}).

The more information the IR SEDs provide, the better the probability distributions are constrained. 
From the analysis performed here and in \citealt{Ramos09a} it appears important to sample the SED
in the wavelength range around 3-4 \micron~and also at $\sim$18 \micron~to constrain the model parameters. 
A detailed study of the influence of different filters/wavelengths in the restriction of the clumpy models parameter space 
will be the subject of a forthcoming paper (Asensio Ramos et al., in prep.).  
The posterior information for the seven Sy1 galaxies is summarized 
in Table \ref{clumpy_parameters}.

\begin{deluxetable*}{lcccccccccccccc}
\tabletypesize{\footnotesize}
\tablewidth{0pt}
\tablecaption{Parameters Derived from the Clumpy Model Fitting}
\tablehead{
\colhead{Galaxy} &  \multicolumn{2}{c}{$\sigma$ (\degr)} &  \multicolumn{2}{c}{$Y$}& \multicolumn{2}{c}{$N_0$} & \multicolumn{2}{c}{$q$} & \multicolumn{2}{c}{$i$ (\degr)} &
\multicolumn{2}{c}{$\tau_{V}$} & \multicolumn{2}{c}{$A_{V}^{LOS}$ (mag)}  \\
 & \colhead{Median} & \colhead{Mode} & \colhead{Median} & \colhead{Mode} &  \colhead{Median} & \colhead{Mode} & \colhead{Median} & \colhead{Mode} & \colhead{Median} & \colhead{Mode} & \colhead{Median} & \colhead{Mode} & \colhead{Median} & \colhead{Mode}}
\startdata
NGC 1097   & 29$\pm^{10}_{8}$ & 28 & 19$\pm^{6}_{8}$ & 23 & $<2$            & 1 & 2.7$\pm^{0.2}_{0.3}$ & 2.9 & 36$\pm^{17}_{19}$ & 43& 42$\pm^{69}_{10}$ & 36 & $<$15 &  1  \\
NGC 1566   & 46$\pm^{12}_{15}$& 52 & 21$\pm$4        & 20 & 2$\pm$1         & 2 & 0.2$\pm$0.2          & 0.1 & 79$\pm^{7}_{15}$  & 86&104$\pm^{28}_{55}$ & 141& 165$\pm^{95}_{100}$ & 160  \\
NGC 6221   & 42$\pm^{15}_{12}$& 36 & 21$\pm^{5}_{6}$ & 24 & 8$\pm^{4}_{3}$  & 7 & 0.8$\pm^{0.7}_{0.5}$ & 0.4 & 59$\pm^{18}_{33}$ & 79& 110$\pm^{23}_{32}$& 128& $<$690&  1  \\  
NGC 6814   & 23$\pm^{13}_{5}$ & 20 & 18$\pm$7        & 23 & 2$\pm$1         & 2 & 1.1$\pm^{0.6}_{0.5}$ & 1.1 & 43$\pm^{27}_{25}$ & 39& 113$\pm^{22}_{30}$& 139& $<$85 &  1  \\  
NGC 7469   & 41$\pm^{17}_{6}$ & 38 & 22$\pm^{5}_{6}$ & 27 & 3$\pm$1         & 3 & 1.9$\pm$0.4          & 1.9 & 84$\pm$2 (fix)    & 84&141$\pm^{6}_{15}$  & 148 & 500$\pm$120 & 430 \\
\hline
NGC 3227   & 36$\pm^{18}_{12}$& 33 & 19$\pm$6	     & 19 & 5$\pm^{4}_{3}$  & 2 & 0.6$\pm^{0.5}_{0.4}$ & 0.5 & 49$\pm^{21}_{26}$ &66 & 118$\pm^{18}_{27}$& 129& $<$240& 1 \\
NGC 4151   & 24$\pm^{17}_{6}$ & 19 & 16$\pm^{8}_{7}$ & 8  & 2$\pm^{2}_{1}$  & 2 & 1.8$\pm$0.6	       & 1.8 & 43$\pm^{18}_{26}$ &57 & 110$\pm^{23}_{26}$& 123& $<$60 & 1 \\
\enddata          
\tablecomments{\footnotesize{Medians and modes of the Sy1 probability distributions. Those presenting single tails 
have been characterized with the mode and upper/lower limit at 68\% confidence.
The models include the intrinsic AGN continuum emission.}}
\label{clumpy_parameters}
\end{deluxetable*}

From the individual fits of the Sy1 galaxies in our sample with clumpy torus models  
we  obtain the following results:

\begin{enumerate}

\item The average number of clouds along an equatorial ray is within the interval $N_0$=[1, 8], 

\item Low values of $\sigma$  are preferred: $\sigma~\simeq$  [25\degr, 45\degr], and intermediate
inclination angles of the torus are found: $i~\simeq$ [35\degr, 85\degr].


\item The radial extent of the torus ($Y$=$R_o$/$R_d$) is weakly constrained within the interval $Y~\simeq$ [15,20]. 

\item Values in the range from $\tau_{V}~\simeq$ [40,140] are found for the optical depth of each cloud for all the galaxies.

\item The radial density profile appears constrained within the interval $q$=[0.2,1.9], with the only exception of
NGC 1097, for which $q$=2.7$\pm^{0.2}_{0.3}$.

\item The 10 \micron~silicate feature appears in shallow emission or absent in the
fitted models with the exception of NGC 6221 MAP model. 
The weak silicate feature arises in the clumpy
models because both illuminated and dark cloud sides
contribute to the observed spectrum (see Section 5.4 in \citealt{Ramos09a} for a detailed discussion on the 
silicate feature modelling).

\item The optical extinction produced by the torus along the LOS results in $A_{V}^{LOS}<690$ mag for all the galaxies.

\item The foreground extinction from the host galaxy, 
which obscures the AGN direct emission in our modelling, results in A$_V$$<$5 mag (see Figures 
\ref{sy1_fits_a} and \ref{sy1_fits_b}). 
The values derived are consistent with those published in the literature, e.g. the nuclear optical extinction 
of A$_V$=3 mag measured from the optical spectrum of NGC 6221 \citep{Levenson01}, A$_V\sim$1 mag determined from NACO/VLT colour 
maps \citep{Prieto05} for NGC 1097, and A$_V$=4.5-4.9 mag reported by \citet{Mundell95} for NGC 3227.

All the above intervals or limits of the parameters correspond to  
median values. We chose the medians instead of the MAPs  because
the former gives a less biased information about the result, 
since it takes into account degeneracies, while the MAP does not.

\end{enumerate}

The clumpy models succesfully reproduce the observed Sy1 SEDs studied here with compatible results
among them. This is indicating that
the NIR and MIR unresolved fluxes employed here are dominated by a combination of 
reprocessed emission from dust in the torus and direct AGN emission.

Our modelling results for NGC 1097 somehow contradict those presented in \citet{Mason07}. The latter authors unsuccessfully 
tried to reproduce the T-ReCS 11.66 and 18.3 \micron~aperture fluxes that they measured for this galaxy using the clumpy models of \citet{Nenkova02}.
The difference with our result is probably due to i) the fact that we obtained unresolved fluxes using PSF subtraction 
over the same images presented in \citet{Mason07}, reducing the potential contamination from star formation; ii)
they did not consider NIR data, but only the MIR fluxes; and finally iii) they faced the degeneracy problem of the clumpy 
models without using any sophisticated tool as e.g. BayesClumpy.


\subsubsection{Sy2 and intermediate-type Seyfert results}
\label{comparison}

As a consequence of the publication of the erratum \citet{Nenkova10}, where the authors report a mistake in the torus emission calculations, 
we repeated all the fits presented in \citealt{Ramos09a} using our updated version of BayesClumpy. 
In order to do a proper comparison with the results for the Sy1 galaxies presented here, we performed the fits 
of the Sy2 and intermediate-type Seyferts considering exactly the same priors as for the Sy1. 
As in \citealt{Ramos09a}, we did not consider the direct AGN contribution for either Sy1.8/1.9 or Sy2.

For the new fits we considered MIR nuclear fluxes from VISIR
when available, in addition to the NIR and MIR fluxes reported in \citealt{Ramos09a}.
We did not include measurements in the VISIR PAH filters (8.59, 11.25 and 11.88 \micron) for those galaxies 
with very intense star formation such as NGC 7582. 
Many of the SEDs in \citealt{Ramos09a} have been also updated with NIR data from recent publications. 
In Table \ref{literature2} (Appendix \ref{appendixB}) we report
the NIR-to-MIR SEDs for all the sample.

In general, the results from the fitting with the most up-to-date version of the interpolated clumpy models, 
which are reported in Table \ref{clumpy_parameters2} 
in Appendix \ref{appendixB}, are compatible with those presented in \citealt{Ramos09a} at the 1-sigma level. 
Indeed, if we compare the results for the five galaxies for which we fitted exactly the same 
SEDs as in the previous work (Circinus, Mrk 573, NGC 1386, NGC 1808, and NGC 1365) we find that they are practically identical. 
The only fits that are completely different from those presented in the previous work correspond to Centaurus A and NGC 3281.
This is a consequence of adding new MIR data from VISIR and/or apriori information for the inclination angle of the torus
(see Appendix \ref{appendixB}).

In Table \ref{types} we show the ranges of the parameters found for the Sy1, intermediate-type Seyferts, and Sy2. 
We have excluded the unreliable fits of NGC 1808 and NGC 7582 as 
we did in \citealt{Ramos09a}. In the case of NGC 1808, due to the intense star formation that is taking place in its 
nuclear region and to the lower spatial resolution IR SED (all from 3-4 m telescopes), it is likely
contaminated with starlight, resulting in its peculiar shape. For NGC 7582, the intense circumnuclear star
formation and the edge-on orientation of the galaxy make it difficult to isolate
the torus emission from that of the host galaxy. In the fit, the silicate feature is predicted in emission, 
while from MIR spectroscopy shows it in strong absorption \citep{Siebenmorgen04}.


\begin{deluxetable*}{lccccccc}
\tablewidth{0pt}
\tablecaption{Ranges of Parameters for Sy1, intermediate-type Seyferts, and Sy2}
\tablehead{
\colhead{Type} & Galaxies & \multicolumn{1}{c}{$\sigma$ (\degr)} &  \multicolumn{1}{c}{$Y$}& \multicolumn{1}{c}{$N_0$} & \multicolumn{1}{c}{$q$} & \multicolumn{1}{c}{$i$ (\degr)} &
\multicolumn{1}{c}{$\tau_{V}$}}
\startdata
Sy1 \& Sy1.5         & 7 & [25, 45] & [15, 20] & [1, 8]  & [0.2, 1.9]\tablenotemark{a}  & [35, 85] & [40, 140] \\
Sy1.8 \& Sy1.9       & 3 & [35, 70] & [20, 25] & [1, 7]  & [0.9, 3.0]  & [25, 85] & [40, 85]  \\
Sy2                  & 9 & [20, 65] & [10, 20] & [6, 14] & [0.0, 3.0]  & [45, 85]  & [5, 95]   \\  
\enddata          
\tablecomments{General ranges of parameters resulting from the fits with the clumpy models for the seven Type-1 Seyferts, the 
three Sy1.8 and Sy1.9 nuclei, and for the nine Sy2 with reliable fits (the extremes of the intervals have been rounded to 
simplify the comparison).
All the above intervals correspond to median values.}
\tablenotetext{a}{With the exception of NGC 1097, for which $q$=2.7$\pm^{0.2}_{0.3}$.}
\label{types}
\end{deluxetable*}


We considered the Sy1.8 and Sy1.9 types as a separate group in between the Sy1 \& Sy1.5 and Sy2 because the IR slopes 
measured from their SEDs are intermediate between those of Sy1 and Sy2, 
as measured for our sample and also as reported in the literature (Section \ref{averageSy1} and references therein). 
By looking at the ranges of parameters for the Sy1.8 and Sy1.9 types, we find similarities with the Sy1 \& Sy1.5
group in $Y$, $N_0$, and $\tau_V$, whereas $\sigma$ and $q$ are  more similar to those of Sy2.

\section{Comparison between Type-1 and Type-2 Seyfert nuclei.}
\label{discussion}

The main aim of this work is to enlarge the number of Type-1 Seyferts in the original sample of \citealt{Ramos09a} in order
to better compare between Type-1 and Type-2 tori under the assumption that the SEDs studied here are torus/AGN dominated. Despite the 
relatively low number of objects considered (7 Type-1
and 9 Type-2 Seyferts\footnote{Here we consider the nine Sy2 in \citealt{Ramos09a} with reliable fits (NGC 1808 and NGC 7582 are excluded).}), 
we find that some of the parameters are significantly different between Sy1 and Sy2.

To take full advantage of the Bayesian approach employed here for the individual fits, the best way to compare the
results for Sy1 and Sy2 galaxies is to derive joint posterior distributions for the full Type-1 and Type-2 datasets respectively.
If $D_i$ contains the observed data from the i-th SED, assuming that the different SEDs are statistically
independent, we can use the Bayes theorem to write the posterior for all galaxies together as:
\begin{equation}
p(\mbox{\boldmath $\theta$}|\{D_i\}) \propto p(\{D_i\}|\mbox{\boldmath $\theta$}) p(\mbox{\boldmath $\theta$}) = 
\prod_{i=1}^N p(D_i|\mbox{\boldmath $\theta$}) p(\mbox{\boldmath $\theta$}),
\end{equation}
where $\mbox{\boldmath $\theta$}=(\sigma,Y,N_0,q,\tau_V,i)$. 

\begin{figure*}[!ht]
\centering
\par{
\includegraphics[width=8.1cm]{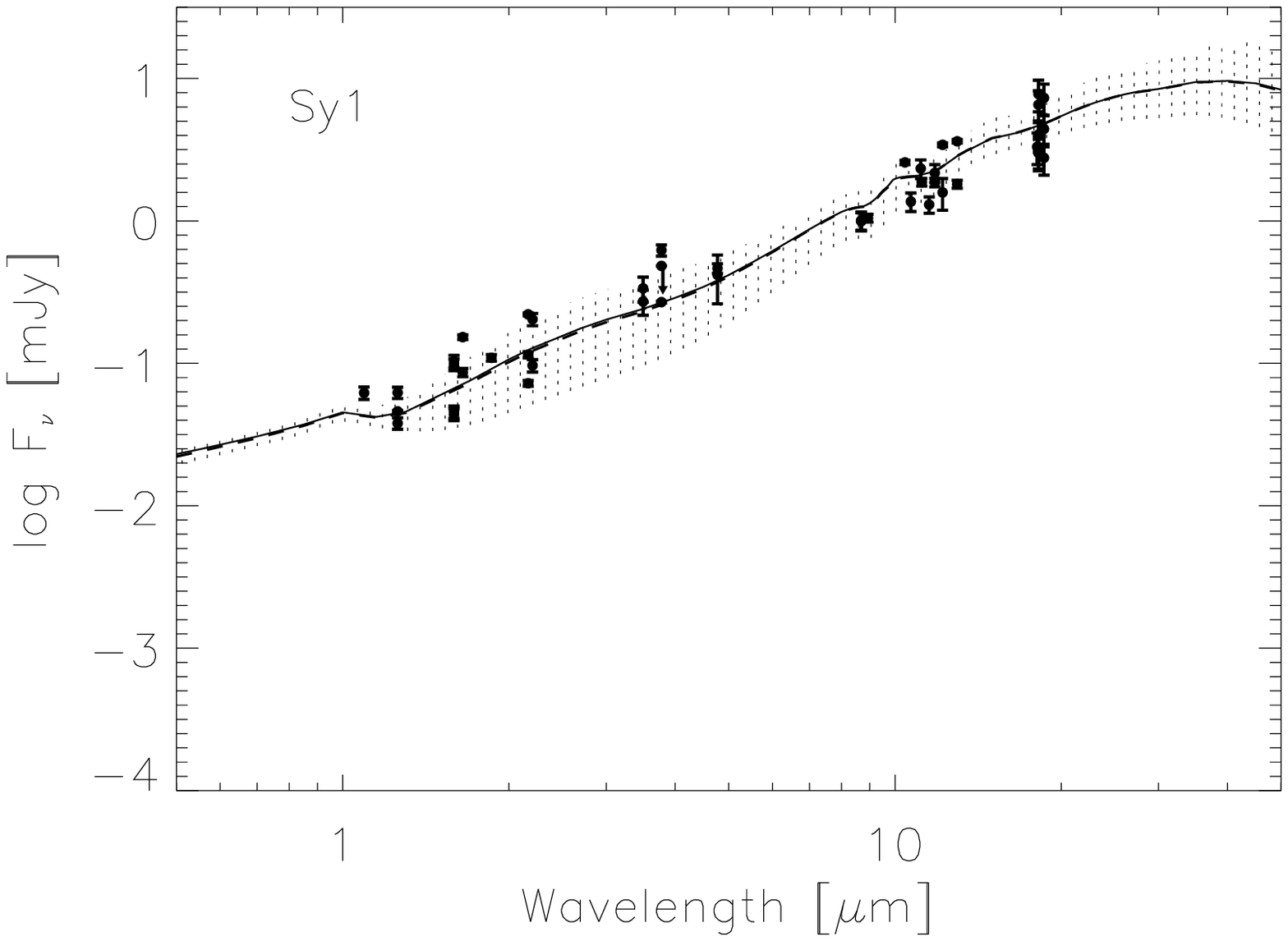}
\includegraphics[width=8.1cm]{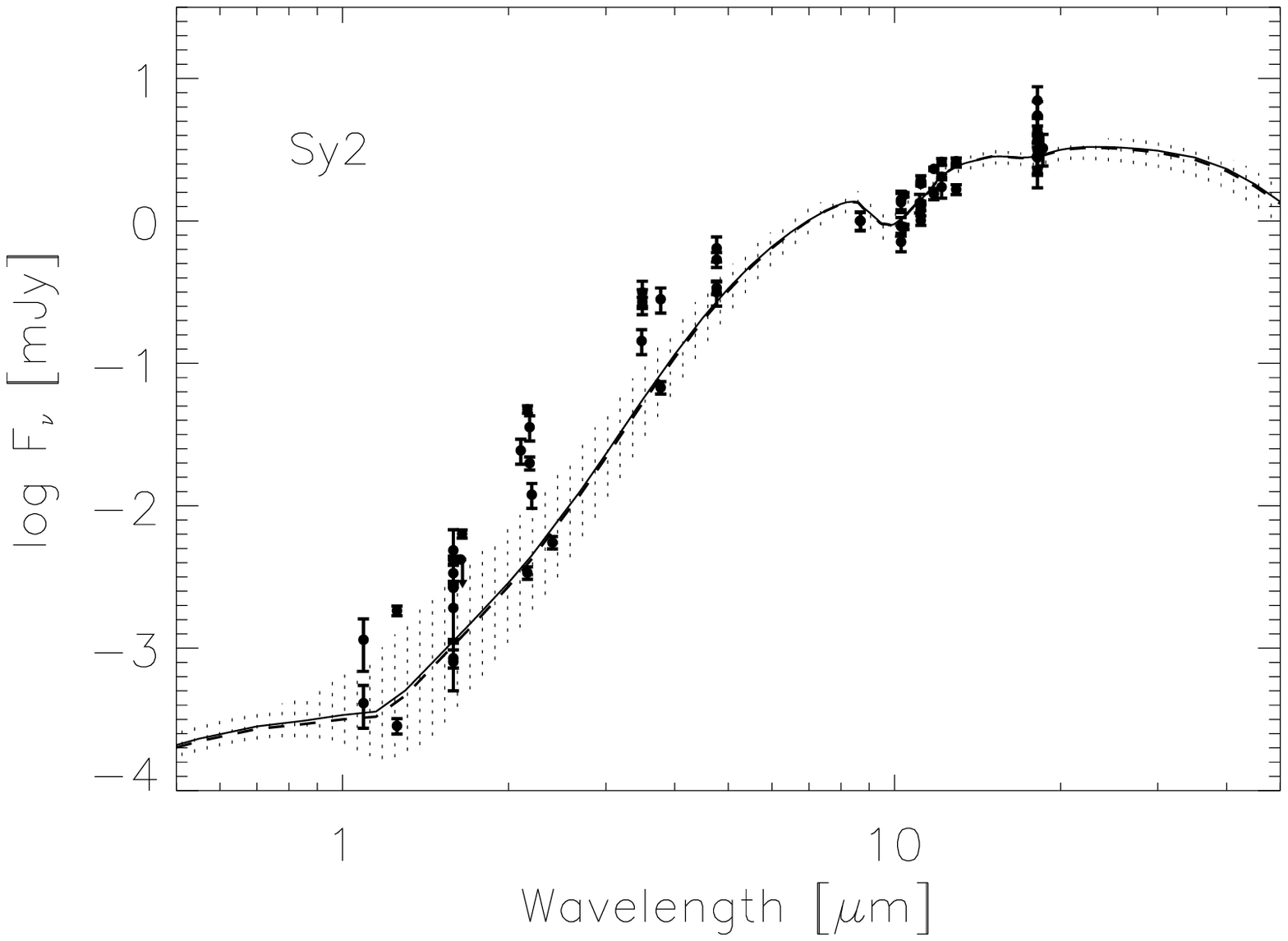}\par}
\caption{\footnotesize{Same as in Figure \ref{sy1_fits_a}, but for the Sy1 (left) and Sy2 SEDs (right) normalized at 8.74 \micron.}
\label{comparison1and2}}
\end{figure*}

Thus, we normalized all the Sy1 SEDs at 8.74 \micron~and fitted them together using BayesClumpy, and we did the same for the Sy2. 
For those galaxies without flux measurements in the Si-2 filter we performed a quadratic interpolation of the SED and used the interpolated
values at 8.74 \micron~to normalize the real data. We did not use the interpolated values in the fits. 
We considered the mean redshift for the Sy1 (z=0.0061$\pm$0.0045) and for the Sy2 (0.0078$\pm$0.0051) in the fits\footnote{Indeed, 
since all the galaxies are local Seyferts, the results are the same if we consider that the SEDs are
rest-frame}.    
In Figure \ref{comparison1and2} we show the Sy1 (left panel) and Sy2 fits (right panel). 
Note that the MAP and median models predict a flat SED with
the silicate feature in weak emission for the Sy1 galaxies, and steeper and with the silicate band in shallow absorption for Sy2. 

\begin{figure*}[!ht]
\centering
{\par
\includegraphics[width=8cm]{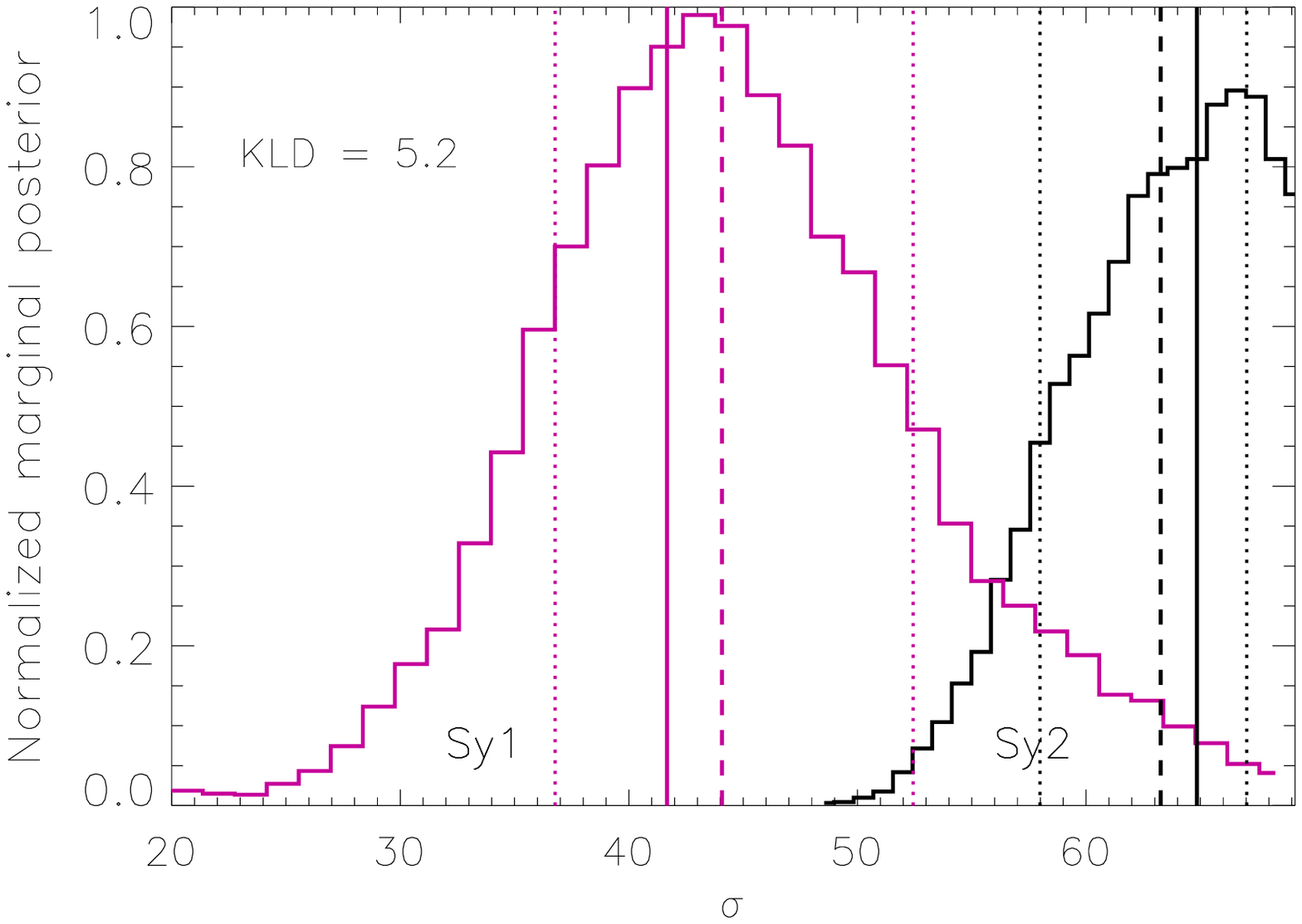}
\includegraphics[width=8cm]{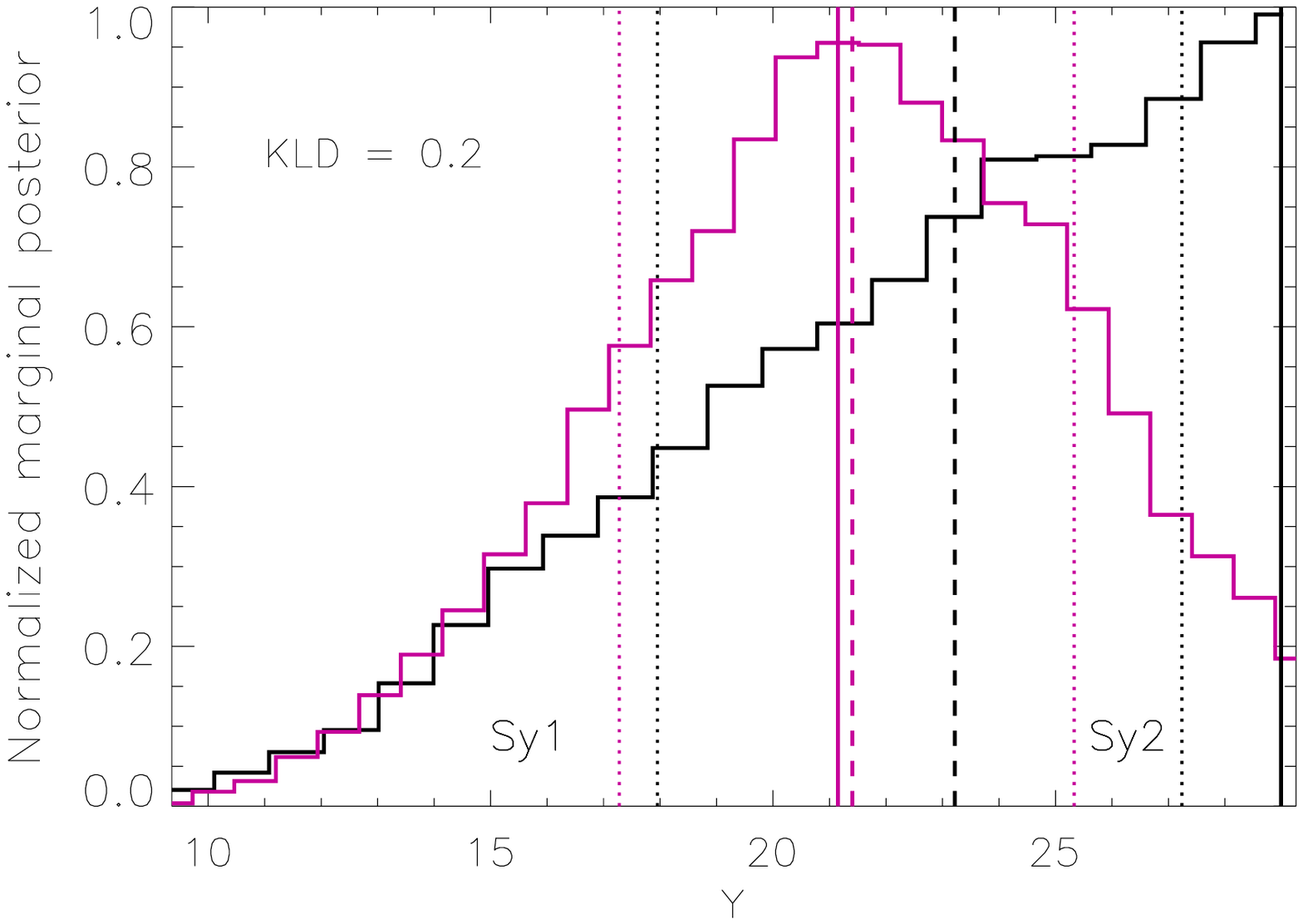}
\includegraphics[width=8cm]{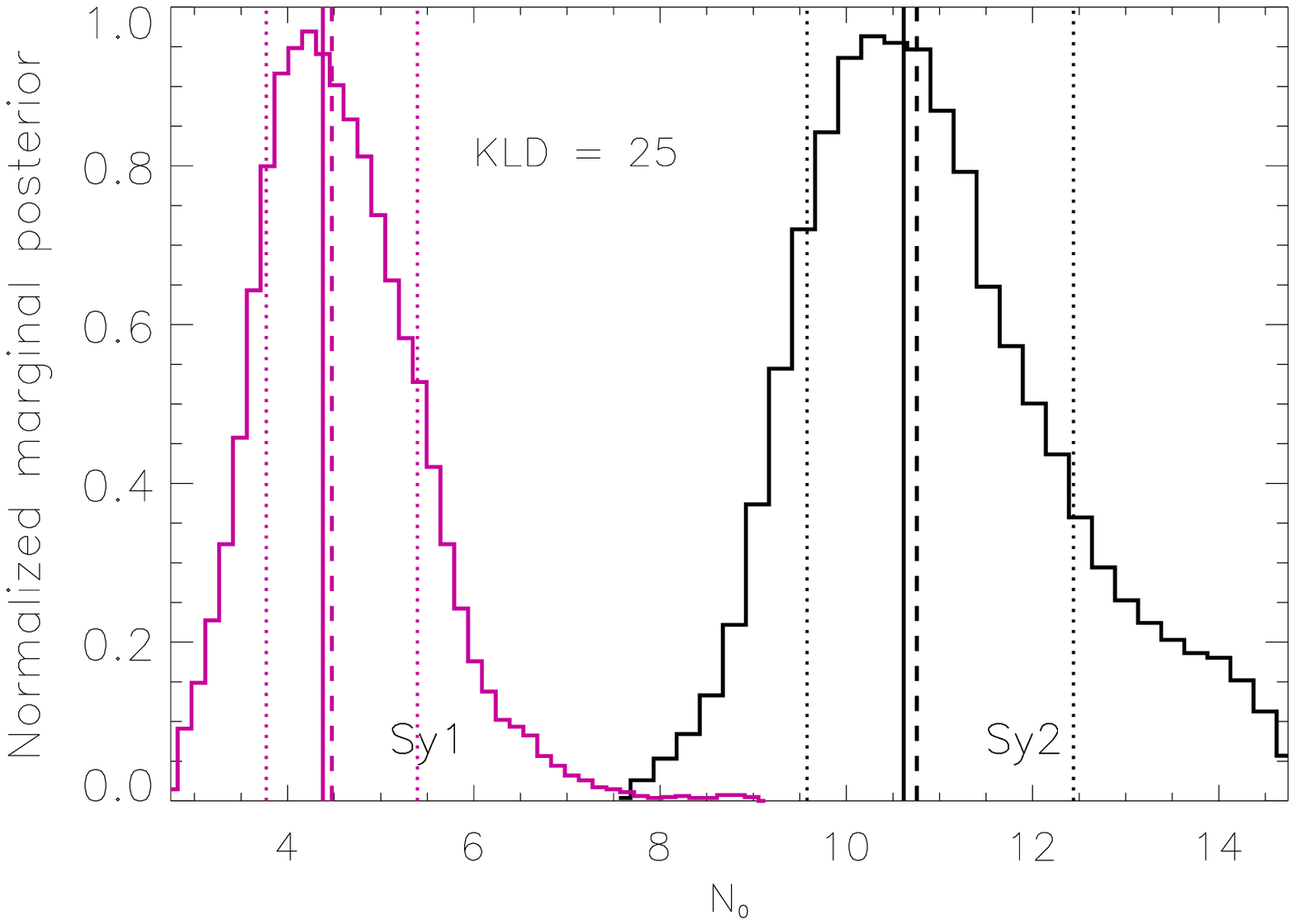}
\includegraphics[width=8cm]{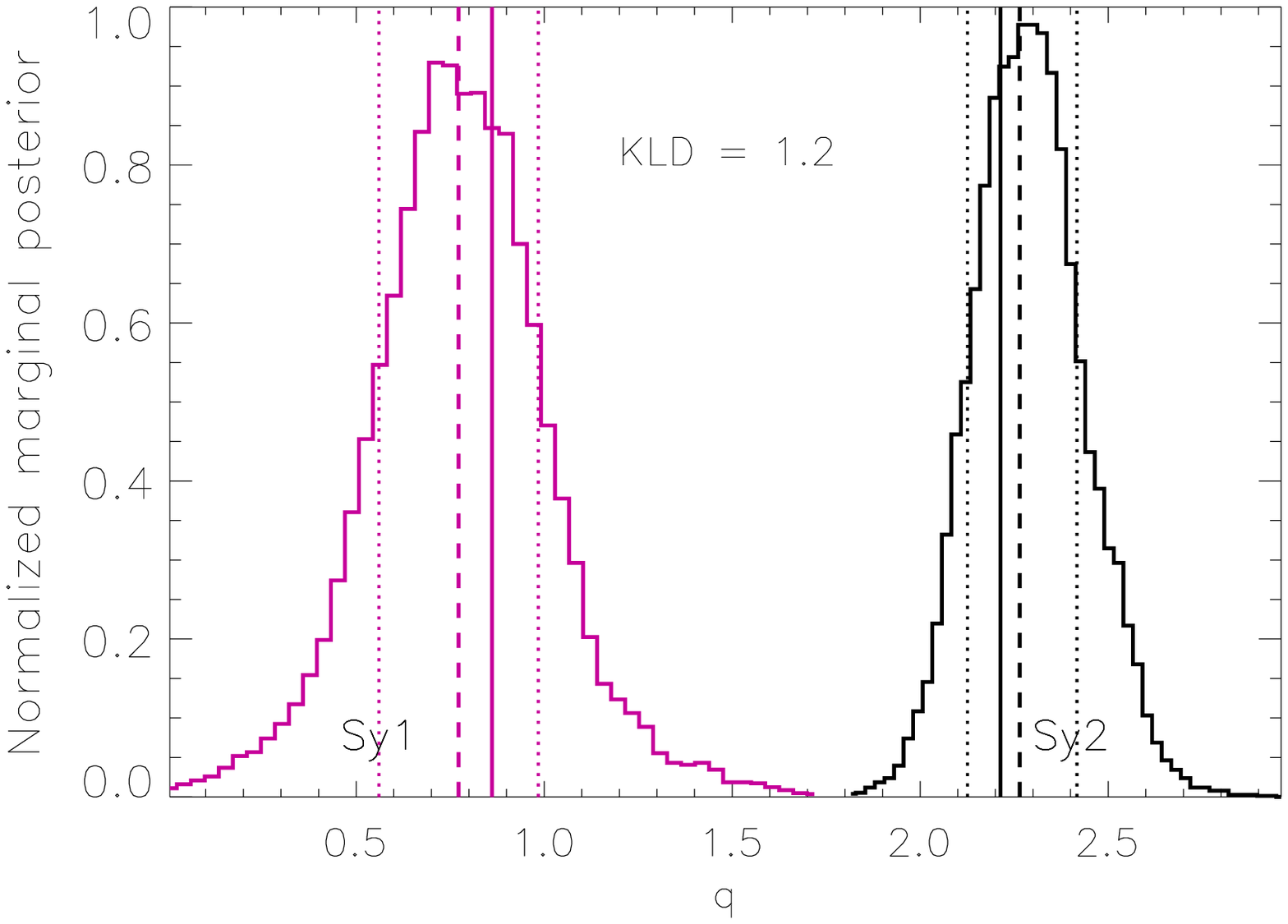}
\includegraphics[width=8cm]{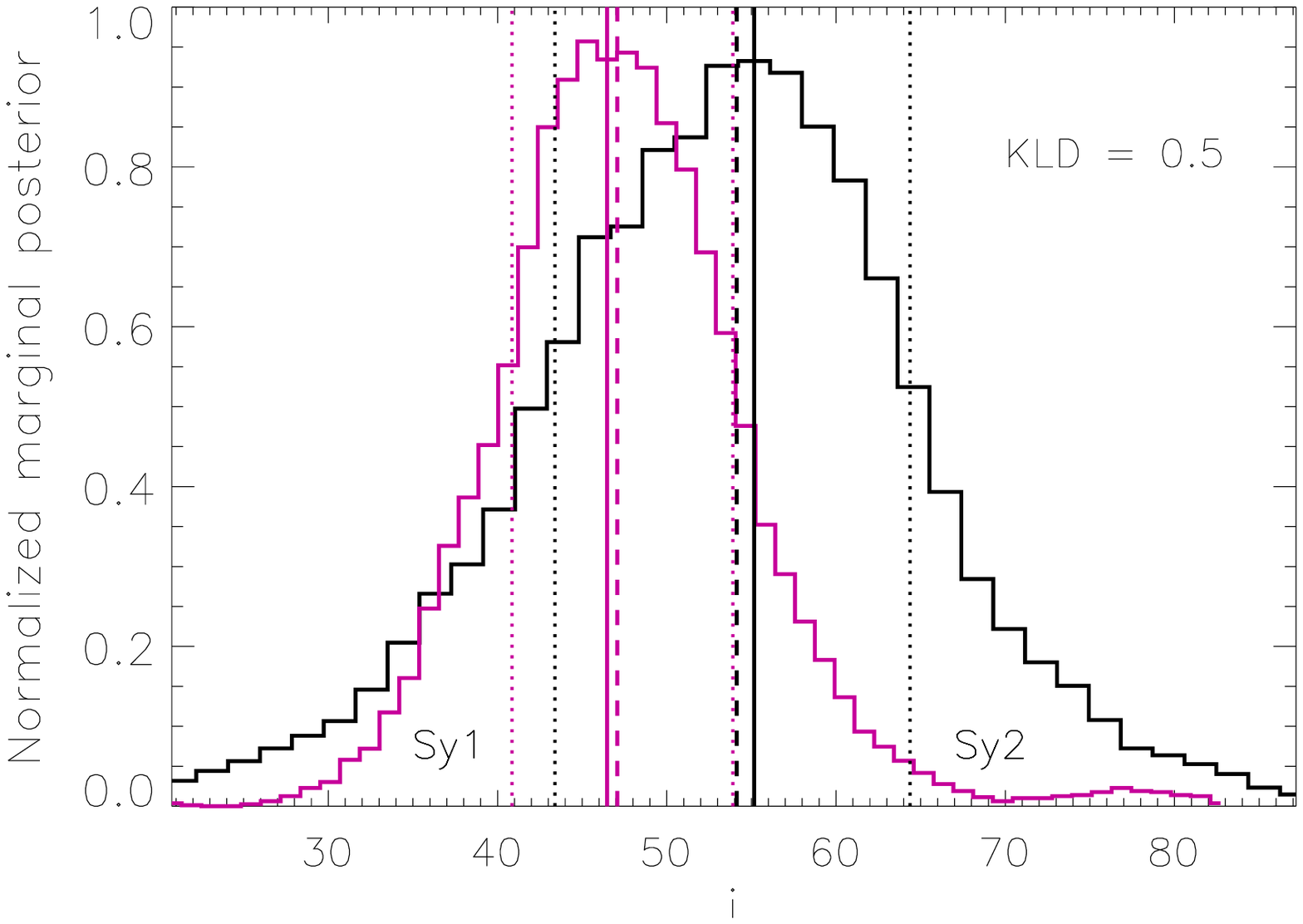}
\includegraphics[width=8cm]{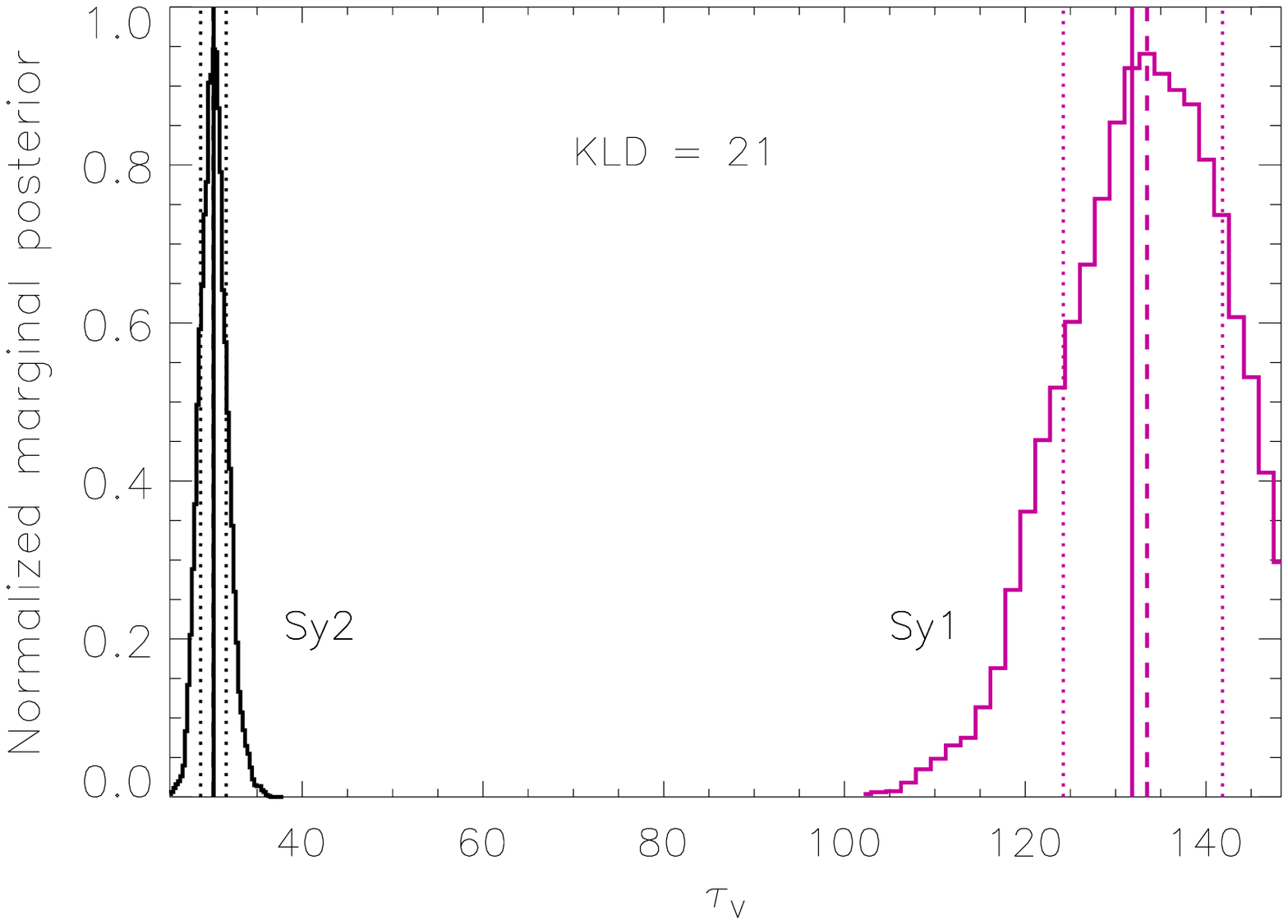}\par}
\caption{\footnotesize{Same as in Figure \ref{ngc1097}, but for the joint Sy1 and Sy2 SEDs. KLD values derived from the comparison between 
Sy1 and Sy2 for each parameter are labelled.}
\label{divergence_plot}}
\end{figure*}

The comparison between the Sy1 and Sy2 posterior distributions is shown in Figure \ref{divergence_plot}. 
From a visual inspection it is clear that the joint posteriors of the parameters N$_0$, $q$, $\tau_V$, and $\sigma$
are completely different between Sy1 and Sy2. There is not overlap between the 
1-sigma intervals. 
In Table \ref{ks} we report the median and mode values of the histograms in Figure \ref{divergence_plot}.

In order to quantify how different the probability distributions are, we calculated the 
Kullback-Leibler divergence (KLD; \citealt{Kullback51}) between the Sy1 and Sy2 posteriors. This divergence takes into
account the full shape of the posterior and it is always
a positive value, and it is equal to zero when two distributions are identical. Therefore, the larger the value of KLD, 
the more different the posteriors. We find KLD$>$1 for $\sigma$ (KLD=5.2), $N_0$ (KLD=25), $q$ (KLD=1.2), and $\tau_V$
(KLD=21). These are indeed the four parameters whose 1-sigma regions do not overlap (see Figure \ref{divergence_plot})
and thus we consider their differences significant between Sy1 and Sy2$\footnote{See further discussion on the Sy2 $q$ parameter results below.}$. 
For both $Y$ and $i$ we find KLD$<$1 and similar median values between Sy1 and Sy2.


\begin{deluxetable*}{lllllllllllll}
\tabletypesize{\footnotesize}
\tablewidth{0pt}
\tablecaption{Statistics of the comparison between Sy1 and Sy2 parameters}
\tablehead{
\colhead{Type} & \multicolumn{2}{c}{$\sigma$ ($\degr$)} & \multicolumn{2}{c}{$Y$} &  \multicolumn{2}{c}{$N_0$} & \multicolumn{2}{c}{$q$} & \multicolumn{2}{c}{$i$ ($\degr$)} &
\multicolumn{2}{c}{$\tau_{V}$} \\
 & \colhead{Sy1} & \colhead{Sy2}  & \colhead{Sy1} & \colhead{Sy2}   & \colhead{Sy1} & \colhead{Sy2} & \colhead{Sy1} & \colhead{Sy2}  & \colhead{Sy1} & \colhead{Sy2} &  \colhead{Sy1} & \colhead{Sy2}}
\startdata
Medians & 44$\pm^{8}_{7}$ & 63$\pm^{4}_{5}$ & 21$\pm$4 & 23$\pm^{4}_{5}$ & 4$\pm$1 & 11$\pm^{2}_{1}$ & 0.8$\pm$0.2 & 2.3$\pm$0.1 & 47$\pm^{7}_{6}$ & 54$\pm^{10}_{11}$  & 133$\pm^{8}_{9}$ & 30$\pm$1	\\
Modes   & 42              & 65              & 21       & 29	         & 4       & 11 	     & 0.9	   & 2.2         & 46		   & 55 		& 132		   & 30 	\\  
\enddata          
\label{ks}
\end{deluxetable*}

Sy1 tori are narrower and have fewer clouds ($\sigma$=44\degr$\pm^{8\degr}_{7\degr}$; N$_0$=4$\pm$1) than those of Sy2 
($\sigma$=63\degr$\pm^{4\degr}_{5\degr}$; N$_0$=11$\pm^{2}_{1}$). The radial density distribution of the clouds is also 
different between the two Seyfert types according to this analysis: in Sy2, the majority of the clumps are distributed very close to the nucleus 
(i.e. steep radial density distribution; $q$=2.3$\pm$0.1) whereas for Sy1 the clouds distribution is flatter ($q$=0.8$\pm$0.2).
On the other hand, the optical depth of the clouds in Sy1 tori is larger ($\tau_V$=133$\pm^{8}_{9}$) than in 
Sy2 ($\tau_V$=30$\pm$1). 


By taking a closer look to the right panel of Figure \ref{comparison1and2}, which corresponds to the Sy2 fit, it is clear that some of the 
data points are underestimated by the model. This happens because of the Circinus SED, which has the shortest errors bars, 
has more weight in the fit than the rest of the SEDs. In order to check that our Sy2 results are not completely biased by Circinus, 
we have repeated the fit excluding it, as well as NGC 1386, which is the least restricted SED in terms of data points, getting
rid of the extremes. From the new fit, we find even larger differences with the Sy1 values of $\sigma$, $N_0$, and $\tau_V$. On the other hand, 
the $q$ joint posterior becomes comparable to that of the Sy1, probably as a consequence of getting rid of the two SEDs fitted with
the largest $q$ values among the Sy2. Considering all the previous we prefer to be cautious about the $q$ parameter, and only consider $\sigma$, 
$N_0$, and $\tau_V$ genuinely different between Sy1 and Sy2.

Interestingly, we find high as well as low values of the inclination angle of the torus for Sy1 and Sy2 (see Table \ref{types}). 
This variety in the $i$ values translates into the similar median values found for the joint Sy1 and Sy2 posterior distributions
(47\degr$\pm^{7\degr}_{6\degr}$ for Sy1 and 54\degr$\pm^{10\degr}_{11\degr}$ for Sy2), which are also intermediate within the considered
prior ($i$=[0\degr,90\degr]). {\it This is telling us that, in the clumpy torus scenario, the classification of a Seyfert galaxy as a Type-1 or Type-2 depends more 
on the intrinsic properties of the torus rather than in its inclination.}

Our results contradict those presented by \citet{Honig10}, who find similar averaged values of $N_0$ for both the Sy1 and Sy2 galaxies
in their sample. However, it is worth mentioning that they fixed low values of the inclination angle of the torus for Sy1 and large for 
Sy2, what can likely have influenced their results.

\begin{figure*}[!ht]
\centering
\includegraphics[width=15cm]{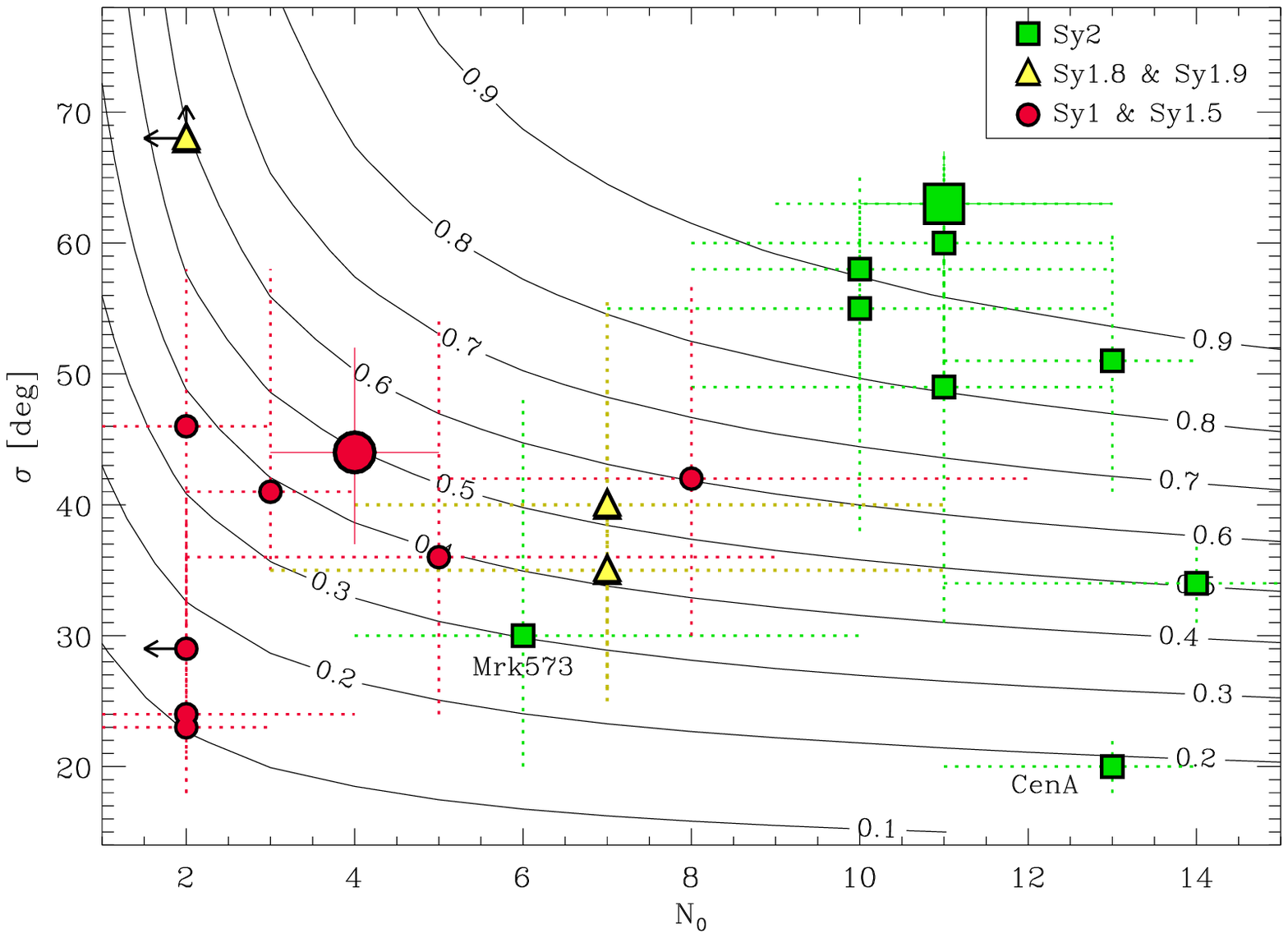}
\caption{\footnotesize{$\sigma$ versus N$_0$ for the individual galaxies. Either median values or upper/lower limits
are taken from the fits presented here. 
Dots correspond to Type-1 Seyferts, triangles to Sy1.8 and Sy1.9, and squares to Sy2. 
Error bars indicate 68\% confidence level around the median.
Note the segregation between Seyfert types, indicating the intrinsic difference between their tori
in terms of covering factor (indicated in contours). The big dot and square correspond to the average $\sigma$ and N$_0$
values for Sy1 and Sy2 from Table \ref{ks}.
\label{covering}}}
\end{figure*}

In Figure \ref{covering} we represent the median values of $\sigma$ and $N_0$ for the different 
Seyfert types over the covering factor contours\footnote{A similar plot showing 
values for Sy2 galaxies from \citealt{Ramos09a} and PG quasars from \citet{Mor09} was shown in the talk by M. Elitzur at the 
Physics of Galactic Nuclei conference held 15-19 June, 2009 at Ringberg Castle. Proceedings published online at 
http://www.mpe.mpg.de/events/pgn09/online$_-$proceedings.html.}. The covering factor is defined as C$_T=1-\int e^{-N_{LOS}(i)}d cos(i)$.
Type-1 nuclei tend to be located within lower C$_T$ contours (C$_T\leq$0.6)
than those of Type-2s, for which C$_T\geq$0.5, with the exception of Centaurus A and Mrk 573\footnote{
As discussed in \citealt{Ramos09a}, the fit of Centaurus A is complicated by the presence of a 
dust lane of A$_V\sim$7-8 mag that is likely affecting the NIR nuclear fluxes, as well as
the possible synchrotron contamination of the MIR fluxes. Mrk~573 has been recently reclassified as an obscured 
Narrow-line Seyfert 1 (NLSy1) based on NIR spectroscopy 
\citep{Ramos08,Ramos09b}. However, here we considered it as a Sy2 
because of the similarity in SED shape with the rest of the sample.}.
We have represented with larger symbols the median values from the joint $\sigma$ and $N_0$ posteriors 
reported in Table \ref{ks}. 

\begin{figure*}[!ht]
\centering
\includegraphics[width=15cm]{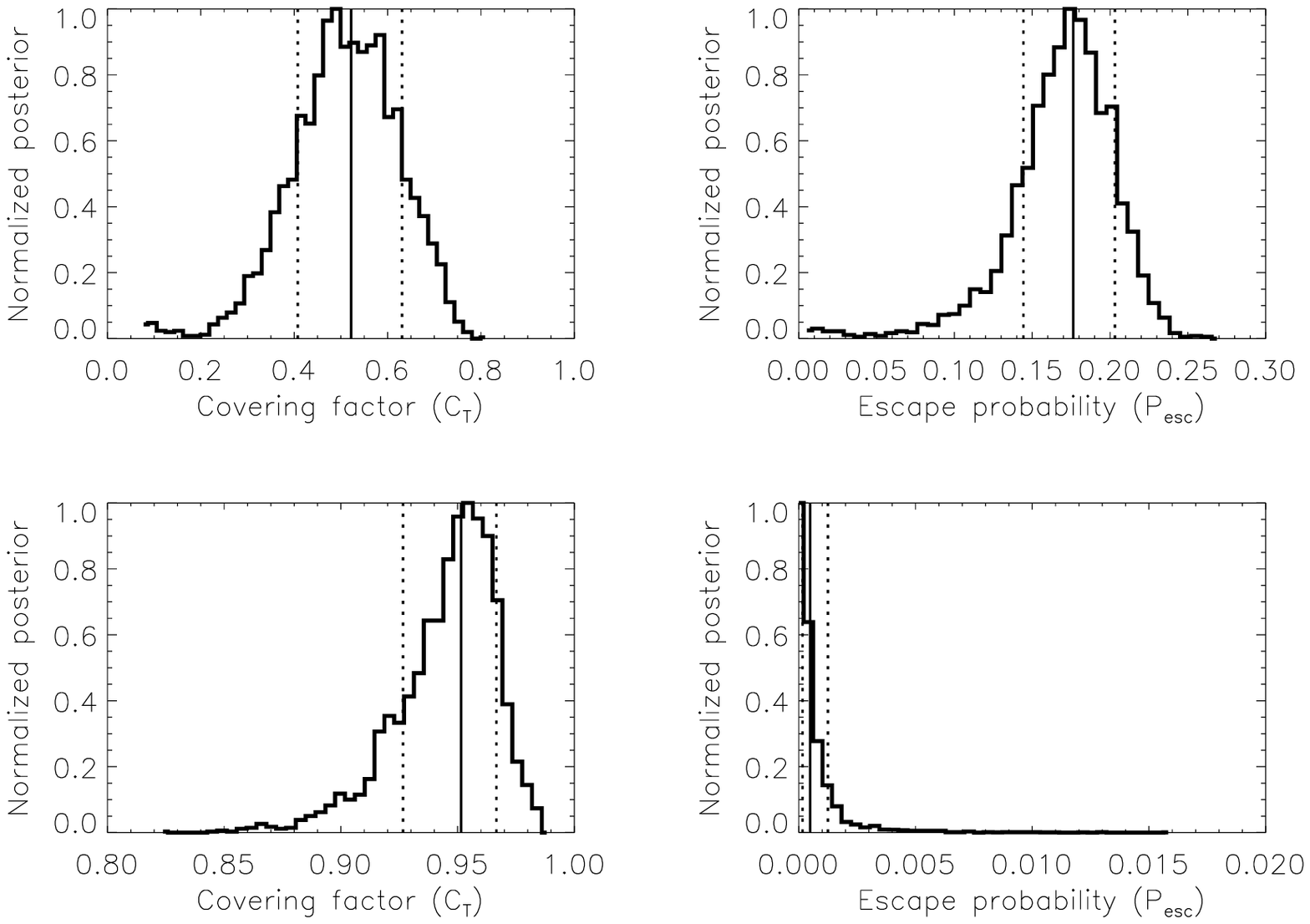}
\caption{\footnotesize{Joint posterior distributions of the torus covering factor (left panels) and escape probability (right panels) for 
Sy1 (top) and Sy2 galaxies (bottom). The values of the Kullback-leibler divergence obtained from the comparison between Sy1 and Sy2 are
KLD=28 for C$_T$ and KLD=29 for P$_{esc}$.
\label{comparison_plot}}}
\end{figure*}

Since the covering factor is a non-linear function of the torus parameters, we took full advantage of our
Bayesian approach and generated joint posterior distributions for C$_T$ from those in Figure  \ref{divergence_plot},
which are shown in the left panel of Figure \ref{comparison_plot}. The median values of the histograms are C$_T$(Sy2)=0.95$\pm$0.02 
and C$_T$(Sy1)=0.5$\pm$0.1. The divergence between the Sy1 and Sy2 C$_T$ posteriors is KLD=28, indicating that the difference 
is significant (the 1-sigma regions do not overlap). Thus, Sy1 tori in our sample have lower C$_T$s than those of Sy2, implying that
they are intrinsically different. 
 
\subsection{Near-infrared Emission and Torus Angular Width}
\label{nirflux}

As reported in Section \ref{averageSy1}, Type-1 Seyferts present characteristically higher H/N
ratios and flatter SEDs than those of Sy2 nuclei. 
The enhancement on the NIR emission of Type-1 AGN is produced by the hot dust from the 
directly-illuminated faces of the clumps in the torus which are close to the central
engine, and also by direct AGN emission (i.e., the tail of the optical/ultraviolet power-law continuum), which
strongly flattens their IR SEDs \citep{Rieke81,Barvainis87}.

Based on the IR SEDs presented here, the relative NIR contribution to the SED is generally correlated with the 
Seyfert type. Sy2 galaxies show lower NIR to MIR ratios (H/N=0.003$\pm$0.002) than Sy1 (0.06$\pm$0.03).


In the context of the clumpy models, the presence of a cloud
along the LOS, which may occur from any viewing angle, results in a Type-2 classification.
{\it Cloud encounters are more probable at large inclination angles ($i$), but there is always
a finite probability for having an unobscured view of the AGN.}
In fact, the likelihood of intercepting a cloud along a LOS depends on the combination of $i$, $N_0$, and
$\sigma$. Thus, the preference for lower values of these parameters (especially $N_0$ and $\sigma$) in
Type-1 Seyferts increases the likelihood of unimpeded views of some 
directly-illuminated cloud faces (i.e., those on the ``back'' side
of the torus) and direct AGN emission, resulting in an increase of the NIR flux.
The latter can be represented in terms of the escape probability 
P$_{esc}$ (see equation 4 in \citealt{Nenkova08a}). For a total number of clouds 
N$_{LOS}$ along a path, P$_{esc}\simeq~exp(-N_{LOS})$ when the clouds are optically thick ($\tau_\lambda>1$).

\begin{figure*}[!ht]
\centering
\includegraphics[width=13cm]{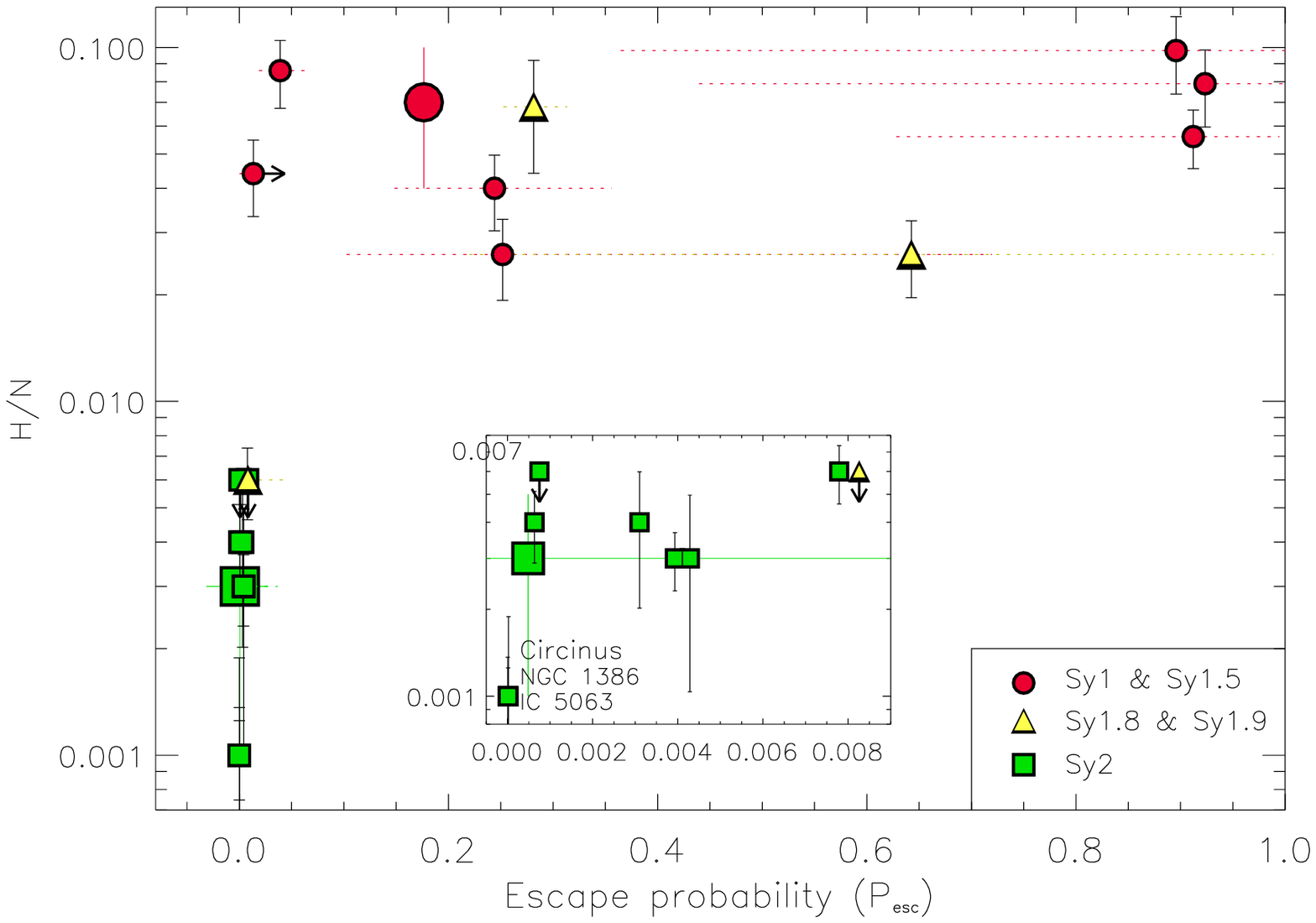}
\caption{\footnotesize{H/N versus P$_{esc}$. Lower values of $N_0$, $\sigma$, and $i$ result in higher P$_{esc}$, 
resulting in higher NIR-to-MIR ratios. The inset shows an amplification of the region occupied by the Sy2. 
Symbols are the same as in Figure \ref{covering}. The big dot and square 
correspond to the median values of P$_{esc}$ for Sy1 and Sy2 from Figure \ref{comparison_plot}, and to the H/N values
of the Sy1 and Sy2 average templates (Table \ref{slopes}).}
\label{slope}}
\end{figure*}

In Figure \ref{slope} we show the dependence of the H/N ratio on the escape probability.
All the Sy2 are in the bottom-left corner of the diagram, whereas the Sy1 have higher
values of the H/N ratio and P$_{esc}$=[1\%,92\%].
Sy1.8 and Sy1.9 galaxies present intermediate values between those of Sy1 and Sy2.  
The derived joint posterior distributions of the escape probabilities for Sy1 and Sy2 (KLD=29 between them) are shown in Figure \ref{comparison_plot}, 
and the median values of the histograms are P$_{esc}$(Sy1) = 18$\pm$3\% and P$_{esc}$(Sy2) = 0.05$\pm^{0.08}_{0.03}$\%.

Thus, while for tori with high values of $i$, $N_0$, and $\sigma$ the probability of having a direct view of the 
AGN is very little, that increases for objects with narrower and less inclined tori, and containing less clumps.
{\it In this work we show for the first time that, in the clumpy torus scenario, the classification as a Type-1 or Type-2 Seyfert depends more on the 
intrinsic properties of the torus, rather than in the inclination angle itself.} The Sy1 galaxies in our sample 
have larger P$_{esc}$ than the Type-2 Seyferts, and as a consequence of that, we detect the broad lines 
in their spectra.

\subsection{AGN Luminosities}
\label{luminosidades}

The clumpy model fits yield the intrinsic bolometric luminosity of 
AGN (L$_{bol}^{AGN}$) by means of the vertical shift applied to match the observational data points.
Combining this value with the torus luminosity (L$_{bol}^{tor}$), obtained by integrating the corresponding model torus emission 
(without the AGN contribution), we derive the reprocessing efficiency (RE) of the torus (L$_{bol}^{tor}$/$L_{bol}^{AGN}$).
The previous values are calculated on the Bayesian framework, by combining the posterior distributions of the 
model parameters. Median values and 1-sigma intervals for Sy1 galaxies are reported in Table \ref{lum}, and those for Sy2
and intermediate-type Seyferts are shown in Table \ref{lum2} (Appendix \ref{appendixB}).

\begin{deluxetable*}{lcccccc}
\tablewidth{0pt}
\tablecaption{Bolometric Luminosity Predictions}
\tablehead{
\colhead{Galaxy}  & \multicolumn{1}{c}{L$_{bol}^{AGN}$}  & \colhead{L$_{bol}^{tor} / L_{bol}^{AGN}$} & 
\colhead{$R_{o}$ (pc)} &  \colhead{L$_{X bol}^{AGN}$} &\colhead{L$_{X bol}^{AGN}$ / L$_{bol}^{AGN}$}  & Ref.}
\startdata
NGC 1097   & 2.4$\pm^{0.8}_{0.4}$  $\times$ 10$^{42}$    &  0.4$\pm$0.1             &    0.4$\pm^{0.1}_{0.2}$  & 1.0 $\times$ 10$^{42}$ & 0.4  &  a \\
NGC 1566   & 5.9$\pm^{2.2}_{1.8}$  $\times$ 10$^{42}$    &  0.4$\pm$0.1             &    0.6$\pm$0.2           & 6.3 $\times$ 10$^{42}$ & 1.1  &  b \\
NGC 6221   & 6.2$\pm^{4.3}_{3.0}$  $\times$ 10$^{42}$    &  0.7$\pm^{0.5}_{0.3}$    &    0.7$\pm^{0.3}_{0.2}$  & 1.3 $\times$ 10$^{43}$ & 2.1  &  c \\
NGC 6814   & 8.1$\pm^{6.0}_{2.4}$  $\times$ 10$^{42}$    &  0.4$\pm$0.2             &    0.7$\pm$0.3           & 2.7 $\times$ 10$^{43}$ & 3.3  &  d \\
NGC 7469   & 3.7$\pm$0.8 $\times$ 10$^{44}$              &  0.5$\pm$0.1             &    5$\pm^{1}_{2}$        & 3.4 $\times$ 10$^{44}$ & 0.9  &  e \\
\hline
NGC 3227   & 1.3$\pm^{1.3}_{0.5}$  $\times$ 10$^{43}$    &  0.7$\pm^{0.5}_{0.3}$    &    0.9$\pm^{0.4}_{0.3}$  & 3.8 $\times$ 10$^{43}$ & 2.9  &  f \\ 
NGC 4151   & 4.6$\pm^{1.5}_{0.9}$  $\times$ 10$^{43}$    &  0.5$\pm$0.1             &    1.4$\pm^{0.8}_{0.6}$  & 1.7 $\times$ 10$^{44}$ & 3.7  &  g \\
\enddata     
\tablecomments{\footnotesize{
    Bolometric luminosities corresponding to the AGN luminosity (L$_{bol}^{AGN}$; in erg~s$^{-1}$).
    Columns 3 and 4 correspond to the RE (L$_{bol}^{tor} / L_{bol}^{AGN}$)
    and the outer radius of the torus calculated using L$_{bol}^{AGN}$ and $Y$.
    Absorption-corrected 2--10 keV X-ray luminosities are taken from
    the literature (references below). L$_{X bol}^{AGN}$ (erg~s$^{-1}$) is derived from 20$\times$L$_{X}^{AGN}$.}}
\tablerefs{\footnotesize{(a) \citet{Terashima02}; (b) \citet{Levenson09}; (c) \citet{Levenson01}; (d) \citet{Gandhi09}; (e) \citet{Nandra07}; (f) \citet{Lamer03}; (g) \citet{Beckmann05}.}}
\label{lum}
\end{deluxetable*}

Sy2 tori in our sample are more efficient reprocessors than Sy1, absorbing and re-emitting the
majority of the intrinsic AGN luminosity in the IR: RE(Sy2)=[0.4, 1.0], with a median value of 0.8 and 
RE(Sy1)=[0.2,0.7], with median of 0.5. It makes no sense to compare the derived joint Sy1 and Sy2 
posterior distributions of RE, because we normalized the SEDs to perform the fits, and consequently the 
derived L$_{bol}^{AGN}$ are meaningless.

We considered a possible dependency of the RE (or alternatively the covering factor; C$_T$) on L$_{bol}^{AGN}$, since 
the amount of incoming radiation from the AGN could possibly have some influence on the reprocessed energy or 
even in the torus properties (e.g., receeding torus scenario; \citealt{Lawrence91}). 
However, we find no relationship between the two quantities in the luminosity range considered. 
This means that the reprocessing efficiency 
depends primarily on the total number of clouds available to absorb the incident radiation, i.e. on the torus covering factor. 
However, the possible dependence of the torus properties on the AGN luminosity considering a broader luminosity range is further investigated
in \citet{Alonso11}.

\begin{figure*}[!ht]
\centering
\includegraphics[width=13cm]{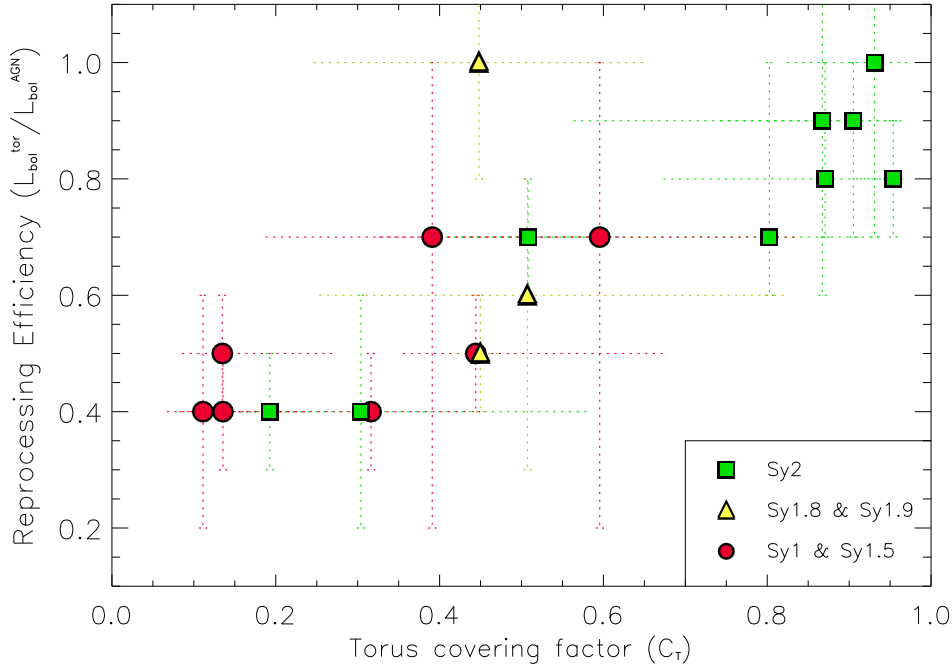}
\caption{\footnotesize{Reprocessing efficiency versus torus covering factor for the individual galaxies. Higher values of C$_T$
translate into more efficient reprocessors. In general, Type-2 tori are more efficient 
than Type-1 tori, with the exceptions of Centaurus A and Mrk 573. Symbols are the same as in Figure \ref{covering}. }
\label{luminosity}}
\end{figure*}

Figure \ref{luminosity} shows that there is a correlation between RE and the torus covering factor for the galaxies in our sample. 
The larger C$_T$ the more efficient reprocessor. 
Type-2 tori have in general larger RE than those of Type-1, with the exceptions of Centaurus A and Mrk 573. 
Despite the relatively small sample, Figure \ref{luminosity} shows a segregation 
between Sy1 and Sy2 galaxies in terms of reprocessing efficiency and covering factor.



{\it If all Seyfert nuclei are identical, as the unified model predicts, only the viewing angle should determine
the classification, not the properties of the torus itself. 
While our results are limited by the small sample analyzed, they suggest instead that the Type-1/Type-2 classification 
depends on the torus intrinsic properties rather than in the mere torus inclination.}

The bolometric luminosity of the intrinsic AGN derived from the fits
(L$_{bol}^{AGN}$; column 2 in Table \ref{lum}) can be directly compared with those from the
absorption-corrected 2-10 keV luminosities compiled from
the literature (L$_{X bol}^{AGN}$; column 6 in Table \ref{lum}), which is an effective proxy for 
the AGN bolometric luminosity \citep{Elvis94}.
To obtain L$_{X bol}^{AGN}$ from the intrinsic 
2-10 keV values we applied a bolometric correction factor of 20 \citep{Elvis94}. 

We find similar values of L$_{Xbol}^{AGN}$ and L$_{bol}^{AGN}$ 
for all the Sy1 (see Table \ref{lum}). In the case of NGC 1097, we expect to have
some contamination from the nuclear starburst (see Section \ref{individual} and \citealt{Mason07}). Thus, if the
L$_{Xbol}^{AGN}$ value represents the AGN contribution-only, $L_{bol}^{AGN}$ may be overestimated because of the 
starburst contribution, thus resulting in a low ratio between the two measurements. 
The good agreement with the observational X-ray measurements
reinforces the results from our SED modelling. The same comparison with X-ray data for Sy2 and intermediate-type Seyferts
is shown in Table \ref{lum2} in Appendix \ref{appendixB}.

\subsection{Torus Size}

In general, the IR SED fitting does not constrain the size of the torus ($Y$) as well as other 
model parameters (see Section 5.3 in \citealt{Ramos09a}). The NIR and MIR observations are sensitive
to the warm dust (located within $\sim$10 pc of the nucleus), which depends on the combination 
of model parameters N$_0$, $q$, and $Y$ \citep{Thompson09}. FIR observations would be more sensitive
to the torus extent independently.  
However, the fits performed here for Type-1 and Type-2 Seyferts 
are consistent with a small torus size, confined to scales
of less than 6 pc (see below). 

Uniform density models require 
the dusty torus to extend over large dimensions, to provide cool dust that produces the IR emission (e.g., \citealt{Granato94}).  
In contrast, in a clumpy distribution, different dust temperatures can coexist 
at the same distance, including cool dust at small radii \citep{Nenkova02}, so large tori, which are inconsistent 
with imaging and interferometry results, are not necessary. Indeed, 
for the Seyfert galaxies considered here, we found $Y$ ranging from 10 to 25 (see Table \ref{clumpy_parameters}), 
showing that small tori can account for the observed IR nuclear emission. 


The outer size of the torus scales with the AGN bolometric luminosity:
$R_{o} = Y R_{d}$, so assuming a dust sublimation temperature of 1500
K, $R_o= 0.4~Y~(L_{bol}^{AGN}/10^{45})^{0.5}$ pc. We derived 
$R_o$ posterior distributions from those of L$_{bol}^{AGN}$ and $Y$ and 
find that all tori in our sample have outer radii smaller than 6 pc
(Tables \ref{lum} and \ref{lum2}), in agreement with MIR
direct imaging of nearby Seyferts \citep{Packham05,Radomski08} and 
also interferometric observations \citep{Jaffe04,Tristram07,Meisenheimer07,Raban09}.  

For example, the estimated outer radius
for NGC 1097 (R$_o$=0.4$\pm^{0.1}_{0.2}$ pc) is in agreement with the value derived from NIR NACO/VLT observations \citep{Prieto05}, 
which placed the radius of the central compact source in r$<$5 pc. \citet{Mason07} also
derived an upper limir of 19 pc for the size of the unresolved component from the MIR images employed in this work for NGC 1097.
For the case of NGC 7469, for which we derive R$_o$=5$\pm^{1}_{2}$ pc, \citet{Tristram09} reported an estimation of the size 
of the dust distribution of 10 pc, based on MIDI/VLT interferometric observations, 
although compromised by the high level of noise and the lack of fringes.

\section{Conclusions}
\label{final}

We present new subarcsecond resolution MIR imaging data at 8.7 and 18.3 $\micron$ for the three Type-1 Seyfert galaxies NGC 6221, 
NGC 6814, and NGC 7469. 
NGC 7469 and NGC 6221 appear extended, with part of this extended emission associated with 
emitting-dust heated by star formation. On the contrary, NGC 6814 lacks of any extended emission. 
Nuclear MIR and NIR fluxes for the three galaxies as well as for NGC 1566, NGC 1097, NGC 3227, and NGC 4151 are reported. 
We construct nuclear SEDs that the AGN dominates and  
fit them with clumpy torus models and a Bayesian approach to derive torus 
parameters. The main results of this work for the individual Sy1 galaxies and from the comparison
with the Sy2 and intermediate-type Seyferts in \citealt{Ramos09a} are summarized as follows:

\begin{itemize}

\item We derived an average Type-1 Seyfert template from the individual Sy1 SEDs, which is flatter
(mean IR slope $\alpha_{IR} = 1.7\pm0.3$) than the Type-2 mean SED 
presented in \citealt{Ramos09a} ($\alpha_{IR} = 3.1\pm0.9$).

\item The NIR-to-MIR flux ratios measured from the individual SEDs are larger for Sy1 (0.07$\pm$0.03) 
than for Sy2 (0.003$\pm$0.002). Indeed, the distributions of NIR-to-MIR values are significantly different 
between the two types at the 100\% confidence level.

\item The interpolated version of the clumpy models of \citet{Nenkova08a,Nenkova08b} employed here 
successfully reproduces the nuclear IR SEDs of Type-1 Seyferts with compatible 
results among them. Consequently, the
observed nuclear IR emission of these galaxies can be accounted for by dust heated by the central engine
and direct AGN emission. 


\item We derive joint posterior distributions for Sy1 and Sy2 and find that the differences in 
N$_0$, $\tau_V$, and $\sigma$ between Type-1 and Type-2 tori are significant according 
to the Kullback-Leibler divergence and the lack of overlap between their 1-sigma confidence intervals. 

\item We find that Sy1 tori are narrower and have fewer clouds than those of Sy2. 
Additionally, the optical depth of the clouds in Sy1 tori is larger than in Sy2.  

\item There is not a clear trend in the values of the inclination angle of the torus for Sy1 and Sy2
(slightly larger values are found for Sy2).

\item The larger the covering factor of the torus, the smaller the likelihood of intercepting a cloud along a LOS. 
In our sample, Seyfert 2 tori have larger covering factors and smaller escape probabilities than those of Seyfert 1.

\item Despite the limited number of galaxies considered, we find that Type-2 tori are in general more efficient 
reprocessors than those of Type-1. Indeed, there is a correlation between the reprocessing efficiency and 
the torus covering factor.

\item For the Seyfert galaxies studied here, we find that tori with outer radii smaller than 6 pc 
can account for the observed NIR/MIR nuclear emission, in agreement with MIR interferometric observations. 

\end{itemize}

{\it Summarizing, we find tantalizing evidence, albeit for a small sample of Seyfert galaxies and under the clumpy torus hypotesis, that the classification 
as a Type-1 and Type-2 depends more on the intrinsic properties of the torus than on its mere inclination towards us.}

\vspace{4cm}

\appendix
\section{Fitting results for Sy1 galaxies}
\label{appendixA}

\begin{figure*}[!ht]
\centering
{\par
\includegraphics[width=5.3cm]{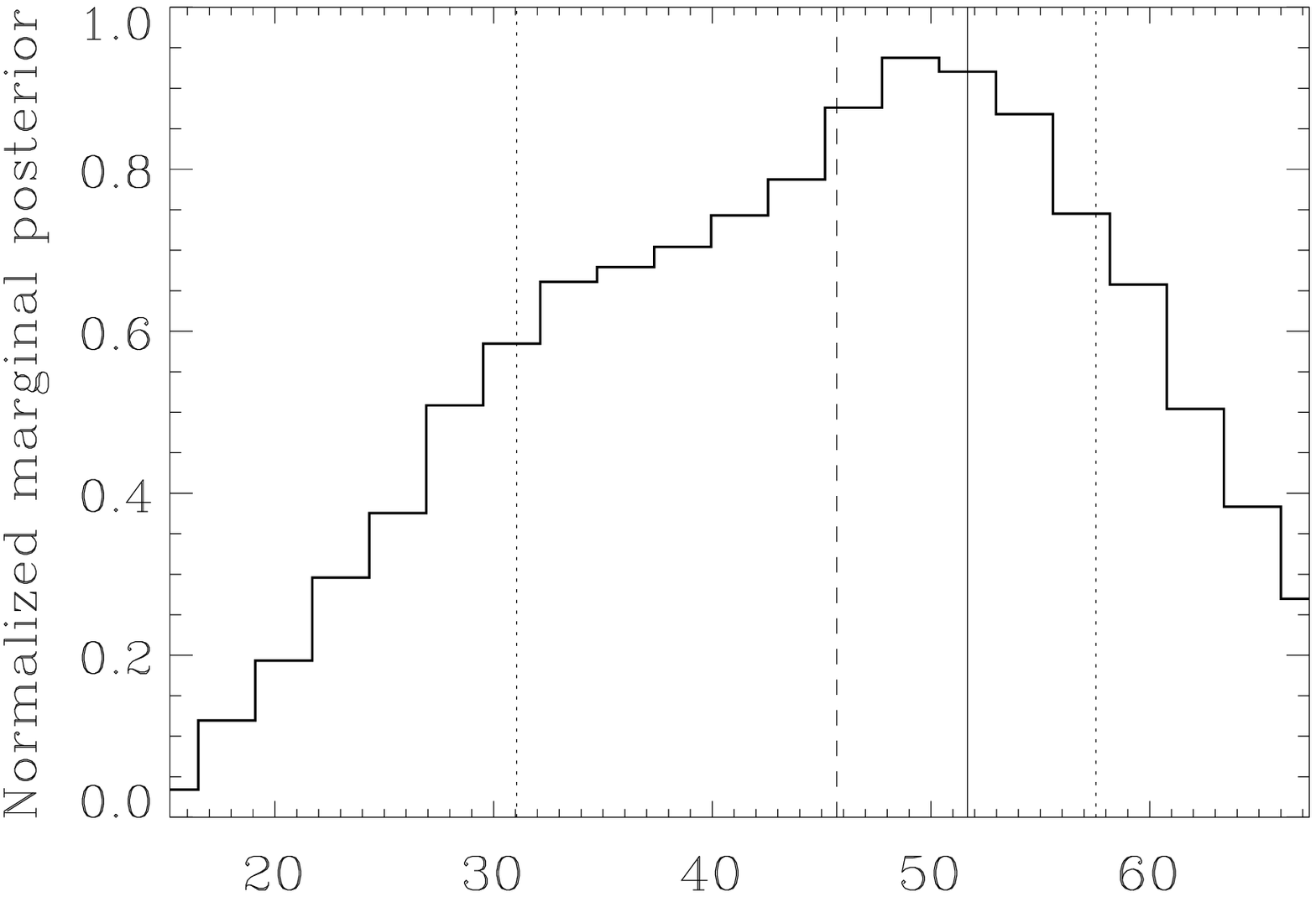}
\includegraphics[width=5.3cm]{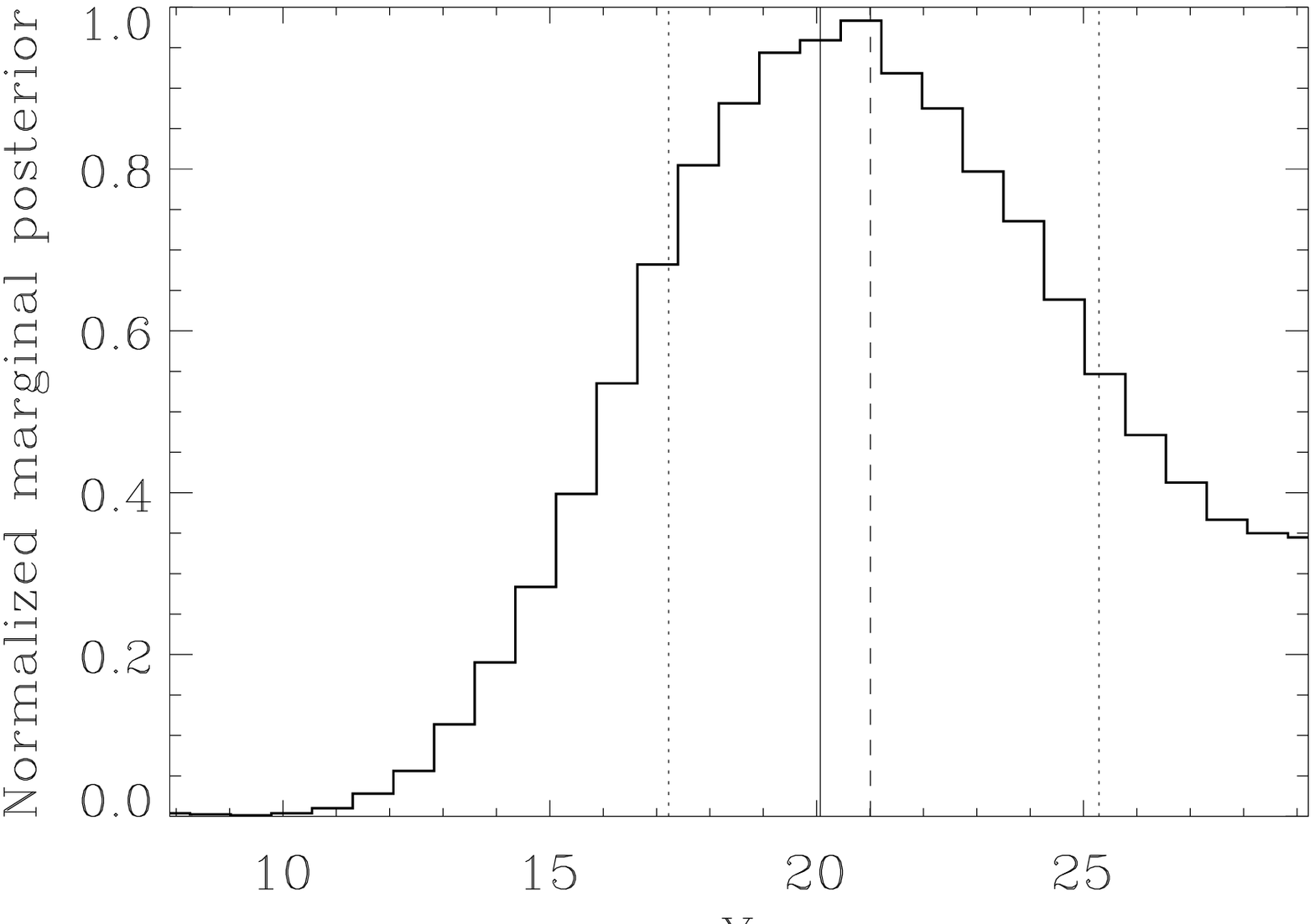}
\includegraphics[width=5.3cm]{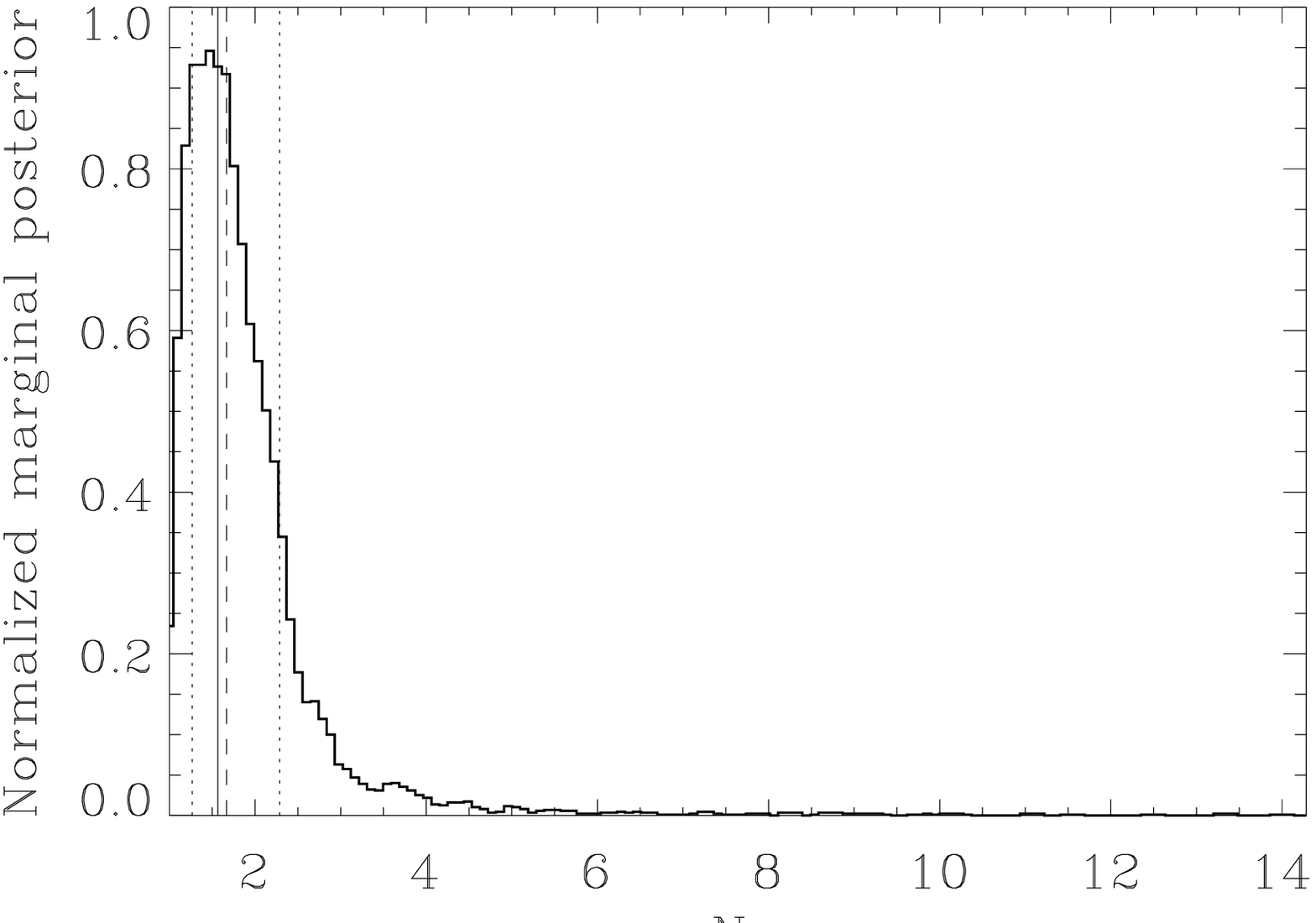}
\includegraphics[width=5.3cm]{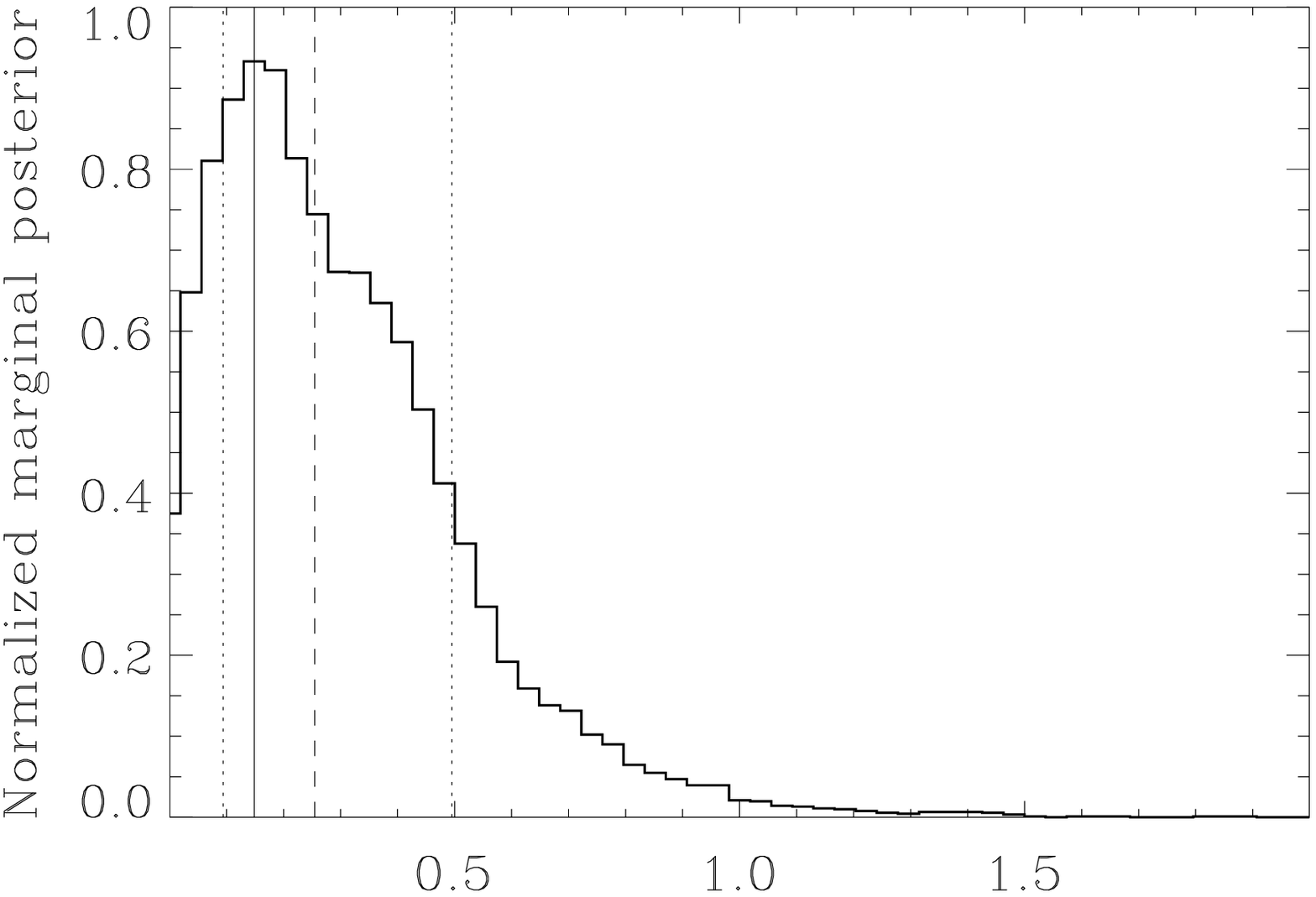}
\includegraphics[width=5.3cm]{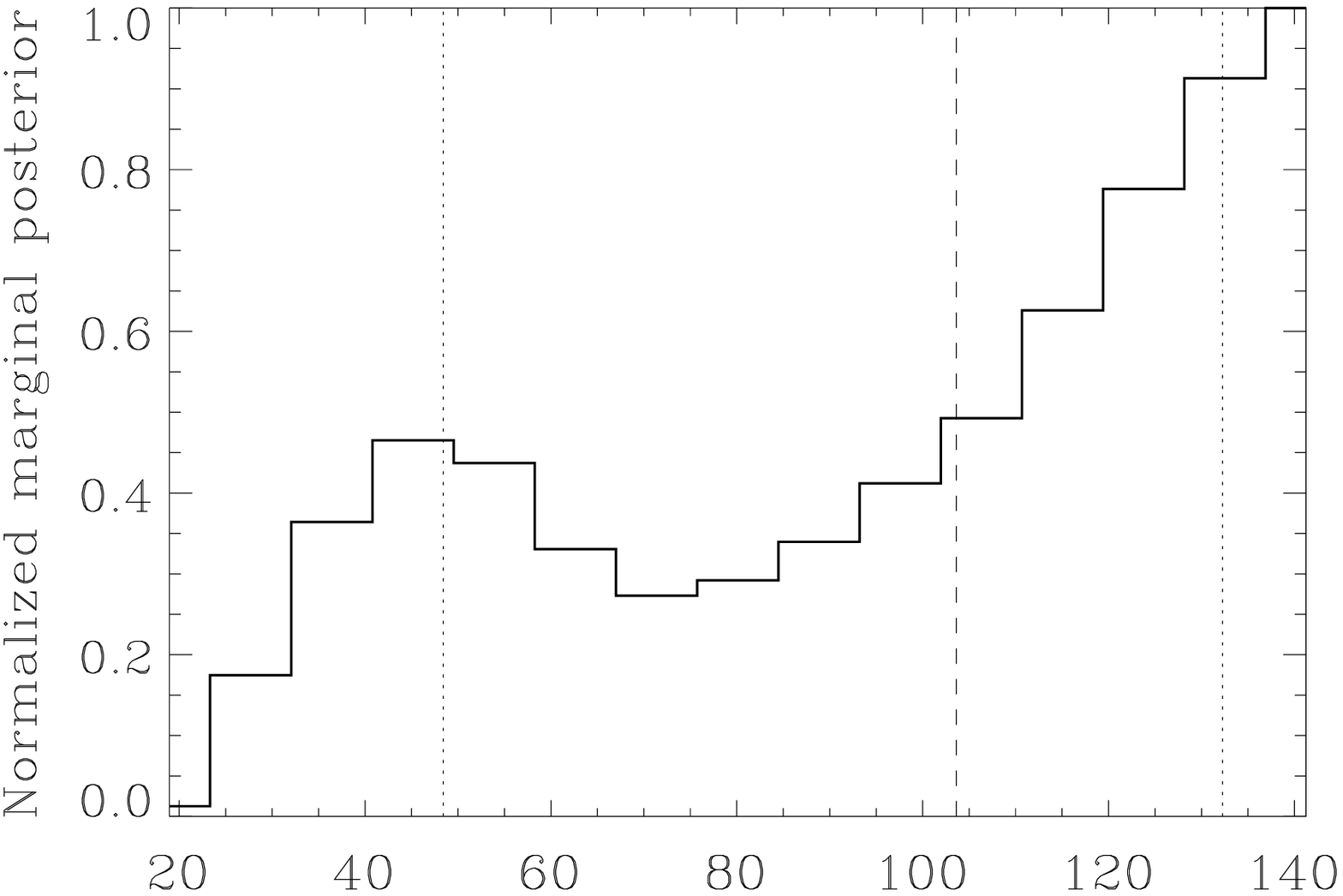}
\includegraphics[width=5.3cm]{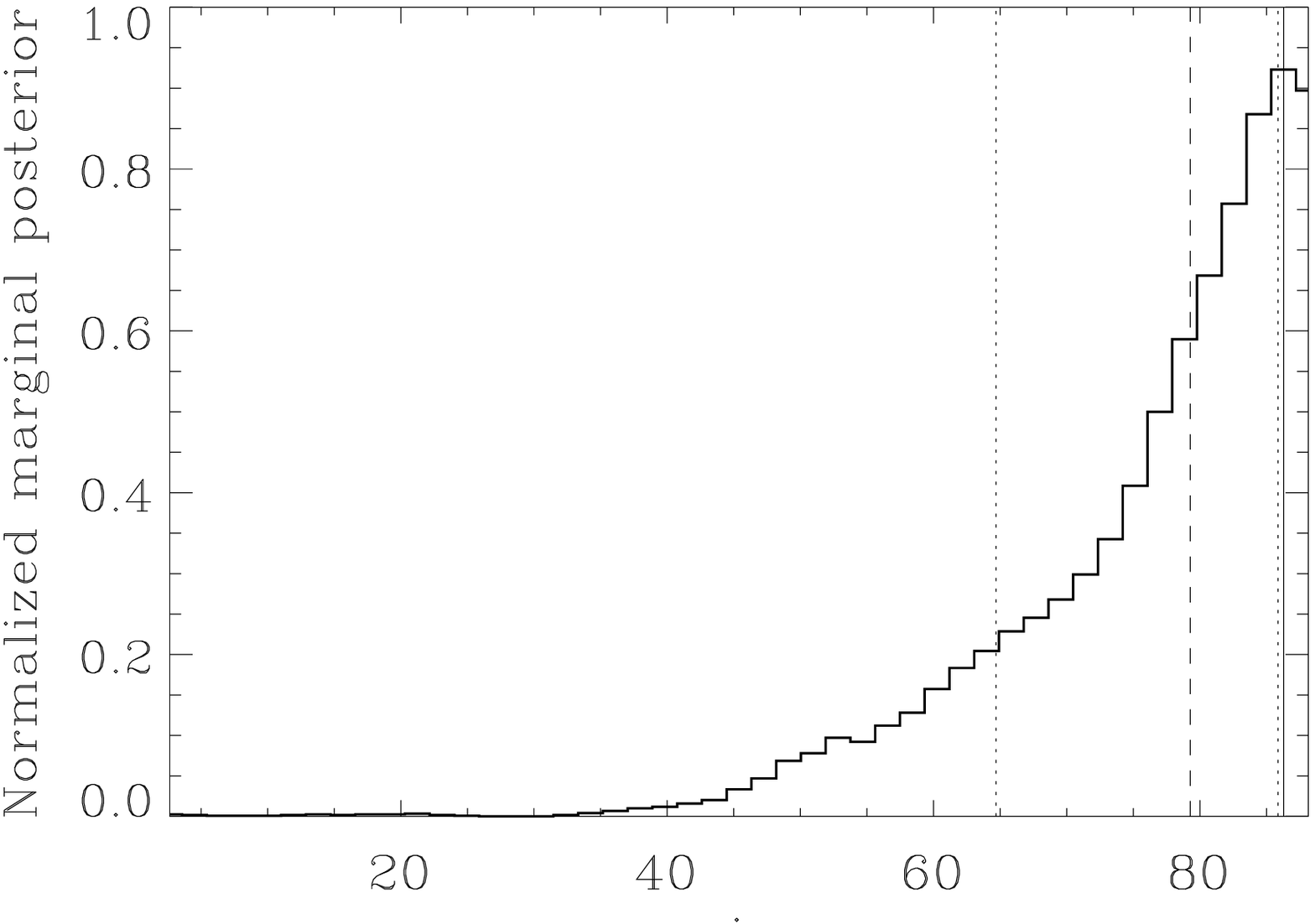}\par}
\caption{\footnotesize{Same as in Figure \ref{ngc1097}, but for the galaxy NGC 1566.}
\label{ngc1566}}
\end{figure*}

\begin{figure*}[!ht]
\centering
{\par
\includegraphics[width=5.3cm]{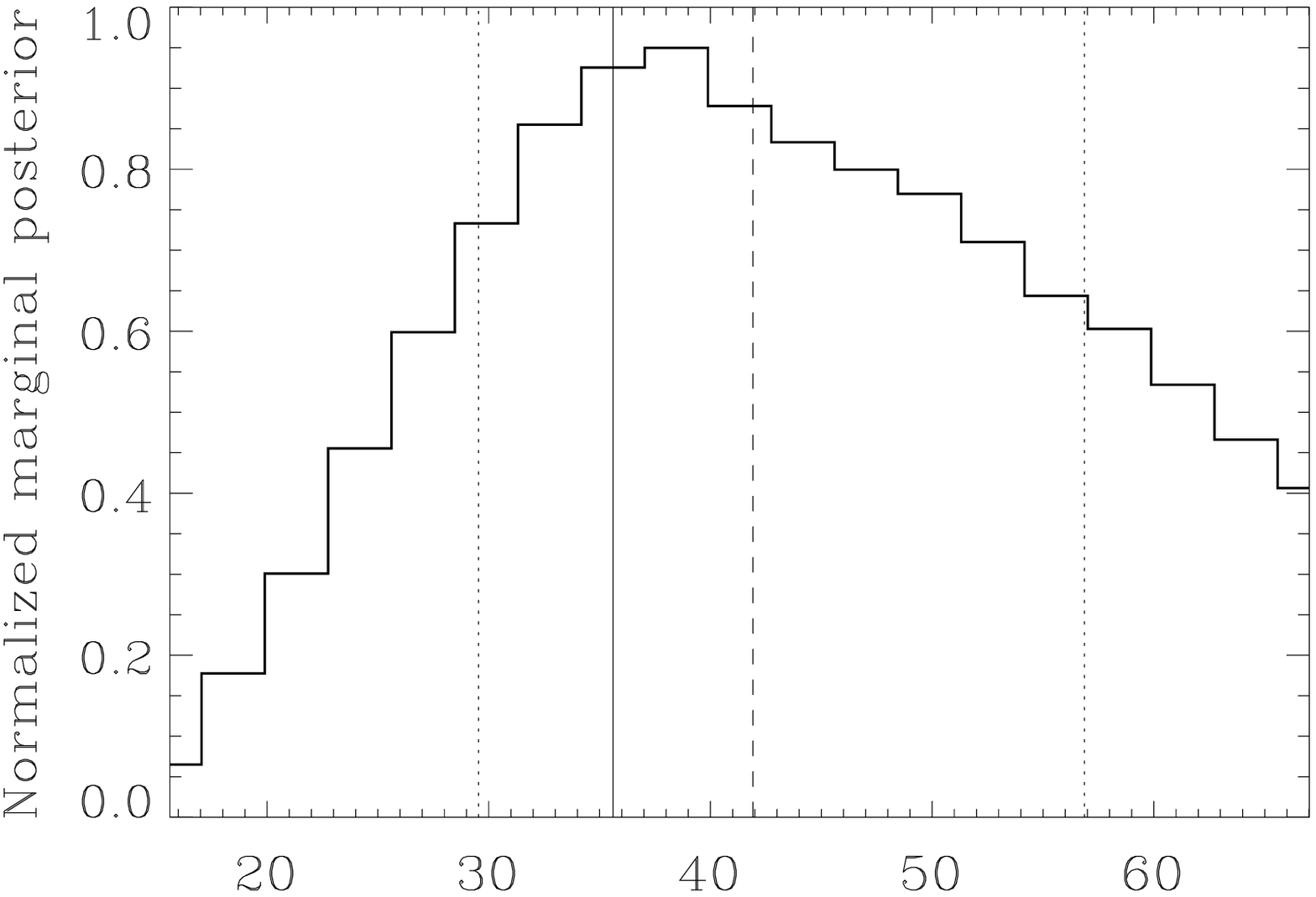}
\includegraphics[width=5.3cm]{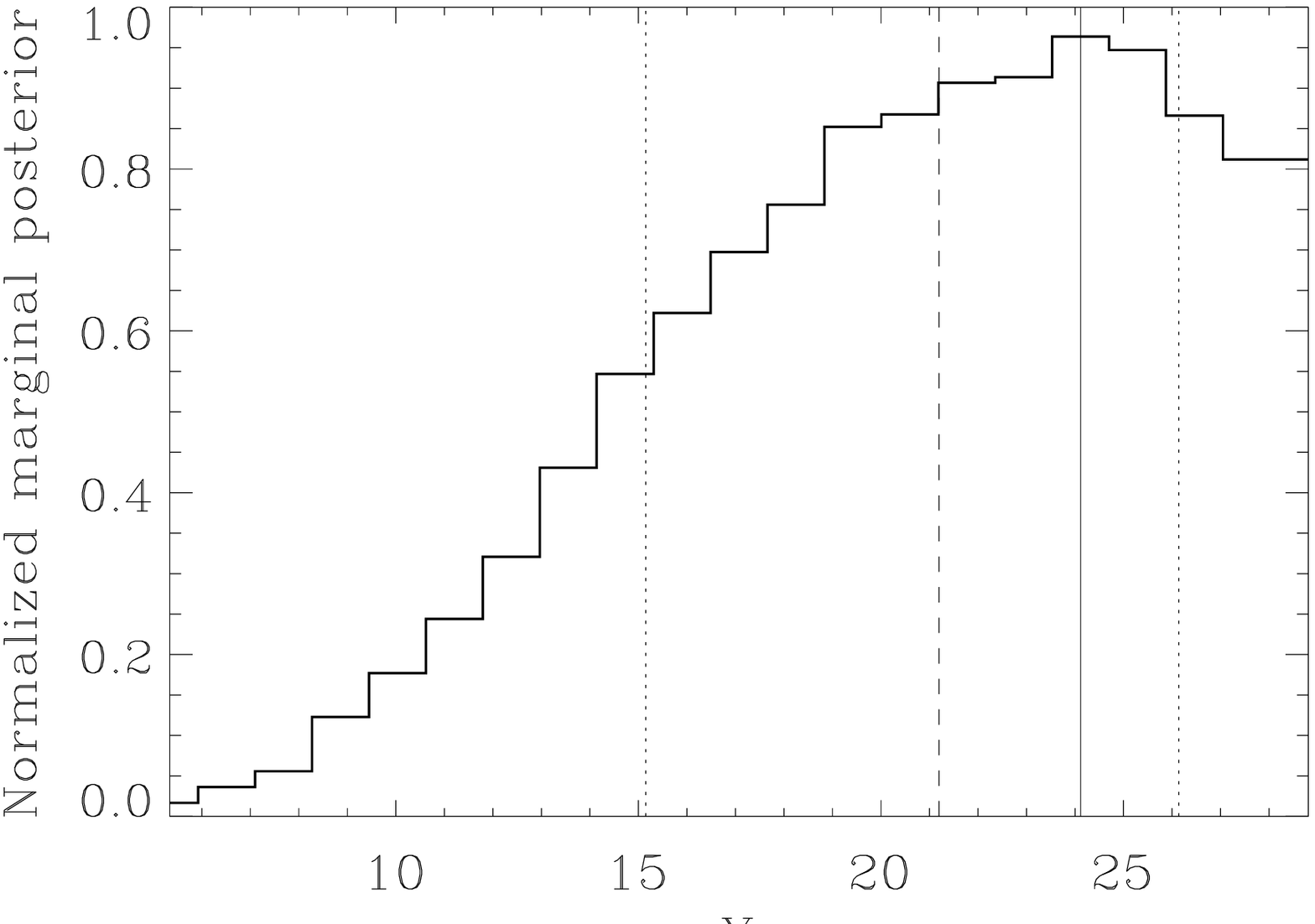}
\includegraphics[width=5.3cm]{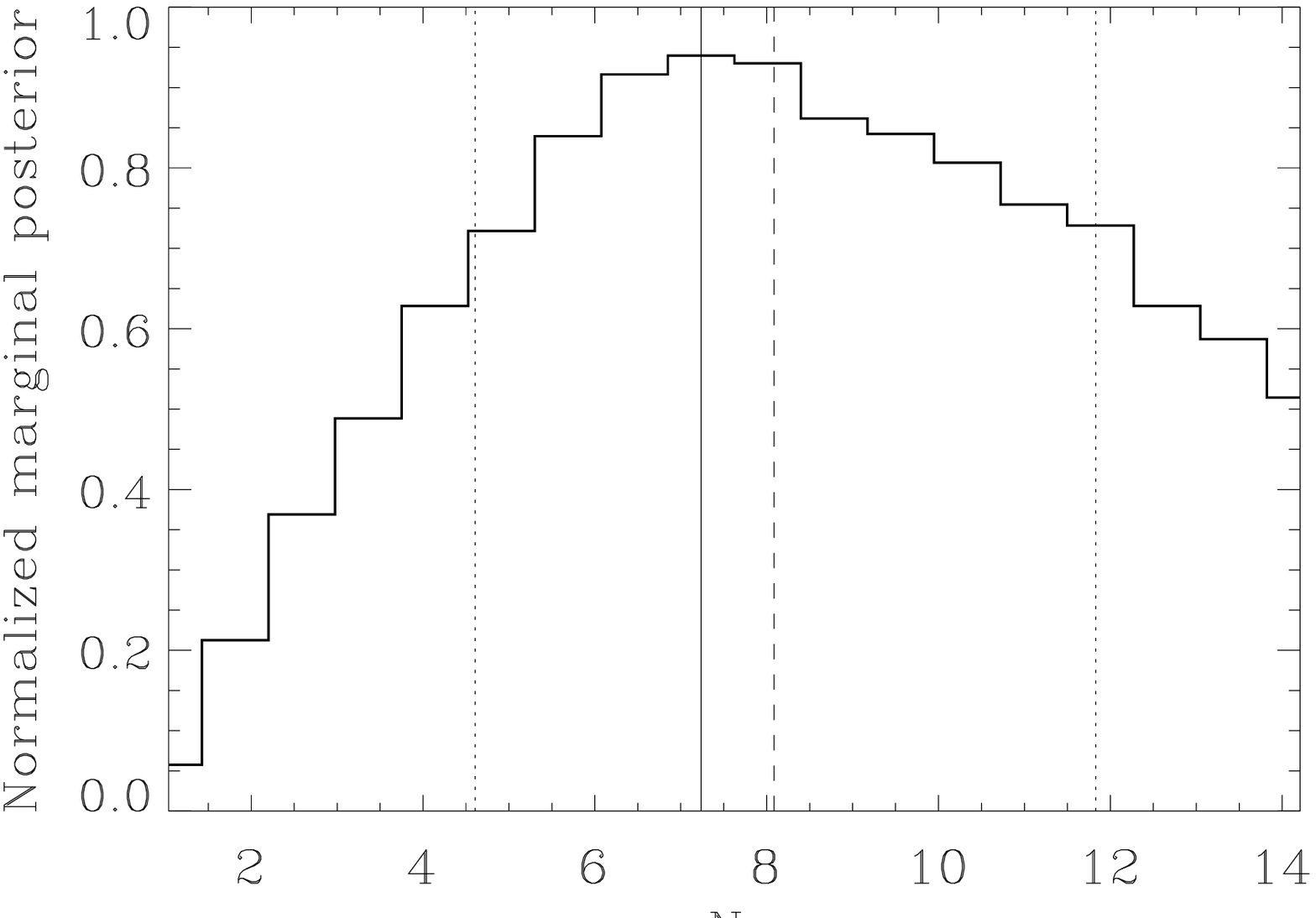}
\includegraphics[width=5.3cm]{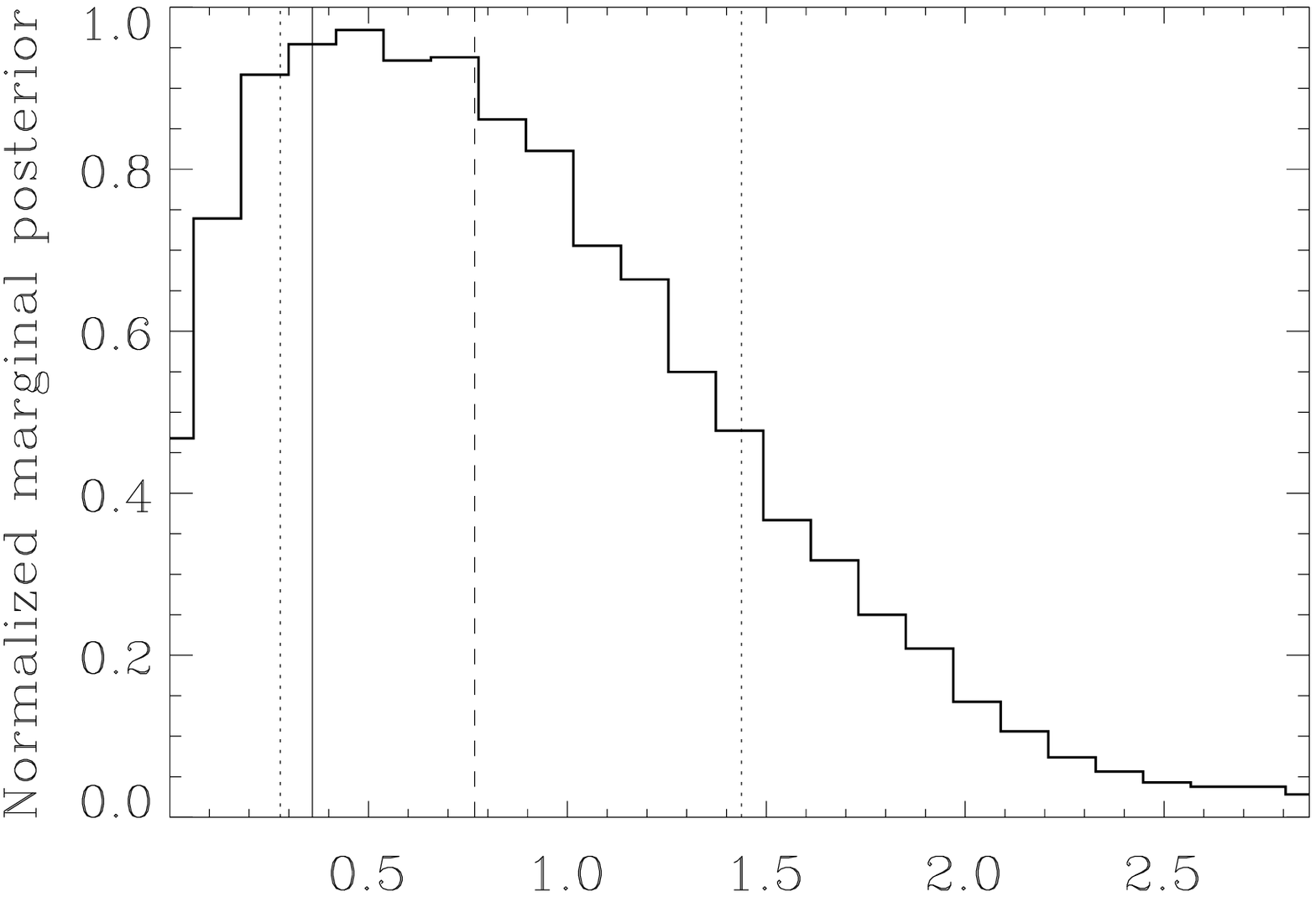}
\includegraphics[width=5.3cm]{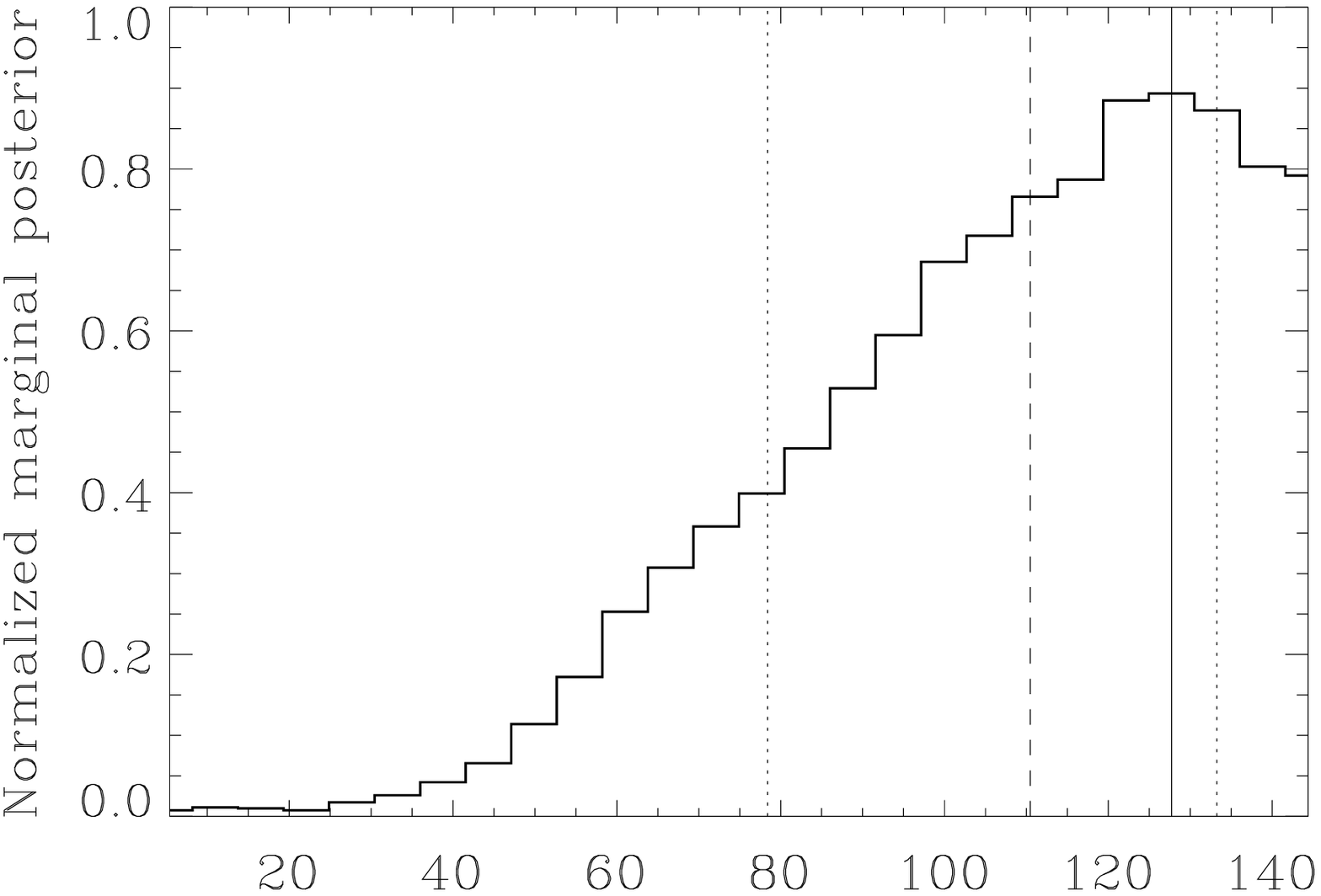}
\includegraphics[width=5.3cm]{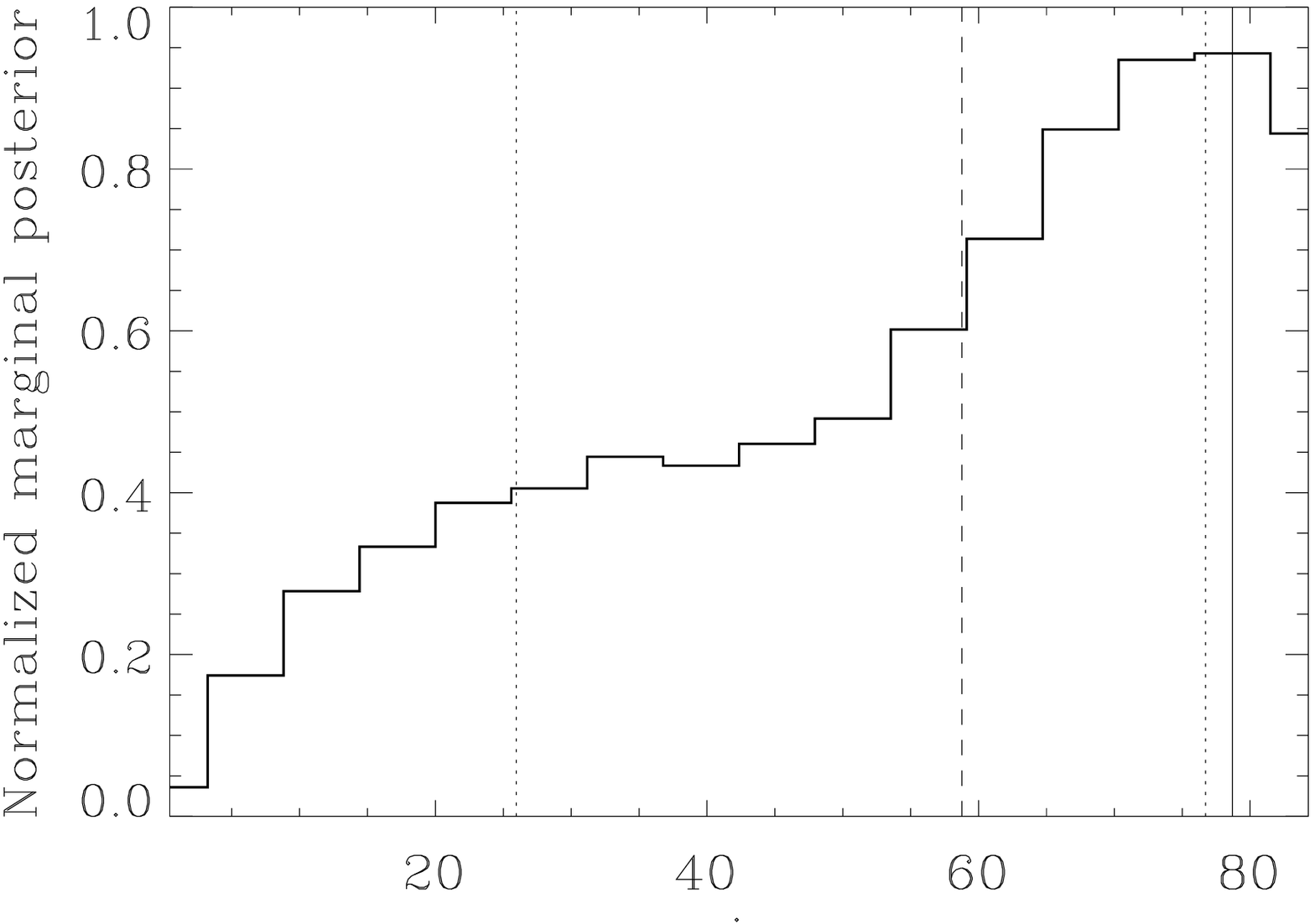}\par}
\caption{\footnotesize{Same as in Figure \ref{ngc1097}, but for the galaxy NGC 6221.}
\label{ngc6221}}
\end{figure*}

\begin{figure*}[!ht]
\centering
{\par
\includegraphics[width=5.3cm]{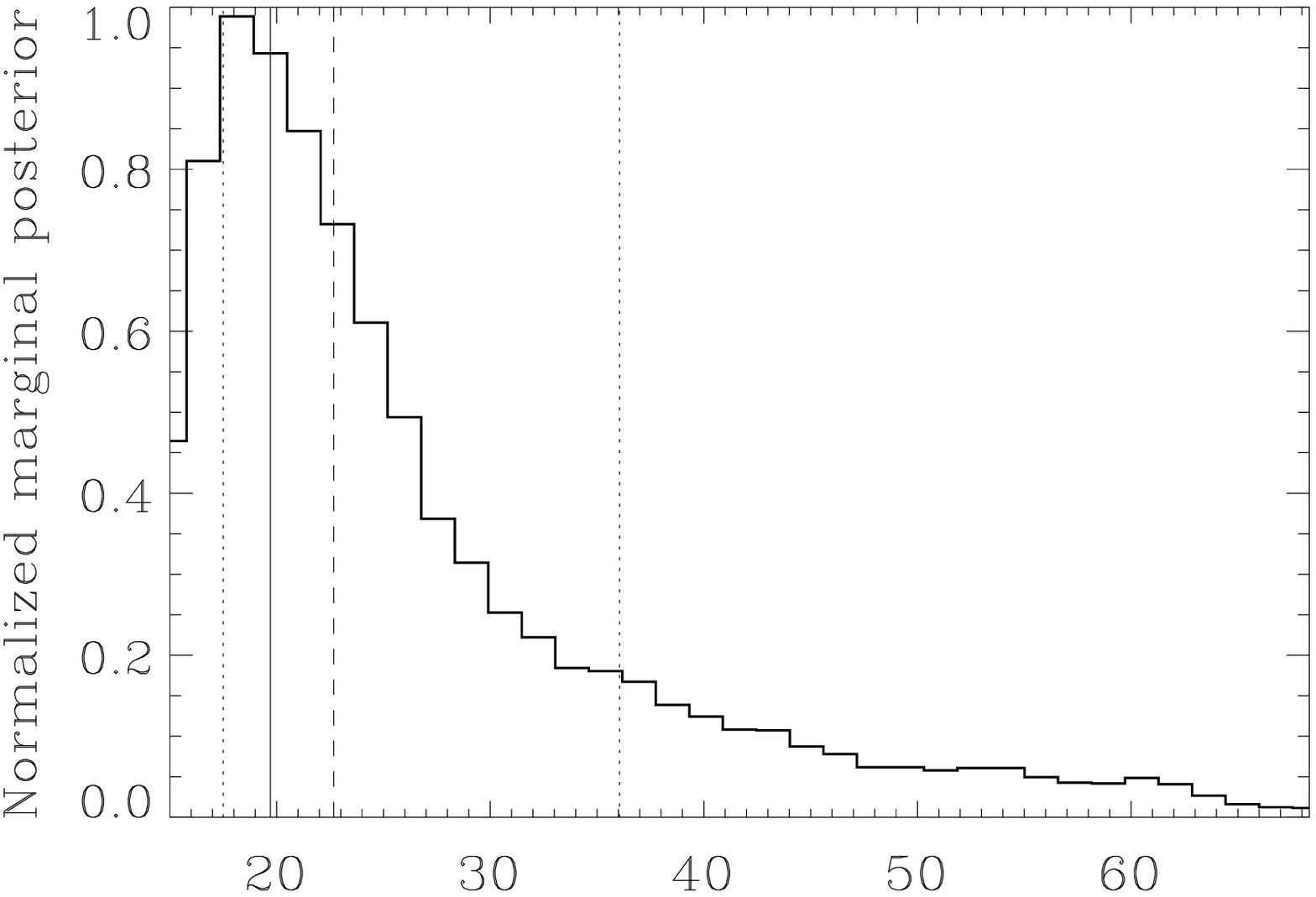}
\includegraphics[width=5.3cm]{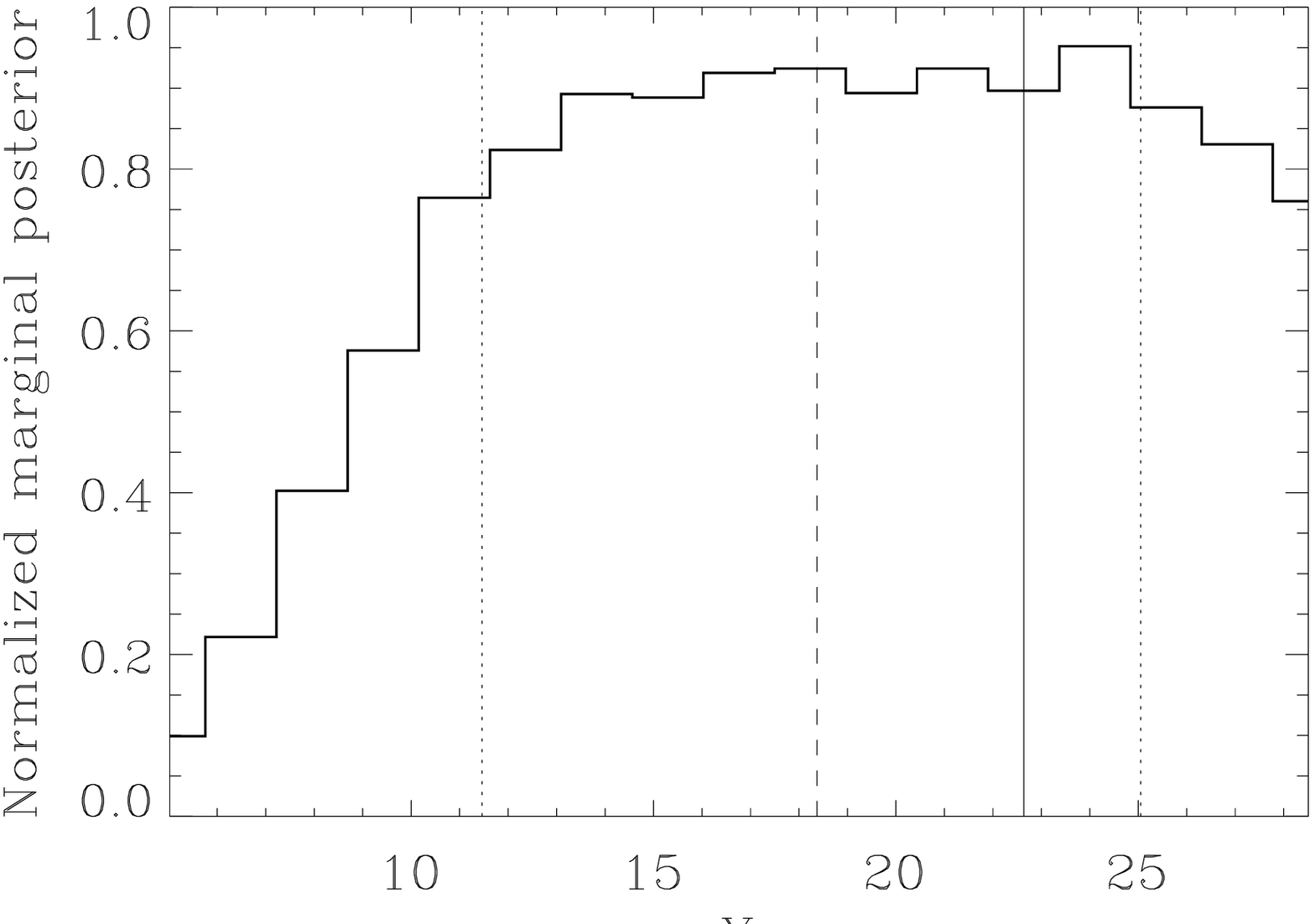}
\includegraphics[width=5.3cm]{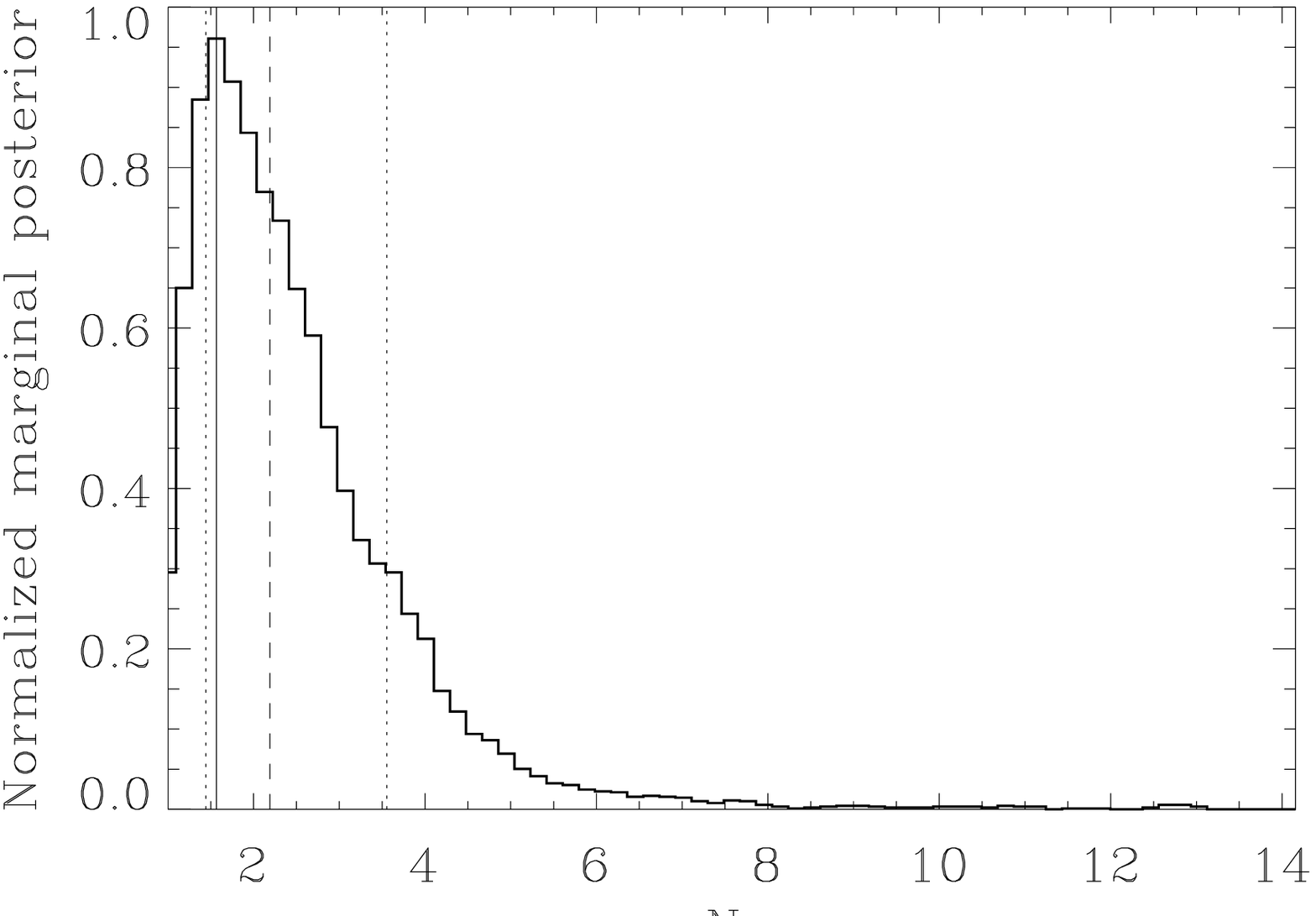}
\includegraphics[width=5.3cm]{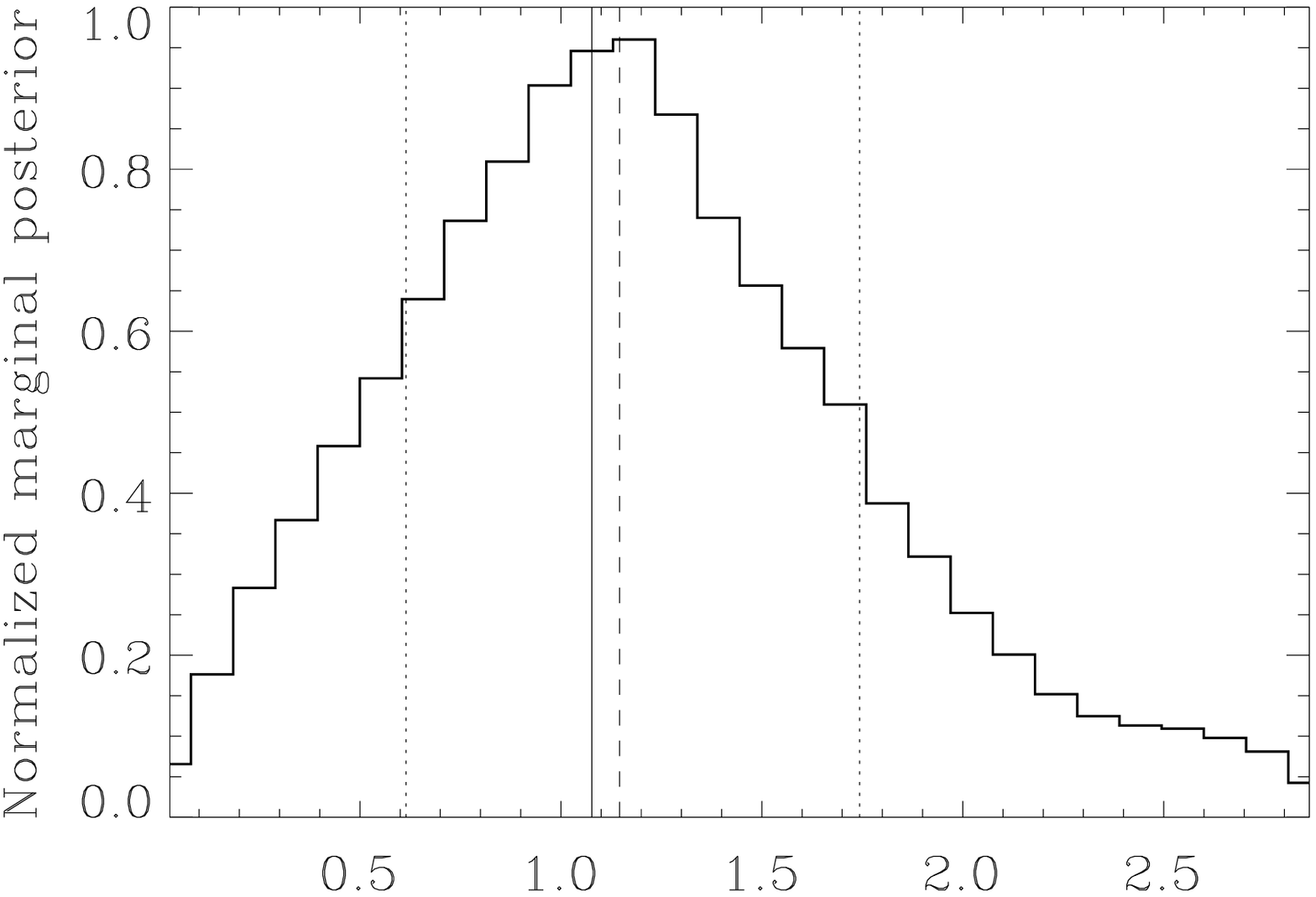}
\includegraphics[width=5.3cm]{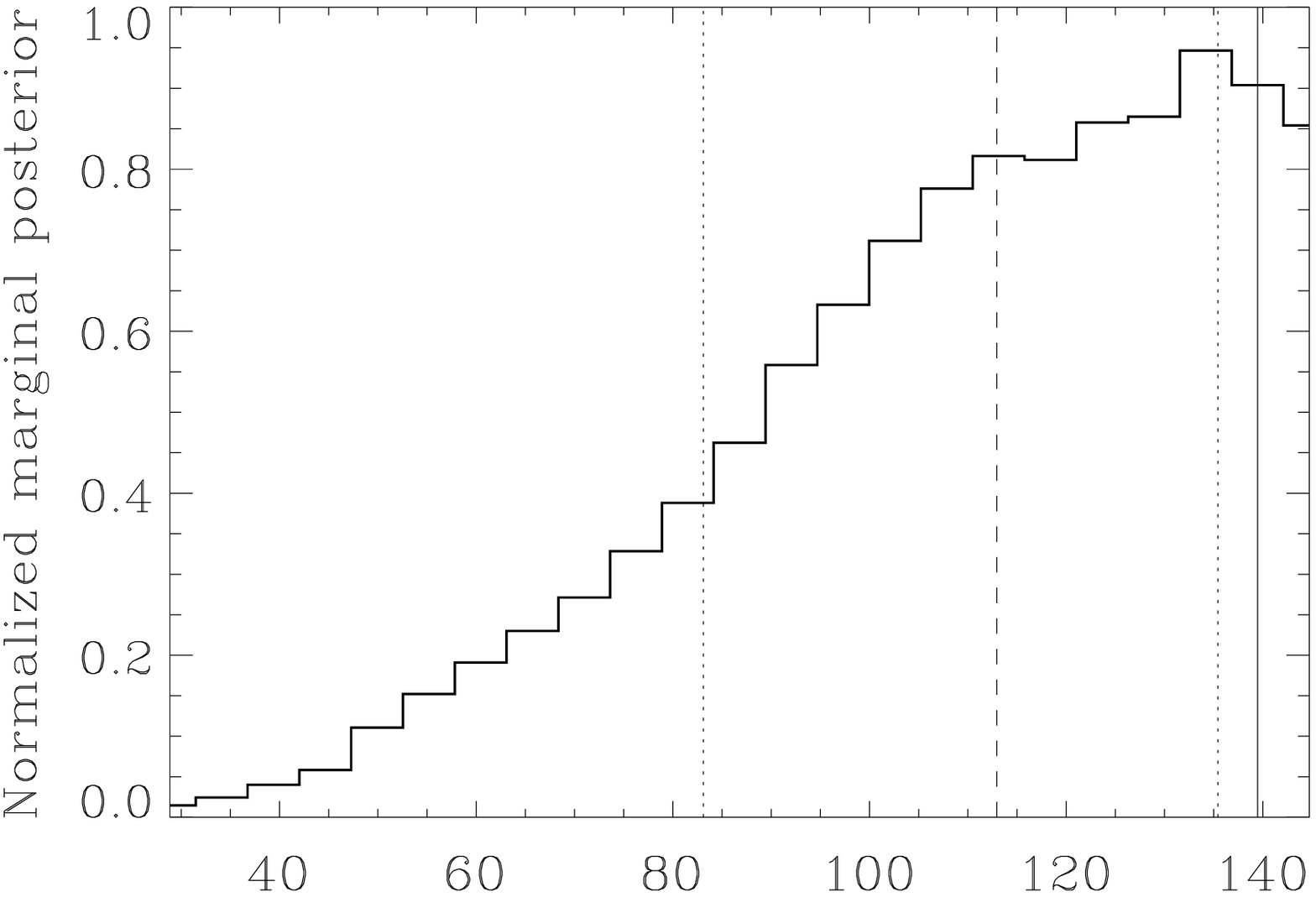}
\includegraphics[width=5.3cm]{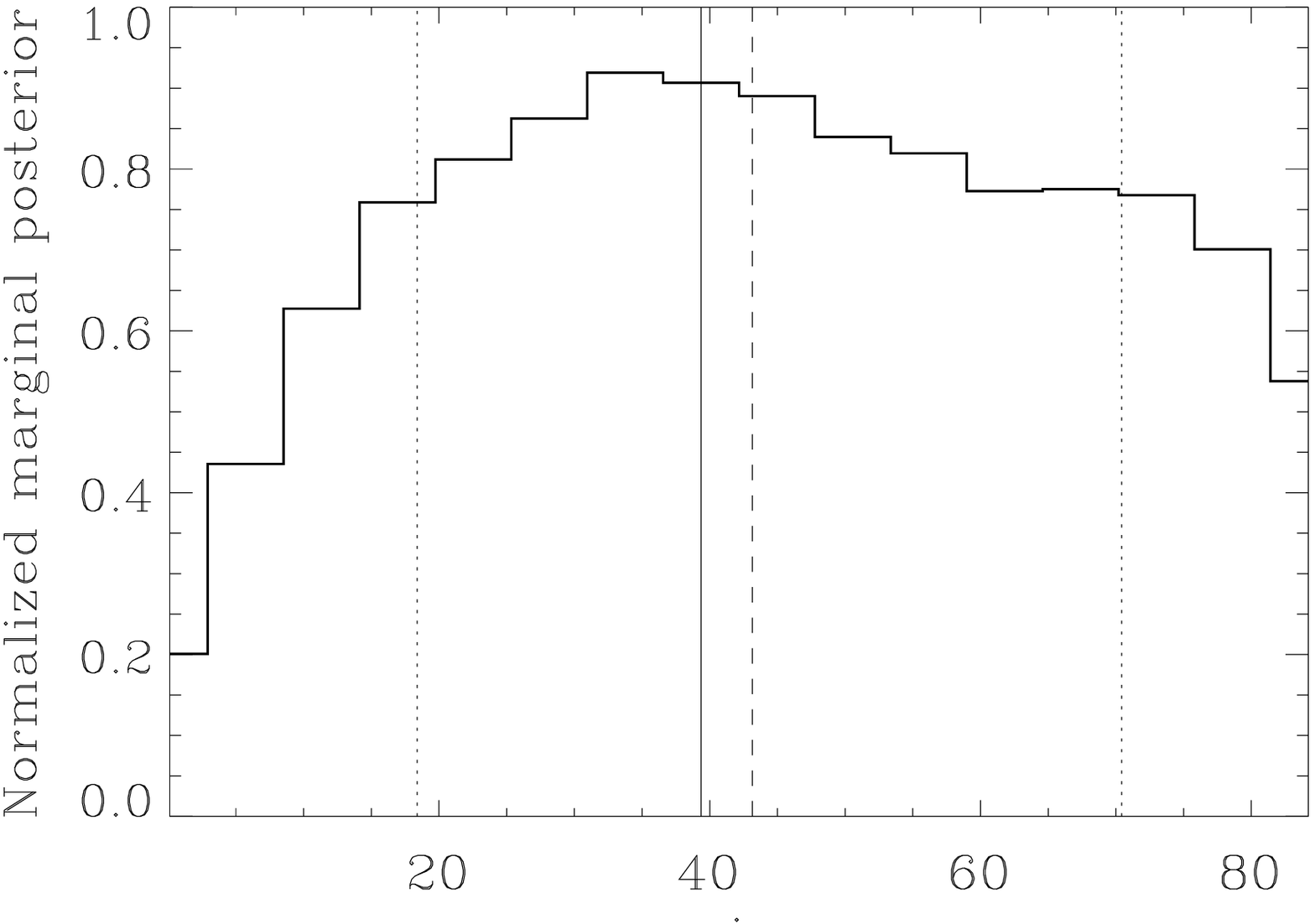}\par}
\caption{\footnotesize{Same as in Figure \ref{ngc1097}, but for the galaxy NGC 6814.}
\label{ngc6814}}
\end{figure*}

\begin{figure*}[!ht]
\centering
{\par
\includegraphics[width=5.3cm]{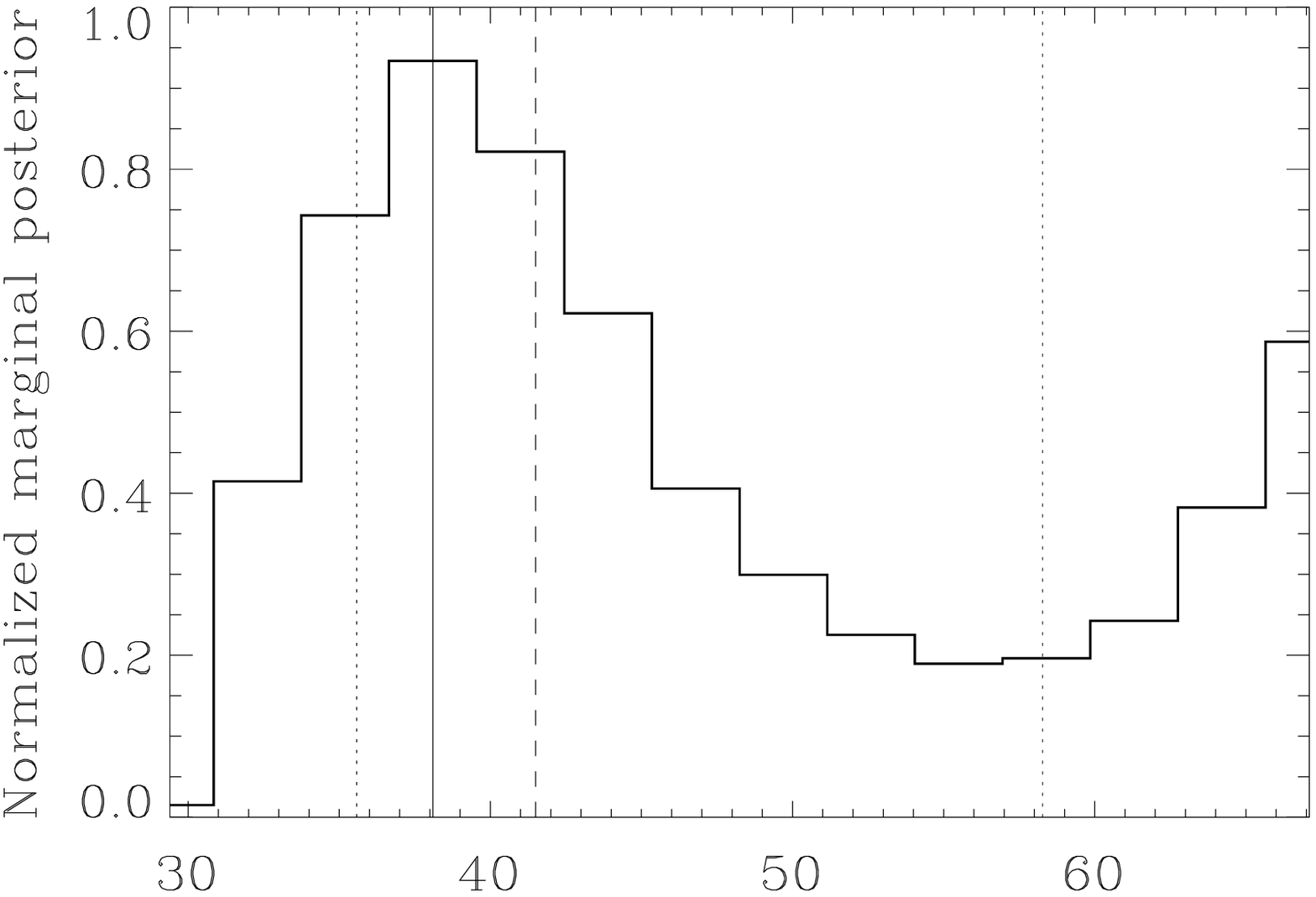}
\includegraphics[width=5.3cm]{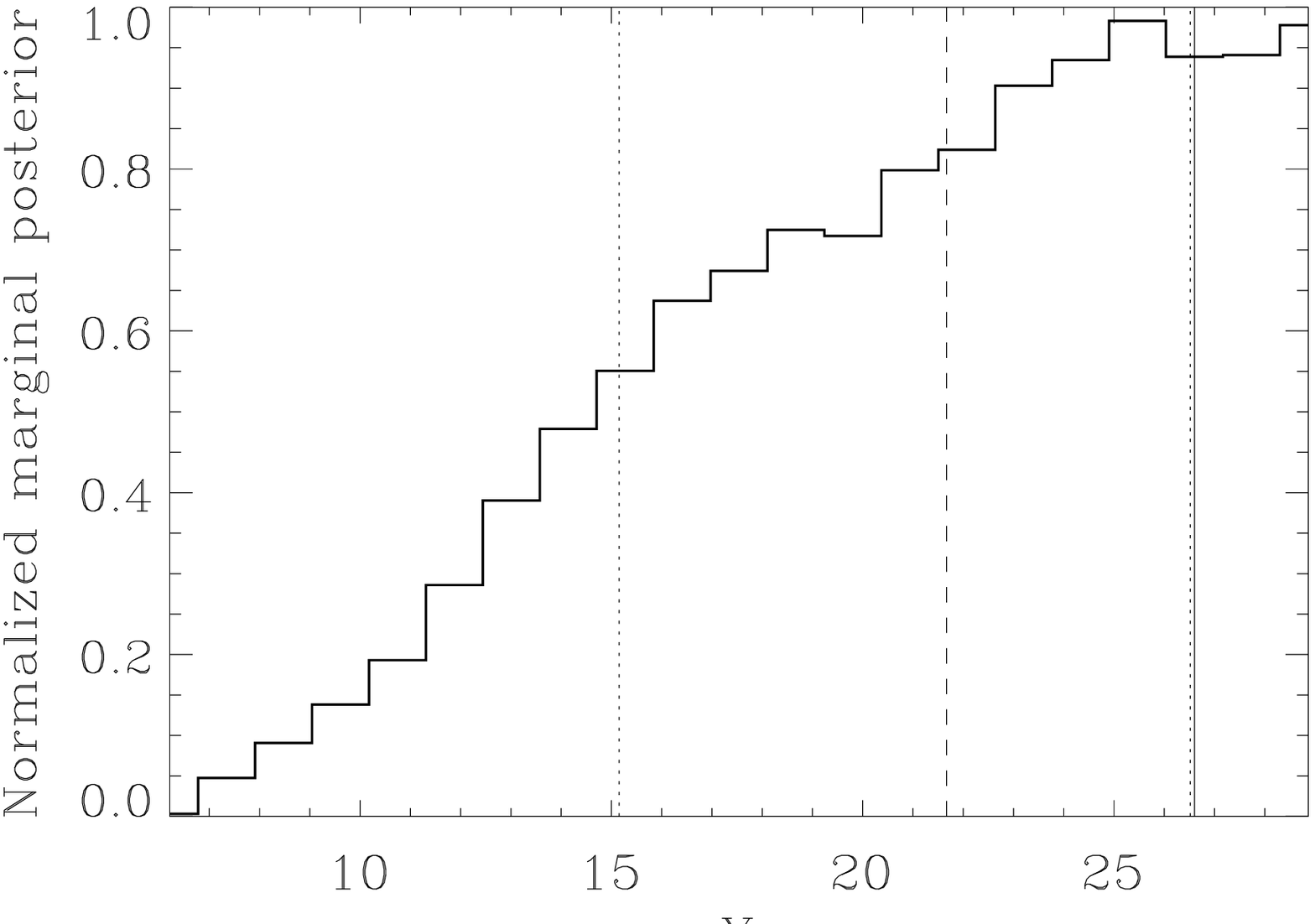}
\includegraphics[width=5.3cm]{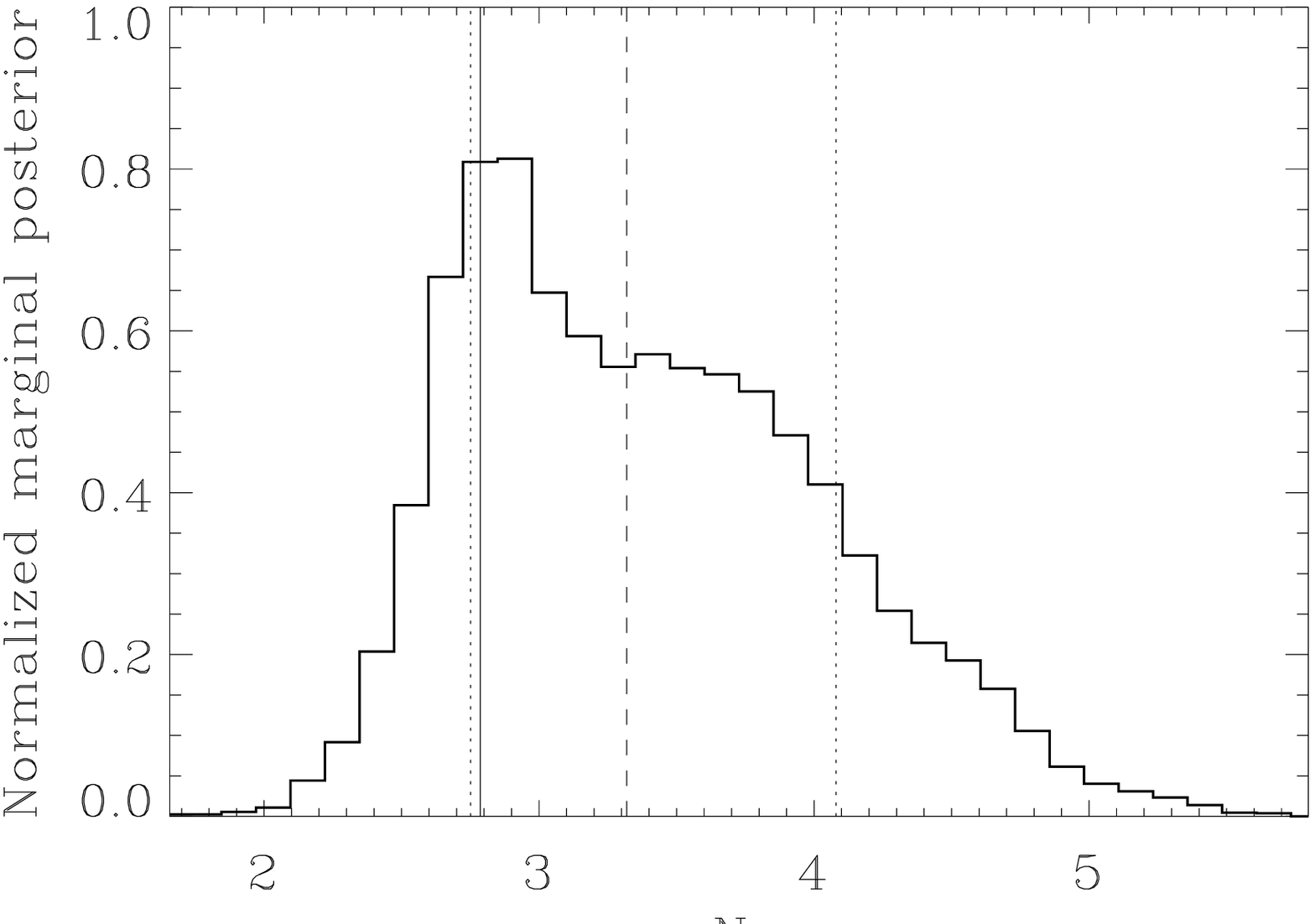}
\includegraphics[width=5.3cm]{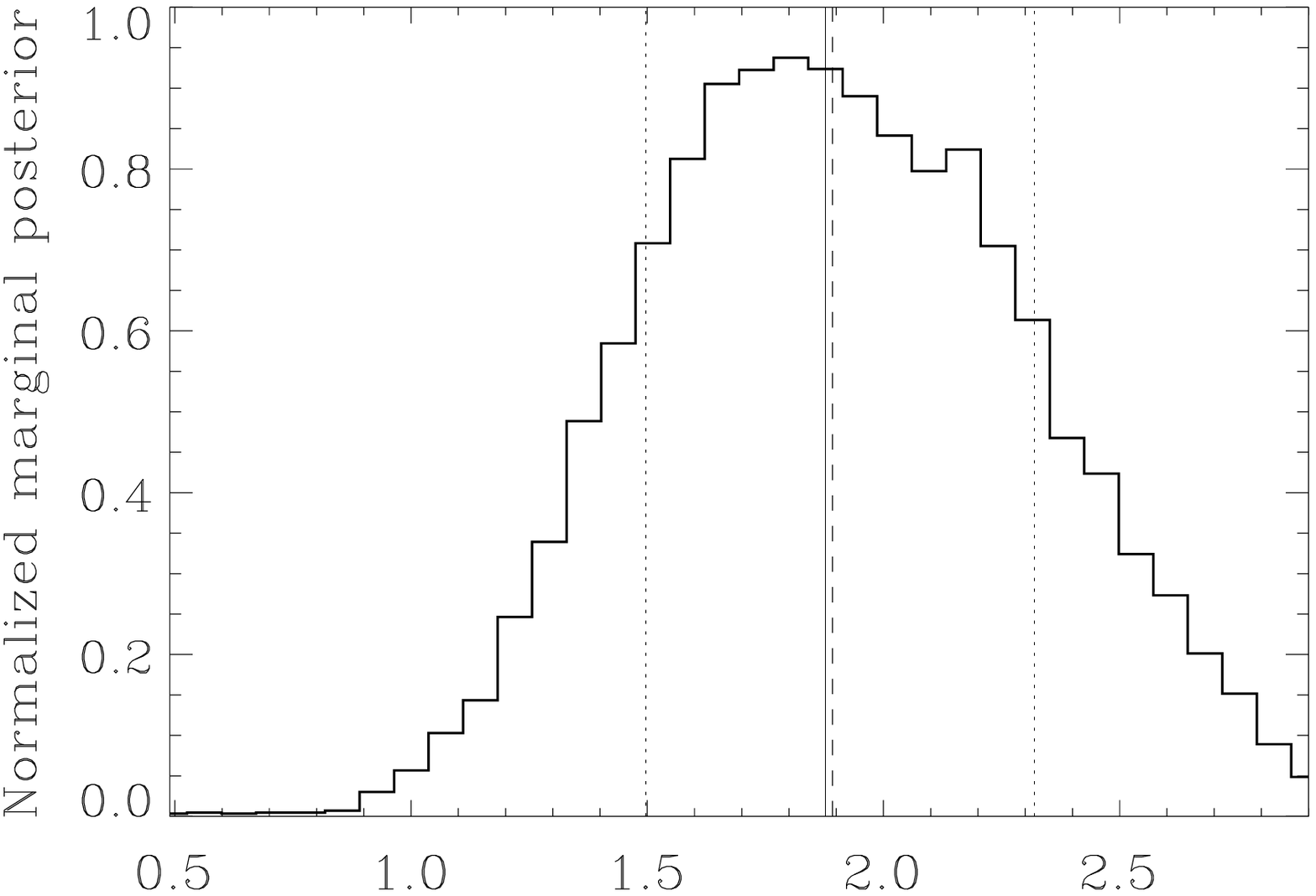}
\includegraphics[width=5.3cm]{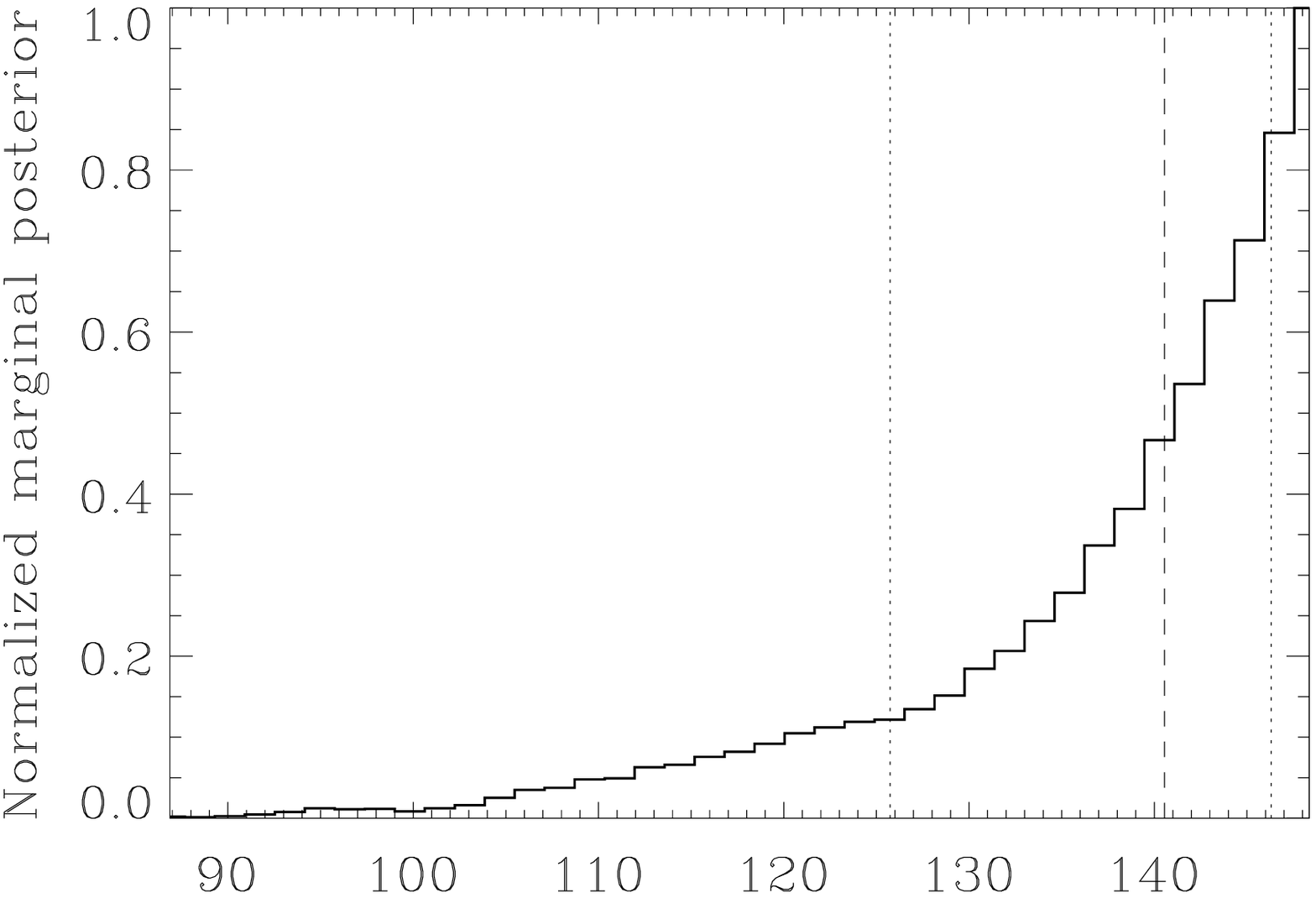}
\includegraphics[width=5.3cm]{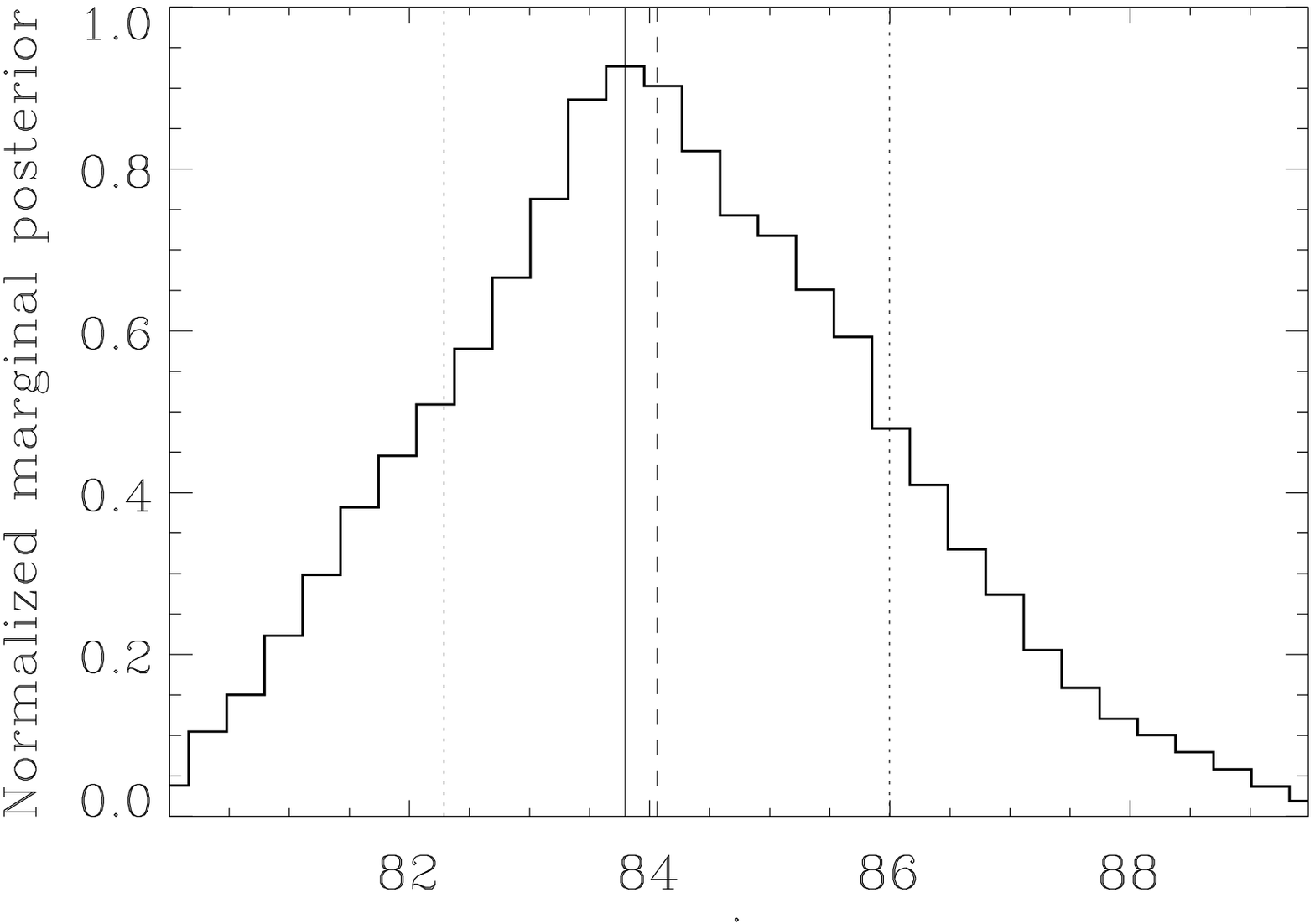}\par}
\caption{\footnotesize{Same as in Figure \ref{ngc1097}, but for the galaxy NGC 7469.}
\label{ngc7469}}
\end{figure*}

\begin{figure*}[!ht]
\centering
{\par
\includegraphics[width=5.3cm]{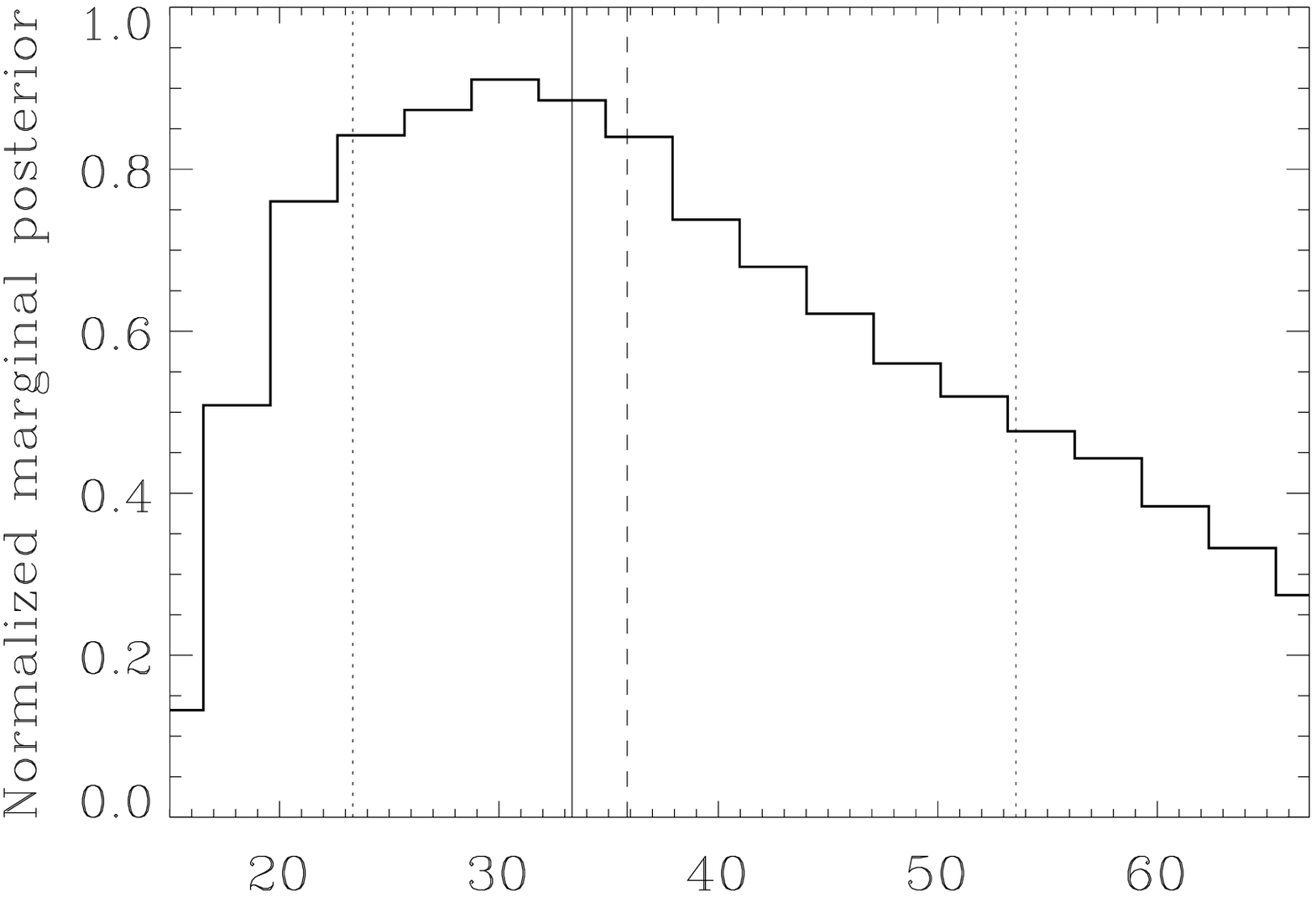}
\includegraphics[width=5.3cm]{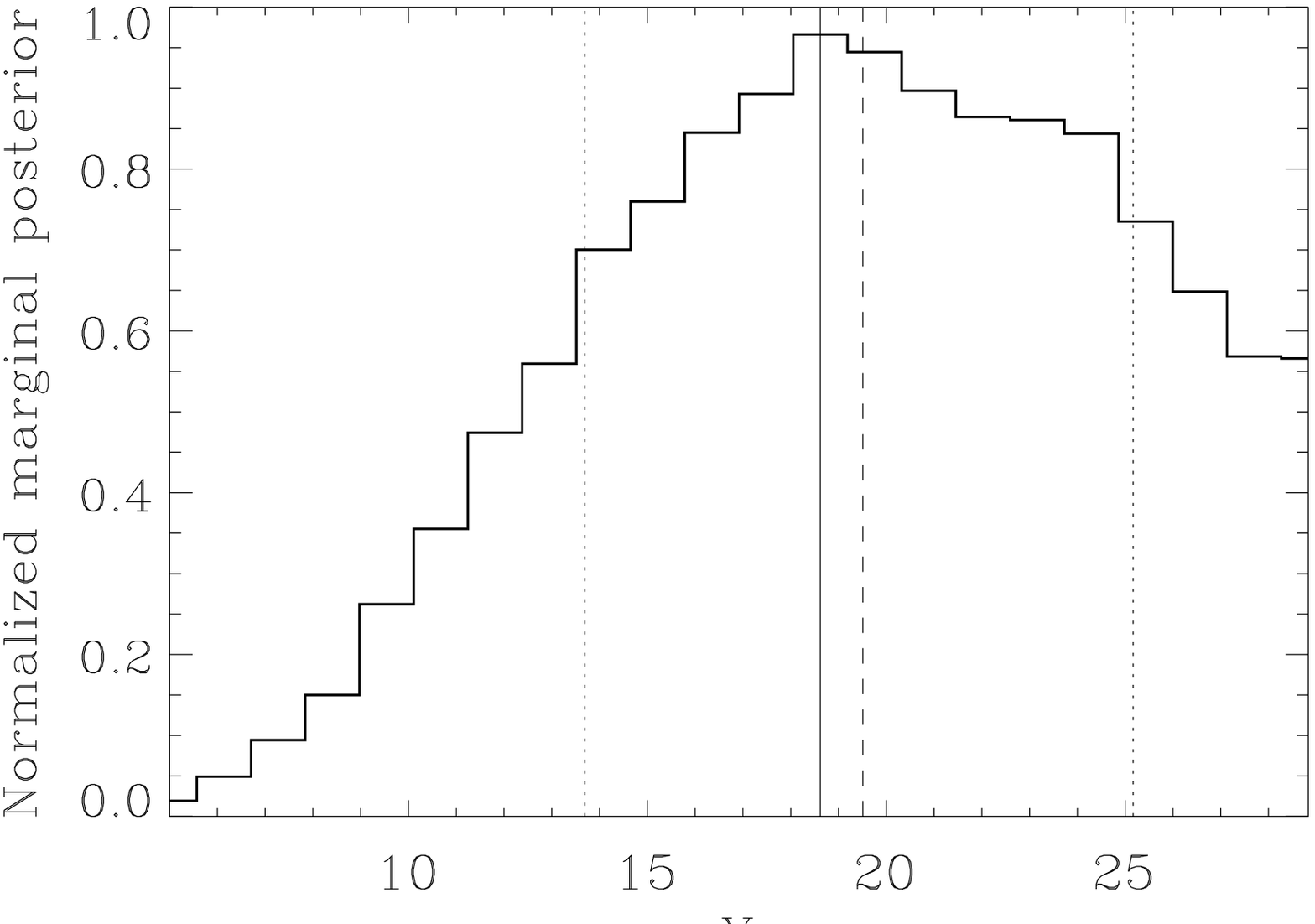}
\includegraphics[width=5.3cm]{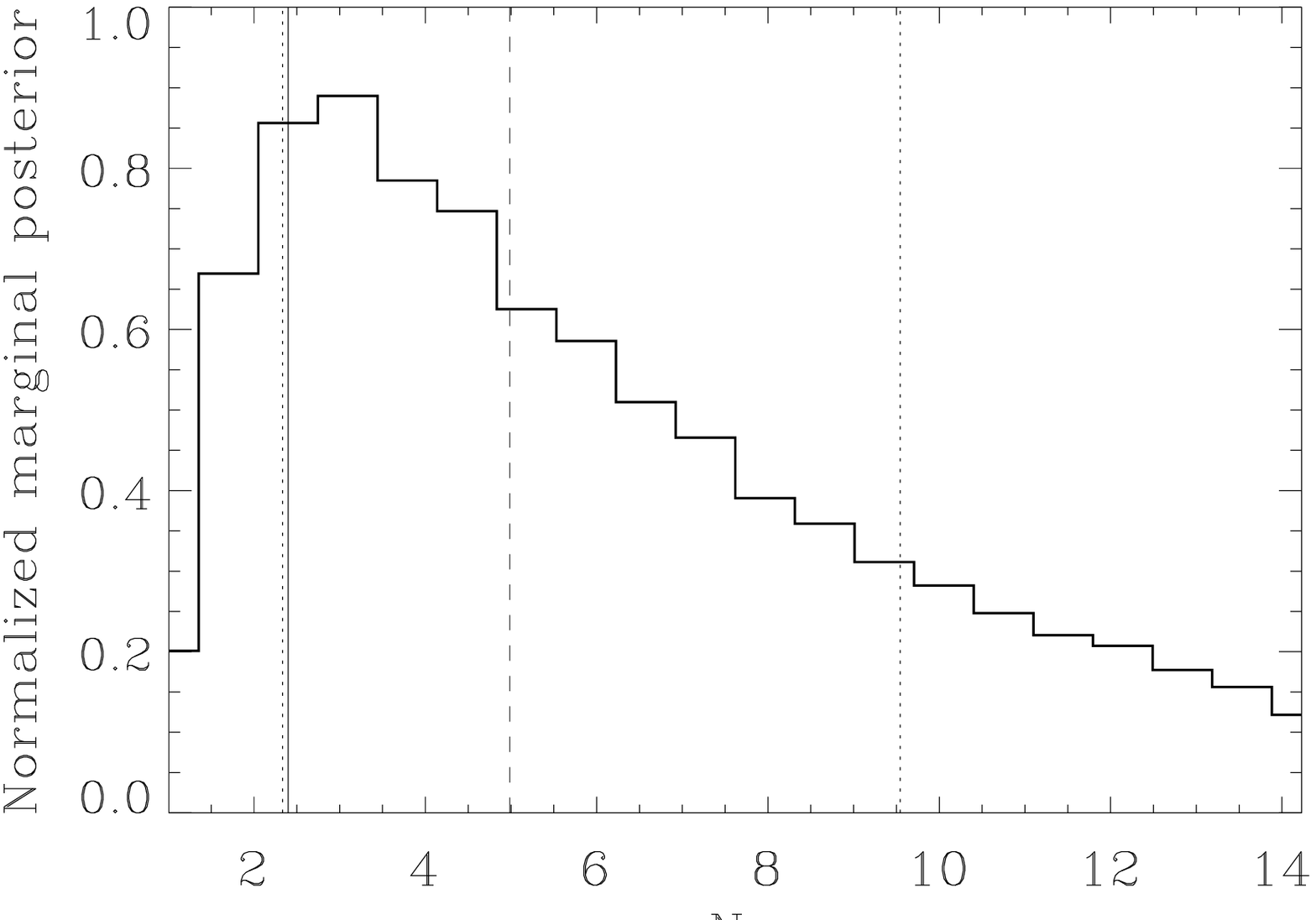}
\includegraphics[width=5.3cm]{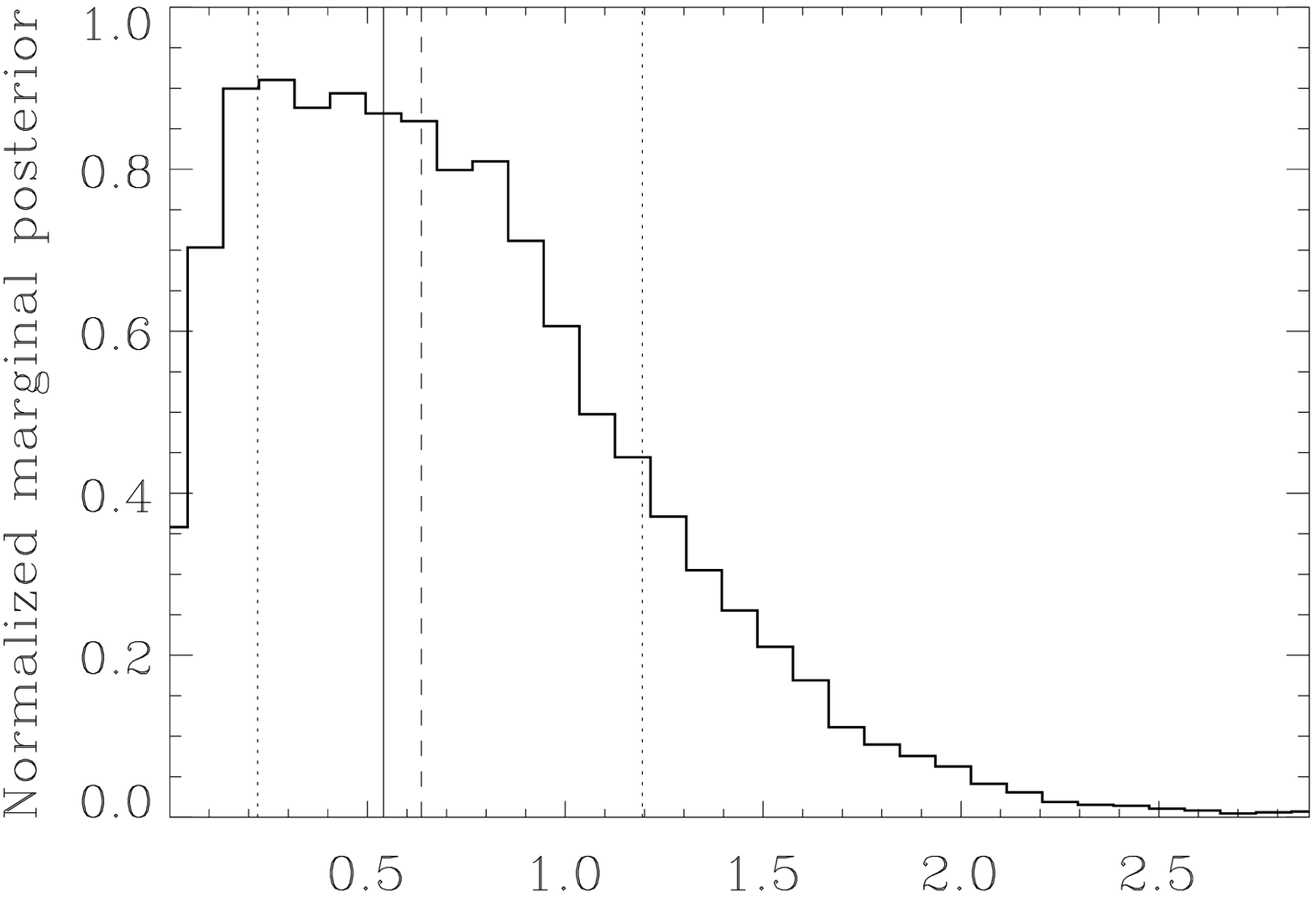}
\includegraphics[width=5.3cm]{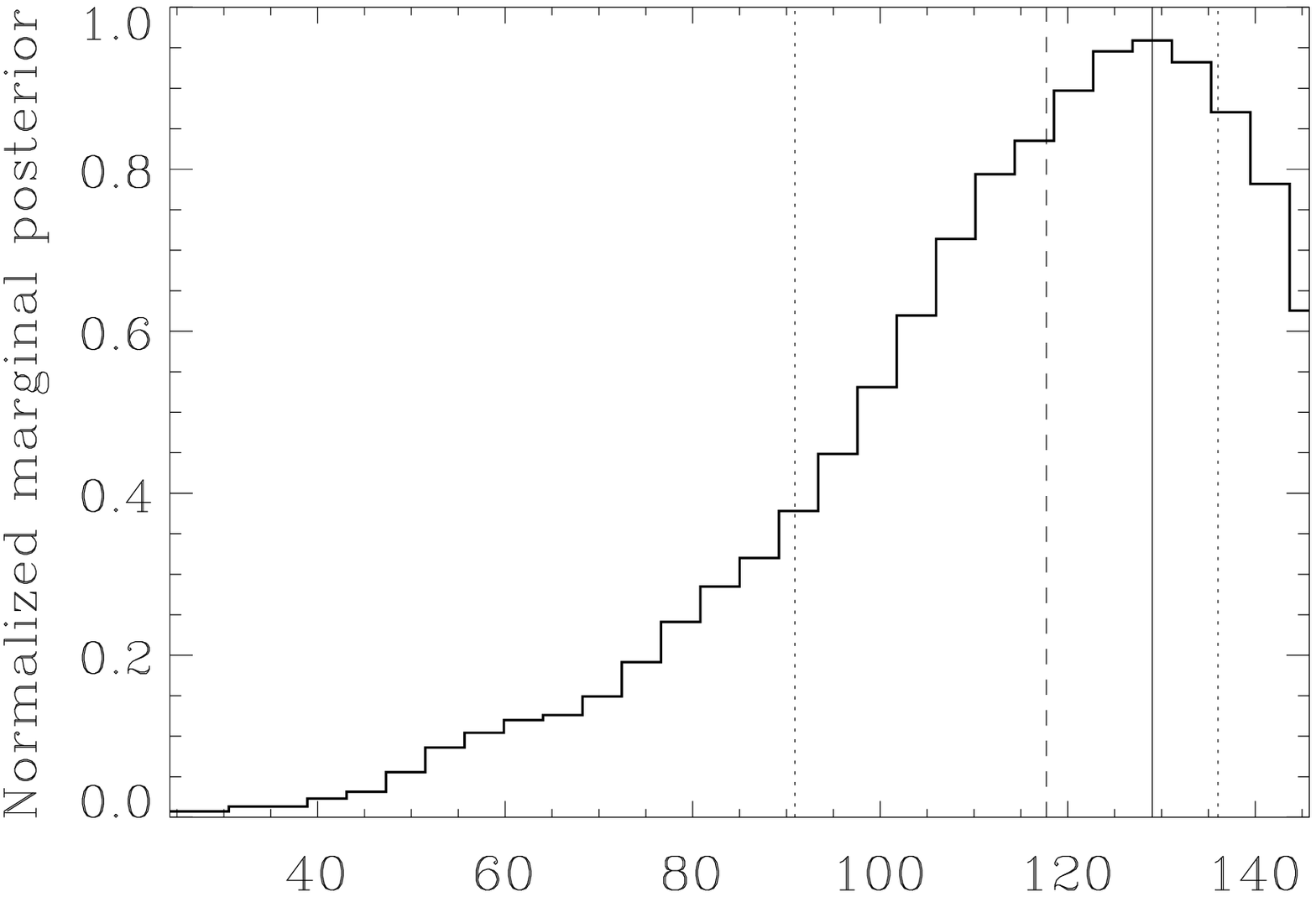}
\includegraphics[width=5.3cm]{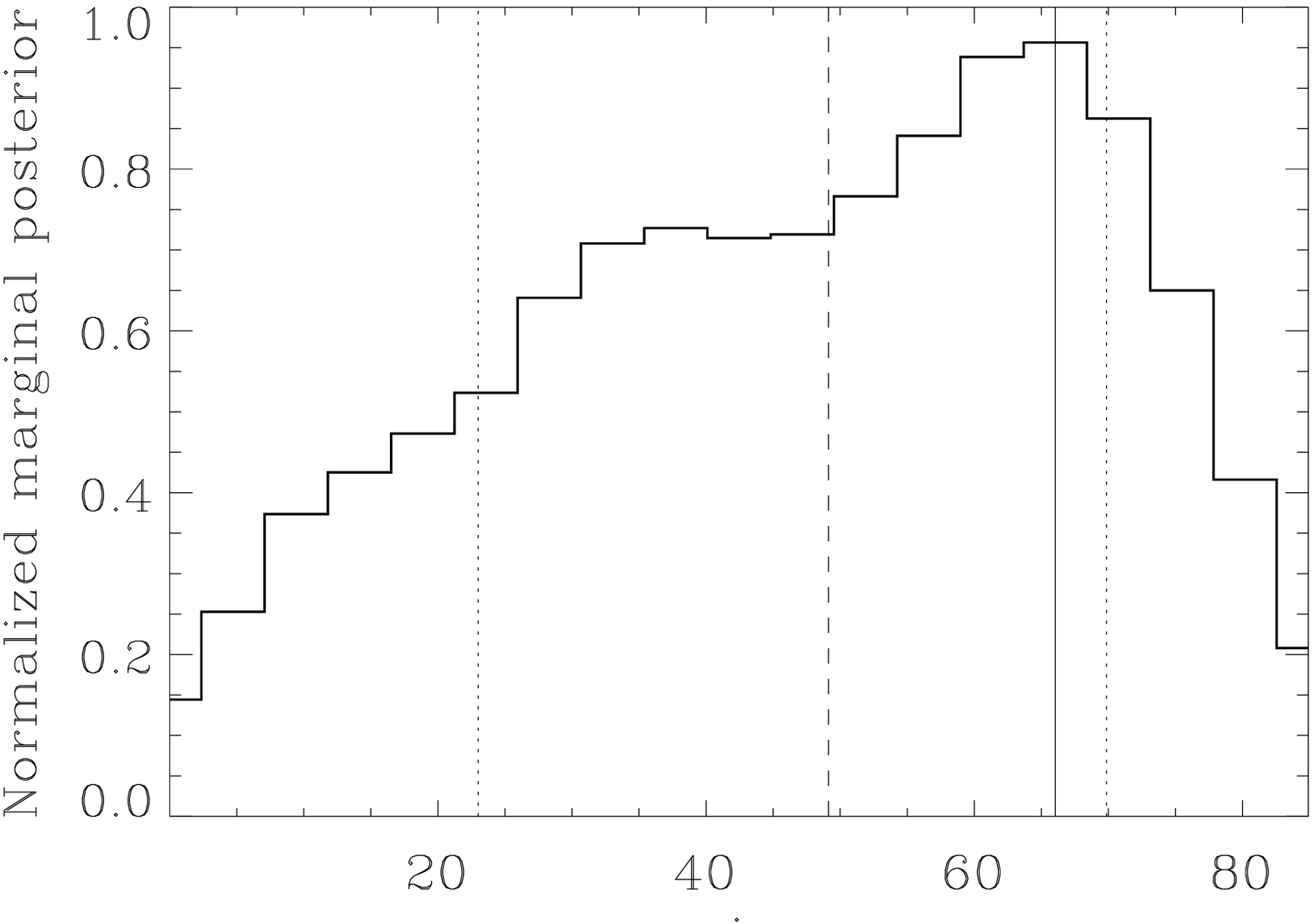}\par}
\caption{\footnotesize{Same as in Figure \ref{ngc1097}, but for the galaxy NGC 3227.}
\label{ngc3227}}
\end{figure*}

\begin{figure*}[!ht]
\centering
{\par
\includegraphics[width=5.3cm]{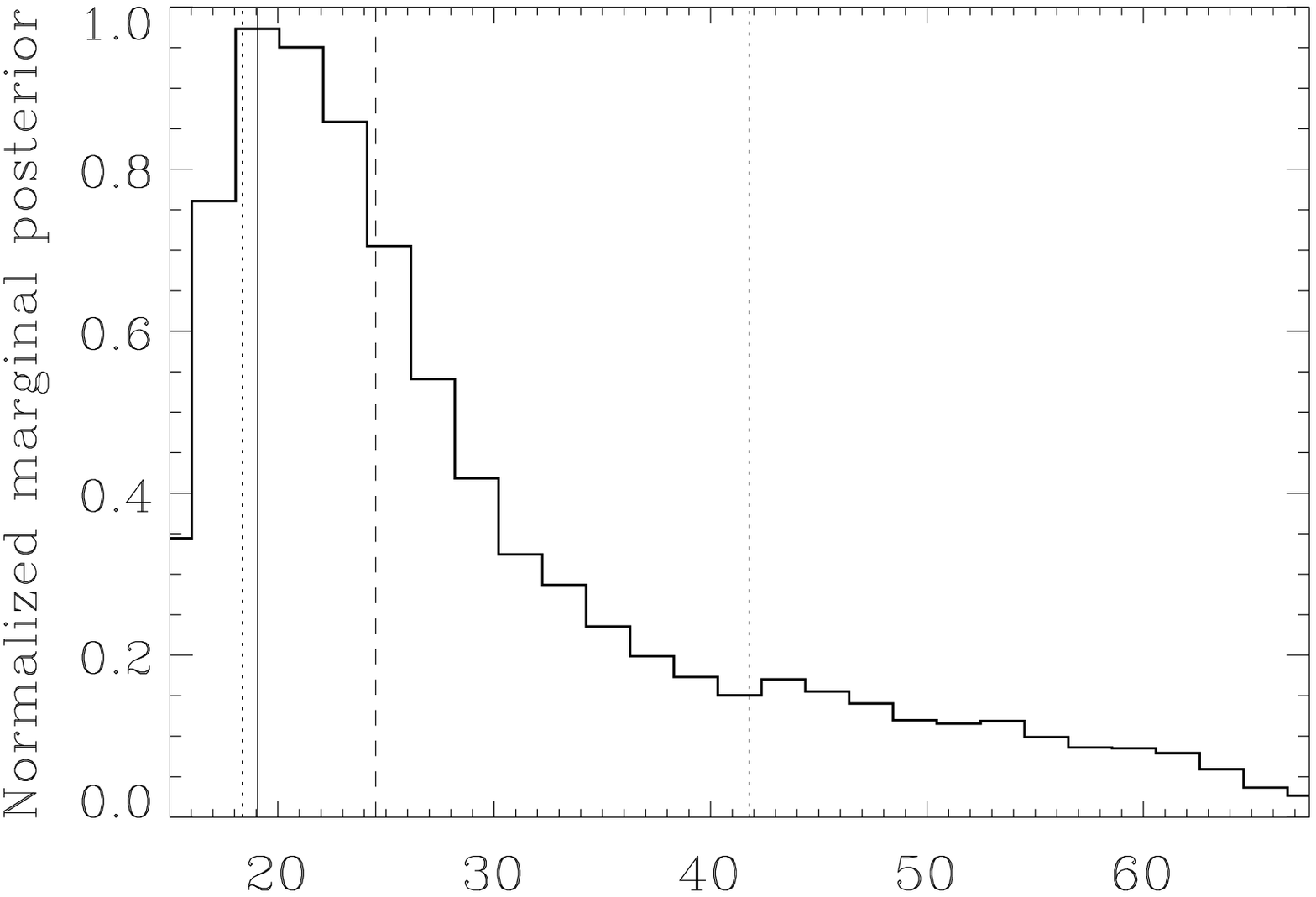}
\includegraphics[width=5.3cm]{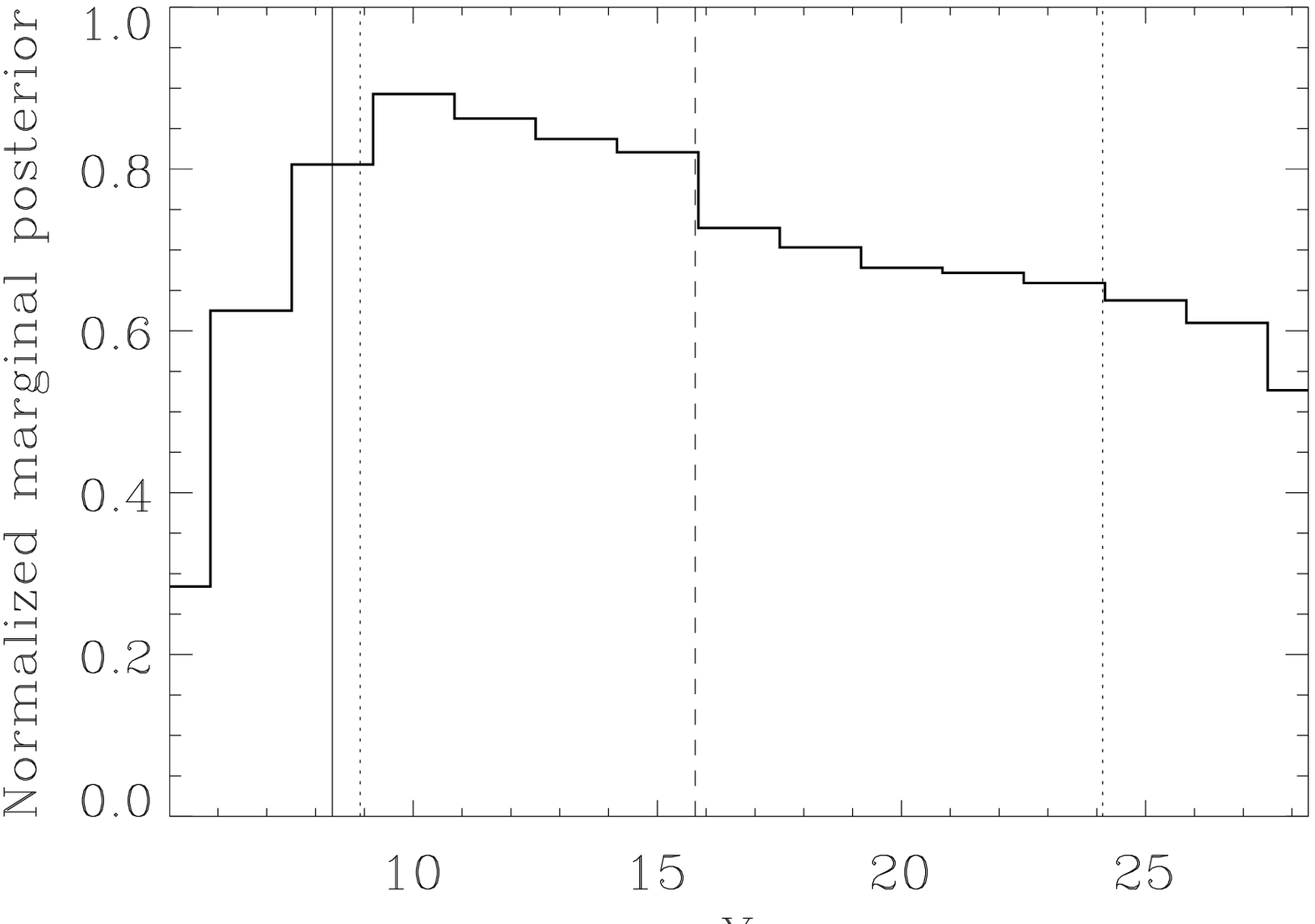}
\includegraphics[width=5.3cm]{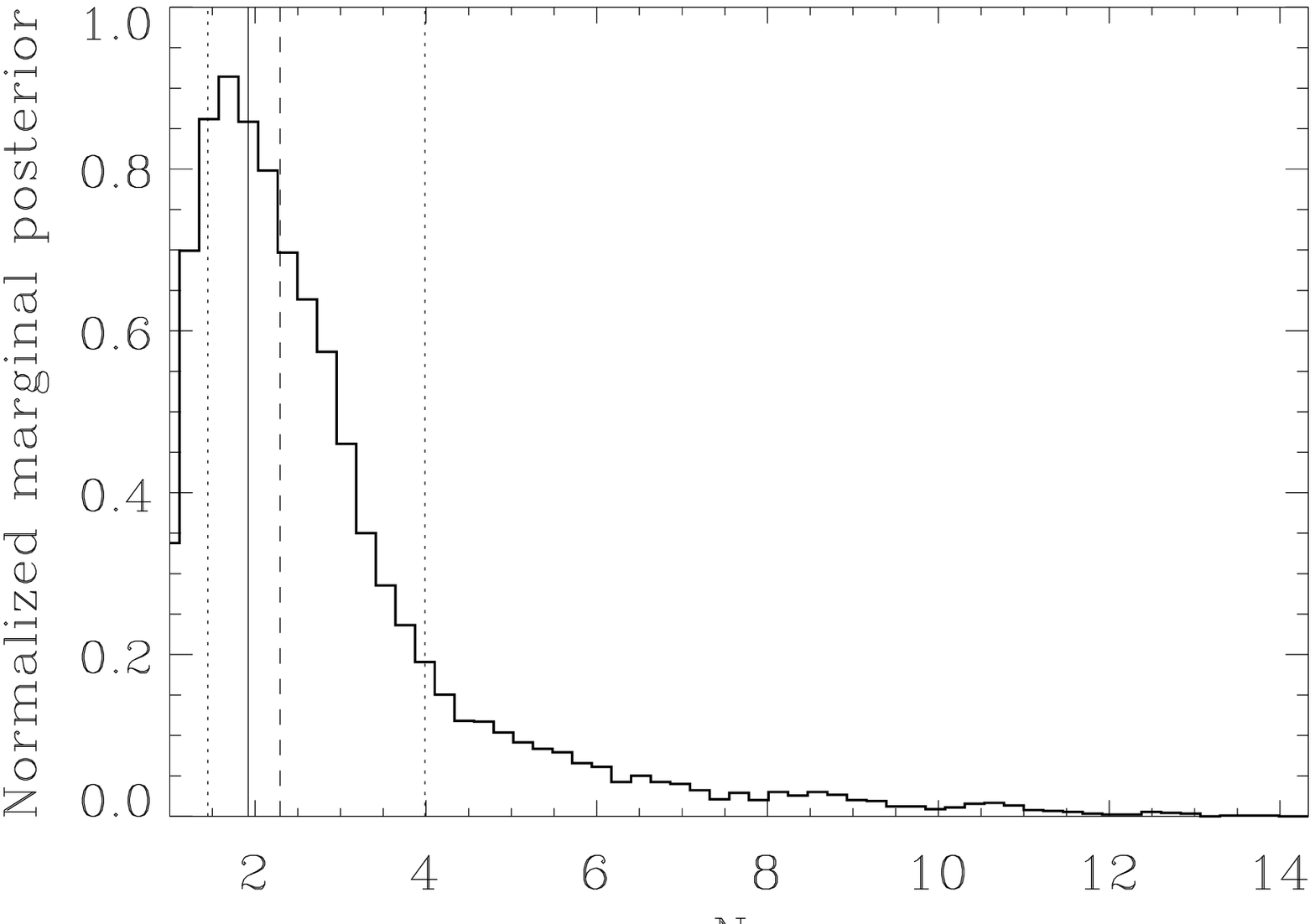}
\includegraphics[width=5.3cm]{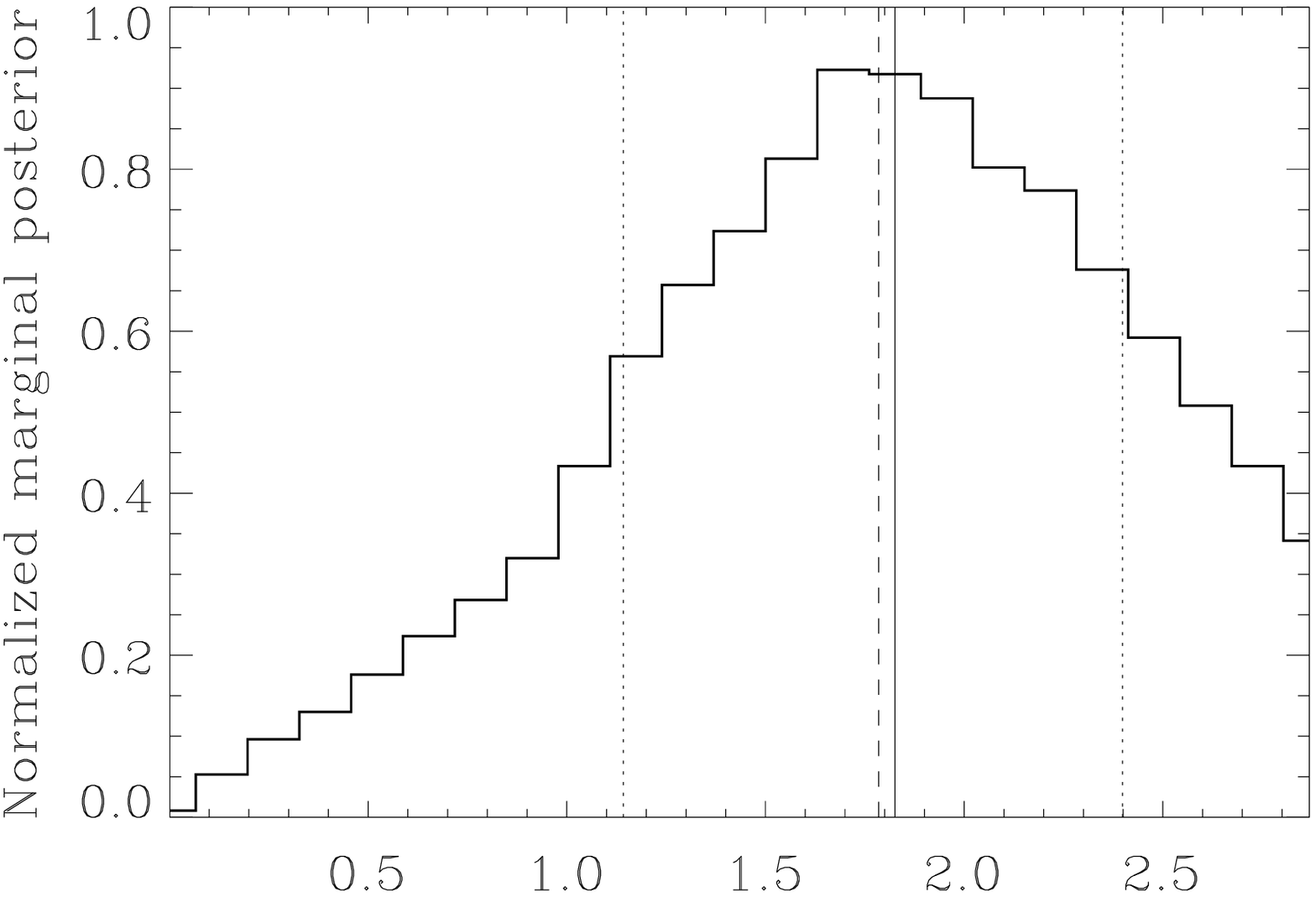}
\includegraphics[width=5.3cm]{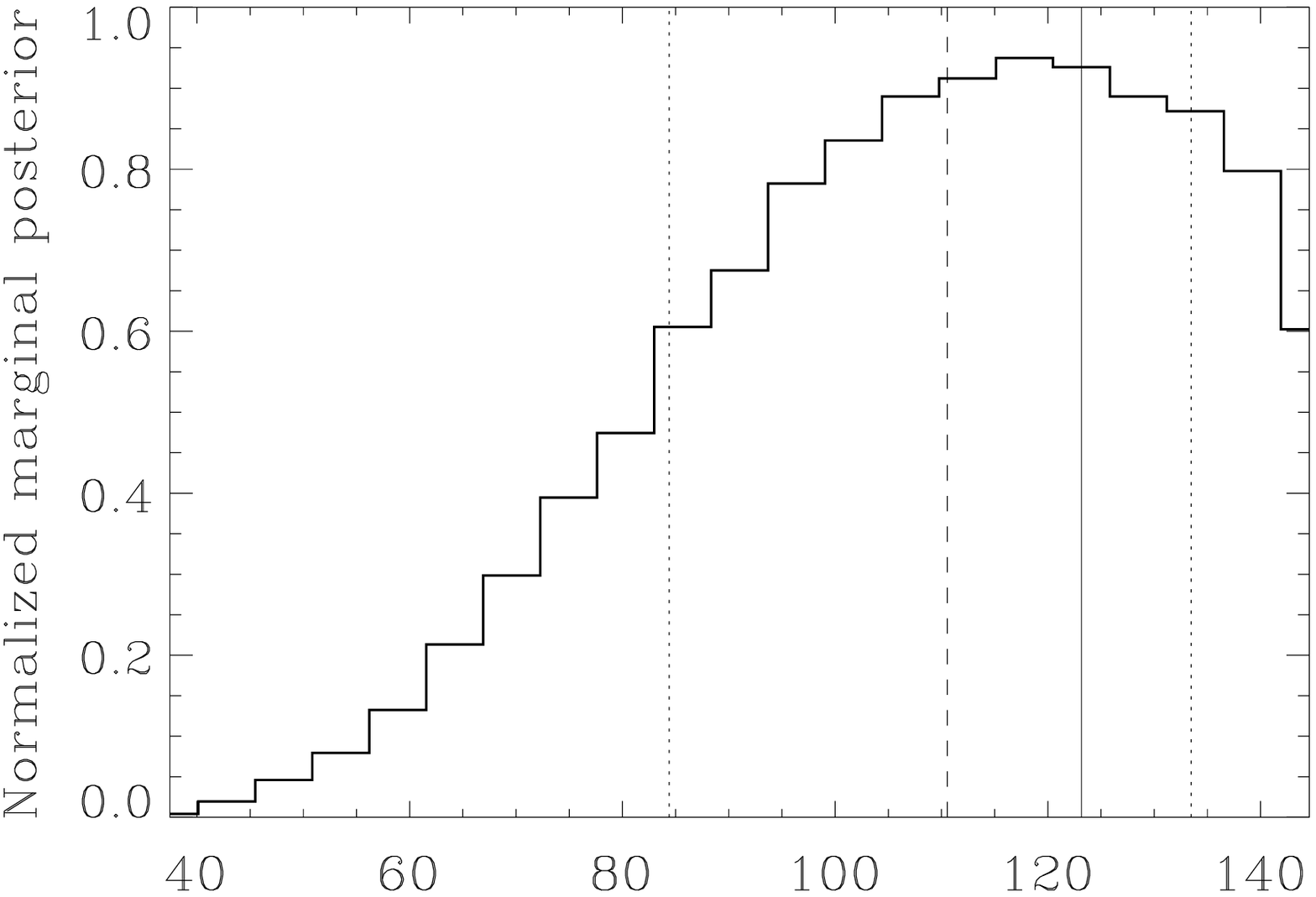}
\includegraphics[width=5.3cm]{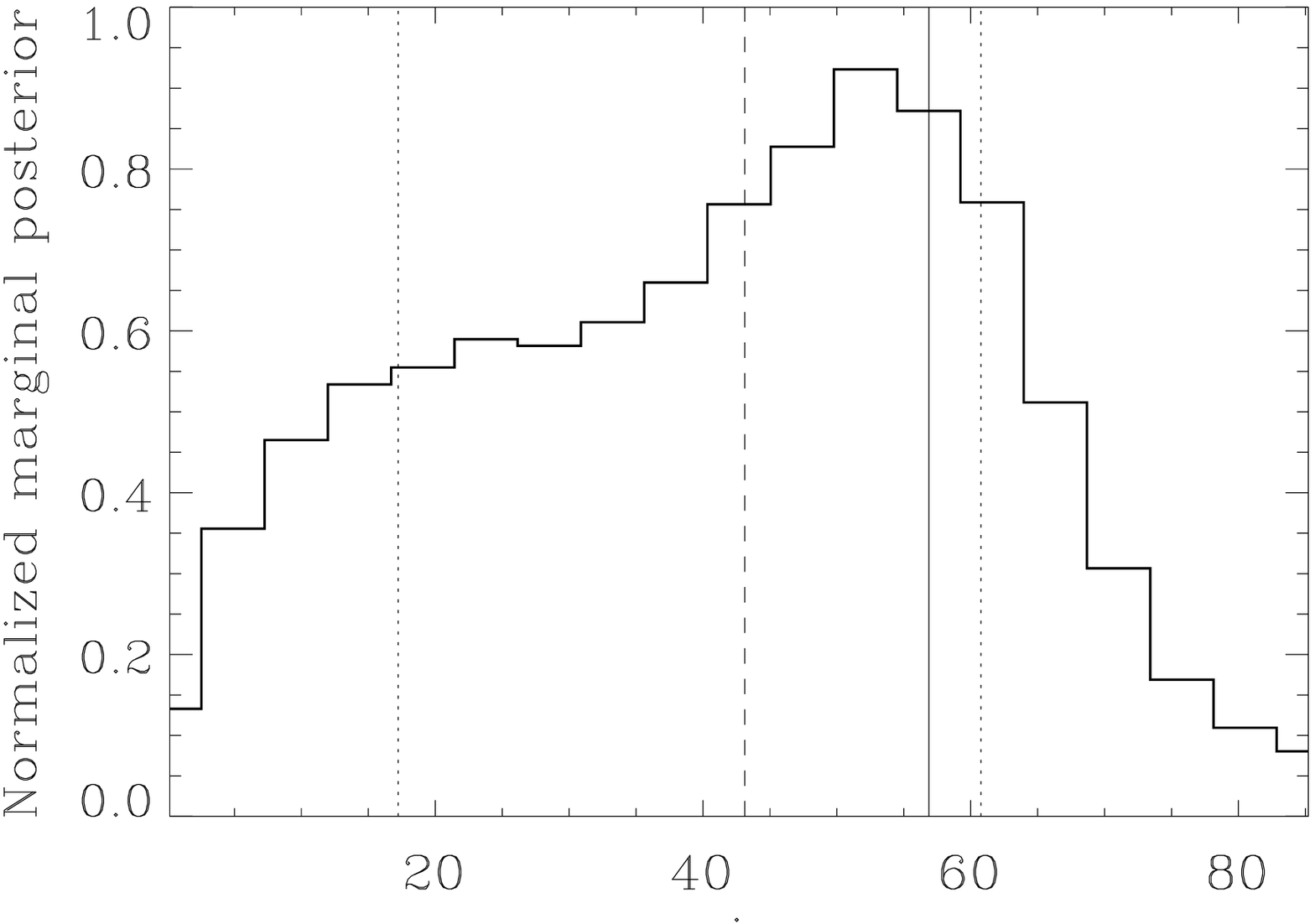}\par}
\caption{\footnotesize{Same as in Figure \ref{ngc1097}, but for the galaxy NGC 4151.}
\label{ngc4151}}
\end{figure*}

Here we include the posterior distributions of the Sy1 galaxies NGC 1566, NGC 6221, NGC 6814, NGC 7469, NGC 3227, and NGC 4151
(Figures \ref{ngc1566} to \ref{ngc4151}).

\begin{deluxetable*}{lllc}
\tablewidth{0pt}
\tablecaption{VISIR MIR fluxes from the literature for the Sy1 galaxies}
\tablehead{
\colhead{Galaxy}  & \colhead{Flux Density (mJy)} & \colhead{Filter(s)} & \colhead{Reference(s)} }
\startdata
NGC 1097    & 28$\pm$7, 49$\pm$12				  & 12.27, 18.72		  & B1,B2    \\ 
NGC 1566    & 63$\pm$9, 128$\pm$32				  & 11.88, 18.72		  & B2       \\
NGC 6814    & 99$\pm$6, 96$\pm$6				  & 11.25,13.04 		  & B3       \\
NGC 7469    & 448$\pm$16, 595$\pm$18, 630$\pm$17, 1270$\pm$317    & 10.49,12.27,13.04,18.72	  & B1,B4,B2 \\
\hline
NGC 3227    & 180$\pm$11, 320$\pm$22				  & 8.99, 11.88 		  & B4       \\ 
\enddata
\tablerefs{(B1) \citet{Horst08}; (B2) \citet{Reunanen10}; (B3) \citet{Gandhi09}; (B4) \citet{Honig10}}
\label{literature1}
\end{deluxetable*}

\clearpage

\section{Fitting results for Sy2 and intermediate-type Seyfert galaxies}
\label{appendixB}

\begin{figure*}[!ht]
\centering
{\par
\includegraphics[width=8cm]{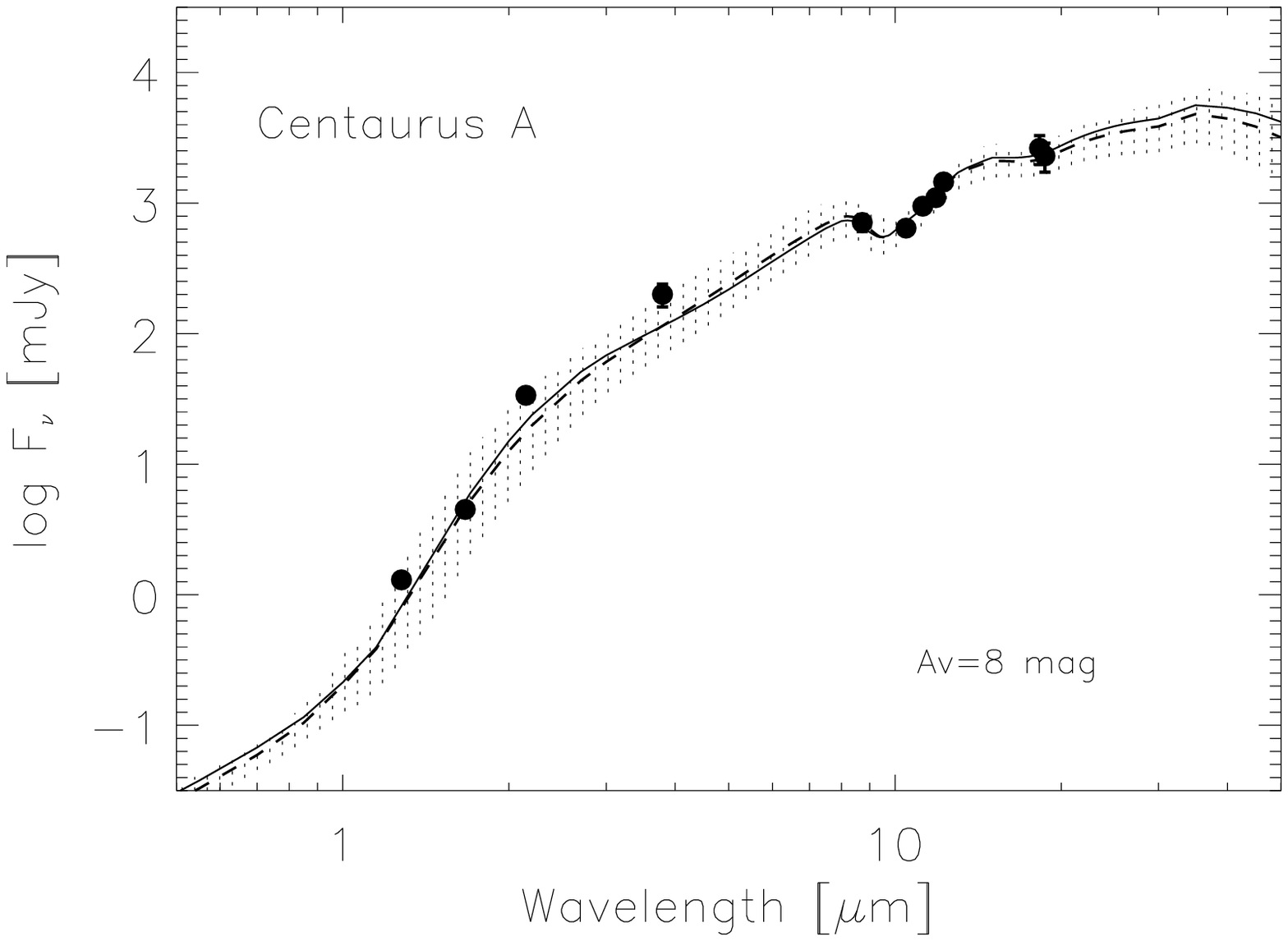}
\includegraphics[width=8cm]{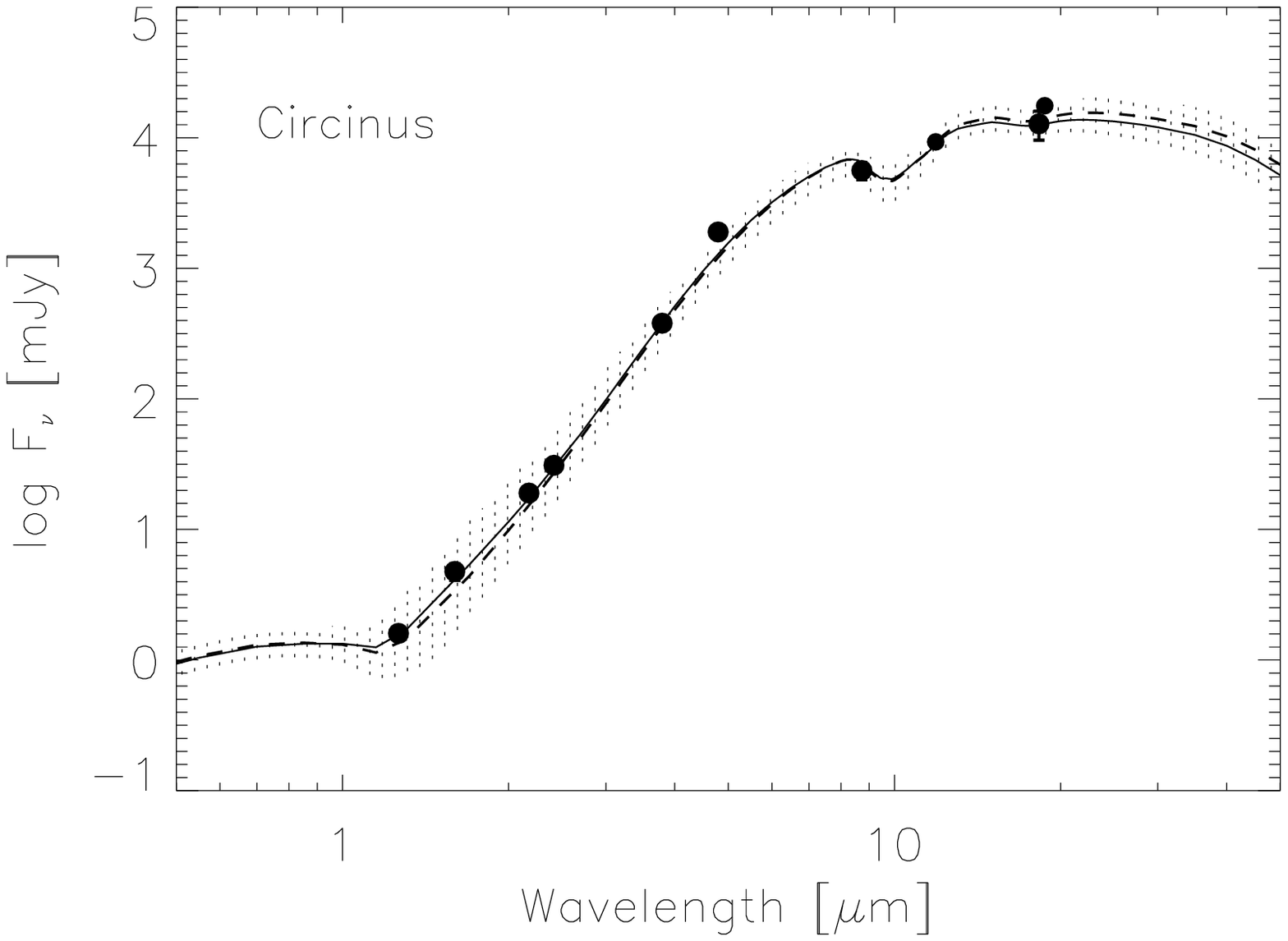}
\includegraphics[width=8cm]{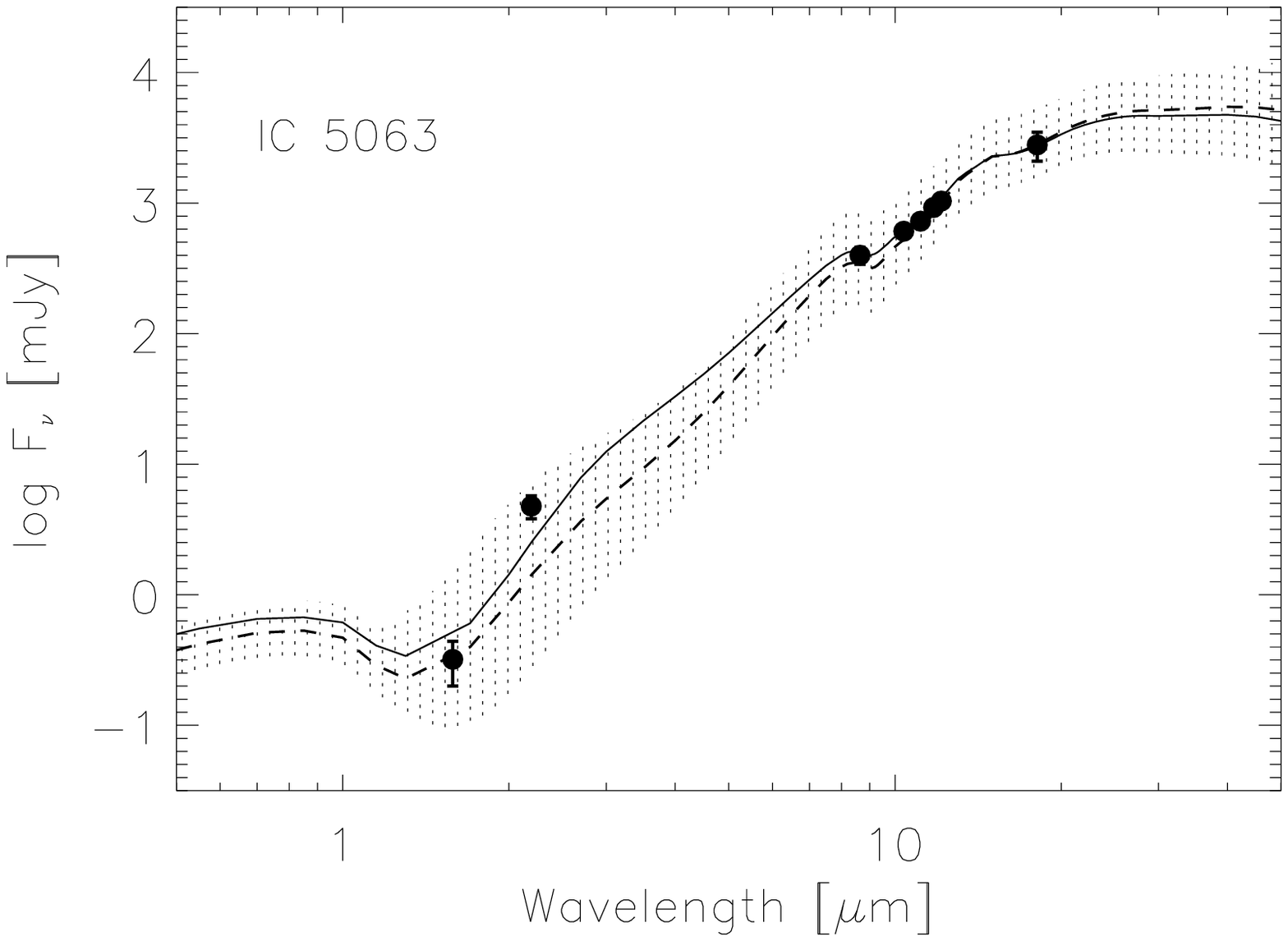}
\includegraphics[width=8cm]{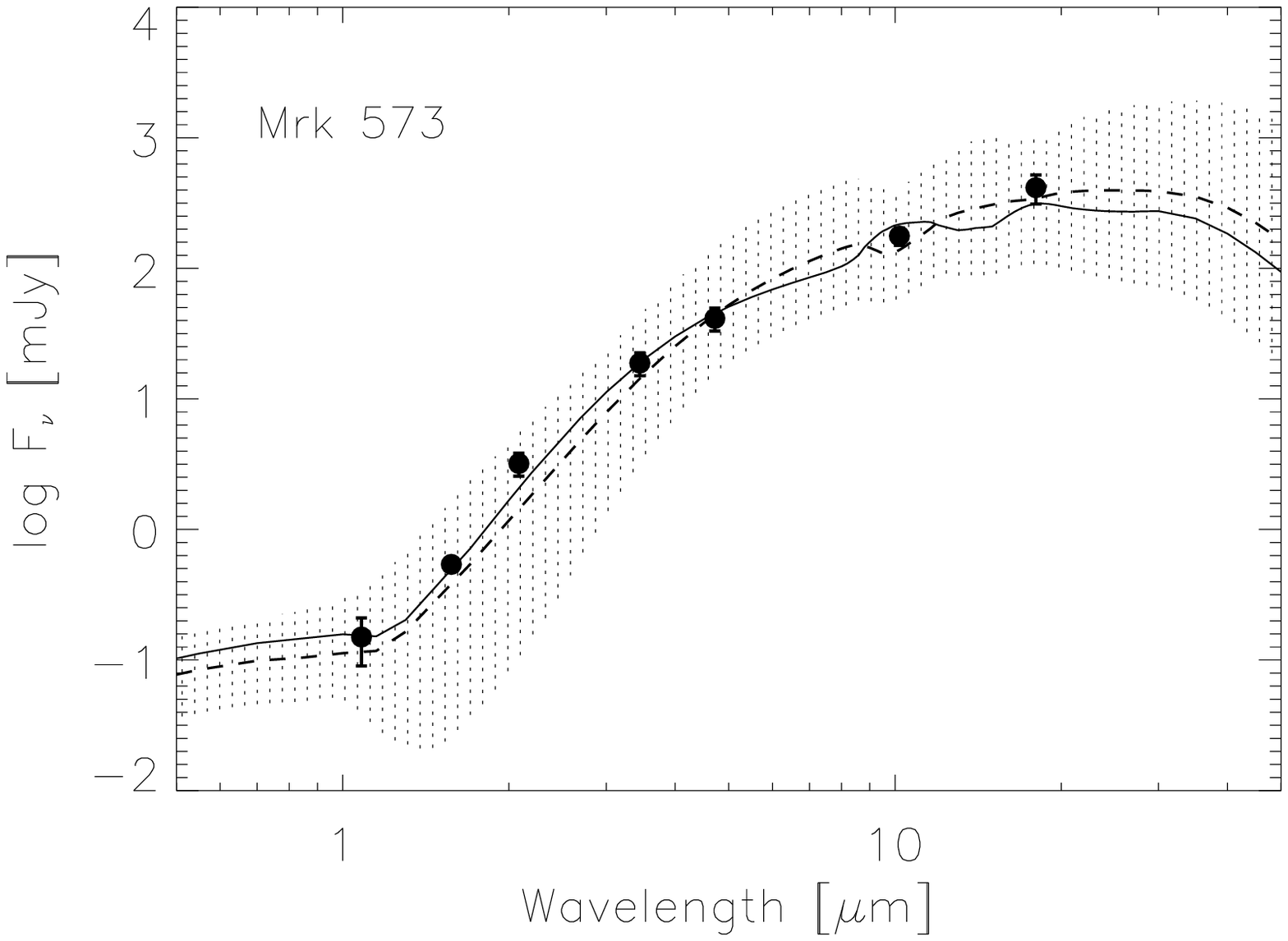}
\includegraphics[width=8cm]{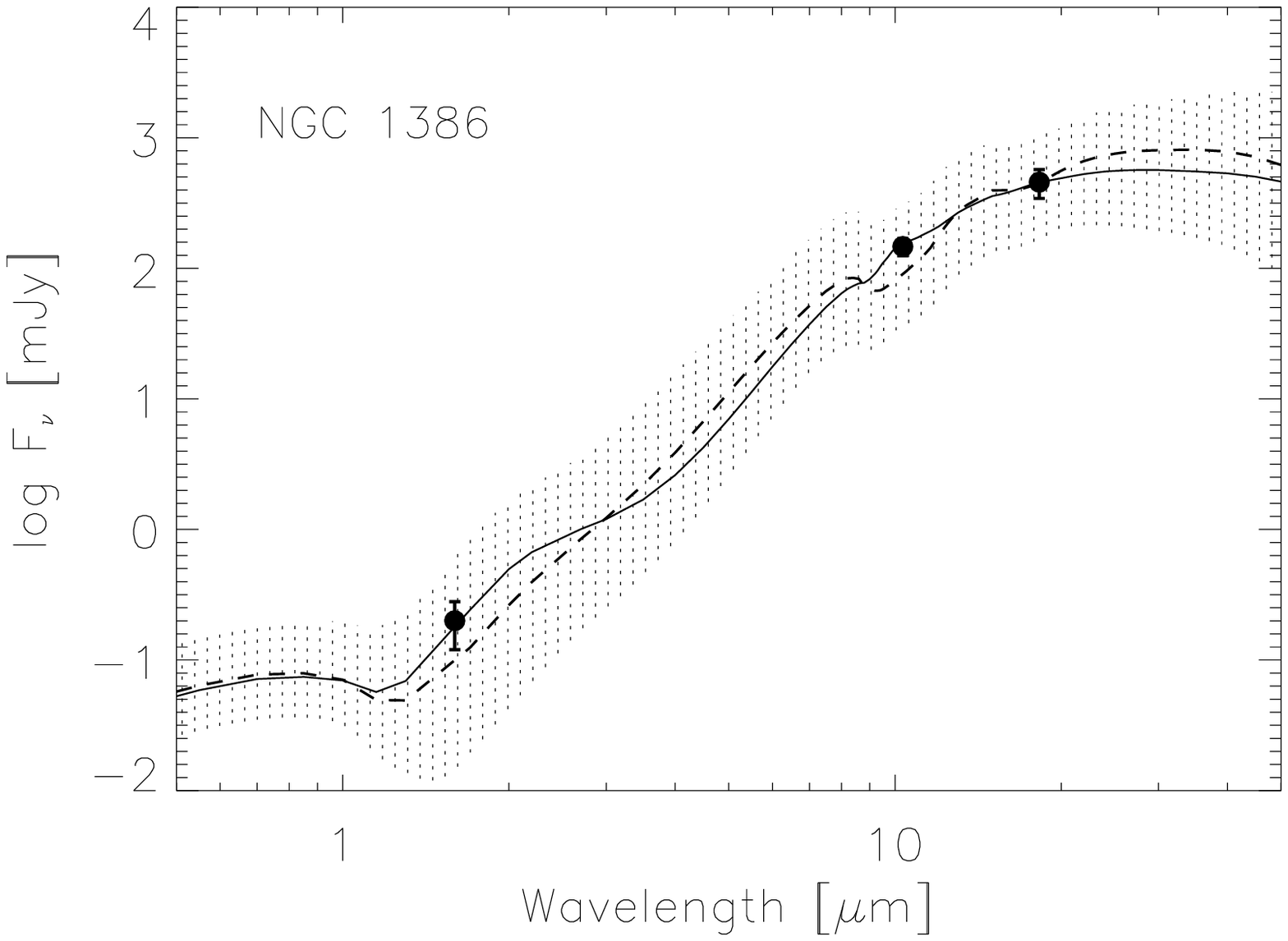}
\includegraphics[width=8cm]{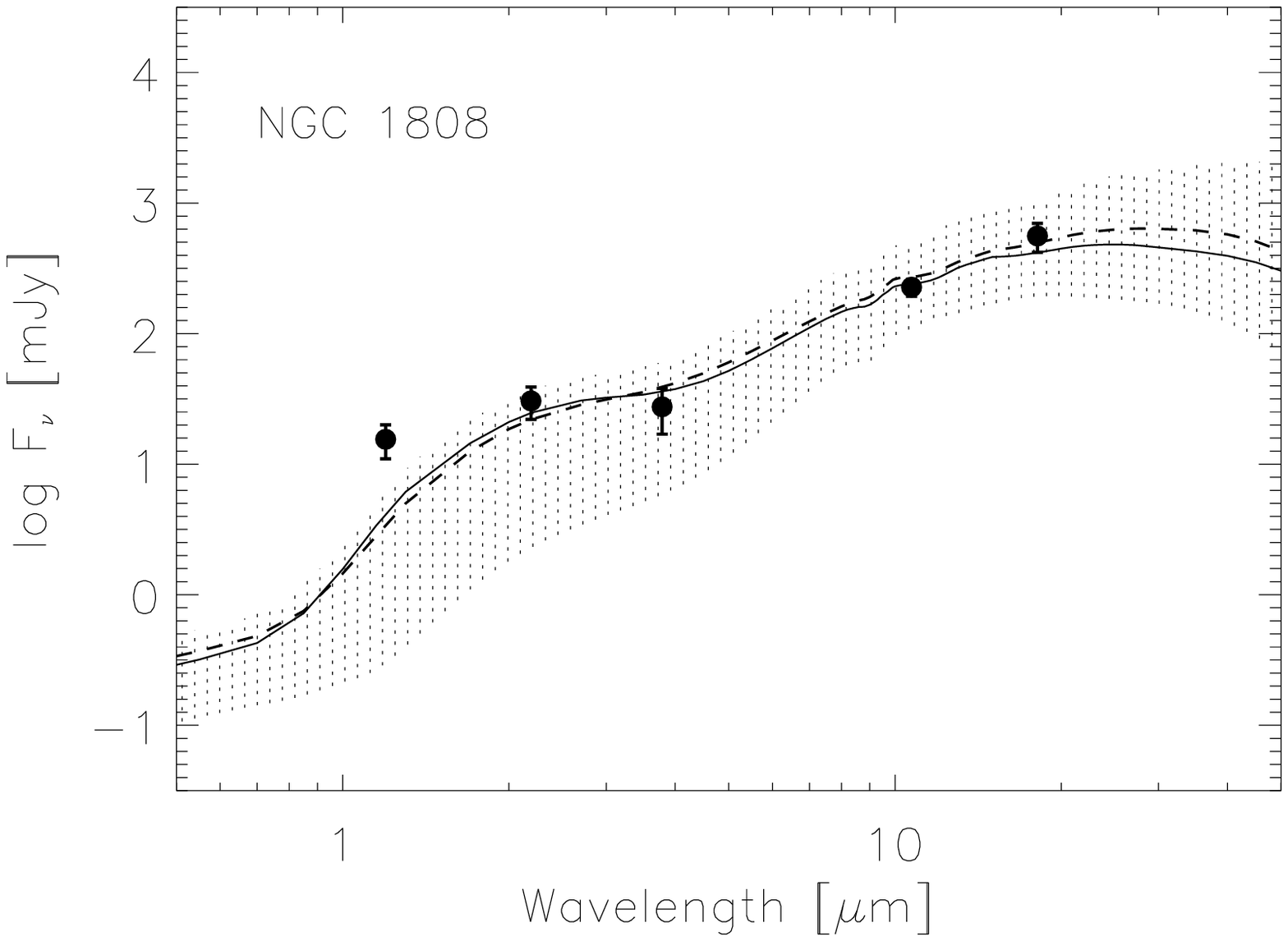}
\includegraphics[width=8cm]{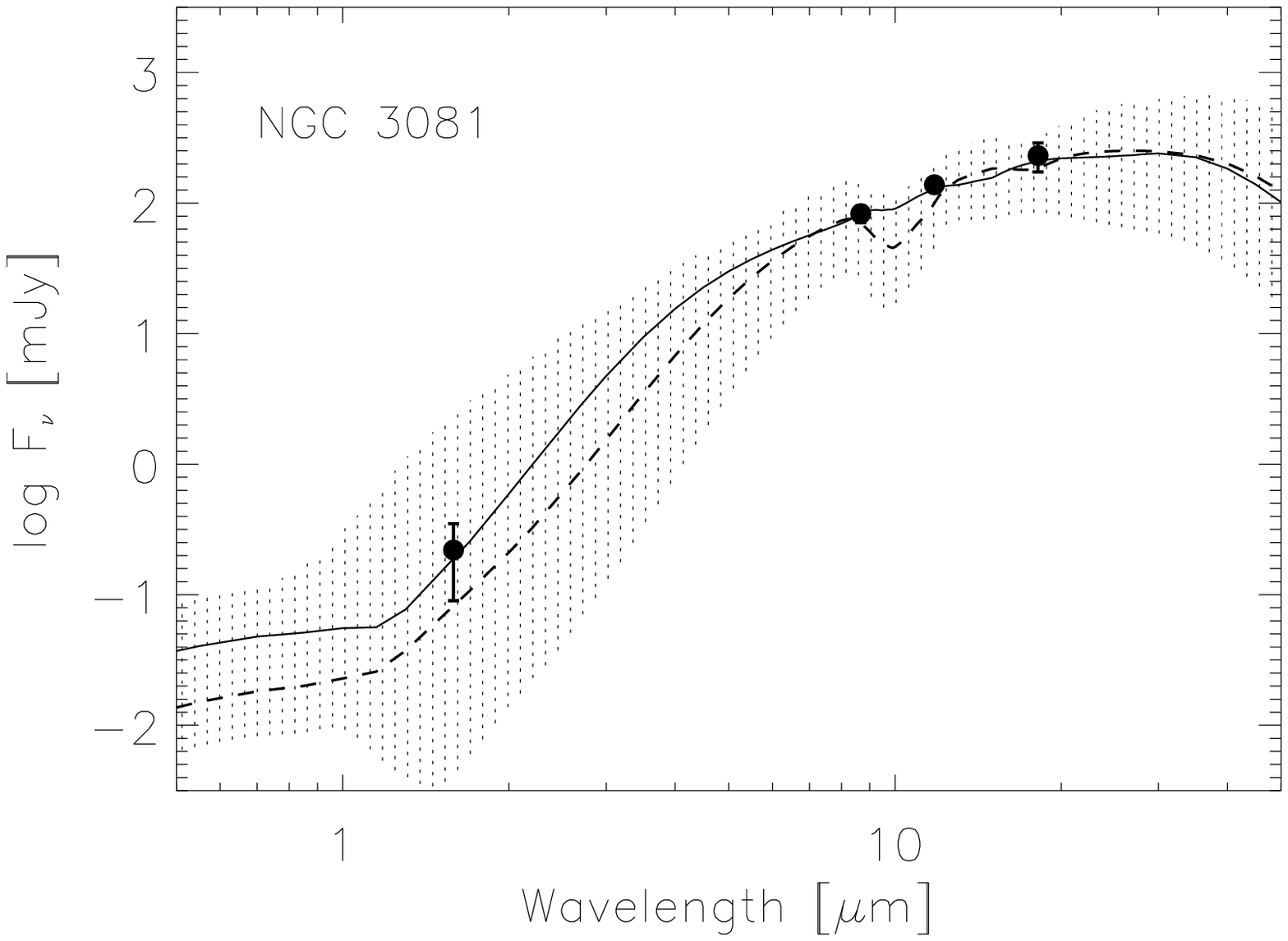}
\includegraphics[width=8cm]{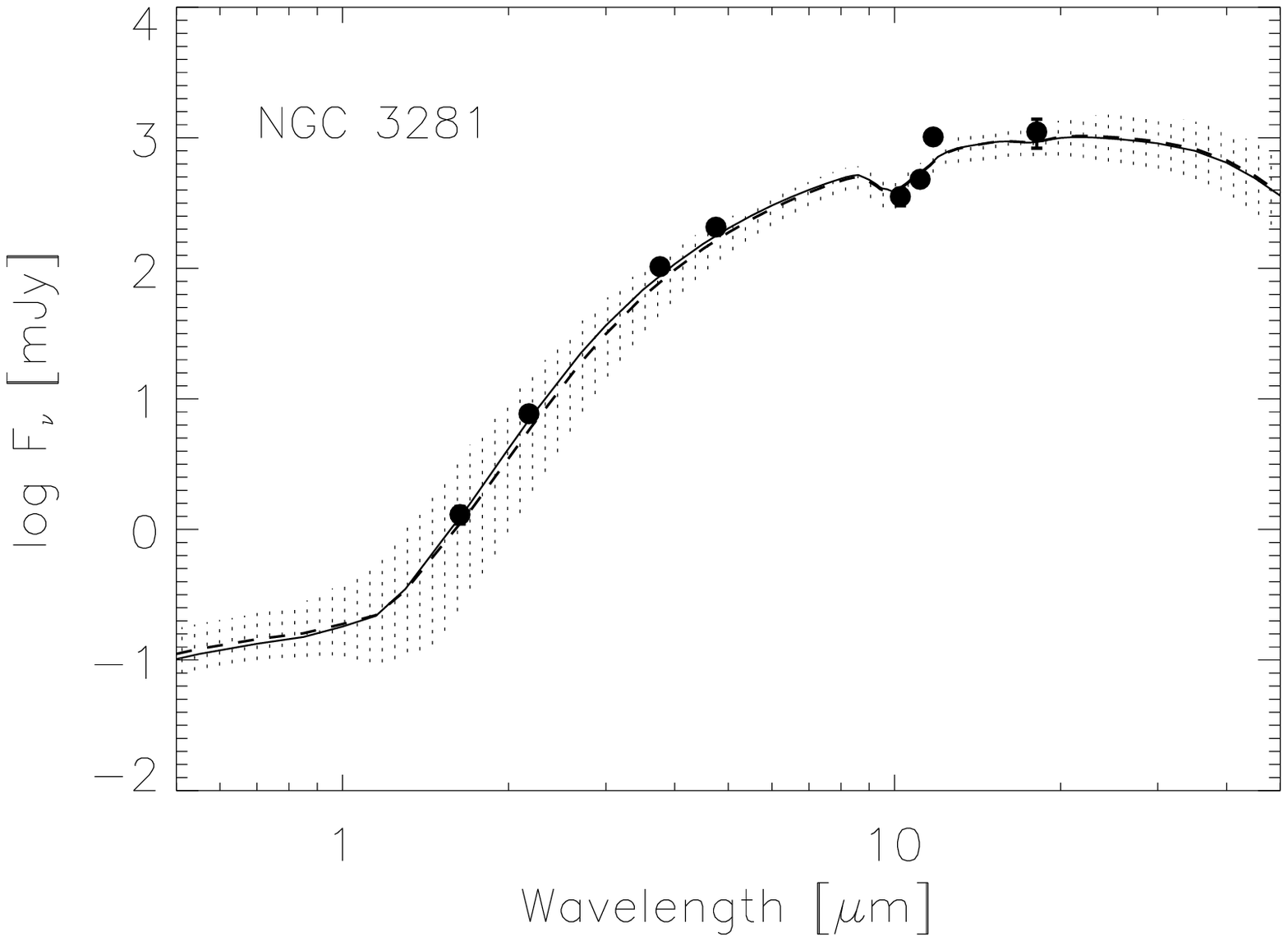}\par}
\caption{\footnotesize{Same as in Figure \ref{sy1_fits_a}, but for the Sy2 galaxies Centaurus A, Circinus, 
IC 5063, Mrk 573, NGC 1386, NGC 1808, NGC 3081, and NGC 3281.}
\label{sed1_appendix}}
\end{figure*}

\begin{figure*}[!ht]
\centering
{\par
\includegraphics[width=8cm]{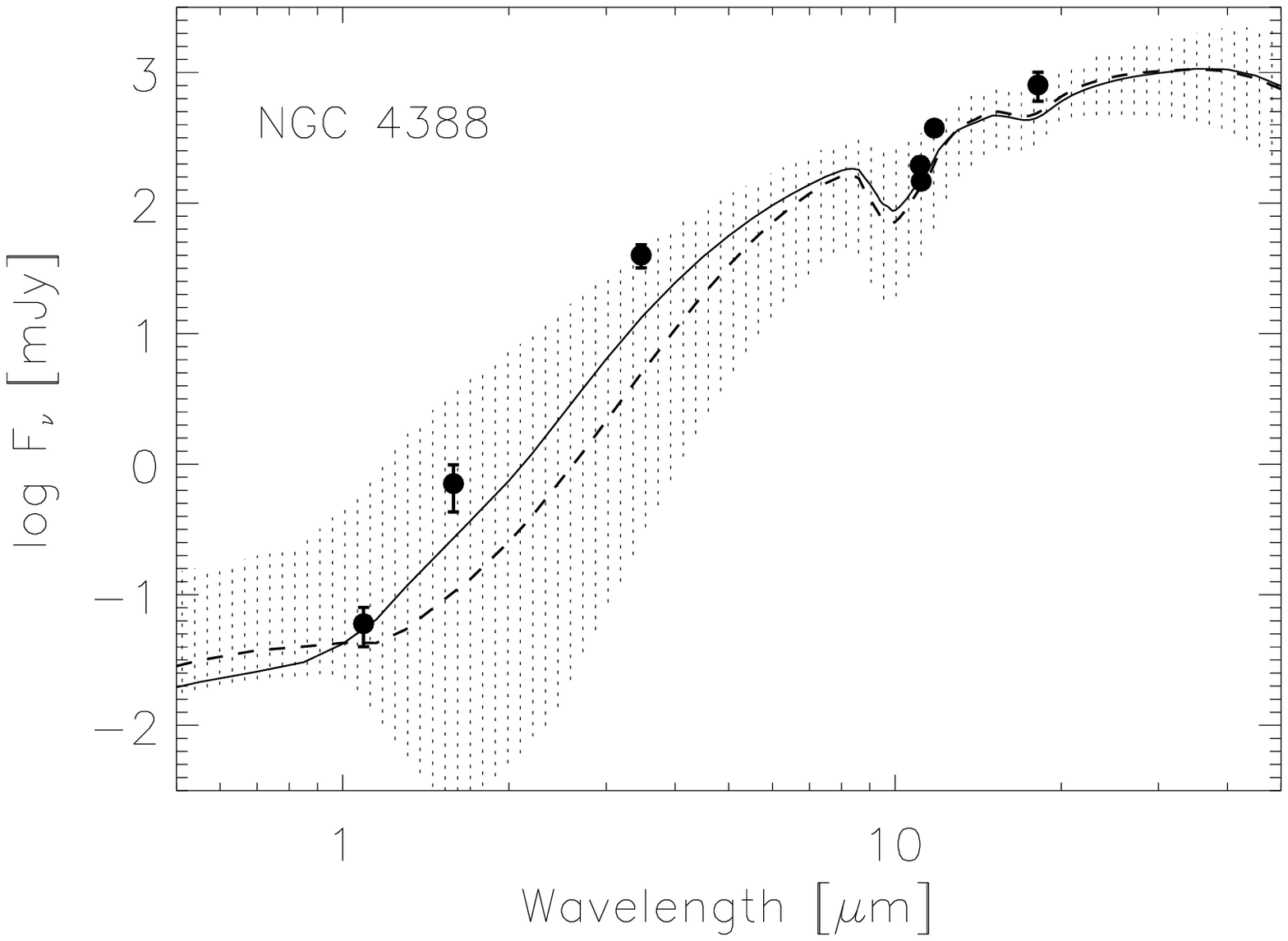}
\includegraphics[width=8cm]{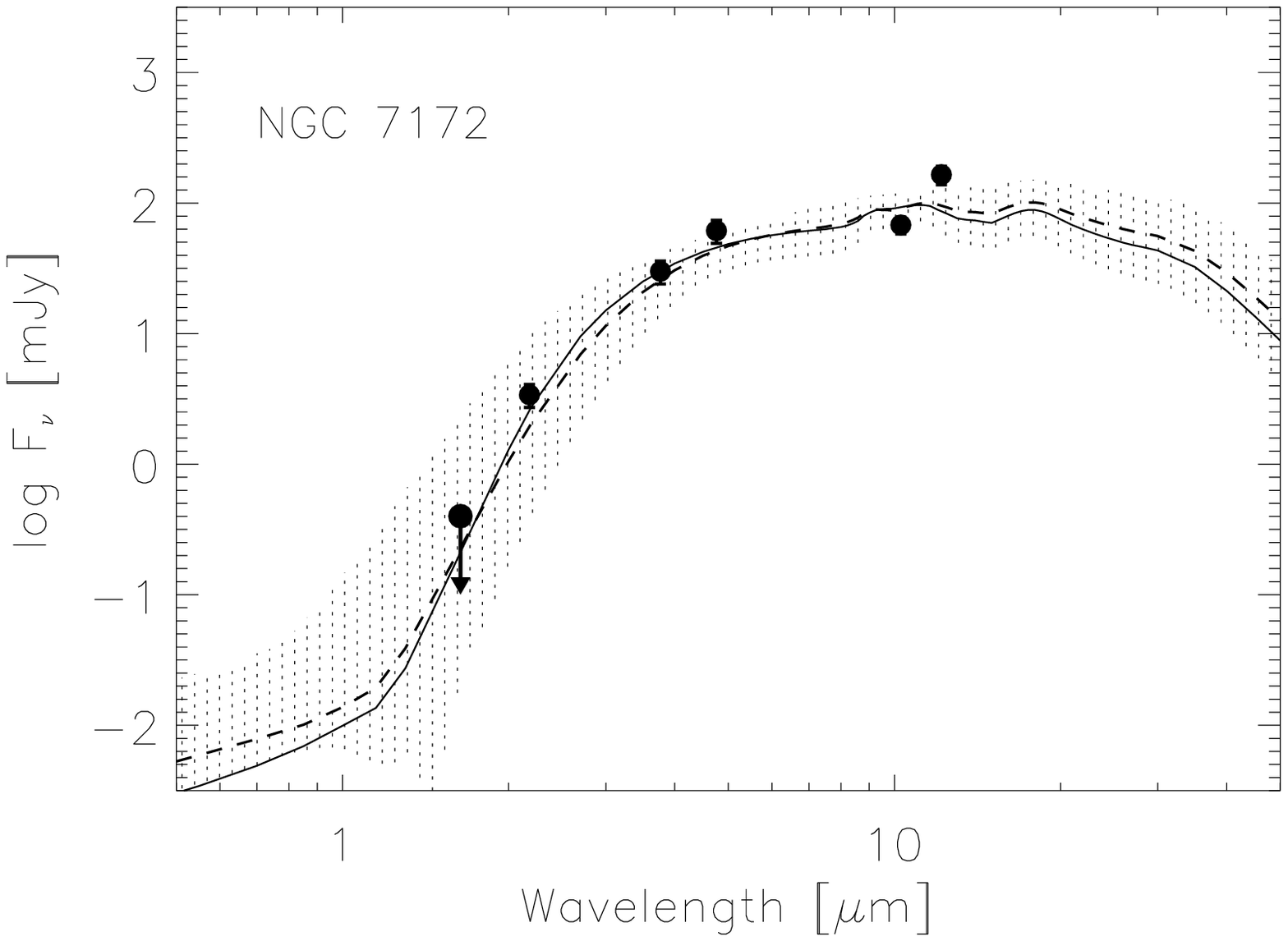}
\includegraphics[width=8cm]{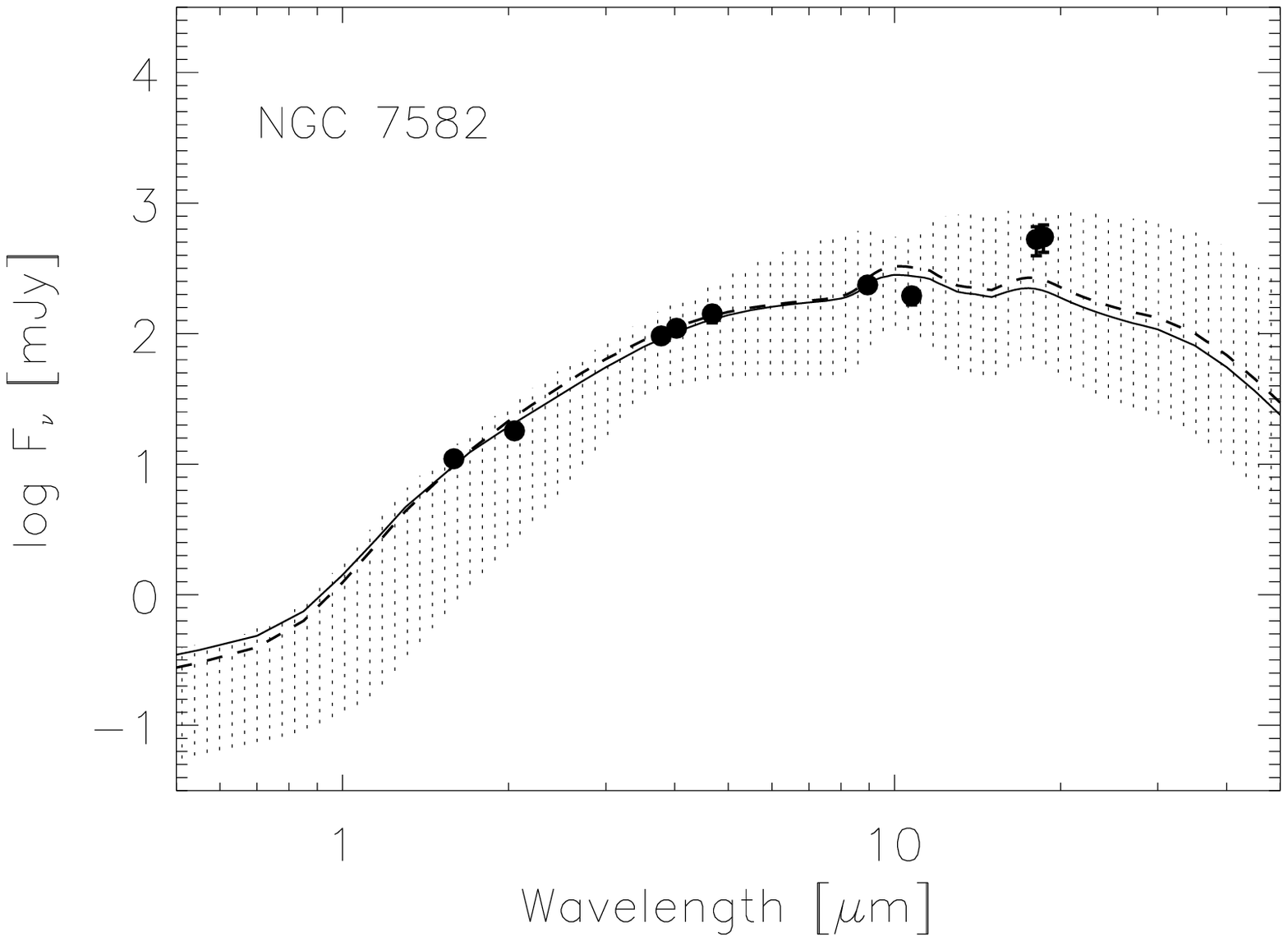}
\includegraphics[width=8cm]{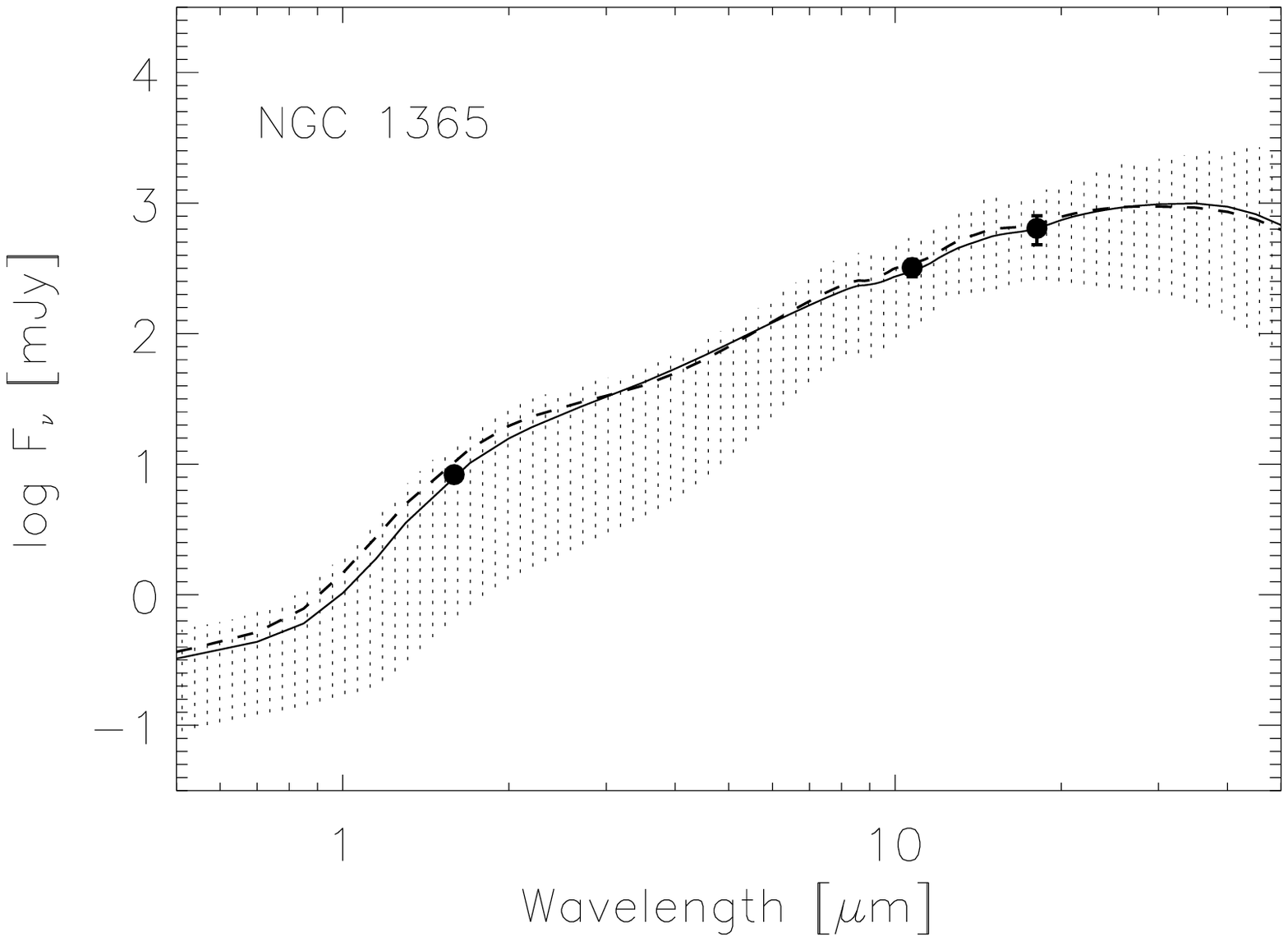}
\includegraphics[width=8cm]{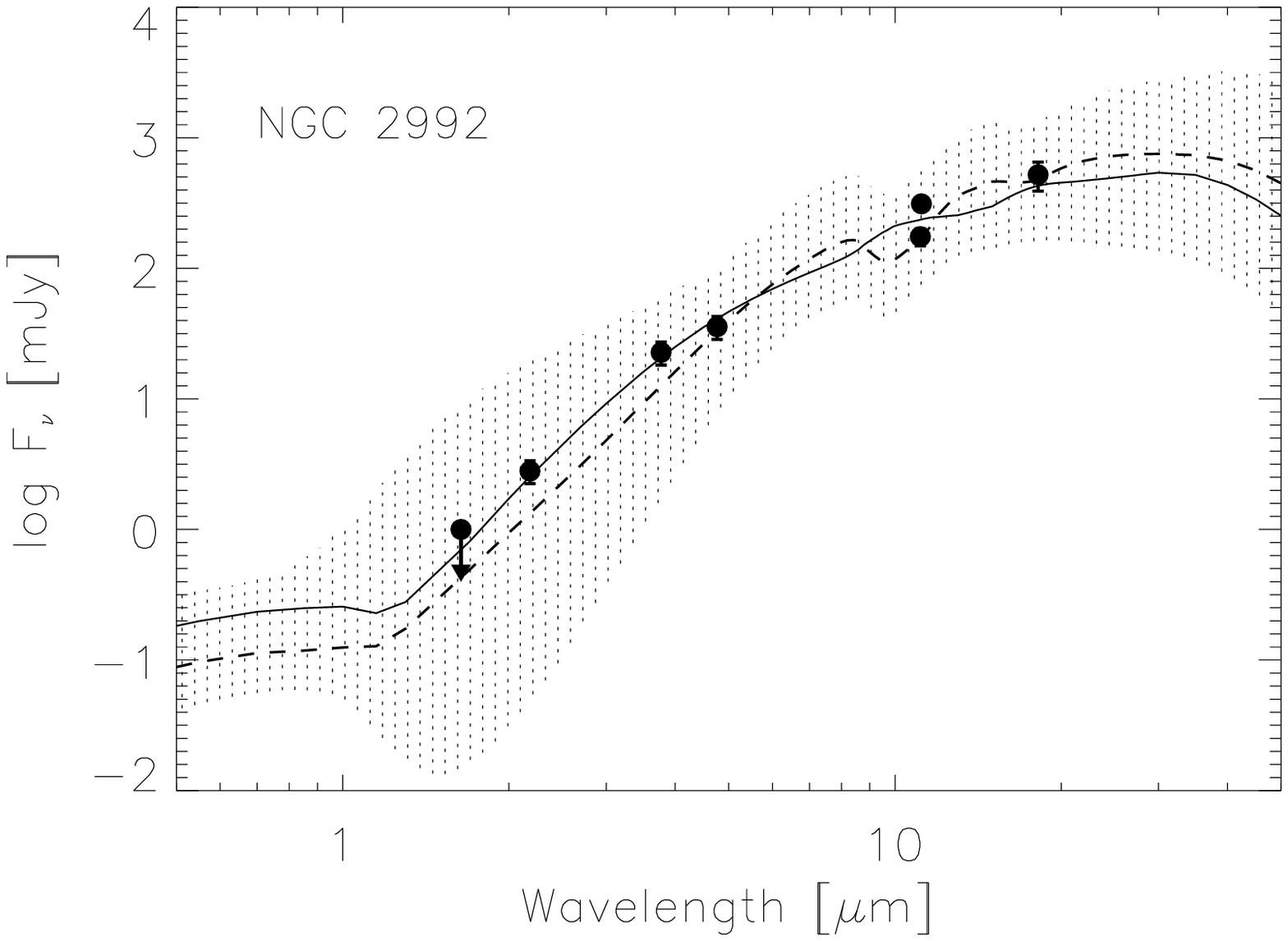}
\includegraphics[width=8cm]{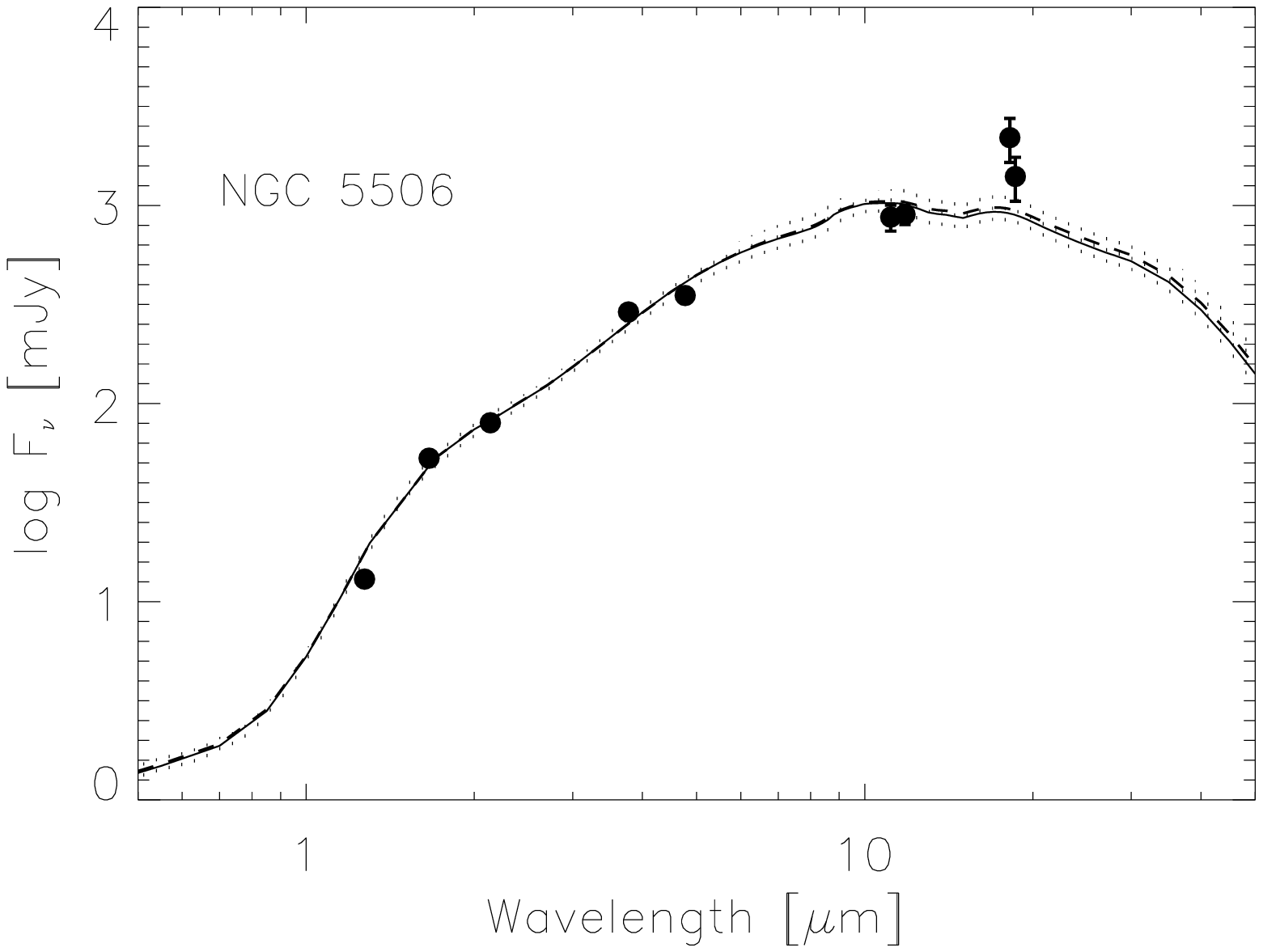}\par}
\caption{\footnotesize{Same as in Figure \ref{sy1_fits_a}, but for the Sy2 galaxies NGC 4388, NGC 7172, and NGC 7582; the
Sy1.8 NGC 1365, and the Sy1.9 galaxies NGC 2992 and NGC 5506.}
\label{sed2_appendix}}
\end{figure*}

\begin{figure*}[!ht]
\centering
{\par
\includegraphics[width=5.3cm]{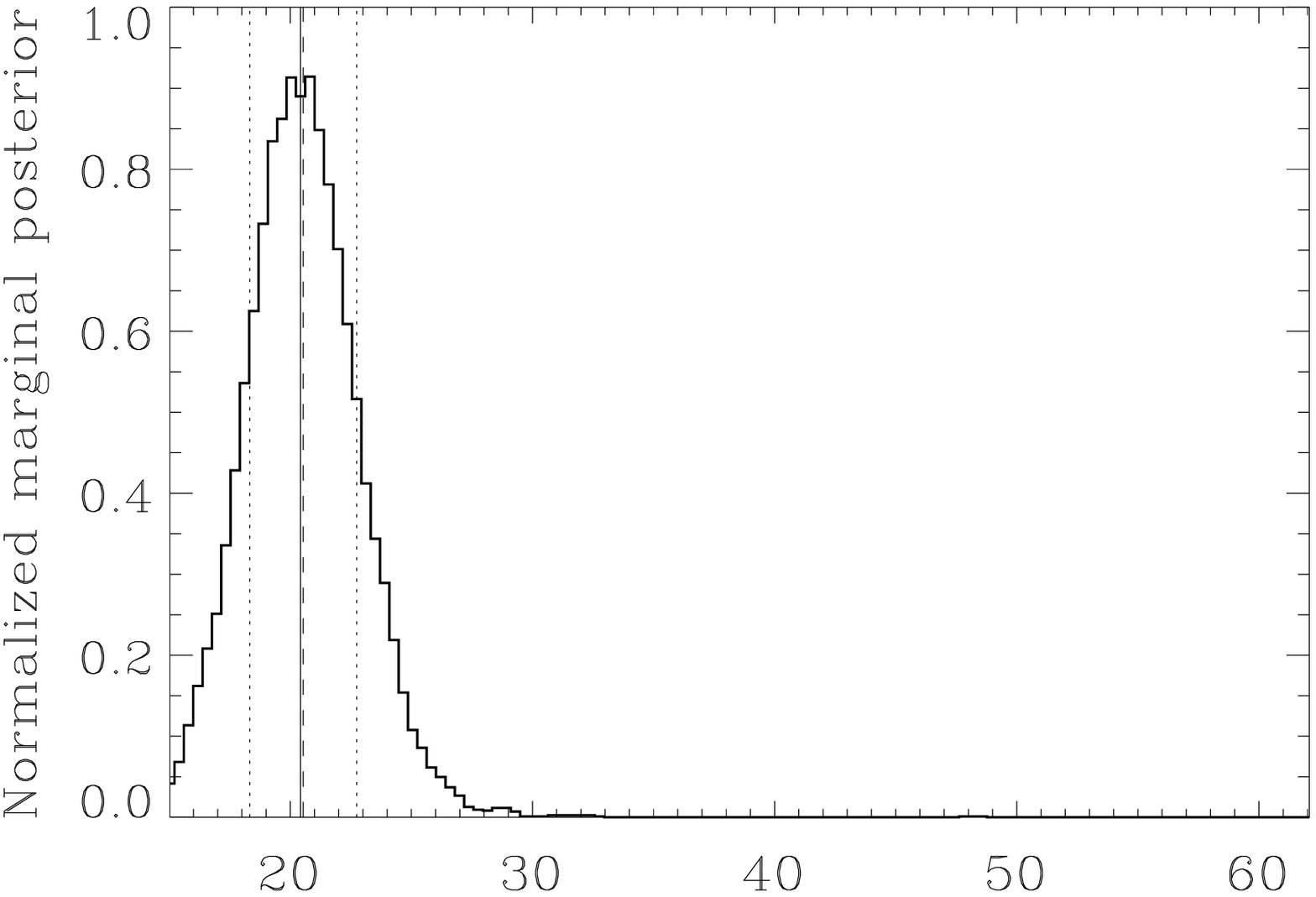}
\includegraphics[width=5.3cm]{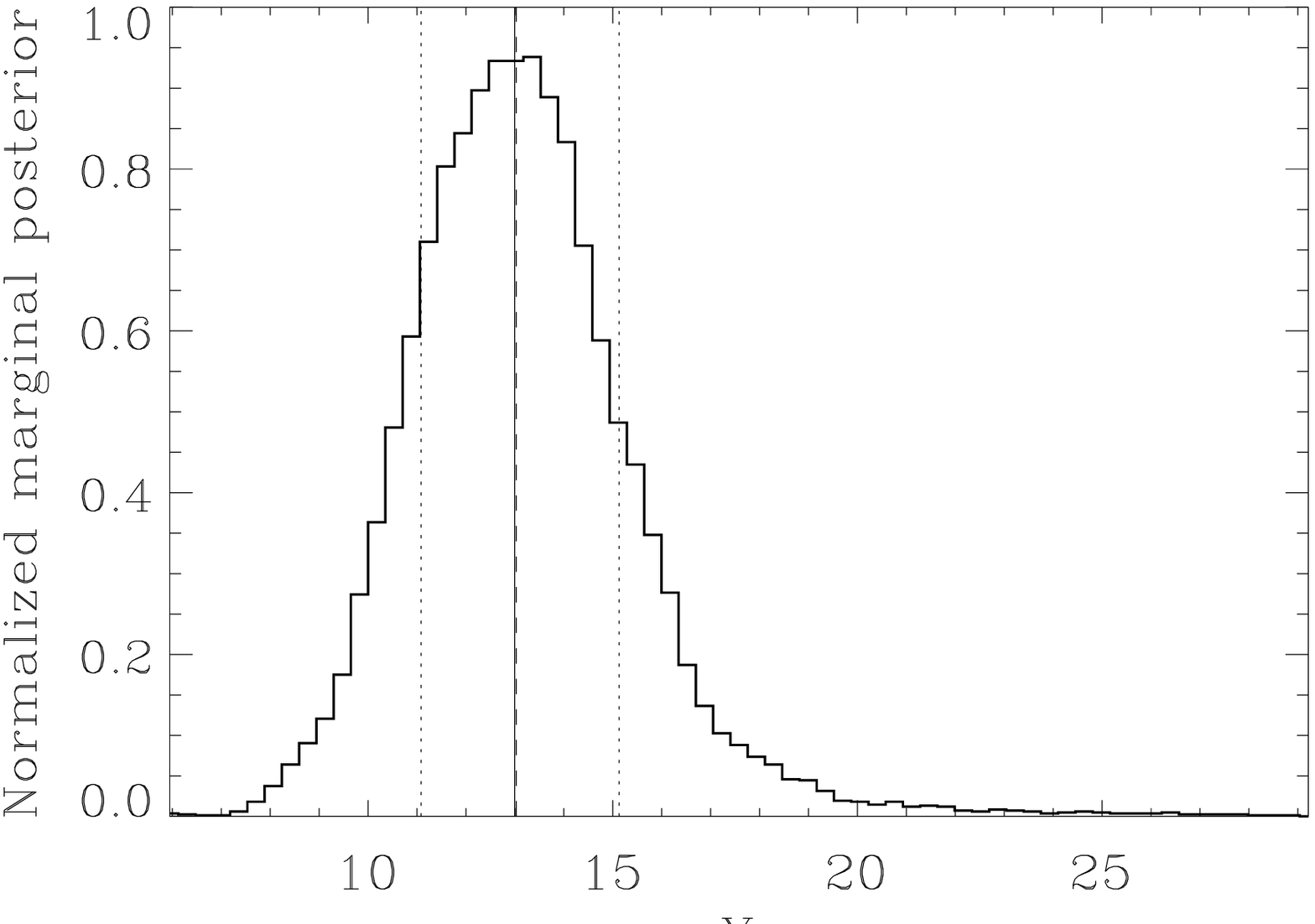}
\includegraphics[width=5.3cm]{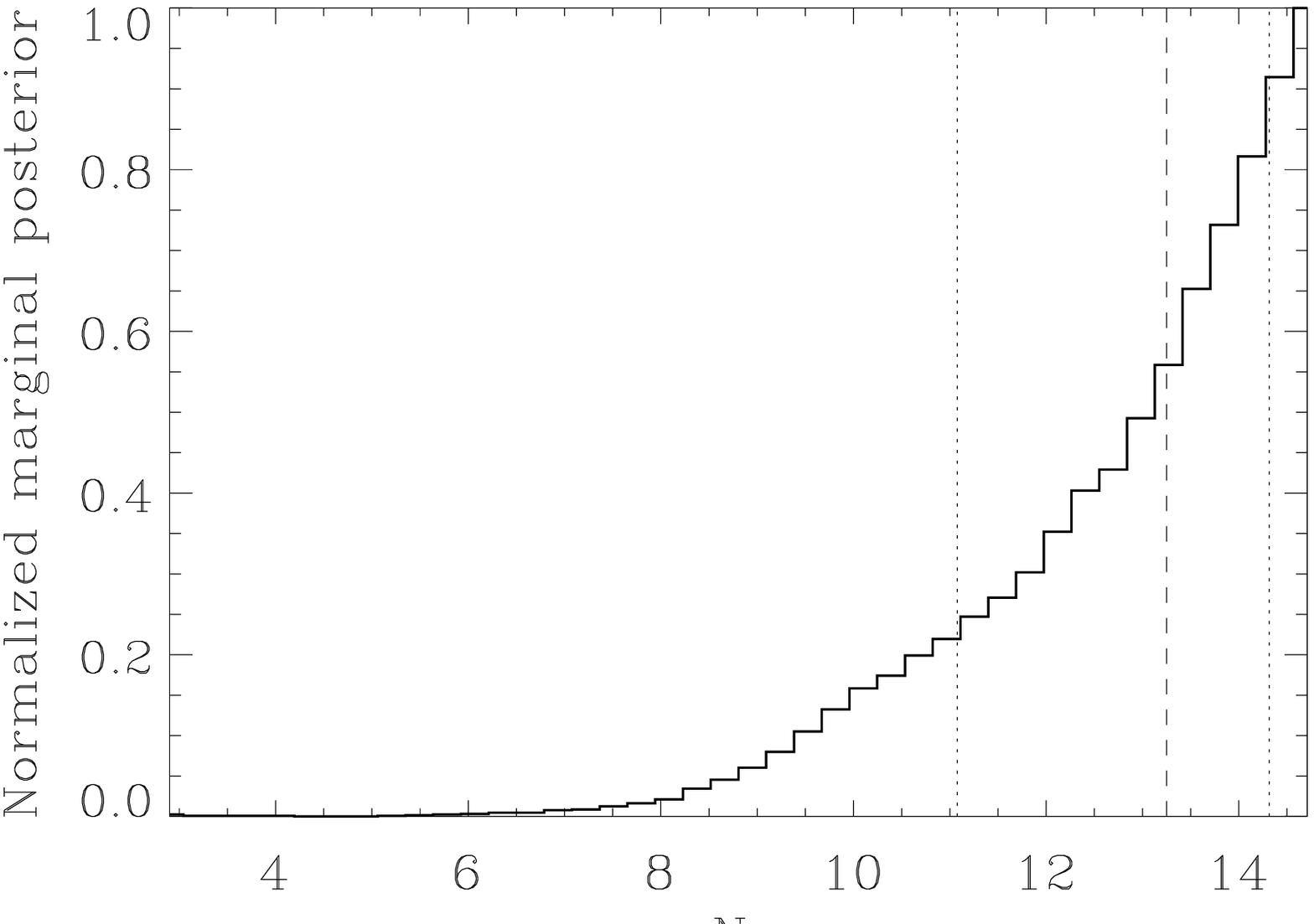}
\includegraphics[width=5.3cm]{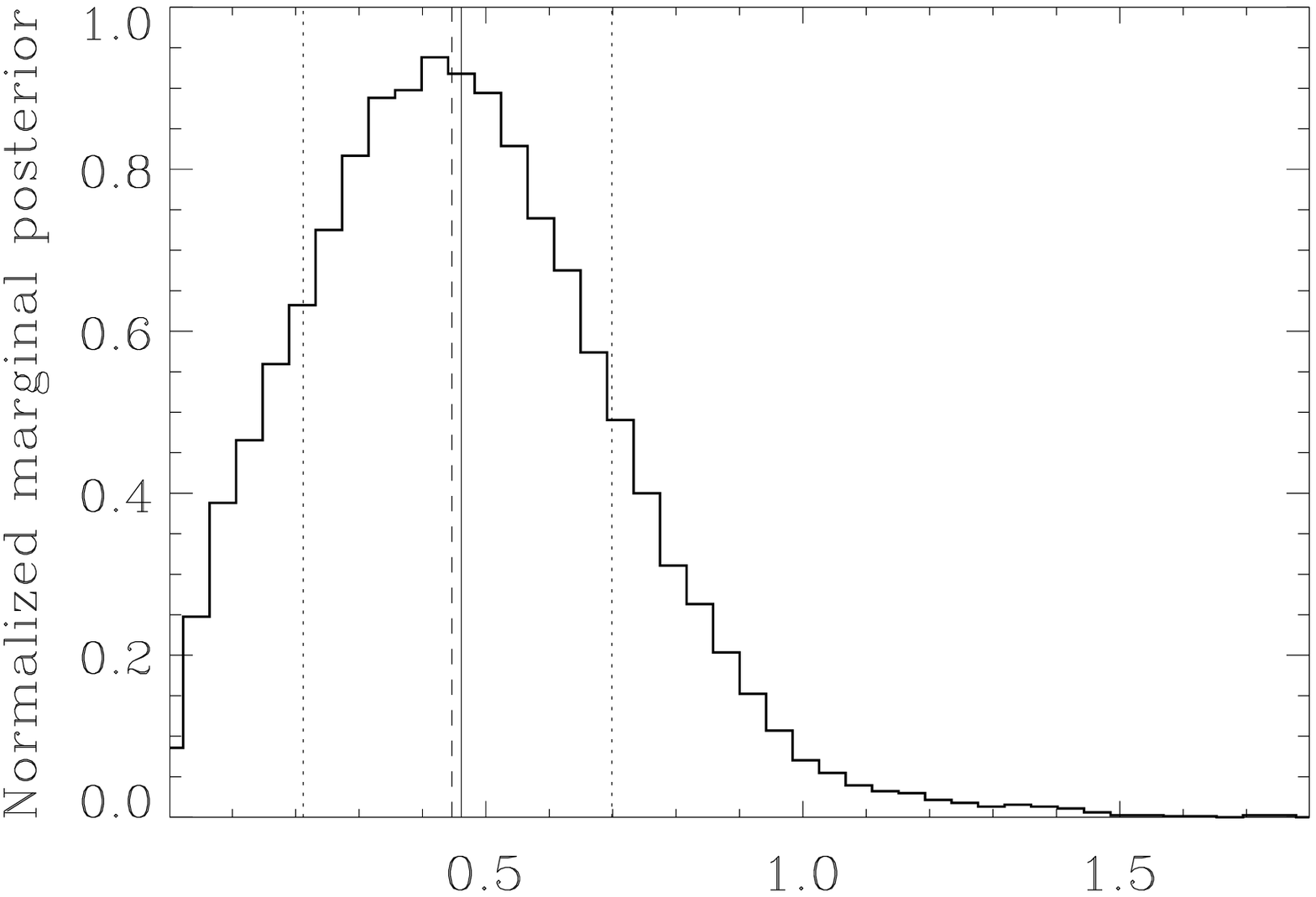}
\includegraphics[width=5.3cm]{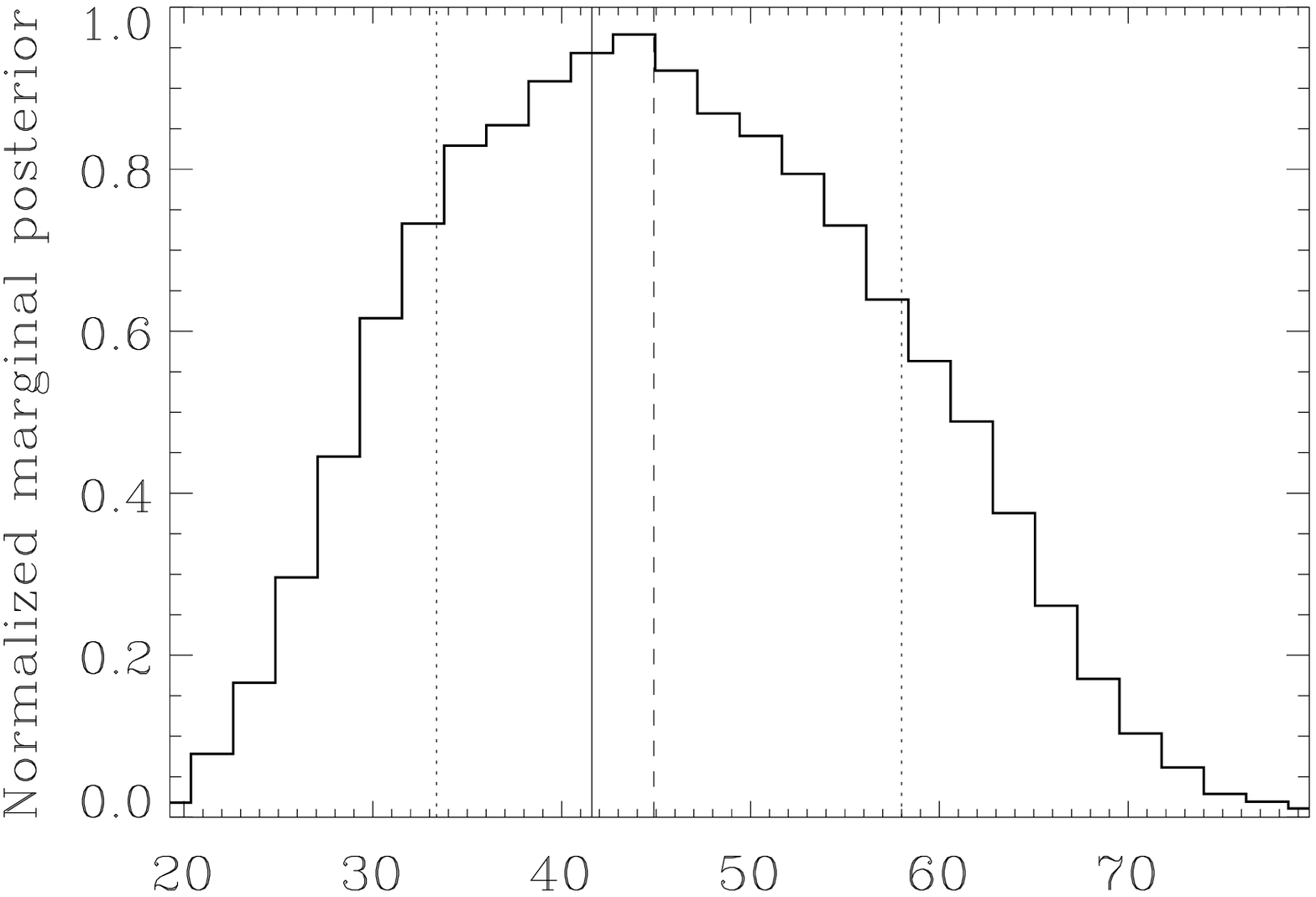}
\includegraphics[width=5.3cm]{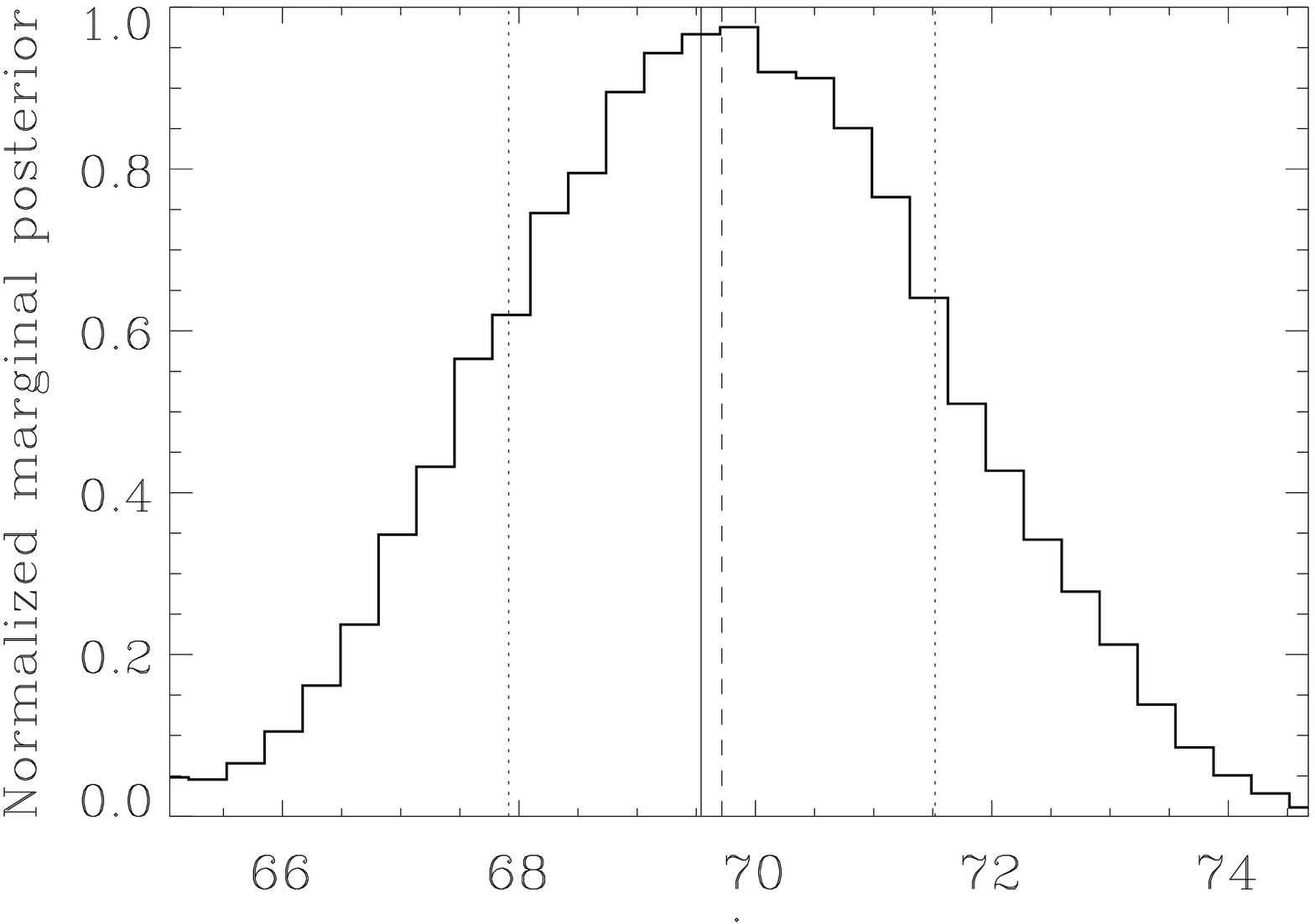}\par}
\caption{\footnotesize{Same as in Figure \ref{ngc1097}, but for the galaxy Centaurus A. See the electronic edition of the Journal for the rest 
of the Sy2 and intermediate-type Seyferts.}
\label{centaurusA}}
\end{figure*}

Here we include the results of the fits of the intermediate-type
Seyferts and the Sy2 galaxies in \citealt{Ramos09a} (Figures \ref{sed1_appendix} and \ref{sed2_appendix})
using the data reported in Table \ref{literature2} and the most up-to-date version of the \citet{Nenkova08a,Nenkova08b} models (see 
erratum \citealt{Nenkova10}). The resulting model parameters for the individual galaxies are shown in Tables \ref{clumpy_parameters2}
and \ref{lum2}. In Figure \ref{centaurusA} we show the posterior distributions from the fit of Centaurus A (see the electronic edition 
of the Journal for the rest of the Sy2 and intermediate-type Seyferts posteriors.).


\begin{deluxetable*}{lcllc}
\tabletypesize{\footnotesize}
\tablewidth{0pt}
\tablecaption{NIR and MIR fluxes from the literature for Sy2 and intermediate-type Seyferts}
\tablehead{
\colhead{Galaxy} & \colhead{Seyfert} & \colhead{Flux Density (mJy)} & \colhead{Filter(s)} & \colhead{Reference(s)} }
\startdata
CenA       &  2   & 1.3$\pm$0.1, 4.5$\pm$0.3, 34$\pm$2, 200$\pm$40 		           & NACO J,H,K,L'		      	 & A1 \\
           &      & 643$\pm$27, 947$\pm$29, 1100$\pm$100, 1451$\pm$73, 2300$\pm$575        & VISIR 10.49,11.25,11.88,12.27,18.72 & B1,B2 \\
Circinus   &  2   & 1.6$\pm$0.2, 4.8$\pm$0.7, 19$\pm$2, 31$\pm$3, 380$\pm$38, 1900$\pm$190 & F160W, NACO J,K,2.42,L,M            & A2 \\
IC 5063	   &  2   & 0.3$\pm$0.1, 4.77$\pm$0.95						   & F160W, F222M		         & A3, A10 \\
           &      & 609$\pm$22, 727$\pm$25, 925$\pm$25, 1036$\pm$56                        & VISIR 10.49,11.25,11.88,12.27       & B3 \\
Mrk 573	   &  2   & 0.15$\pm$0.06, 0.54$\pm$0.04, 3.2$\pm$0.6, 18.8$\pm$3.8, 41.3$\pm$8.3  & F110W,F160W, NSFCam K',L,M          & A4 \\ 
NGC 1386   &  2   & 0.2$\pm$0.1 						           & F160W			         & A3 \\
NGC 1808   &  2   & 15.5$\pm$4.5, 30.5$\pm$8.5, 27.5$\pm$10.5			           & ISAAC J,Ks,L'		         & A5 \\
NGC 3081   &  2   & 0.22$\pm$0.13						           & F160W			         & A3 \\
           &      & 138$\pm$11                                                             & VISIR 13.04                         & B4 \\
NGC 3281   &  2   & 1.3$\pm$0.2, 7.7$\pm$0.8, 103$\pm$9, 207$\pm$25		           & IRAC-1 H,K, IRCAM3 L',M	         & A6 \\
           &      & 481$\pm$24, 1016$\pm$52                                                & VISIR 11.25,13.04                   & B4 \\
NGC 4388   &  2   & 0.06$\pm$0.02, 0.71$\pm$0.28, 40$\pm$8			           & F110W,F160W, NSFCam L	         & A4 \\ 
           &      & 147$\pm$11, 375$\pm$28                                                 & VISIR 11.25,13.04                   & B4 \\
NGC 7172   &  2   & $<$0.4, 3.4$\pm$0.7, 30$\pm$6, 61$\pm$12			           & IRCAM3 H,K,L',M		         & A7 \\
           &      & 165$\pm$27                                                             & VISIR 12.27                         & B1 \\
NGC 7582   &  2   & 11$\pm$1, 18$\pm$2, 96$\pm$10, 110$\pm$11, 142$\pm$21 		   & F160W, NACO 2.06,L,4.05, ISAAC M	 & B2,A8 \\
           &      & 236$\pm$27, 550$\pm$130						   & VISIR 8.99, 18.72                   & B3,B2 \\ 
\hline
NGC 1365   & 1.8  & 8.3$\pm$0.8                                    		           & F160W			         & A9 \\
NGC 2992   & 1.9  & $<$1, 2.8$\pm$0.6, 22.7$\pm$4.5, 35.7$\pm$7.1  		           & IRCAM3 H,K,L',M		         & A7 \\
           &      & 312$\pm$31                       					   & VISIR 11.25                         & B5 \\ 
NGC 5506   & 1.9  & 13$\pm$1, 53$\pm$1, 80$\pm$1, 290$\pm$10, 351$\pm$29             	   & NACO J,H,K,L', NSFCam M  	         & B2,A11 \\
           &      & 900$\pm$100, 1400$\pm$350  						   & VISIR 11.88, 18.72                  & B2 \\ 
\enddata
\tablecomments{\scriptsize{Ground-based instruments and telescopes are: NACO and ISAAC on the 8 m VLT,  NSFCam on the 3 m NASA IRTF, IRCAM3 on the  3.8 m UKIRT, and IRAC-1 on the 2.2 m ESO telescope. 
Measurements in the F110W, F160W, and F222M filters are from NICMOS on HST.}}
\tablerefs{\scriptsize{(A1) \citet{Meisenheimer07}; (A2) \citet{Prieto04}; (A3) \citet{Quillen01}; (A4) \citet{Alonso03}; (A5) \citet{Galliano08}; (A6) \citet{Simpson98};
(A7) \citet{Alonso01}; (A8) \citet{Prieto02}; (A9) \citet{Carollo02}; (A10) \citet{Kulkarni98}; (A11) \citet{Ward87}; 
(B1) \citet{Horst08}; (B2) \citet{Prieto10}; (B3) \citet{Honig10}; (B4) \citet{Gandhi09}; (B5) \citet{Haas07}}}
\label{literature2}
\end{deluxetable*}

\begin{deluxetable*}{llclclclclclclc}
\tabletypesize{\scriptsize}
\tablewidth{0pt}
\tablecaption{Parameters from the fits for Sy2 and intermediate-type Seyferts}
\tablehead{
\colhead{Galaxy} &  \multicolumn{2}{c}{$\sigma$} & \multicolumn{2}{c}{$Y$} & \multicolumn{2}{c}{$N_0$} & \multicolumn{2}{c}{$q$} & \multicolumn{2}{c}{$i$} &
\multicolumn{2}{c}{$\tau_{V}$} & \multicolumn{2}{c}{$A_V^{LOS}$}  \\
&  \colhead{Med} & \colhead{Mode} & \colhead{Med} & \colhead{Mode} &  \colhead{Med} & \colhead{Mode} & \colhead{Med} & \colhead{Mode} & \colhead{Med} & 
\colhead{Mode} & \colhead{Med} & \colhead{Mode} & \colhead{Med} & \colhead{Mode}}
\startdata
CenA       &   20$\pm$2 	   & 20   & 13$\pm$2	     & 13    & 13$\pm^{1}_{2}$ & 15    & 0.4$\pm$0.2	      & 0.5 & 70$\pm$2  	& 69& 45$\pm^{13}_{11}$ & 42 & 235$\pm^{110}_{90}$& 200 \\
Circinus   &   63$\pm^{4}_{7}$     & 68   & 20$\pm^{6}_{10}$ & 27    & 11$\pm$2        & 11    & 2.6$\pm^{0.2}_{0.4}$ & 2.7 & 85$\pm$2 (fix)	& 85& 31$\pm$2  	& 31 & 365$\pm^{70}_{55}$ & 340 \\
IC 5063	   &   51$\pm$10	   & 56   & 12$\pm^{4}_{3}$  & 10    & 13$\pm^{1}_{2}$ & 14    & 0.8$\pm^{0.9}_{0.6}$ & 0.3 & 81$\pm^{5}_{8}$	& 85& 90$\pm$19 	& 96 & 1135$\pm^{315}_{240}$ & 990 \\
Mrk 573	   &   30$\pm^{18}_{10}$   & 23   & 17$\pm$8	     & 25    & 6$\pm^{4}_{2}$  & 5     & 0.9$\pm^{0.9}_{0.6}$ & 0.6 & 85$\pm$2 (fix)	& 85& 72$\pm^{41}_{33}$ & 53 & 155$\pm^{100}_{80}$& 105  \\
NGC 1386   &   49$\pm^{11}_{18}$   & 52   & 17$\pm^{8}_{7}$  & 17    & 11$\pm^{2}_{3}$ & 13    & 1.5$\pm^{0.8}_{0.9}$ & 2.3 & 85$\pm$2 (fix)	& 85& 95$\pm^{51}_{66}$ & 51 & 805$\pm^{645}_{440}$& 425 \\  
NGC 1808   &   26$\pm^{12}_{7}$    & 20   & 18$\pm^{7}_{8}$  & 18    & 8$\pm$4         & 12    & 1.3$\pm^{0.9}_{0.8}$ & 0.9 & 41$\pm^{16}_{24}$ & 53& 111$\pm^{23}_{31}$& 122& $<$100 & 1  \\
NGC 3081   &   55$\pm^{8}_{17}$    & 61   & 18$\pm^{7}_{8}$  & 22    & 10$\pm$3        & 12    & 1.4$\pm$0.9	      & 2.3 & 54$\pm^{21}_{31}$ & 79& 29$\pm^{23}_{12}$ & 21 & $<$230 & 70 \\	    
NGC 3281   &   34$\pm$3 	   & 34   & 10$\pm$2	     & 9     & 14$\pm^{1}_{3}$ & 15    & $<$0.2 	      & 0.0 & 63$\pm^{4}_{2}$ (a)& 60& 15$\pm$2 	& 14 & 120$\pm^{20}_{15}$ & 120  \\
NGC 4388   &   60$\pm^{6}_{10}$    & 67   & 20$\pm^{5}_{6}$  & 22    & 11$\pm^{2}_{3}$ & 14    & 0.5$\pm^{0.6}_{0.3}$ & 0.2 & 45$\pm^{25}_{22}$ & 28& 31$\pm^{16}_{8}$  & 25 & $<$270  & 110 \\
NGC 7172   &   58$\pm^{7}_{11}$    & 66   & 14$\pm^{9}_{6}$  & 8     & 10$\pm^{3}_{2}$ & 9     & 1.9$\pm^{0.7}_{1.0}$ & 2.5 & 63$\pm^{16}_{26}$ & 77& 7$\pm^{2}_{1}$	& 6  & 50$\pm$15 & 50 \\
NGC 7582   &   34$\pm^{20}_{13}$   & 23   & 18$\pm^{7}_{8}$  & 18    & 3$\pm^{4}_{2}$  & 2     & 2.3$\pm^{0.4}_{0.6}$ & 2.6 & 43$\pm^{19}_{22}$ & 45& 16$\pm^{7}_{5}$	& 13 & $<$15	  & 1 \\  
\hline
NGC 1365   &   35$\pm^{14}_{10}$   & 32   & 18$\pm$7	     & 22    & 7$\pm$4         & 4     & 1.1$\pm^{0.8}_{0.7}$ & 0.7 & 27$\pm^{19}_{16}$ & 19& 86$\pm^{36}_{38}$ & 96 & $<$110	  & 1 \\ 
NGC 2992   &   40$\pm^{16}_{14}$   & 28   & 18$\pm$7	     & 24    & 7$\pm^{4}_{3}$  & 5     & 0.9$\pm^{1.0}_{0.7}$ & 0.2 & 75$\pm^{8}_{22}$  & 78& 39$\pm^{12}_{10}$ & 34 & 210$\pm^{145}_{105}$ & 235 \\
NGC 5506   &   $>$68		   & 69   & 23$\pm^{4}_{6}$  & 26    & $<$2	       & 1     & $>$2.9 	      & 3.0 & 85$\pm$2 (fix)	& 85& 42$\pm$3  	& 43		  & 60$\pm$8 & 60 \\ 
\enddata          
\tablecomments{\footnotesize{For the galaxies Circinus, Mrk 573, NGC 1386, and NGC 5506 the $i$ parameter has been introduced as a Gaussian prior into the computations, centered at 85\degr with a 
width of 2\degr, based on other observations. 
Probability distributions presenting a single tail have been characterized with the mode and upper/lower limits at 68\% confidence.
Sy1.8 and Sy1.9 have been fitted with the geometry corresponding to torus emission-only.  }}
\tablenotetext{a}{\footnotesize{The inclination angle of the torus was introduced as a uniform prior $i$=[60\degr,90\degr] following \citet{Storchi92}.}}
\label{clumpy_parameters2}
\end{deluxetable*}

\begin{deluxetable*}{llccccc}
\tablewidth{0pt}
\tablecaption{Bolometric Luminosity Predictions}
\tablehead{
\colhead{Galaxy}  & \multicolumn{1}{c}{L$_{bol}^{AGN}$}  & \colhead{L$_{bol}^{tor} / L_{bol}^{AGN}$} & 
\colhead{$R_{o}$ (pc)} &  \colhead{L$_{X bol}^{AGN}$} &\colhead{L$_{X bol}^{AGN}$ / L$_{bol}^{AGN}$} & \colhead{Ref.}}
\startdata
Centaurus A & 4.7$\pm^{0.8}_{0.7}$  $\times$ 10$^{42}$   &  0.4$\pm$0.1           &    0.4$\pm$0.1           & 1.3 $\times$ 10$^{43}$ &   3    &  a  \\
Circinus    & 1.0$\pm^{0.2}_{0.1}$  $\times$ 10$^{43}$   &  0.8$\pm$0.1           &    0.8$\pm^{0.3}_{0.4}$  & 1.2 $\times$ 10$^{43}$ &   1.2  &  b  \\
IC 5063     & 2.7$\pm^{1.2}_{0.7}$  $\times$ 10$^{44}$   &  0.8$\pm^{0.1}_{0.2}$  &    2.5$\pm^{1.3}_{0.8}$  & 1.7 $\times$ 10$^{44}$ &   0.6  &  c  \\
Mrk 573	    & 1.5$\pm^{1.1}_{0.6}$  $\times$ 10$^{44}$   &  0.4$\pm$0.2           &    2.8$\pm^{1.4}_{1.2}$  & 4.4 $\times$ 10$^{44}$ &   3    &  d  \\
NGC 1386    & 2.8$\pm^{2.9}_{0.8}$  $\times$ 10$^{42}$   &  0.7$\pm^{0.1}_{0.3}$  &    0.4$\pm$0.2           & 1.3 $\times$ 10$^{43}$ &   5    &  e  \\
NGC 1808    & 3.1$\pm^{1.4}_{0.8}$  $\times$ 10$^{42}$   &  0.8$\pm^{0.4}_{0.2}$  &    0.4$\pm$0.2           & 2.2 $\times$ 10$^{41}$ &   0.1  &  f  \\
NGC 3081    & 7.6$\pm^{4.6}_{2.0}$  $\times$ 10$^{42}$   &  0.9$\pm$0.3           &    0.7$\pm$0.3           & 1.0 $\times$ 10$^{44}$ &   13   &  d  \\
NGC 3281    & 1.1$\pm^{0.2}_{0.1}$  $\times$ 10$^{44}$   &  0.7$\pm$0.1           &    1.3$\pm$0.3           & 3.0 $\times$ 10$^{44}$ &   3    &  g  \\
NGC 4388    & 2.3$\pm^{1.1}_{0.6}$  $\times$ 10$^{43}$   &  1.0$\pm$0.3           &    1.3$\pm$0.4           & 1.5 $\times$ 10$^{44}$ &   6    &  h  \\
NGC 7172    & 8.4$\pm$1.6           $\times$ 10$^{42}$   &  0.9$\pm^{0.2}_{0.1}$  &    0.5$\pm^{0.3}_{0.2}$  & 1.1 $\times$ 10$^{44}$ &   13   &  i  \\
NGC 7582    & 1.4$\pm^{0.4}_{0.2}$  $\times$ 10$^{43}$   &  0.6$\pm$0.1           &    0.8$\pm$0.4           & 9.8 $\times$ 10$^{43}$ &   7    &  j  \\  
\hline	      	 		   
NGC 1365    & 8.1$\pm^{2.9}_{1.4}$  $\times$ 10$^{42}$   &  1.0$\pm^{0.3}_{0.2}$  &    0.7$\pm$0.3           & 3.0 $\times$ 10$^{43}$ &   4    &  k  \\
NGC 2992    & 3.0$\pm^{2.5}_{1.2}$  $\times$ 10$^{43}$   &  0.6$\pm^{0.3}_{0.2}$  &    1.3$\pm^{0.7}_{0.6}$  & 3.0 $\times$ 10$^{43}$ &   1.0  &  l  \\
NGC 5506    & 9.3$\pm^{0.4}_{0.2}$  $\times$ 10$^{43}$   &  0.5$\pm$0.1           &    2.9$\pm^{0.4}_{0.7}$  & 2.2 $\times$ 10$^{44}$ &   2    &  m  \\
\enddata     
\tablecomments{Same as in Table \ref{lum}, but for the Sy2 and intermediate-type Seyfert galaxies in \citealt{Ramos09a}.}
\tablerefs{(a)
\citet{Markowitz07}; (b) \citet{Soldi05}; (c) \citet{Turner97}; (d) \citealt{Ramos09a}; (e) \citet{Levenson06}; (f) \citet{Jimenez05}; (g)
\citet{Vignali02}; (h) \citet{Elvis04}; (i) \citet{Awaki06}; (j) \citet{Turner00}; (k) \citet{Risaliti05}; (l) \citet{Yaqoob97};
(m) \citet{Lamer00}; (n) \citet{Lamer03}; (o) \citet{Beckmann05}.}
\label{lum2}
\end{deluxetable*}

As mentioned in Section \ref{comparison}, the results presented in this Appendix are compatible in general with those presented in \citealt{Ramos09a} 
at the 1-sigma level. For the majority of the galaxies, we have included IR data from recent publications. In particular, the use of VISIR MIR data 
have been key in predicting the silicate feature in absorption in the case of Centaurus A\footnote{In order to account for the extinction caused by 
the dust lane of A$_V$ $\sim$ 7-8 mag in Centaurus A, we have considered the \citet{Chiar06} law in the fit fixing A$_V$=8 mag.} and NGC 3281, 
and in emission in NGC 3227. 

In the case of Centaurus A, the fit presented in \citealt{Ramos09a} predicted the silicate feature in emission, 
in clear contradiction with MIR spectroscopic data \citep{Siebenmorgen04,Meisenheimer07}. By including the VISIR data for this galaxy, the fitted 
models predict a broad silicate feature, in agreement with the observations. Additionally, we have also introduced the inclination angle of the
torus as a gaussian prior centred in 70$\degr$ as infered from the inclination of the inner radio jet of Centaurus A \citep{Tingay98}. This value 
is also coincident with the 66$\degr$ inclination angle of the obscuring structure derived from MIR interferometric observations \citep{Burtscher10}.

For the Sy2 galaxy NGC 3281 we also have considered a uniform prior in the range [60\degr,90\degr] for the inclination angle of the torus, based on 
the inclination of the ionization cone axis derived in \citet{Storchi92} from optical imaging and spectroscopic observations, which implies
a nearly edge-on orientation of the obscuring structure.

\clearpage

\acknowledgments

C.R.A. and J.R.E. acknowledge the Spanish Ministry of Science and Innovation (MICINN) through 
Consolider-Ingenio 2010 Program grant CSD2006-00070: First Science with the GTC 
(http://www.iac.es/consolider-ingenio-gtc/).
C.R.A. ackowledges financial support from STFC PDRA (ST/G001758/1). 
A.A.H acknowledges support from the Spanish Plan Nacional del Espacio under grant
ESP2007-65475-C02-01 and Plan Nacional de Astronom\' ia y Astrof\' isica under grant AYA2009-05705-E.
A.A.R. acknowledges the Spanish Ministry of Science and Innovation (MICINN) through project AYA2007-63881.
A.M.P.G. acknowledges the Spanish Ministry of Science and Innovation (MICINN) through project AYA2008-06311-C02-01.
C.P. acknowledges support from the NSF under grant number 0904421.

This work is based on observations obtained at the Gemini Observatory, which is operated by the
Association of Universities for Research in Astronomy, Inc., under a cooperative agreement
with the NSF on behalf of the Gemini partnership: the National Science Foundation (United
States), the Science and Technology Facilities Council (United Kingdom), the
National Research Council (Canada), CONICYT (Chile), the Australian Research Council
(Australia), Minist\'{e}rio da Ci\^{e}ncia e Tecnologia (Brazil), and Ministerio de Ciencia, Tecnolog\' ia e Innovaci\'on Productiva (Argentina). 
The Gemini programs under which the data were obtained are GS-2005B-DD, GS-2005B-Q-29, and GS-2009B-Q-43.

This work is based on observations made with the NASA/ESA Hubble Space Telescope, and obtained 
from the Hubble Legacy Archive, which is a collaboration between the Space Telescope Science Institute 
(STScI/NASA), the Space Telescope European Coordinating Facility (ST-ECF/ESA) and the Canadian Astronomy 
Data Centre (CADC/NRC/CSA).

This research has made use of the NASA/IPAC Extragalactic Database (NED) which is 
operated by the Jet Propulsion Laboratory, California Institute of Technology, under 
contract with the National Aeronautics and Space Administration.


We finally acknowledge useful comments from the anonymous referee.

\end{document}